\documentstyle[12pt,dbl12,rac,epsfig,amscd,amsmath]{report}
\newcommand{\re}{\mbox{Re}}

\newcommand{\wtd}{\widetilde}
\newcommand{\be}{\begin{equation}}
\newcommand{\ee}{\end{equation}}
\newcommand{\ba}{\begin{eqnarray}}
\newcommand{\ea}{\end{eqnarray}}
\newcommand{\notE}{E\kern-0.6em\hbox{/}\kern0.05em}
\newcommand{\notEt}{E_{T}\kern-1.21em\hbox{/}\kern0.45em}
\newcommand{\notsusy}{SUSY\kern-1.21em\hbox{/}\kern0.45em}

\newcommand{\bea}{\begin{eqnarray}}
\newcommand{\eea}{\end{eqnarray}}
\newcommand{\nnmb}{\nonumber}

\def\4vol{{\int d^4x \sqrt{-g}}}

\newcommand{\nc}{\newcommand}

\nc{\nt}{\tilde{N}} \nc{\ra}{\rightarrow}
\nc{\lsim}{\begin{array}{c}\,\sim\vspace{-21pt}\\< \end{array}}
\nc{\gsim}{\begin{array}{c}\sim\vspace{-21pt}\\> \end{array}}
\nc{\tnt}{\tilde{N}} \nc{\tst}{\tilde{t}}
\nc{\beq}{\begin{equation}} \nc{\eeq}{\end{equation}}
\nc{\gev}{\mbox{GeV}} \nc{\tev}{\mbox{TeV}} \nc{\eev}{\mbox{eV}}
\nc{\db}{\Delta b} \nc{\met}{\,/\!\!\!\!E_T}
\nc{\mpt}{\,/\!\!\!p_T} \nc{\nfive}{N_{{\bf 5}\oplus\bar{\bf 5}}}
\nc{\five}{{\bf 5}\oplus\bar{\bf 5}}

\nc{\LL}{L} \nc{\vv}{\tilde{v}} \nc{\bit}{\begin{itemize}}
\nc{\eit}{\end{itemize}}

\begin{document}
\titlepage{Connecting String/$M$ Theory to the           %  Mandatory
  Electroweak Scale and to LHC Data      }                  % -- Title, in capitals
    {Piyush Kumar   }                                 % -- Your name
    {Doctor of Philosophy}                          % -- Name of the degree
    {Physics}                % -- Area
    {2007}                                          % -- Year
    {Professor Gordon L. Kane,     Chairperson \\       % -- Committee members:
     Professor Dante E. Amidei    \\                       % -- The Chairperson first,
     Professor Douglas O. Richstone   \\                        % -- Others alphabetically
     Associate Professor James D. Wells  \\
     Assistant Professor Leopoldo A. Pando Zayas}
\unnumberedpage                             % Mandatory
\copyrightpage{Piyush Kumar}                   % A copyright page (unmembered);  Optional
\initializefrontsections                    % Important for right format

\dedicationpage{TO MY PARENTS $\&$ MY WIFE
           } % Optional
\startacknowledgementspage                  % Optional
I would like to express my heartfelt gratitude to my advisor Prof.
Gordon Kane. I am really grateful to him for accepting me as a
student when I transferred to Michigan, for his continuous
encouragement and for inspiring me by his infectious enthusiasm
about physics. I would also like to thank other professors - most
notably Bobby Acharya, Joseph Lykken and James Wells, for the
various illuminating discussions and collaborations I had with
them which helped me gain insight and perspective about various
branches of theoretical high-energy physics.

I have also benefitted a great amount by interactions with fellow
students and postdocs in the particle theory group. I would like
to thank them, particularly Konstantin Bobkov, Jacob Bourjaily,
Joshua Davis, David Morrissey, Jing Shao, Manuel Toharia, Diana
Vaman and Ting Wang, for increasing my understanding of many
topics as well as offering constructive criticism.

Finally, I would like to thank my family. My mother and father,
for everything they did for me and for having so much confidence
in their eldest son. My brother and sister, for always looking up
to their elder brother. And my wife, Kriti, for her unconditional
love, care and support as well as for being my best friend.

%\startprefacepage                           % Optional

\tableofcontents                     % Mandatory
\listoffigures                       % Necessary if you have two or more figures
\listoftables                        % Necessary if you have two or more tables
%\def\anythingtopic{FLOWCHARTS}       % Define the topic first.
%\listofanythings{LIST OF FLOWCHARTS} % Necessary if you have two or more 'anythings'
%\listofmaps                          % Necessary if you have two or more maps
\listofappendices                    % Necessary if you have two or more appendices
%\defineanythingtopic{Flowchart}      % Define 'anything' to be 'Flowchart'.
\startthechapters                    % Important for right format.
\chapter{Introduction}\label{intro}
We are about to enter a unique era in high energy physics. For the
first time in history, the energy and technology required to probe
the electroweak scale in a controlled manner is at hand. The Large
Hadron Collider (LHC), which will supposedly turn on by the end of
this year, will be able to achieve this. We have many reasons to
believe that the mysteries of the electroweak scale (the Standard
Model) will be revealed at the LHC.

Even though this thesis will focus on forthcoming data from the
LHC, we are very fortunate that forthcoming data from many other
closely related fields in particle physics and cosmology will
complement and supplement data from the LHC. For example, major
clues about extensions to the Standard Model can come from
indirect information from rare decays, magnetic moments, proton
decay and particularly WIMP\footnote{Weakly Interacting Massive
Particle} detection in Dark Matter detection experiments. In
addition, any extension to the Standard Model of particle physics
with an underlying microscopic origin will also affect
astrophysical/cosmological observables. Recent data from WMAP,
Supernovae and Galactic Clusters as well as forthcoming data from
PLANCK, LIGO/VIRGO and LISA will further constrain various
approaches aiming to explain the origin of the electroweak scale.
Therefore, in my opinion, it is fair to say that we are on the
threshold of a unique data-rich era.

Assuming one obtains data confirming the existence of new physics
beyond the Standard Model, we, as theorists, would have to grapple
with the following questions : \begin{itemize} \item What is the
broad framework for new physics? \item What is the spectrum of
particles and the effective theory at $\sim$ TeV scale within this
broad framework? \item What is the structure of the underlying
deeper, short distance, theory? \end{itemize}

These questions are collectively known as the ``\emph{Inverse
Problem}". The first two questions pertaining to the Inverse
Problem above have been receiving more and more attention in the
past few years. However, the third question - the \emph{deeper}
Inverse Problem, has not even been addressed in a systematic way.
This is hardly surprising, as this is arguably the most
challenging problem in fundamental physics. The goal of this
thesis is to explore the third question and try to get insights
about addressing the deeper Inverse Problem in a meaningful way.
In this thesis, the nature of the underlying theory will be
assumed to be string/$M$ theory. This is because of the following
reasons. At present, string theory is the only known consistent
theory of quantum gravity, at least at the perturbative level. But
for the purposes of this thesis, more importantly, it is the only
known ultra-violet complete theory which can naturally give rise
to effective four dimensional theories of particle physics with
the most important features of the Standard Model, namely,
non-abelian gauge fields and chiral fermions. In addition, string
theory can address \emph{all} open questions in particle physics
and cosmology within its framework and hopefully, solve them.
Therefore, it seems reasonable to assume the existence of such a
theory providing an underlying microscopic theoretical structure
to our universe.

Even if one assumes the existence of string theory as the
fundamental microscopic theory, it is still a herculean task to
solve the deeper Inverse Problem, to say the least. One has to
first explore this question carefully and identify approaches in
which the question can be addressed meaningfully. In my opinion,
the first steps towards addressing the deeper Inverse Problem are:
\begin{itemize} \item String/$M$ theory
constructions should be developed enough to make contact with low
energy physics. \item Various specific classes of constructions,
with ``reasonable assumptions", should be systematically analyzed
to the extent that predictions for real experimental observables,
say, signatures at the LHC, can be made. \item Effort should be
made to identify experimental observables which probe and are
sensitive to the properties of the underlying microscopic
construction, or equivalently, different microscopic constructions
should be distinguished on the basis of experimental observables.
\item The origin of distinguishibility should be understood in
terms of the structure of the theory. \item This program should be
complemented with a bottom-up program - that of finding the
effective theory which explains all data. A combination of
top-down and bottom-up approaches will be much more powerful than
either.
\end{itemize}

In this work, all steps mentioned above will be examined and
studied. The thesis is organized as follows. In chapter
\ref{string-pheno}, the motivation for and importance of string
phenomenology will be described in detail. Chapter \ref{hierarchy}
will try to explain the hierarchy problem, which is one of the
most important problems in particle physics, within the context of
field theory as well as string theory. Then, in chapter
\ref{topdownstringpheno}, two particular examples of string/$M$
theory compactifications will be analyzed, with a particular
emphasis on their predictions for low-energy physics (of the order
of the electroweak scale). In chapter \ref{distinguishing}, it
will be shown that many string motivated constructions can be
distinguished on the basis of patterns of signatures at the LHC
and the origin of distinguishibility can also be explained on the
basis of the underlying theoretical structure of the
constructions. Finally, a more bottom-up approach to the Inverse
Problem, {\it viz.} to go from data to theory in a more
model-independent way will be studied in chapter \ref{lowtohigh}.
This will be followed by conclusions in chapter \ref{conclusions}.

\chapter{Why is it important to do String Phenomenology?}
\label{string-pheno}

Before moving on to discuss more technical aspects of the
dissertation in subsequent chapters, it is worthwhile to review
the current status of string phenomenology, its importance and its
role in the future.

At first glance, making reliable predictions from string theory
might seem a long shot. A background independent non-perturbative
formulation of string theory is not at hand at present. One does
not even have a microscopic non-perturbative definition of string
theory in general backgrounds such as backgrounds with
Ramond-Ramond (RR) fluxes or cosmological backgrounds. There are
also no hints of a deep underlying ``vacuum selection principle"
which would uniquely predict the properties of the vacuum which we
live in, in a natural way. In fact, recent developments point in
the opposite direction -- towards a vast multitude of
possibilities for four dimensional string vacua. This vast
multitude of possibilities has been given the name - ``Landscape".
The extent of the landscape of four dimensional string vacua is
not known. In addition to the well known examples of Calabi-Yau
compactifications, many other kinds of compactifications have been
studied in the literature, some of them recently -- such as
generalized K\"{a}hler compactifications, non-K\"{a}hler
compactifications like $G_2$ compactifications, etc. and
compactifications with non-geometric fluxes. In such a situation,
the goal of making reliable predictions for low energy physics
from string theory appears to be quite challenging.

Therefore, in my opinion, the situation warrants a pragmatic
approach if one is still interested in connecting string theory to
real observable physics. In fact, developments in the last twenty
years and particularly in the last five years or so, actually give
us a clue as to what the pragmatic approach might be. Even though
we may not have a good understanding of the \emph{entire} $M$
theory landscape in all its glory, we have gained a lot of
understanding about different corners of $M$ theory, such as
weakly and strongly coupled heterotic string theory, Type IIA and
IIB string theories and $M$ theory on $G_2$ manifolds. Detailed
studies of these corners have shown that string theory has the
ability to address \emph{all} issues in particle physics and
cosmology. For example, the origin of forces and matter, in
particular non-abelian gauge fields and chiral fermions, can be
explained. The origin of more than one flavor and hierarchical
yukawa couplings can also be explained in the various corners,
albeit in different ways. Heterotic string constructions and $M$
theory constructions can naturally give rise to gauge coupling
unification at a high scale. In Type II constructions, gauge
coupling unification is less natural, but it is possible to
construct models in which the gauge couplings unify. Model
building in heterotic and type II string theories is a healthy
area of research with many semi-realistic examples, and new
approaches to model building are emerging. Moreover, in recent
years, there has been a lot of progress in understanding
\emph{dynamical} issues - such as the stabilization of
moduli\footnote{moduli are effective four dimensional scalar
fields which characterize the size and shape of the internal
manifold in a string compactification; astrophysical observations
require that these scalars be sufficiently massive.},
supersymmetry breaking and generation of hierarchy between the
Planck and electroweak scales. Regarding connection to cosmology,
many approaches to achieving inflation in string theory have been
proposed in the literature. Many of these issues will be analyzed
in detail in subsequent chapters in the context of specific
string/$M$ theory constructions.

A pragmatic approach, therefore, in my opinion as well as of many
other people is to systematically study models in \emph{all}
corners of the entire $M$ theory landscape (where it is possible
to do so) in a way such as to connect to real physics observables
like collider (LHC) observables, dark matter (DM) observables and
inflationary observables to name a few, and then use data to gain
insights about the nature of the underlying theory. Developments
in the last few years have actually made it possible to address
each of these issues in a reliable way. In a string/$M$ theory
construction, all such observables come from the \emph{same}
underlying microscopic physics, implying that forthcoming data has
a great potential to constrain and favor or rule out specific
classes of constructions in string/$M$ theory.

In the absence of a breakthrough in discovering a deep dynamical
principle which uniquely selects the vacuum we live in, science
should proceed in small systematic steps, which makes the
pragmatic approach described above a sensible one to pursue. Of
course, a breakthrough in some of the conceptual issues mentioned
in the previous paragraphs would sharpen the approach further and
make it even more useful.

Hoping that the case for string phenomenology has been made, the
subsequent chapters will deal with various aspects of string
phenomenology. After a detailed explanation of the hierarchy
problem and its importance, two particular string/$M$ theory
constructions will be studied so as to connect them to observable
physics. Then, a general approach which allows us to distinguish
different string constructions on the basis of their predictions
for pattern of experimental observables, such as LHC signatures,
will be described. This will be followed by a discussion of some
issues in going from low-scale data to a high scale theory in a
more model independent way (although still within the framework of
supersymmetry).

\chapter{The Hierarchy Problem and Supersymmetry
Breaking} \label{hierarchy}

In this chapter, we would like to explain the Hierarchy Problem in
detail, which is the most important problem in particle physics at
present. This chapter is organized as follows. The nature of the
problem will be first elucidated in a simple manner. We will then
describe the paradigm of low energy supersymmetry which is perhaps
the most appealing solution to the problem. Finally, since our
interest lies in connecting String/$M$ theory to the (observed)
electroweak scale, we will discuss the various approaches to the
problem in String/$M$ theory. It turns out that the issue of
supersymmetry breaking is intimately connected to the Hierarchy
Problem, which will also be explained.

To begin, let's first state the Hierarchy Problem. The Hierarchy
problem is actually two separate problems:
\begin{itemize}
\item What is the origin of the electroweak scale and why is it so
much smaller than the fundamental mass scale $M_{planck}$?

\item Since the higgs mass parameter in the Standard Model is not
protected by any symmetry, why is the higgs mass of the order of
the electroweak scale instead of some very high cutoff scale even
though it is disturbingly sensitive to almost any new physics in
any imaginable extension of the Standard Model?
\end{itemize}

The first part of the Hierarchy Problem is known as the Gauge
Hierarchy Problem while the second part is known as the Technical
Hierarchy Problem. As seen from above, the Gauge Hierarchy Problem
is extremely simple to state and understand conceptually, but
incredibly challenging to answer in a compelling way from an
underlying theory. This would be explained in more detail later in
this chapter. Let us now try to understand the technical hierarchy
problem. The electrically neutral part of the Standard Model Higgs
field is a complex scalar $H$ with a classical potential \be V =
m_H^2 |H|^2 + {\lambda} |H|^4\> . \label{higgspotential} \ee The
Standard Model requires a non-vanishing vacuum expectation value
(VEV) for $H$ at the minimum of the potential. This will occur if
$\lambda > 0$ and $m_H^2 < 0$, resulting in $\langle H \rangle =
\sqrt{-m_H^2/2\lambda}$. Since we know experimentally that
$\langle H \rangle$ is approximately 174 GeV, from measurements of
the properties of the weak interactions, it must be that $m_H^2$
is very roughly of order $-$(100 GeV)$^2$. The problem is that
$m_H^2$ receives enormous quantum corrections from the virtual
effects of every particle that couples, directly or indirectly, to
the Higgs field. For example, we have a correction to $m_H^2$ from
a loop containing a Dirac fermion $f$ with mass $m_f$. If the
Higgs field couples to $f$ with a term in the Lagrangian -
$\lambda_f H \bar{f} f$, then the corresponding Feynman diagram
yields a correction \be \Delta m_H^2 \>=\> -{|\lambda_f|^2\over 8
\pi^2} \Lambda_{\rm UV}^2 + \ldots . \label{quaddiv1} \ee Here
$\Lambda_{\rm UV}$ is an ultraviolet momentum cutoff used to
regulate the loop integral; it should be interpreted as at least
the energy scale at which new physics enters to alter the
high-energy behavior of the theory. The ellipses represent terms
proportional to $m_f^2$, which grow at most logarithmically with
$\Lambda_{\rm UV}$ (and actually differ for the real and imaginary
parts of $H$). Each of the leptons and quarks of the Standard
Model can play the role of $f$; for quarks, eq.~(\ref{quaddiv1})
should be multiplied by 3 to account for color. The largest
correction comes when $f$ is the top quark with $\lambda_f\approx
1$. The problem is that if $\Lambda_{\rm UV}$ is of order
$M_{Planck}$, say, then this quantum correction to $m_H^2$ is some
30 orders of magnitude larger than the required value of $m_H^2
\sim -(100$ GeV$)^2$. This is only directly a problem for
corrections to the Higgs scalar boson squared mass, because
quantum corrections to fermion and gauge boson masses do not have
the direct quadratic sensitivity to $\Lambda_{\rm UV}$ found in
eq.~(\ref{quaddiv1}).  However, the quarks and leptons and the
electroweak gauge bosons $Z^0$, $W^\pm$ of the Standard Model all
obtain masses from $\langle H \rangle$, so that the entire mass
spectrum of the Standard Model is directly or indirectly sensitive
to the cutoff $\Lambda_{\rm UV}$.

Furthermore, there are contributions similar to
eq.~(\ref{quaddiv1}) from the virtual effects of any arbitrarily
heavy particles that might exist, and these involve the masses of
the heavy particles, not just the cutoff.

For example, suppose there exists a heavy complex scalar particle
$S$ with mass $m_S$ that couples to the Higgs with a Lagrangian
term $ -\lambda_S |H|^2 |S|^2$. This gives a correction \be \Delta
m_H^2 \>=\> {\lambda_S\over 16 \pi^2} \left [\Lambda_{\rm UV}^2 -
2 m_S^2 \> {\rm ln}(\Lambda_{\rm UV}/m_S) + \ldots \right ].
\label{quaddiv2} \ee

This problem arises even if there is no direct coupling between
the Standard Model Higgs boson and the unknown heavy particles.
For example, suppose there exists a heavy fermion $F$ that, unlike
the quarks and leptons of the Standard Model, has vector-like
quantum numbers and therefore gets a large mass $m_F$ without
coupling to the Higgs field. [In other words, an arbitrarily large
mass term of the form $m_F \overline F F$ is not forbidden by any
symmetry, including weak isospin $SU(2)_L$.] In that case, no
one-loop diagram like (\ref{quaddiv1}) exists for $F$.
Nevertheless there will be a correction to $m_H^2$ as long as $F$
shares some gauge interactions with the Standard Model Higgs
field; these may be the familiar electroweak interactions, or some
unknown gauge forces that are broken at a very high energy scale
inaccessible to experiment. This would give rise to a contribution
: \ba \Delta m_H^2 \>=\> C_H T_F \left ( {g^2 \over 16 \pi^2}
\right )^2 \left [ a \Lambda_{\rm UV}^2 + 24 m_F^2 \>{\rm ln}
(\Lambda_{\rm UV}/m_F) + \ldots \right ], \label{quaddiv3} \ea
where $C_H$ and $T_F$ are group theory
factors\footnote{Specifically, $C_H$ is the quadratic Casimir
invariant of $H$, and $T_F$ is the Dynkin index of $F$ in a
normalization such that $T_F=1$ for a Dirac fermion (or two Weyl
fermions) in a fundamental representation of $SU(n)$.} of order 1,
and $g$ is the appropriate gauge coupling.

Therefore, the important point is that these contributions to
$\Delta m_H^2$ are sensitive both to the largest masses and to the
ultraviolet cutoff in the theory, presumably of order
$M_{Planck}$. Thus, the ``natural" squared mass of a fundamental
Higgs scalar, including quantum corrections, should be more like
$M_{Planck}^2$ than the experimentally favored value. Even very
indirect contributions from Feynman diagrams with three or more
loops can give unacceptably large contributions to $\Delta m_H^2$.
The argument above applies not just for heavy particles, but for
arbitrary high-scale physical phenomena such as condensates or
additional compactified dimensions.

It could be that there is no fundamental Higgs boson, as in
technicolor models, top-quark condensate models, and models in
which the Higgs boson is composite. Or it could be that the
ultimate ultraviolet cutoff scale is much lower than the Planck
scale. These ideas are certainly worth exploring, although they
often present difficulties in their simplest forms. But, if the
Higgs boson is a fundamental particle, as we will assume in this
work henceforth, and if there really is physics far above the
electroweak scale which will also be assumed, then we have two
remaining options: either we must make the rather bizarre
assumption that there do not exist {\it any} high-mass particles
or effects that couple (even indirectly or extremely weakly) to
the Higgs scalar field, or else some striking cancellation is
needed between the various contributions to $\Delta m_H^2$.

\section{Low Energy Supersymmetry}

Theories with ``low energy supersymmetry" have emerged as the
strongest candidates for physics beyond the SM. By ``low energy
supersymmetry", one means that supersymmetry remains an unbroken
symmetry at very low energies compared to the fundamental scale
$M_{Planck};$ it is somehow broken at a scale such that it gives
rise to masses of extra particles which are required by
supersymmetry, to be of the order of the TeV scale, so as to solve
the (Technical) Hierarchy Problem (this will become clear soon).
There are strong reasons to expect that low energy supersymmetry
is the probable outcome of experimental and theoretical progress
and that it will soon be directly confirmed by experiment.  In the
simplest supersymmetric world, each particle has a {\it
superpartner} which differs in spin by $1/2$ and is related to the
original particle by a supersymmetry transformation. Since
supersymmetry relates the scalar and fermionic sectors, the chiral
symmetries which protect the masses of the fermions also protect
the masses of the scalars from quadratic divergences, leading to
an elegant resolution of the hierarchy problem. Comprehensive
reviews of supersymmetry from a particle physics and
phenomenological perspective can be found in \cite{Martin:1997ns}.

Historically though, supersymmetry had been proposed entirely from
a mathematical and formal perspective. It was found that the
Coleman-Mandula Theorem \cite{Coleman:1967ad}  for interacting
quantum field theories could be generalized if one postulates a
fermionic symmetry which connects bosons to fermions and vice
versa. This is known as the Haag-Lopuszanski theorem
\cite{Haag:1974qh}. Thus, before its good phenomenological
properties were realized, supersymmetry was studied purely as a
formal theory in the 1970s. Supersymmetry is also a critical
ingredient in the microscopic formulation of String theory. It
also turns out that many solutions of string theory give rise to
low energy supersymmetry, as will be discussed in detail in
chapter \ref{topdownstringpheno}. It is therefore, remarkable that
a symmetry which was proposed entirely from a formal point of view
has the potential to solve many problems in particle physics as
well.

Supersymmetry must be a broken symmetry, because exact
supersymmetry dictates that every superpartner is degenerate in
mass with its corresponding SM particle, a possibility which is
decisively ruled out by experiment. Possible ways to achieve a
spontaneous breaking of supersymmetry breaking depend on the form
of the high energy theory. In many ways, it is not surprising that
supersymmetry breaking is not yet understood --- the symmetry
breaking was the last thing understood for the Standard Model too
(assuming it is indeed understood).

An important clue as to the nature of supersymmetry breaking can
be obtained by returning to the motivation provided by the
hierarchy problem. Supersymmetry forced us to introduce two
complex scalar fields for each Standard Model Dirac fermion, which
is just what is needed to enable a cancellation of the
quadratically divergent $(\Lambda_{\rm UV}^2)$ pieces of
eqs.~(\ref{quaddiv1}) and (\ref{quaddiv2}). This sort of
cancellation also requires that the associated dimensionless
couplings should be related (for example $\lambda_S =
|\lambda_f|^2$). The necessary relationships between couplings
indeed occur in unbroken supersymmetry. In fact, unbroken
supersymmetry guarantees that the quadratic divergences in scalar
squared masses must vanish to all orders in perturbation
theory.\footnote{A simple way to understand this is to recall that
unbroken supersymmetry requires the degeneracy of scalar and
fermion masses. Radiative corrections to fermion masses are known
to diverge at most logarithmically in any renormalizable field
theory, so the same must be true for scalar masses in unbroken
supersymmetry.} Now, if broken supersymmetry is still to provide a
solution to the hierarchy problem even in the presence of
supersymmetry breaking, then the relationships between
dimensionless couplings that hold in an unbroken supersymmetric
theory must be maintained. Otherwise, there would be quadratically
divergent radiative corrections to the Higgs scalar masses of the
form \be \Delta m_H^2 = {1\over 8\pi^2} (\lambda_S -
|\lambda_f|^2) \Lambda_{\rm UV}^2 + \ldots .
\label{eq:royalewithcheese} \ee We are therefore led to consider
``soft" supersymmetry breaking. This means that the effective
Lagrangian of the MSSM can be written in the form \be \mathcal{L}
= \mathcal{L}_{\rm SUSY} + \mathcal{L}_{\rm soft}, \ee where
$\mathcal{L}_{\rm SUSY}$ contains all of the gauge and Yukawa
interactions and preserves supersymmetry invariance, and
$\mathcal{L}_{\rm soft}$ violates supersymmetry but contains only
mass terms and coupling parameters with {\it positive} mass
dimension. Without further justification, soft supersymmetry
breaking might seem like a rather arbitrary requirement.
Fortunately, theoretical models for supersymmetry breaking do
indeed yield effective Lagrangians with just such terms for
$\mathcal{L}_{\rm soft}$. If supersymmetry is broken in this way,
the superpartner masses can be lifted to a phenomenologically
acceptable range. Furthermore, the scale of the mass splitting
should be of order the $Z$ mass to TeV range because it can be
tied to the scale of electroweak symmetry breaking.

Thus, we see that low energy supersymmetry provides an elegant
solution to the Technical Hierarchy Problem. As we will see
shortly, it also mitigates the Gauge Hierarchy Problem by breaking
the electroweak symmetry radiatively through logarithmic running,
which explains the large number $\sim 10^{13}$. Apart from the
Hierarchy problem, low energy supersymmetry has had many other
``successes" as well:

\begin{itemize}
\item {\it Radiative electroweak symmetry breaking.} With
plausible boundary conditions at a high scale (certain couplings
such as the top quark Yukawa of $O(1)$ and no bare Higgs mass
parameter $\mu$ in the superpotential), low energy supersymmetry
can provide the explanation of the origin of electroweak symmetry
breaking \cite{Ibanez:1982fr, Alvarez-Gaume:1981wy, Inoue:1982ej}.
To oversimplify a little, the SM effective Higgs potential has the
form $V=m^2h^2+\lambda h^4$. First, supersymmetry requires that
the quartic coupling $\lambda$ is a function of the $U(1)_Y$ and
$SU(2)$ gauge couplings $\lambda=({g'}^2+g^2)/2$.  Second, the
$m^2$ parameter runs to negative values at the electroweak scale,
driven by the large top quark Yukawa coupling. Thus the ``Mexican
hat'' potential with a minimum away from $h =0$ is derived rather
than assumed.  As is typical for progress in physics, this
explanation is not from first principles, but it is an explanation
in terms of the next level of the effective theory which depends
on the crucial assumption that the ${\cal L}_{soft}$ mass
parameters have values of order the electroweak scale.  Once
superpartners are discovered, the question of supersymmetry
breaking must be answered in any event and it is a genuine success
of the theory that whatever explains supersymmetry breaking is
also capable of resolving the crucial issue of $SU(2)\times U(1)$
breaking.

\item {\it Gauge coupling unification.} In contrast to the SM, the
MSSM allows for the unification of the gauge couplings, as first
pointed out in the context of GUT models by
\cite{Dimopoulos:1981yj,Dimopoulos:1981zb,Sakai:1981gr}. The
extrapolation of the low energy values of the gauge couplings
using renormalization group equations and the MSSM particle
content shows that the gauge couplings unify at the scale $M_G
\simeq 3 \times 10^{16}$ GeV
\cite{Giunti:ta,Amaldi:1991cn,Langacker:1991an,Ellis:1990wk}.
Gauge coupling unification and electroweak symmetry breaking
depend on essentially the same physics since each needs the soft
masses and $\mu$ to be of order the electroweak scale.

\item {\it Cold dark matter.} In supersymmetric theories, the
lightest superpartner (LSP) can be stable.  This stable
superpartner provides a nice cold dark matter candidate
\cite{Pagels:ke,Goldberg:1983nd}.  Simple estimates of its relic
density are of the right order of magnitude to provide the
observed amount. LSPs were noticed as good candidates before the
need for nonbaryonic cold dark matter was established.

\end{itemize}

\section{String/$M$ Theory and the Hierarchy Problem}

As mentioned in the previous subsection, low energy supersymmetry
alone can only mitigate the Gauge Hierarchy Problem, but it cannot
solve it. An explanation of the \emph{origin} of the electroweak
scale has to come from an underlying microscopic theory which
incorporates both non-abelian gauge theories and gravitation, like
String/$M$ theory. This subsection is devoted to the various
approaches in String/$M$ Theory to the Hierarchy Problem.

The particle spectrum of string theory consists of a finite number
of massless states and an infinite number of massive states
characterized by the string scale. For a phenomenological
description of the consequences of string theory for low-energy
physics, it should not be necessary to describe the dynamics of
massive states. Formulating an effective theory based entirely on
fields corresponding to massless (light) degrees of freedom is the
most natural thing to do in such a situation. Such a description
is useful not only for a phenomenological analysis, but also for
addressing certain theoretical issues, such as the occurrence of
anomalies.

In principle, it must be possible to describe string theory by a
classical action $S(\phi,\Phi)$, $\phi$ denoting the light degrees
of freedom and $\Phi$ denoting the heavy degrees of freedom. One
could then imagine integrating out the heavy fields $\Phi$ from
the action and obtain a low-energy effective action for the light
fields $S_{eff}(\phi)$. However, at present, the exact string
theory action $S(\phi,\Phi)$ is not known (even at the classical
level). Therefore, it is not possible to construct the low-energy
effective action for the light fields. What is usually done is to
study string $S$-matrix elements and construct a classical action
for the massless fields that reproduces them. Such an action is
extremely useful since it can be written as a systematic expansion
in number of derivatives, the higher derivatives being unimportant
at low energies.

Since string theory and $M$ theory live in ten and eleven
dimensions respectively, in order to connect to four dimensional
physics, one needs to compactify ten or eleven dimensions to four
and construct solutions of the compactified equations of motion.
Since supersymmetry makes small masses stable against radiative
corrections (for example, it makes the Higgs mass natural) in an
elegant way, one wants to compactify to four dimensions so as to
preserve $\mathcal{N}$=1 supersymmetry in four dimensions. The
requirement of $\mathcal{N}$=1
supersymmetry\footnote{Compactifications preserving more
supersymmetries in four dimensions are uninteresting
phenomenologically as they do not give rise to chiral fermions.}
in four dimensions is also useful from a technical point of view,
as it is much easier to find solutions to the equations of
motion\footnote{Compactifications satisfying supersymmetry
conditions (which are first order equations) automatically obey
the equations of motion and are also stable against quantum
corrections.}. Restricting oneself to $\mathcal{N}$=1
compactifications does not guarantee low energy supersymmetry in
the sense of giving rise to superpartners of $\mathcal{O}$(TeV),
since supersymmetry can still be broken at around the
compactification scale, which is near the string scale or the
eleven dimensional Planck scale (typically much above the TeV
scale).

Therefore, one has to find mechanisms within $\mathcal{N}$=1
compactifications to generate or at least accommodate a large
hierarchy. If one wants a high string scale or eleven dimensional
Planck scale ($\geq M_{unif}$), one mechanism to generate
hierarchies is by strong gauge dynamics in the hidden sector. This
works for many regions of the entire $M$ theory moduli space --
weakly \cite{Dine:1985rz} and strongly coupled
\cite{Horava:1996vs} heterotic string theory, type IIA string
theory \cite{Blumenhagen:2005mu} and $M$ theory on $G_2$ manifolds
\cite{Acharya:2006ia}. Keeping the string scale high, a second
mechanism is to utilize the discretuum of flux vacua obtained in
flux compactifications of Type IIB string theory and obtain a
small scale by tuning the flux superpotential to be very small in
Planck units \cite{Giddings:2001yu,Kachru:2003aw}. A third way of
obtaining a small scale is to relax the requirement of a high
string scale, making it sufficiently small\footnote{The precise
value will depend on explicit constructions.}
\cite{Balasubramanian:2005zx}. Finally, it turns out that Type IIB
flux compactifications cause \emph{warping} of the extra
dimensions which can also give rise to the observed hierarchy of
scales \cite{Verlinde:1999fy}. In this work, we will analyze the
consequences of many of the above mechanisms in detail in later
sections.

\chapter{Top-Down String Phenomenology}
\label{topdownstringpheno}

As mentioned in the previous subsection, in order to connect
string/$M$ theory to four dimensional physics, we are interested
in compactifications of string/$M$ theory to four dimensions with
$\mathcal{N}$=1 supersymmetry.

In general, string compactifications fall into two general
categories---one based on free or solvable world-sheet CFTs and
the second based on compactification on a smooth compact manifold.
Compactifications based on free conformal field theories (CFTs)
are characterized by singular spaces called orbifolds.
Compactifications based on smooth manifolds require the smooth
manifold to satisfy certain conditions so as to have
$\mathcal{N}$=1 supersymmetry in four dimensions. These are known
as Calabi-Yau manifolds. On the other hand, phenomenologically
interesting $M$ theory compactifications with $\mathcal{N}$=1
supersymmetry in four dimensions are characterized by
\emph{singular} manifolds of $G_2$ holonomy. Specific kinds of
singularities are required to obtain non-abelian gauge groups and
chiral fermions. It is also important to realize that different
regions of the $M$ theory moduli space are connected to each other
through a web of dualities \cite{Witten:1995ex}.

There are two main approaches to string phenomenology within
(ten-dimensional) string theory. Historically, the first approach
is concerned with the $E_8 \times E_8$ heterotic string
constructions. The $E_8 \times E_8$ theory is interesting because
it can produce $\mathcal{N}$=1 supersymmetry in four dimensions
and it also has gauge fields which can generate chiral fermions in
four dimensions. The most promising compactifications of the
heterotic string giving rise to a semi-realistic spectrum and
interactions are orbifold compactifications in the weakly coupled
regime \cite{Dixon:1986qv}. These have the advantage that CFT
techniques can be used to compute the complete massless spectrum,
as well as many of their interactions. Perturbative heterotic
string compactifications on Calabi-Yau manifolds give a less
detailed but global picture. Compactifications of the strongly
coupled heterotic string have also been constructed
\cite{Witten,BD}.

The other approach to string phenomenology is more recent. It was
realized in the mid-1990s that type I, IIA and IIB string theories
are actually different states in a single theory, which also
includes states containing general configurations of D-branes
(boundaries of open strings). This, together with the
understanding of dualities, has led to a deeper understanding of
type I/II, $\mathcal{N}$=1, four-dimensional vacua. The most
promising models for phenomenological purposes in this approach
are type II orientifold compactifications. Conformal field theory
techniques in the open string sectors, which end on D-branes,
allow for exact constructions of consistent $\mathcal{N}$=1,
four-dimensional chiral models with non-Abelian gauge symmetry on
type II orientifolds. Within this framework, chiral matter can
appear on the worldvolume of D-branes at orbifold singularities
and/or at the intersections of D-branes in the internal space (in
type IIA picture). The intersecting D-brane configurations also
have a T-dual description (in type IIB) in terms of D-branes with
open string 2-form fluxes on them. As in the heterotic case, type
II compactifications on Calabi-Yau manifolds are useful for a
global picture.

Finally, one can study compactifications in eleven-dimensional $M$
theory. It is believed that the different ten-dimensional string
theories are particular limits of a deeper eleven dimensional
theory, known as $M$ theory \cite{Witten:1995ex}. Even though a
quantum description of $M$ theory is not available at present, its
low energy limit is described by eleven dimensional supergravity
which is well understood. In the $M$ theory approach,
phenomenologically interesting compactifications on manifolds with
$G_2$ holonomy (for $\mathcal{N}$=1 supersymmetry) require the
presence of appropriate gauge and conical singularities. At
present, it has not been possible to contruct a physically
interesting global compactification in this approach, because of
its mathematical complexity. However, the existence of these
compactifications is guaranteed by dualities with $E_8 \times E_8$
heterotic string theory and type IIA string theory
\cite{Acharya:2001gy,Acharya:1998pm,Atiyah:2001qf}. Also, local
constructions with phenomenologically interesting gauge groups and
chiral spectrum have been constructed in the literature
\cite{Acharya:2004qe}.

The first step towards obtaining a low energy description of
String/$M$ theory compactifications is to derive the spectrum of
massless particles. As mentioned above, heterotic and type II
compactifications on orbifolds and orientifolds respectively have
the advantage that CFT techniques can be employed to compute the
complete massless spectrum. Therefore, a great amount of work on
these compactifications has been done in the literature. However,
since any given string compactification has to satisfy many
stringy consistency conditions (such as the tadpole cancellation
conditions), it is quite challenging to construct a global model
with a massless spectrum which has three families, is MSSM-like,
and does not have fractionally electrically charged states or
charged chiral exotics.

Once the massless spectrum is determined, one has to construct the
four dimensional low-energy effective action consistent with the
symmetries of the theory. One first obtains the low-energy
effective action in the \emph{field theory approximation}. This
means that the compactification radius is assumed to be larger
than the string length or the eleven dimensional Planck length, so
that we can restrict attention to massless fields in the ten or
eleven dimensional theory. One can then calculate higher order
corrections to this approximation. We are interested in theories
with $\mathcal{N}$=1 supersymmetry in four dimensions, which
combined with gravity gives rise to $\mathcal{N}$=1 supergravity.
The vacuum structure of $\mathcal{N}$=1 supergravity in four
dimensions is specified completely by three functions---the
(holomorphic) gauge kinetic function $(f)$ which determines the
gauge couplings, the (holomorphic) superpotential $(W)$ which
determines the Yukawa couplings and the K\"{a}hler potential $(K)$
which is a general function of all the four dimensional chiral
superfields and determines the K\"{a}hler metric for the matter
fields among other things. The effects of higher order corrections
to the field theory approximation can be incorporated within the
$\mathcal{N}$=1 supergravity formalism by including them as
corrections to the above three functions - $K, W$ $\&$ $f$. These
functions depend non-trivially on the closed string moduli which
characterize the size and shape of the extra dimensions, and the
matter fields. Deducing the dependence of the three functions
$K,W\,$ and $f$ on the moduli is an important task. This can be
done by the calculation of various string scattering amplitudes.
Alternatively, part of the effective action can also be determined
by a dimensional reduction of ten dimensional supergravity. In the
four dimensional supergravity theory, these moduli are classically
represented as \emph{massless} chiral superfields. This is
disastrous for two reasons: a) All four dimensional masses and
couplings are functions of the moduli. So, unless the moduli are
stabilized, the masses and couplings cannot be predicted. b)
Massless scalars have been ruled out by cosmological and
astrophysical observations. Therefore, one has to stabilize the
moduli and make them massive. In addition, as emphasized in the
previous chapter, one also has to break supersymmetry and generate
the hierarchy between the Planck and electroweak scales. Thus,
starting from a string/$M$ theory compactification, it is an
extremely daunting task to get a realistic matter spectrum,
stabilize the moduli and break supersymmetry in a controlled
manner in such a way as to generate the hierarchy.

In the following, we will focus on two corners of the $M$ theory
moduli space - that of type IIA string theory on toroidal
orientifolds and $M$ theory on $G_2$ manifolds. We will see that
issues related to model-building such as constructing the massless
chiral matter spectrum can be better understood in the type IIA
picture while issues related to supersymmetry breaking and moduli
stabilization can be better understood in the $M$ theory picture.
We will start with the type IIA constructions and then move on to
$M$ theory on $G_2$ manifolds.

\section{Type IIA Intersecting D-brane Constructions}

This section is devoted to the detailed study of a particular
class of models based on type $II$ string theory compactifications
on toroidal orientifolds with D\textit{p}-branes wrapping
intersecting cycles on the compact space. This approach to string
model building is distinguished by its computability and
simplicity, together with very appealing phenomenological
possibilities. In these models, gauge interactions are confined to
D-branes. Chiral fermions are open strings which are stretched
between two intersecting branes. They are localized at the brane
intersections. If certain conditions are satisfied, there will be
massless scalars associated with the chiral fermions such that we
have $\mathcal{N}$=1 supersymmetry in the effective field theory.
Because of these attractive features, intersecting brane model
building has drawn considerable attention in recent years and
several semi-realistic models with an MSSM like spectrum have been
constructed \cite{realistic}.

To test these approximate clues and to begin to test string
theory, only reproducing the SM particle content is not enough.
Numerical predictions must be made. In addition, a successful
theory should not just explain existing data, it must also make
predictions which can be tested in future experiments. For the
brane models, if supersymmetry exists and is softly broken, soft
SUSY breaking terms have to be calculated and tested by future
experimental measurements. A fair amount of work on the low-energy
effective action of intersecting D-brane models has been done. The
stability of these kind of models has been discussed in
\cite{Blumenhagen:2001te}. The issues of tree level gauge
couplings, gauge threshold corrections and gauge coupling
unification has been addressed in
\cite{Shiu:1998pa,Cvetic:2002qa,Lust:2003ky,Antoniadis:2000en,
Cremades:2002te,Blumenhagen:2003jy,Berg:2004ek}. Yukawa couplings
and higher point scattering have been studied in
\cite{Cremades:2003qj,Cremades:2004wa,Cvetic:2003ch,Abel:2003yx}.
Some preliminary results for the K\"{a}hler metric have been
obtained in \cite{Kors:2003wf}. A more complete derivation of the
K\"{a}hler metric directly from open/closed string scattering
amplitudes has been done in \cite{Lust:2004cx,Bertolini:2005qh},
which we use in this section.

At present, the closely related issues of moduli stabilization and
supersymmetry breaking such as to give rise to low energy
supersymmetry have not been understood well enough in these
compactifications. These can be better addressed in flux
compactifications of type IIB string theory, which are T-dual to
these compactifications.  In this section, we have taken a
phenomenological approach, parametrizing the effects of
supersymmetry breaking in a self-consistent way and examining the
consequences. Even though the supersymmetry breaking effects have
not been derived from first principles, it should still be
preferred to a blind parameterization of the mechanism of SUSY
breaking and its transmission to the observable sector. In the
absence of a complete supersymmetry-breaking model, such a
parameterization, in terms of non-zero $F$-terms with the
assumption of vanishing vacuum energy is useful as it gives us
some idea about the low energy consequences of these
constructions. Our main goal here is to use the results of
\cite{Lust:2004cx} to calculate and analyze effective low energy
soft supersymmetry breaking terms. We also look at some of their
dark matter applications.

\subsection{General construction of intersecting brane models.}
\label{idb:set:gen}
%%% ----------------------------------------------------------------------
In this section, we will briefly review the basics of constructing
these models.  More comprehensive treatments can be found in
\cite{Aldazabal:2000cn,Aldazabal:2001jmp,Blumenhagen:2000jhep,
Blumenhagen:2001jhep,Ott:2003yv}. The setup is as follows - we
consider type $IIA$ string theory compactified on a six
dimensional manifold $\mathcal{M}$. It is understood that we are
looking at the large volume limit of compactification, so that
perturbation theory is valid. In general, there are $K$ stacks of
intersecting D6-branes filling four dimensional Minkowski
spacetime and wrapping internal homology 3-cycles of
$\mathcal{M}$. Each stack $P$ consists of $N_P$ coincident D6
branes whose worldvolume is $M_4 \times {\Pi}_P$, where ${\Pi}_P$
is the corresponding homology class of each 3-cycle. The closed
string degrees of freedom reside in the entire ten dimensional
space, which contain the geometric scalar moduli fields of the
internal space besides the gravitational fields. The open string
degrees of freedom give rise to the gauge theory on the D6-brane
worldvolumes, with gauge group ${\Pi}_P \,U(N_{P})$. In addition,
there are open string modes which split into states with both ends
on the same stack of branes as well as those connecting different
stacks of branes. The latter are particularly interesting. If for
example, the 3-cycles of two different stacks, say ${\Pi}_P$ and
${\Pi}_Q$ intersect at a single point in $\mathcal{M}$, the lowest
open string mode in the Ramond sector corresponds to a chiral
fermion localized at the four dimensional intersection of $P$ and
$Q$ transforming in the bifundamental of $U(N_{P}) \times
U(N_{Q})$ \cite{Berkooz:1996npb}. The net number of left handed
chiral fermions in the $a b$ sector is given by the intersection
number $I_{P Q} \equiv [{\Pi}_{P}] \cdot [{\Pi}_{Q}]$.

The propagation of massless closed string RR modes on the compact
space $\mathcal{M}$ under which the D-branes are charged, requires
some consistency conditions to be fulfilled. These are known as
the $RR$ tadpole-cancellation conditions, which basically means
that the net $RR$ charge of the configuration has to vanish
\cite{Uranga:2001npb}. In general, there could be additional RR
sources such as orientifold planes or background fluxes. So they
have to be taken into account too. Another desirable constraint
which the models should satisfy is $\mathcal{N}$=1 supersymmetry.
Imposing this constraint on the closed string sector requires that
the internal manifold $\mathcal{M}$ be a Calabi-Yau manifold. We
will see shortly that imposing the same constraint on the open
string sector leads to a different condition.

A technical remark on the practical formulation of these models is
in order. Till now, we have described the construction in type
$IIA$ string theory. However, it is also possible to rephrase the
construction in terms of type $IIB$ string theory. The two
pictures are related by T-duality. The more intuitive picture of
type $IIA$ intersecting D-branes is converted to a picture with
type $IIB$ D-branes having background magnetic fluxes on their
world volume. It is useful to remember this equivalence as it
turns out that in many situations, it is more convenient to do
calculations in type $IIB$.

Most of the realistic models constructed in the literature involve
toroidal (${T^6}$) compactifications or orbifold/orientifold
quotients of those. In particular, orientifolding introduces O6
planes as well as mirror branes wrapping 3-cycles which are
related to those of the original branes by the orientifold action.
For simplicity, the torus (${T^6}$) is assumed to be factorized
into three 2-tori, i.e ${T^6}$ = $T^2 \times T^2 \times T^2$. Many
examples of the above type are known, especially with those
involving orbifold groups - i) $Z_2 \, \times \,Z_2$
\cite{Cvetic:2001tj} ii) $Z_4 \, \times \, Z_2$
\cite{Honecker:2003npb}, iii) $Z_4$ \cite{Blumenhagen:2003jhep},
iv) $Z_6$ \cite{Honecker:2004th}, etc.

%%% ----------------------------------------------------------------------
\subsection{A local MSSM-like model}\label{idb:set:model}
%%% ----------------------------------------------------------------------
In order to make contact with realistic low energy physics while
keeping supersymmetry intact, we are led to consider models which
give rise to a chiral spectrum close to that of the MSSM. In any
case, it is a useful first step to analyze. It has been shown
that this requires us to perform an orientifold twist. A stack of
$N_P$ D6 branes wrapping a 3-cycle not invariant under the
orientifold projection will yield a $U(N_P)$ gauge group,
otherwise we get a real $(SO(2N_P))$ or pseudoreal $(USp\,(2N_P))$
gauge group.

Using the above fact, the brane content for an MSSM-like chiral
spectrum with the correct intersection numbers has been presented
in \cite{Cremades:2003qj}. Constructions with more than four
stacks of branes can be found in \cite{Kokorelis:2003jr}. In the
simplest case, there are four stacks of branes which give rise to
the initial gauge group : $U(3)_a \times Sp(2)_b \times U(1)_c
\times U(1)_d$, where $a,b,c\,\&\,d$ label the different stacks.
The intersection numbers $I_{PQ} = [{\Pi}_P] \cdot [{\Pi}_Q]$
between a D6-brane stack $P$ and a D6-brane stack $Q$ is given in
terms of the 3-cycles $[{\Pi}_P]$ and $[{\Pi}_Q]$, which are
assumed to be factorizable.
\begin{equation}
[{\Pi}_P] \equiv
[(n_P^1,m_P^1)\otimes(n_P^2,m_P^2)\otimes(n_P^3,m_P^3)]
\end{equation}
where $(n_{P}^i,m_{P}^i)$ denote the wrapping numbers on the
$i^{th}$ 2-torus.The $O6$ planes are wrapped on 3-cycles :
\begin{equation}
[{\Pi}_{O6}] = \bigotimes_{r=1}^3[(1,0)]^r
\end{equation}
\begin{table}
\begin{center}
\begin{tabular}[c]{|c|c|c|c|c|c|}
\hline Stack  & Number of Branes & Gauge Group &
$(n_{\alpha}^1,m_{\alpha}^1)$ & $(n_{\alpha}^2,m_{\alpha}^2)$ &
$(n_{\alpha}^3,m_{\alpha}^3)$\\
\hline $Baryonic$ & $N_a =3$ & $U(3)= SU(3) \times U(1)_a$ &
$(1,0)$ & $(1/{\rho},3{\rho})$ & $(1/{\rho},-3{\rho})$
\\
$Left$ & $N_b =1$ & $USp(2)\cong SU(2)$ & $(0,1)$ & $(1,0)$ & $(0,-1)$\\
$Right$ & $N_c =1$ & $U(1)_c$ & $(0,1)$ & $(0,-1)$ & $(1,0)$\\
$Leptonic$ & $N_d =1$ & $U(1)_d$ & $(1,0)$ & $(1/{\rho},3{\rho})$
& $(1/{\rho},-3{\rho})$
\\
\hline
\end{tabular}
\end{center}
\caption {Brane content for an MSSM-like spectrum. The mirror
branes $a^*,b^*,c^*,d^*$ are not shown. $\rho$ can take values 1,
1/3. For concreteness, we take $\rho =1$ for calculating the soft
terms. However, the parameter space for the soft terms remains the
same for both $\rho =1$ and $\rho=1/3$.}\label{idb:tab:wr}
\end{table}
Note that for stack $b$, the mirror brane $b^*$ lies on top of
$b$. So even though $N_b = 1$, it can be thought of as a stack of
two D6 branes, which give an $USp(2) \cong SU(2)$ group under the
orientifold projection.

The brane wrapping numbers are shown in Table \ref{idb:tab:wr} and
the chiral particle spectrum from these intersecting branes are
shown in Table \ref{idb:tab:spec}.
\begin{table}[ht]
\begin{center}
\begin{tabular}{|c|c|c|c|c|c|c||c|}\hline
fields& sector& I& $SU(3)_c\times SU(2)_L$& $U(1)_a$&
  $U(1)_c$&$U(1)_d$&$U(1)_Y$ \\ \hline
$Q_L$& $(a,b)$& 3& $(3,2)$  & 1& 0& 0& 1/6 \\ \hline $U_R$&
$(a,c)$& 3& $(3,1)$  &-1& 1& 0&-2/3 \\ \hline $D_R$& $(a,c^*)$& 3&
$(3,1)$&-1&-1& 0& 1/3 \\ \hline
  $L$& $(d,b)$& 3& $(1,2)$  & 0& 0& 1&-1/2 \\ \hline
$E_R$& $(d,c^*)$& 3& $(1,1)$& 0&-1&-1& 1   \\ \hline $N_R$&
$(d,c)$& 3& $(1,1)$  & 0& 1&-1& 0   \\ \hline $H_u$& $(b,c)$& 1&
$(1,2)$  & 0&-1& 0& 1/2 \\ \hline $H_d$& $(b,c^*)$& 1& $(1,2)$& 0&
1& 0&-1/2 \\ \hline
\end{tabular}
\end{center}
\caption{The MSSM spectrum from intersecting branes. The
hypercharge normalization is given by $Q_Y=
\frac{1}{6}Q_a-\frac{1}{2}Q_c-\frac{1}{2}Q_d$. }
\label{idb:tab:spec}
\end{table}

\subsubsection{Getting the MSSM}
The above spectrum is free of chiral anomalies. However, it has an
anomalous $U(1)$ given by $U(1)_a$ + $U(1)_d$. This anomaly is
canceled by a generalized Green-Schwarz mechanism
\cite{Aldazabal:2001jmp}, which gives a Stuckelberg mass to the
$U(1)$ gauge boson. The two nonanomalous $U(1)$s are identified
with $(B-L)$ and the third component of right-handed weak isospin
$U(1)_R$ \cite{Cremades:2003qj}. In orientifold models, it could
sometimes happen that some nonanomalous $U(1)$s also get a mass by
the same mechanism, the details of which depend on the specific
wrapping numbers. It turns out that in the above model, two
massless $U(1)$s survive. In order to break the two $U(1)$s down
to $U(1)_Y$, some fields carrying non-vanishing lepton number but
neutral under $U(1)_Y$ are assumed to develop vevs. This can also
be thought of as the geometrical process of brane recombination
\cite{Cremades:2002jhep}.

\subsubsection{Global embedding and supersymmetry breaking}

As can be checked from Table 1, the brane content by itself does
not satisfy the $RR$ tadpole cancellation conditions :
\begin{equation}
\sum_{\alpha}
([{\Pi}_{\alpha}]+[{\Pi}_{\alpha^*}])=32\,[{\Pi}_{O6}]
\end{equation}
\noindent Therefore, this construction has to be embedded in a
bigger one, with extra $RR$ sources included. There are various
ways to do this such as including hidden D-branes or adding
background closed string fluxes in addition to the open string
ones. As a bonus, this could also give rise to spontaneous
supersymmetry breaking. With extra D-branes, one might consider
the possibility of gaugino condensation in the hidden sector
\cite{Cvetic:2003yd}. Alternatively, one could consider turning on
background closed string $NS$-$NS$ and $RR$ fluxes which generate
a non-trivial effective superpotential for moduli, thereby
stabilizing many of them.

In this work, we will leave open the questions of actually
embedding the above model in a global one and the mechanism of
supersymmetry breaking. We shall assume that the embedding has
been done and also only \emph{parametrize} the supersymmetry
breaking, in the spirit of
\cite{Nilles:1983ge,Kaplunovsky:1993rd,Brignole:1997dp}. We are
encouraged because there exists a claim of a concrete mechanism
for the global embedding of (the T-dual of) this model as well as
supersymmetry breaking \cite{Marchesano:2004yq}.

\subsubsection{Exotic matter and $\mu$
problem}\label{idb:set:model:exo}

The above local model is very simple in many respects, especially
with regard to gauge groups and chiral matter. However, it also
contains exotic matter content which is non-chiral. These
non-chiral fields are related to the untwisted open string moduli
- the D-brane positions and Wilson lines. The presence of these
non-chiral fields is just another manifestation of the old moduli
problem of supersymmetric string vacua. However, it has been
argued \cite{Gorlich:2004qm} that mass terms for the above moduli
can be generated by turning on a $F$- theory 4-form flux. One then
expects that a proper understanding of this problem will result in
a stabilization of all the moduli. As explained in
\cite{Marchesano:2004yq}, there could be $\mathcal{N}$=1
embeddings of this local model in a global construction. This
requires additional D-brane sectors and background closed string
3-form fluxes. The other D-brane sectors add new gauge groups as
well as chiral matter, some of which could be charged under the
MSSM gauge group. This may introduce chiral exotics in the
spectrum, an undesirable situation. However, many of these exotics
uncharged under the MSSM gauge group can be made to go away by
giving vevs to scalars parametrizing different flat directions. In
this work, we assume that there exists an embedding such that
there are no chiral exotics charged under the MSSM. Such exotics
can cause two types of problems. It is of course essential that no
states exist that would already have been observed. It seems
likely that can be arranged. In addition, states that would change
the RGE running and details of the calculations have to be taken
into account eventually.

The higgs sector in the local model arises from strings stretching
between stacks $b$ and $c$. However, the net chirality of the $bc$
sector is zero, since the intersection number $I_{bc}$ is zero.
The higgs sector in the above model has a $\mu$ term, which has a
geometrical interpretation. The real part of the $\mu$ parameter
corresponds to the separation between stacks $b$ and $c$ in the
first torus, while the imaginary part corresponds to a Wilson line
phase along the 1-cycle wrapped on the first torus. These
correspond to flat directions of the moduli space. Adding
background closed string fluxes may provide another source of
$\mu$ term \cite{Camara:2003ku}, which will lift the flat
direction in general. Thus, the effective $\mu$ term relevant for
phenomenology is determined by the above factors and the problem
of obtaining an electroweak scale $\mu$ term from a fundamental
model remains open. In this work, therefore, we will not attempt
to calculate $\mu$, and fix it by imposing electroweak symmetry
breaking (EWSB). It is important to study further the combined
effect of the several contributions to $\mu$ and to EWSB.

\subsubsection{Type IIA - type IIB equivalence}
As mentioned earlier, it is useful to think about this model in
terms of its T-dual equivalent. In type $IIB$, we are dealing with
D9 branes wrapped on $T^2 \times T^2 \times T^2$ with an open
string background magnetic flux $\mathcal{F}^j$ turned on.
Therefore the D9-branes have in general mixed Dirichlet and
Neumann boundary conditions. The flux has two parts - one coming
from the antisymmetric tensor $(b^j)$ and the other from the gauge
flux $(F^j)$ so that :
\begin{equation}
{\mathcal{F}}^j = b^j + 2\pi{\alpha}'\,F^j
\end{equation}
\noindent The above compactification leads to the following closed
string K\"{a}hler and complex structure moduli, each of which are
three in number for this model:
\begin{equation}
T^{'j} = b^j + iR_1^{'j}R_2^{'j}\,\sin({\alpha'}^j); \;\;\;
U^{'j}= \frac{R_2^{'j}}{R_1^{'j}}\, e^{i{\alpha'}^j};\;\;j=1,2,3.
\end{equation}

\noindent where $R_1^{'j}$ and $R_2^{'j}$ are lengths of the basis
lattice vectors characterizing the torus $T^{2,j}$ and
${\alpha}^j$ is the angle between the two basis vectors of the
torus $T^{2,j}$. By performing a T-duality in the $y$ direction of
each torus $T^{2,j}$, the D9 brane with flux ${\cal F}^j$ is
converted to a D6 brane with an angle ${\theta}^j$ with respect to
the x-axis. This is given by \cite{Ardalan:1998ce}:
\begin{equation}
\tan(\pi\,{\theta}^j)=\frac{f^j}{\mathrm{Im}(T^{'j})}
\end{equation}
\noindent where $f^j$ is defined by the quantization condition for
the net 2-form fluxes ${\cal F}^j$ as
\begin{equation}
f^j \equiv \frac{1}{(2\pi)^2{\alpha}'} \int_{T^{2,j}} \, {\cal
F}^j = \frac{m^j}{n^j}; \;\; m^j,n^j \,\in\,\mathbf{Z},
\end{equation}
\noindent Using the above equation and the relation between the
type $IIA$ and type $IIB$: \ba T^j &=&
-\frac{{\alpha}'}{U^{'j}},\;\;\;U^j =
-\frac{{\alpha}'}{T^{'j}},\;\;\;j=1,2,3. \\
R^j_1&=&R_1^{'j}, \qquad R^j_2=\frac{1}{R_2^{'j}} \ea \noindent we
get the corresponding type $IIA$ relation:
\begin{equation}
\tan(\pi{\theta}^j) = \frac{m^j}{n^j}
\frac{|U^j|^2}{\mathrm{Im}\,(U^j)}; \;\; j=1,2,3. \label{idb:eq:a}
\end{equation}

\noindent The unprimed symbols correspond to the type IIA version
while the primed ones to the type IIB.

\subsubsection{$\mathcal{N}$=1 \textit{SUSY}}

We now look at the $\mathcal{N}$=1 supersymmetry constraint on the
open string sector.  In type $IIA$, this leads to a condition on
the angles ${\theta}^j$ \cite{Berkooz:1996npb}:
\begin{eqnarray}
{\theta}^1+{\theta}^2+{\theta}^3 &=& 0\;\; \mathrm{mod} \;\;2
\;\;\;\;
\mathrm{or} \nonumber \\
\sum_{j=1}^3 \frac{m^j}{n^j} \frac{|U^j|^2}{\mathrm{Im}\,(U^j)}
&=& \prod_{j=1}^3 \frac{m^j}{n^j}
\frac{|U^j|^2}{\mathrm{Im}\,(U^j)}
\end{eqnarray}

\noindent which after T-duality leads to a condition on the fluxes
in type $IIB$.

%%% ----------------------------------------------------------------------
\subsection{Low energy effective action and soft terms}
\label{idb:set:soft}
%%% ----------------------------------------------------------------------
We now analyze the issue of deriving the four dimensional
$\mathcal{N}$=1 low energy effective action of these intersecting
brane models. In the type $IIB$ picture, this has been done in
\cite{Grimm:2004uq,Jockers:2004yj} by Kaluza Klein reduction of
the Dirac-Born-Infeld and Chern-Simons action. The effective
action can also be obtained by explicitly computing disk
scattering amplitudes involving open string gauge and matter
fields as well as closed string moduli fields and deducing the
relevant parts of the effective action directly. This has been
done in \cite{Lust:2004cx}. We will follow the results of
\cite{Lust:2004cx} in our analysis.

The $\mathcal{N}$=1 supergravity action thus obtained is encoded
by three functions, the K\"{a}hler potential $K$, the
superpotential $W$ and the gauge kinetic function $f$
\cite{Cremmer:1982en}. Each of them will depend on the moduli
fields describing the background of the model. One point needs to
be emphasized. When we construct the effective action and its
dependence on the moduli fields, we need to do so in terms of the
moduli $s$, $t$ and $u$ in the field theory basis, in contrast to
the $S$, $T$ and $U$ moduli in the string theory basis
\cite{Lust:2004cx}. In type $IIA$, the real part of the field
theory $s$, $u$ and $t$ moduli are related to the corresponding
string theory $S$, $U$ and $T$ moduli by :
\begin{eqnarray}
\mathrm{Re}\,(s)& =&
\frac{e^{-{\phi}_4}}{2\pi}\,\left(\frac{\sqrt{\mathrm{Im}\,U^{1}\,
\mathrm{Im}\,U^{2}\,\mathrm{Im}\,U^3}}{|U^1U^2U^3|}\right)
\nonumber \\
\mathrm{Re}\,(u^j)& =&
\frac{e^{-{\phi}_4}}{2\pi}\left(\sqrt{\frac{\mathrm{Im}\,U^{j}}
{\mathrm{Im}\,U^{k}\,\mathrm{Im}\,U^l}}\right)\;
\left|\frac{U^k\,U^l}{U^j}\right| \qquad
(j,k,l)=(\overline{1,2,3})
\nonumber \\
\re(t^j)&=&\frac{i\alpha'}{T^j} \label{idb:eq:moduli}
\end{eqnarray}

\noindent where $j$ stands for the $j^{th}$ 2-torus. The above
formulas can be inverted to yield the string theory $U$ moduli in
terms of the field theory moduli $s$ and $u$.
\begin{equation}
\frac{|U^j|^2}{\mathrm{Im}\,(U^j)} = \sqrt{
\frac{\mathrm{Re}\,(u^k)\,\mathrm{Re}\,(u^l)}{\mathrm{Re}\,(u^j)\mathrm{Re}\,
(s)}}\qquad  (j,k,l)=(\overline{1,2,3}) \label{eq:b}
\end{equation}

\noindent The holomorphic gauge kinetic function for a D6 brane
stack $P$ is given by :
\begin{eqnarray}
f_P &=& \frac{1}{\kappa_P}
(n_P^1\,n_P^2\,n_P^3\,s-n_P^1\,m_P^2\,m_P^3\,u^1-n_P^2\,m_P^1\,m_P^3\,u^2-
n_P^3\,m_P^1\,m_P^2\,u^3)
\nonumber \\
g_{D6_P}^{-2} &=& |\mathrm{Re}\,(f_P)|\label{idb:eq:gkf}
\end{eqnarray}
\noindent The extra factor ${\kappa}_P$ is related to the
difference between the gauge couplings for $U(N_P)$ and
$Sp(2N_P),\,SO(2N_P)$. ${\kappa}_P =1$ for $U(N_P)$ and
${\kappa}_P =2$ for $Sp(2N_P)$ or $SO(2N_P)$
\cite{Klebanov:2003my}.

\noindent The SM hypercharge $U(1)_Y$ gauge group is a linear
combination of several $U(1)$s: \be
Q_Y=\frac{1}{6}Q_a-\frac{1}{2}Q_c-\frac{1}{2}Q_d. \ee Therefore
the gauge kinetic function for the $U(1)_Y$ gauge group is
determined to be\cite{Blumenhagen:2003jy}: \be
f_Y=\frac{1}{6}f_{D_a}+\frac{1}{2}f_{D_c}+\frac{1}{2}f_{D_d}. \ee
The K\"{a}hler potential to the second order in open string matter
fields is given by :
\begin{eqnarray}
K(M,\bar{M},C,\bar{C}) = \hat{K}(M,\bar{M}) +  \sum_{untwisted\,
i,j} \tilde{K}_{C_{i}\bar{C}_j}(M,\bar{M}) C_i\bar{C}_j +
\sum_{twisted, \, \theta}
\tilde{K}_{C_{\theta}\bar{C}_{\theta}}(M,\bar{M})
C_{\theta}\bar{C}_{\theta}
\end{eqnarray}

\noindent where $M$ collectively denote the moduli; $C_i$ denote
untwisted open string moduli which comprise the D-brane positions
and the Wilson line moduli which arise from strings with both ends
on the same stack; and $C_{\theta}$ denote twisted open string
states arising from strings stretching between different stacks.

The open string moduli fields could be thought of as matter fields
from the low energy field theory point of view. The untwisted open
string moduli represent non-chiral matter fields and so do not
correspond to the MSSM. For the model to be realistic, they have
to acquire large masses by some additional mechanism, as already
explained in section \ref{idb:set:model:exo}.

Let's now write the K\"{a}hler metric for the twisted moduli
arising from strings stretching between stacks $P$ and $Q$, and
comprising $1/4$ BPS brane configurations. In the type IIA
picture, this is given by \cite{Cvetic:2003ch,Lust:2004cx}:
\begin{equation}
\tilde{K}_{C_{\theta_{PQ}}\bar{C}_{\theta_{PQ}}} \equiv
\tilde{K}_{PQ}= e^{{\phi}_4}\,\left(e^{{\gamma}_E\,\sum_{j=1}^{3}
{\theta}^j_{PQ}} \,\prod_{j=1}^3 \; \left[
\sqrt{\frac{{\Gamma}(1-{\theta}^j_{PQ})}{{\Gamma}({\theta}^j_{PQ})}}\;\;
(t^j+\bar{t}^j)^{-{\theta}^j_{PQ}}\right]\right)\label{idb:eq:kahler}
\end{equation}
\noindent where ${\theta}^j_{PQ} = {\theta}^j_{P}-{\theta}^j_{Q}$
is the angle between branes in the $j^{th}$ torus and ${\phi}_4$
is the four dimensional dilaton. From (\ref{idb:eq:moduli}),
${\phi}_4$ can be written as
$(\mathrm{Re}(s)\,\mathrm{Re}(u_1)\,\mathrm{Re}(u_2)\,\mathrm{Re}(u_3))^{-1/4}$.
String-loop corrections to the K\"{a}hler metric have been
calculated for some orientifolds in \cite{Berg:2005ja}. The above
K\"{a}hler metric depends on the field theory dilaton and complex
structure moduli $u^j$ through ${\phi}_4$ and ${\theta}_{PQ}$. It
is to be noted that (\ref{idb:eq:kahler}) is a product of two
factors, one which explicitly depends on the field theory $s$ and
$u$ moduli ($e^{{\phi}_4}$), and the other which implicitly
depends on the $s$ and $u$ moduli (through the dependence on
${\theta}^j_{PQ}$). Thus, $\tilde{K}_{PQ}$ can be symbolically
written as :\begin{equation} \tilde{K}_{PQ} =
e^{{\phi}_4}\,\tilde{K}^0_{PQ} \label{idb:eq:metric}
\end{equation}

\noindent The K\"{a}hler metric for $1/2$ BPS brane configurations
is qualitatively different from that of the $1/4$ BPS brane
configurations mentioned above. Generically, these cases arise if
both branes $P$ and $Q$ have a relative angle
${\theta}^j_{PQ}=0,1$ in the same complex plane $j$. It is
worthwhile to note that the higgs fields in Table
\ref{idb:tab:spec} form a $1/2$ BPS configuration and are
characterized by the following K\"{a}hler metric
\cite{Lust:2004fi}:
\begin{equation} \tilde{K}^{higgs}_{PQ} =
\left((s+\bar{s})(u^1+\bar{u}^1)(t^2+\bar{t}^2)(t^3+\bar{t}^3)\right)^{-1/2}
\label{idb:eq:higgs}
\end{equation}

\noindent An important point about the above expressions needs to
be kept in mind. These expressions for the K\"{a}hler metric has
been derived using \emph{tree level} scattering amplitudes and
with Wilson line moduli turned \emph{off}. Carefully taking the
Wilson lines into account as in \cite{Cremades:2004wa}, we see
that the K\"{a}hler metric has another multiplicative factor which
depends on the Wilson line moduli as well as $t$ moduli in type
$IIA$. If the Wilson line moduli do not get a vev, then our
analysis goes through. However, if they do, they change the
dependence of the metric on the $t$ moduli. As will be explained
later, we only choose the $u$ moduli dominated case for our
phenomenological analysis, so none of our results will be
modified.

The superpotential is given by:
\begin{equation}
W = \hat{W} + \frac{1}{2} {\mu}_{\alpha \beta}(M)\,
C^{\alpha}\,C^{\beta}+ \frac{1}{6}\,Y_{\alpha \beta
\gamma}(M)\,C^{\alpha}\,C^{\beta}\,C^{\gamma}+... \label{eq:W}
\end{equation}
In our phenomenological analysis, we have not included the Yukawa
couplings for simplicity. But as explained later, in the $u$
moduli dominant SUSY breaking case, the soft terms are independent
of the Yukawa couplings and will not change the phenomenology.

\subsubsection{Soft terms in general soft broken $\mathcal{N}$=1, $D=4$
supergravity Lagrangian}

From the gauge kinetic function, K\"{a}hler potential and the
superpotential, it is possible to find formulas for the
\emph{normalized} soft parameters - the gaugino mass parameters,
mass squared parameter for scalars and the trilinear parameters
respectively. These are given by \cite{Brignole:1997dp}:
\begin{eqnarray}
M_P &=& \frac{1}{2\,\mathrm{Re}\,f_P}\, (F^M\,\partial_M\,f_P) \nonumber \\
m_{PQ}^2 &=& (m_{3/2}^2 + V_0) - \sum_{M,N}\, \bar{F}^{\bar{M}}
F^N\,{\partial}_{\bar{M}}\,{\partial}_{N}\,\log({\tilde{K}}_{PQ}) \nonumber \\
A_{PQR} &=&
F^M[\hat{K}_M+{\partial}_M\,\log(Y_{PQR})-{\partial}_{M}\,
\log(\tilde{K}_{PQ}\tilde{K}_{QR}\tilde{K}_{RP})]\label{idb:eq:soft}
\end{eqnarray}

\noindent For our purposes, $P, Q$ and $R$ denote brane stacks. So
$M_P$ denotes the gaugino mass parameter arising from stack $P$;
$m_{PQ}^2$ denotes mass squared parameters arising from strings
stretching between stacks $P$ and $Q$ and $Y_{PQR}$ denotes Yukawa
terms arising from the triple intersection of stacks $P$, $Q$ and
$R$. The terms on the RHS without the indices $P$, $Q$ and $R$ are
flavor independent. Also, $M$ and $N$ run over the closed string
moduli. $F^M$ stands for the auxiliary fields of the moduli in
general. Supersymmetry is spontaneously broken if these fields get
non-vanishing vevs. It is assumed here that the auxiliary fields
$D$ have vanishing vevs. Their effect on the soft terms can be
calculated as in \cite{Kawamura:1996ex}, which we assume to be
zero. These formulas have been derived for the case when the
K\"{a}hler metric for the observable (MSSM) fields is diagonal and
universal in flavor space. In principle, there are also
off-diagonal terms in the K\"{a}hler metric. They relate twisted
open string states at different intersections and therefore are
highly suppressed. We neglect the off- diagonal terms in our
study. If the seperations between the intersections are very
small, the off-diagonal terms or non-universal diagonal terms may
have observable effects leading to interesting flavor physics.

The effective $\mathcal{N} = 1$, $4\,d$ field theory is assumed to
be valid at some high scale, presumably the string scale. The
string scale in our analysis is taken to be the unification scale
$( \sim 2 \times 10^{16} \; \mathrm{GeV})$. We then need to use
the renormalization group equations (RGE) to evaluate these
parameters at the electroweak scale. In this work, as mentioned
before, it is assumed that the non-chiral exotics have been made
heavy by some mechanism and there are no extra matter fields at
any scale between the electroweak scale and the unification scale.
This is also suggested by gauge coupling unification at the
unification scale.

One might wonder whether including the Yukawas in the analysis may
lead to significant modifications in the spectrum at low energies
because of their presence in the formulas for the soft terms
(\ref{idb:eq:soft}). However, this does not happen. This is
because the Yukawa couplings ($Y_{\alpha\beta\gamma}$) appearing
in the soft terms are \emph{not} the physical Yukawa couplings
($Y^{\mathrm{ph}}_{\alpha\beta\gamma}$). The two are related by:
\be Y^{\mathrm{ph}}_{\alpha\beta\gamma} = Y_{\alpha\beta\gamma}\,
\frac{\hat{W}^*}{|\hat{W}|}\,e^{\hat{K}/2}\,
(\tilde{K}_{\alpha}\tilde{K}_{\beta}\tilde{K}_{\gamma})^{-1/2} \ee
The Yukawa couplings ($Y_{\alpha\beta\gamma}$) between fields
living at brane intersections in intersecting D-brane models arise
from worldsheet instantons involving three different boundary
conditions \cite{Aldazabal:2000cn}. These semi-classical instanton
amplitudes are proportional to $e^{-A}$ where $A$ is the
worldsheet area. They have been shown to depend on the K\"{a}hler
($t$) moduli (complexified area) and the Wilson line moduli
\cite{Cremades:2003qj} in type $IIA$. Although the physical
Yukawas ($Y^{\mathrm{ph}}_{\alpha\beta\gamma}$) depend on the $u$
moduli through their dependence on the K\"{a}hler potential, the
fact that $Y_{\alpha\beta\gamma}$ do not depend on the $u$ moduli
in type $IIA$ ensures that in the $F^u$ dominant supersymmetry
breaking case, the soft terms are independent of
$Y_{\alpha\beta\gamma}$.

Thus our analysis is similar in spirit to those in the case of the
heterotic string, where dilaton dominated supersymmetry breaking
and moduli dominated supersymmetry breaking are analyzed as
extreme cases. It should be remembered however, that T-duality
interchanges the field theory ($t$) and ($u$) moduli. Thus what we
call $u$ moduli in type $IIA$, become $t$ moduli in type $IIB$ and
vice versa. In a general situation, in which the $F$-terms of all
the moduli get vevs, the situation is much more complicated and a
more general analysis needs to be done. This is left for the
future.

\subsubsection{Soft terms in intersecting brane models}
We calculate the soft terms in type IIA picture. As already
explained, we assume that only $F$-terms for $u$ moduli get
non-vanishing vevs. In terms of the goldstino angle $\theta$ as
defined in \cite{Brignole:1997dp}, we have $\theta = 0$. We assume
the cosmological constant is zero and introduce the following
parameterization of $F^{u^i}$: \be
F^{u^i}=\sqrt{3}m_{3/2}(u^i+\bar u^i)\Theta_{i}e^{-i\gamma_i}
\qquad \mbox{for } i=1,2,3 \label{idb:eq:Fu} \ee with
$\sum|\Theta_i|^2=1$. To calculate the soft terms, we need to know
the derivatives of the K\"{a}hler potential with respect to $u$.
For a chiral field $C$ arising from open strings stretched between
stacks $P$ and $Q$, we denote its K\"{a}hler potential as
$\tilde{K}_{PQ}$. From (\ref{idb:eq:metric}), we see that their
derivatives with respect to $u^i$ are \ba \frac{\partial
\log{\tilde{K}_{PQ}}}{\partial u^i}&=& \sum_{j=1}^3\frac {\partial
\log{\tilde{K}^0_{PQ}}}{\partial\theta^j_{PQ}}
\frac{\partial\theta^j_{PQ}}{\partial u^i} + \frac{-1}{4(u^i+\bar{u}^i)}\\
\frac{\partial^2 \log{\tilde{K}_{PQ}}}{\partial u^i\partial\bar
u^j}&=& \sum_{k=1}^3\left(\frac{\partial
\log{\tilde{K}^0_{PQ}}}{\partial\theta^k_{PQ}}
\frac{\partial^2\theta^k_{PQ}}{\partial u^i\partial\bar u^j}+
\frac{\partial^2\log \tilde{K}^0_{PQ}}{\partial (\theta^k_{PQ})^2}
\frac{\partial\theta^k_{PQ}}{\partial u^i}
\frac{\partial\theta^k_{PQ}}{\partial \bar
u^j}+\frac{{\delta}_{ij}}{4\,(u^i+\bar{u}^i)^2}\right) \ea From
the K\"{a}hler potential in eq.(\ref{idb:eq:kahler}), we have \ba
\Psi(\theta^j_{PQ})&\equiv&\frac {\partial
\log{\tilde{K}^0_{PQ}}}{\partial\theta^j_{PQ}}=
\gamma_E+\frac{1}{2}\frac{d}{d{\theta}^j_{PQ}}\,\ln{\Gamma(1-\theta^j_{PQ})}-
\frac{1}{2}\frac{d}{d{\theta}^j_{PQ}}\,\ln{\Gamma(\theta^j_{PQ})}-\log(t^j+\bar t^j) \label{eq:idb:gamma}\\
\Psi'(\theta^j_{PQ})&\equiv& \frac{\partial^2\log
\tilde{K}^0_{PQ}} {\partial(\theta^j_{PQ})^2}=
\frac{d\Psi(\theta^j_{PQ})}{d \theta^j_{PQ}} \ea The angle
$\theta^j_{PQ}\equiv\theta^j_P-\theta^j_Q$ can be written in terms
of $u$ moduli as: \be \tan(\pi\theta^j_P)=\frac{m^j_P}{n^j_P}
\sqrt{\frac{\re u^k \re u^l}{\re u^j\re s}} \qquad\mbox{where }
(j,k,l)=(\overline{1,2,3}) \ee Then we have \be
{\theta}^{j,k}_{PQ} \equiv (u^k+\bar u^k)\,\frac{\partial
\theta^j_{PQ}}{\partial u^k}= \left\{\begin{array}{cc}
 \left[-\frac{1}{4\pi}
 \sin(2\pi\theta^j)
 \right]^P_Q & \mbox{ when }j=k  \vspace*{0.6cm} \\
 \left[\frac{1}{4\pi}
\sin(2\pi\theta^j)
 \right]^P_Q & \mbox{ when }j\neq k
\end{array}\right.\label{idb:eq:dthdu}
\ee where we have defined
$[f(\theta^j)]^P_Q=f(\theta^j_P)-f(\theta^j_Q)$. The second order
derivatives are \be {\theta}^{j,k\bar{l}}_{PQ} \equiv (u^k+\bar
u^k)(u^l+\bar u^l)\,\frac{\partial^2 \theta^j_{PQ}}{\partial
u^k\partial\bar u^l}= \left\{\begin{array}{cc} \frac{1}{16\pi}
  \left[ \sin(4\pi\theta^j)+4\sin(2\pi\theta^j)
 \right]^P_Q &
   \mbox{when }j=k=l  \vspace*{0.6cm} \\
 \frac{1}{16\pi}  \left[
 \sin(4\pi\theta^j)-4\sin(2\pi\theta^j)
 \right]^P_Q &
   \mbox{when }j\neq k=l  \vspace*{0.6cm} \\
 -\frac{1}{16\pi}\left[
 \sin(4\pi\theta^j)
 \right]^P_Q &
   \mbox{ when }j=k\neq l\mbox{ or } j=l\neq k \vspace*{0.4cm} \\
 \frac{1}{16\pi}\left[
\sin(4\pi\theta^j)
 \right]^P_Q &
   \mbox{when }j\neq k\neq l\neq j
\end{array}\right.\label{idb:eq:dth2du}
\ee Now, one can substitute
eq.(\ref{idb:eq:Fu}-\ref{idb:eq:dth2du}) to the general formulas
of soft terms in eq.(\ref{idb:eq:soft}). The formulas for the soft
parameters in terms of the wrapping numbers of a general
intersecting D-brane model are given by:

\begin{itemize}
\item Gaugino mass parameters \be M_P=\frac{-\sqrt{3}m_{3/2}}{\re
f_P}\sum_{j=1}^3 \left(\re u^j\,\Theta_j\,
e^{-i\gamma_j}\,n^j_Pm^k_Pm^l_P\right) \qquad
(j,k,l)=(\overline{1,2,3}) \label{eq:idb:gaugino}\ee As an
example, the $M_{\tilde{g}}$ can be obtained by putting $P = a$
and using the appropriate wrapping numbers, as in Table
\ref{idb:tab:wr}.

\item Trilinears : \begin{eqnarray}
A_{PQR}&=&-\sqrt{3}m_{3/2}\sum_{j=1}^3 \left[
\Theta_je^{-i\gamma_j}\left(1+(\sum_{k=1}^3
 \theta_{PQ}^{k,j}\Psi(\theta^k_{PQ})-\frac{1}{4})+(\sum_{k=1}^3
 \theta_{RP}^{k,j}\Psi(\theta^k_{RP})-\frac{1}{4})
\right)\right] + \nonumber \\
& &
\frac{\sqrt{3}}{2}m_{3/2}{\Theta}_{1}e^{-i{\gamma}_1}\end{eqnarray}
This arises in general from the triple intersections $PQ$, $QR$
and $RP$, where $PQ$ and $RP$ are 1/4 BPS sector states and $QR$
is a 1/2 BPS state. $QR$ being the higgs field, has a special
contribution to the trilinear term compared to $PQ$ and $RP$. So
as an example, $A_{Q_LH_uU_R}$ can be obtained as a triple
intersection of $ab$, $bc$ and $ca$, as seen from Table
\ref{idb:tab:spec}.

\item Scalar mass squared (1/4 BPS) : \be
m^2_{PQ}=m_{3/2}^2\left[1-3\sum_{m,n=1}^3
\Theta_m\Theta_ne^{-i(\gamma_m-\gamma_n)}\left(
\frac{{\delta}_{mn}}{4}+ \sum_{j=1}^3 (\theta^{j,m\bar
n}_{PQ}\Psi(\theta^j_{PQ})+
 \theta^{j,m}_{PQ}\theta^{j,\bar n}_{PQ}\Psi'(\theta^j_{PQ}))\right)
\right] \ee By using the definitions in eq.(\ref{idb:eq:dthdu},
\ref{idb:eq:dth2du}), we see that $\theta^{j,k}_{PQ}$ and
$\theta^{j,k\bar l}_{PQ}$ do not depend on  the vevs of $u$
moduli: $u^j$. As an example, the squark mass squared
$m^2_{\tilde{Q}}$ can be obtained by putting $P,Q = a,b$; as can
be seen again from Table \ref{idb:tab:spec}.
\end{itemize}

\noindent We can now use the wrapping numbers in Table
\ref{idb:tab:wr} to get explicit expressions for the soft terms
for the particular model in terms of the $s,t$ and $u$ moduli and
the parameters $m_{3/2}$, $\gamma_i$ and $\Theta_i$, $i=1,2,3$.
The expressions for the trilinear $A$ parameters and scalar mass
squared parameters (except those for the up and down type higgs)
are provided in the Appendix \ref{idbsoft}. Using Table
\ref{idb:tab:wr}, the formula for the gaugino mass parameters and
the mass squared parameters for the up and down higgs are given
by:

\begin{itemize}
\item Gaugino mass parameters: \ba
%M_{\tilde B}&=&\frac{3\sqrt{3}}{5}m_{3/2}(3\re u_1\Theta_1
% e^{-i\gamma_1}+ \Theta_3e^{-i\gamma_3}) \\
%M_{\tilde W}&=&\sqrt{3}m_{3/2}\Theta_2 e^{-i\gamma_2} \\
%M_{\tilde g}&=&\frac{9\sqrt{3}}{2}m_{3/2}
% \re u_1\Theta_1 e^{-i\gamma_1}
M_{\wtd B}&=&3\sqrt{3}m_{3/2}
 \frac{12e^{-i\gamma_1}\,\re u_1\,\Theta_1
 +e^{-i\gamma_3}\,\re u_3\,\Theta_3}
 {3\re u_3+4\re s+ 36 \re u_1} \\
M_{\wtd W}&=&\sqrt{3}m_{3/2}\,e^{-i\gamma_2}\,\Theta_2 \\
M_{\wtd g}&=&\sqrt{3}m_{3/2}\frac{9e^{-i\gamma_1}\,\re
u_1\,\Theta_1}
 {\re s+9\re u_1}
\ea It is important to note that there is no gaugino mass
degeneracy at the unification scale, unlike other models such as
mSUGRA. This will lead to interesting experimental signatures.

\item Higgs mass parameters \ba m^2_{H_u} = m^2_{H_d} =
m^2_{3/2}\,(1-\frac{3}{2}\,|{\Theta}_1|^2)\ea

\noindent The Higgs mass parameters are equal because they are
characterized by the same K\"{a}hler metric
$\tilde{K}^{higgs}_{PQ}$ in (\ref{idb:eq:higgs}).

\end{itemize}

\noindent For completeness, we would also like to list the
relative angles between a brane $P$ and the orientifold plane on
the $j$th torus are denoted by $\theta^j_P$ in Table
\ref{idb:tab:angl}. The soft terms depend on these angles.

\begin{table}
\begin{center}
\begin{tabular}[h]{|c|c|c|c|c|c|}
\hline Stack  & Number of Branes & Gauge Group & $({\theta}^1_P)$
&
$({\theta}^2_P)$ & $({\theta}^3_P)$\\
\hline $Baryonic$ & $N_a =3$ & $U(3)= SU(3) \times U(1)_a$ &
$0$ & $\alpha$ & $-\alpha$ \\
$Left$ & $N_b =1$ & $USp(2)\cong SU(2)$ & $\frac{1}{2}$
 & $0$ & $-\frac{1}{2}$\\
$Right$ & $N_c =1$ & $U(1)_c$ & $\frac{1}{2}$ &
 $-\frac{1}{2}$ & $0$\\
$Leptonic$ & $N_d =1$ &$U(1)_d$ &$0$ & $\alpha$ & $-\alpha$
\\
\hline
\end{tabular}
\end{center}
\caption{Angles between a brane $P$ and the orientifold plane on
the $j$th torus: $\theta^j_P$.} \label{idb:tab:angl}
\end{table}

\vspace{0.3cm}

\noindent The only non-trivial angle is $\alpha$, which is given
by \be \tan (\pi\alpha) =3\,\rho^2\sqrt{\frac{\re u^1\re u^3}{\re
u^2 \re s}}. \ee as can be seen from
eq.(\ref{idb:eq:a},\ref{eq:b}) and Table \ref{idb:tab:wr}.

We would now like to compare our setup and results for the soft
terms with those obtained in \cite{Lust:2004dn}. The setup
considered in \cite{Lust:2004dn} is very similar to ours, but with
a few differences. It is a type $IIB$ setup with three stacks of
D7 branes wrapping four cycles on $T^6$. The last stack is
equipped with a non-trivial open string flux, to mimic the
intersecting brane picture in type $IIA$, as explained in section
3.4. There is also a particular mechanism of supersymmetry
breaking through closed string 3-form fluxes. Thus, there is an
explicit superpotential generated for the closed string moduli,
which leads to an explicit dependence of the gravitino mass
$(m_{3/2})$, $F^{s,t^i,u^i}$ and the cosmological constant $(V)$
on the moduli $s$, $t^i$ and $u^i$. The cosmological constant is
zero if the goldstino angle ($\theta$) is zero which is the same
in our case. It turns out that using these formulas, in order for
the gravitino mass to be small, the string scale is sufficiently
low for reasonable values ($\mathcal{O}$(1) in string units) of
the flux. We have not assumed any particular mechanism of
supersymmetry breaking, so we do not have an explicit expression
for $(m_{3/2})$, $F^{s,t^i,u^i}$ and $(V)$ in terms of the moduli
and have taken the string scale to be of the order of the
unification scale.

The model considered in \cite{Lust:2004dn} considers non-zero
(0,3) and (3,0) form fluxes only, which leads to non-vanishing
$F^{s}$ and $F^{t^i}$. In the T-dual version, this means that
$F^{s}$ and $F^{u^i}$ are non-zero, which is the case we examined
in detail. However, in \cite{Lust:2004dn}, an isotropic
compactification is considered, while we allow a more general
situation.

For the calculation of soft terms, we have used the updated form
of the 1/4 BPS sector K\"{a}hler metric as in \cite{Font:2004cx},
which we have also explicitly checked. In \cite{Lust:2004dn}, the
\emph{un-normalized} general expression for calculating the soft
terms has been used following \cite{Kaplunovsky:1993rd}, whereas
we use the \emph{normalized} general expression for the soft terms
in eq.(\ref{idb:eq:soft}) \cite{Brignole:1997dp}. In contrast to
\cite{Lust:2004dn} which has a left-right symmetry, we have also
provided an expression for the Bino mass parameter, since we have
the SM gauge group (possibly augmented by $U(1)'s$) and the exact
linear combination of $U(1)'s$ giving rise to $U(1)_Y$ is known.

%%% ----------------------------------------------------------------------
\subsection{Some phenomenological implications}
\label{idb:set:phen}
%%% ----------------------------------------------------------------------
Using the formulas for the soft terms given in the previous
section, we can study some aspects of the phenomenology of the
model that has the brane setup shown in Table \ref{idb:tab:wr}.

Although ideally the theory should generate $\mu$ of order the
soft terms and $\tan\beta$ should be calculated, that is not yet
possible in practice as explained before. Therefore we will not
specify the mechanism for generating the $\mu$ term. We will take
$\tan\beta$ and $m_Z$ as input and use EWSB to fix $\mu$ and $b$
terms.

Unlike the heterotic string models where the gauge couplings are
automatically unified\footnote{But usually unified at a higher
scale than the true GUT scale.}, generic brane models don't have
this nice feature. This is simply because in brane models
different gauge groups originate from different branes and they do
not necessarily have the same volume in the compactified space.
Therefore to ensure gauge coupling unification at the scale
$M_X\approx 2 \times 10^{16}$GeV, the vev of some moduli fields
need to take certain values. In our models, the gauge couplings
are determined according to eq.(\ref{idb:eq:gkf}). Thus the
unification relations \be g_s^2=g_w^2=\frac{5}{3}g_Y^2\approx 0.5.
\label{idb:eq:guni} \ee lead to  three conditions on the four
variables: $\re\, s$ and $\re\, u^i$ where $i=1,2,3$. One of the
solutions is \be \re s=2-9\re u_1 \qquad \re u_2=4 \qquad \re
u_3=4. \ee It's interesting to note that $\mathcal{N}$=1 SUSY
condition actually requires $\re\,u_2=\re\, u_3$. Therefore
although at first sight it seems that the three gauge couplings
are totally unrelated in brane models, in this case requiring
$\mathcal{N}$=1 SUSY actually guarantees one of the gauge coupling
unification conditions \cite{Blumenhagen:2003jy}.

After taking into account the gauge coupling unification
constraint, the undetermined parameters we are left with are
$m_{3/2}$, $\tan\beta$, $\re u^1$, $\re t^2$, $\re t^3$,
$\Theta_i$ and $\gamma_i$, where $i=1,2,3$. The phases
$\gamma_{1,2,3}$ enter both gaugino masses and trilinears and in
general can not be rotated away, leading to EDMs and a variety of
other interesting phenomena for colliders, dark matter, etc. in
principle. However, for simplicity, in this work we set all phases
to be zero: $\gamma_{1,2,3}=0$. The only dependence on $\re\, t^2$
and $\re\, t^3$ is in the scalar mass squared terms and is
logarithmic. For simplicity, we set them equal: $\re t^2=\re t^3$.
Using the relation $\sum|\Theta_i|^2=1$, we can eliminate the
magnitude of one of the $\Theta_i$s but its sign is free to vary.
Thus  we are left with the following six free parameters and two
signs: \be m_{3/2},\hspace*{0.3cm}\tan\beta, \hspace*{0.3cm}\re
u^1, \hspace*{0.3cm}\re t^2, \hspace*{0.3cm}\Theta_1,
\hspace*{0.3cm}\Theta_2, \hspace*{0.3cm}\mbox{sign}(\mu),
\hspace*{0.3cm} \mbox{sign}(\Theta_3). \ee Instead of scanning the
full parameter space, we show here three representative models
which correspond to three interesting points in the parameter
space. In order to reduce fine tuning, we require that the gluino
be not too heavy. We also require that the higgs mass is about 114
GeV, that all other experimental bounds are satisfied and that the
universe is not overclosed. These requirements strongly constrains
the free parameters. The parameters of these three points are
shown in Table \ref{idb:tab:para}.
\begin{table}[h]
\begin{center}
\begin{tabular}{|c|c|c|c|c|c|c|c|c|c|} \hline
LSP &$m_{3/2}$ & $\tan\beta$&$\re u^1$&$\re t^2$&$\Theta_1$&
$\Theta_2$
 & $\mbox{sign}(\mu)$ & $\mbox{sign}(\Theta_3)$ &
  $\Omega^{\mbox{thermal}}_{\wtd N_1}$ \\ \hline
$\wtd W$  & 1500& 20& 0.025& 0.01& 0.50& 0.060& $+$& $-$& $\sim
0$\\ \hline $\wtd H$  & 2300& 30& 0.025& 0.01& -0.75& 0.518& $+$&
$+$& $\sim 0$\\ \hline $\wtd B$-$\wtd H$ mixture  &
               2300& 30& 0.025& 0.01& -0.75& 0.512& $+$& $+$& $\sim 0.23$ \\ \hline
\end{tabular}
\end{center}
\caption{Parameter choices for three particular models. All masses
are in GeV. We set $\re t^3=\re t^2$. $|\Theta_3|$ will be fixed
by the condition $\sum|\Theta_i|^2=1$. $\re s$, $\re u^2$ and $\re
u^3$ are determined by requiring gauge coupling unification. }
\label{idb:tab:para}
\end{table}

Using the values of the moduli, one can calculate the string scale
$M_S$. It is indeed between the unification scale and the Planck
scale. Notice that since $t\sim 1/T$, we are in the large radius
limit of compactification and perturbation theory holds good.

From the parameters shown in Table \ref{idb:tab:para}, we can
calculate the soft terms at high scale. They are shown in Table
\ref{idb:tab:hsoft}.
\begin{table}[h]
\begin{center}
\begin{tabular}{|c|c|c|c|c|c|} \hline
 & $M_1$ & $M_2$ & $M_3$ &  $A_t$& $m_0$ \\ \hline
$\wtd W$ LSP &-1288& 156& 146 & -728 & $\sim m_{3/2}$  \\ \hline
$\wtd H$ LSP &849&2064& -336 & 633 & $\sim m_{3/2}$   \\ \hline
$\wtd B$-$\wtd H$ &
               866& 2040& -336& 640 & $\sim m_{3/2}$   \\ \hline
\end{tabular}
\end{center}
\caption{Soft terms at the unification scale. The input parameters
for calculating the soft terms are shown in Table
{\ref{idb:tab:para}}. $m_0$ denotes the average of scalar masses.
In both models, they are roughly the gravitino mass. The sign of
the trilinears is according to the convention used by SUSPECT. It
should be kept in mind that this is opposite to the convention
used in supergravity formulas.} \label{idb:tab:hsoft}
\end{table}
We use SUSPECT \cite{Djouadi:2002ze} to run the soft terms from
the high scale to the weak scale and calculate the sparticle
spectrum. Most string-based models that have been analyzed in such
detail have had Bino LSPs. The three models we examine give wino,
higgsino and mixed bino-higgsino LSP, respectively. The gluino
masses\footnote{The radiative contributions to the gluino pole
mass from squarks are included.} in the three models are ${\cal
O}(500\mbox{GeV})$. They are significantly lighter than the
gluinos in most of existing supergravity and superstring based
MSSM models. These three models satisfy all current experimental
constraints. If the LSP is a pure bino, usually the relic density
is too large to satisfy the WMAP\cite{Eguchi:2002dm}
 constraint. For the third model, the LSP is a
mixture of bino and higgsino such that its thermal relic density
is just the measured $\Omega_{CDM}$. For the first two models,
LSPs annihilate quite efficiently such that their thermal relic
density is negligible. Thus if only thermal production is taken
into account, the LSP can not be the cold dark matter (CDM)
candidate. But in general, there are non-thermal production
mechanisms for the LSP, for example the late gravitino
decay\cite{Kawasaki:1994af} or Q-ball decays
\cite{Fujii:2001xp,Kasuya:1999wu}. These non-thermal production
mechanisms have uaually been neglected but could have important
effects on predicting the relic density of the LSP. Since at
present, it is not known whether non-thermal production mechansims
were relevant in the early universe, we have presented examples of
both possibilities.

Late gravitino decay actually can not generate enough LSPs to
explain the observed $\Omega_{CDM}$ for the first two models. This
is because the gravitino decays after nucleosynthesis. Thus to
avoid destroying the successful nucleosynthesis predictions, the
gravitino should not be produced abundantly during the reheating
process. Therefore, LSPs from the gravitino decay can not be the
dominant part of CDM.

Another mechanism for non-thermal production is by the late Q-ball
decay. Q-balls are non-topological solitons\cite{Coleman:1985ki}.
At a classical level, their stability is guaranteed by some
conserved charge, which could be associated with either a global
or a local $U(1)$ symmetry. At a quantum level, since the field
configuration corresponding to Q-balls does not minimize the
potential globally and in addition Q-balls by definition do not
carry conserved topological numbers, they will ultimately decay.
Q-balls can be generated in the Affleck-Dine mechanism of
baryogenesis\cite{Dine:1995kz}. Large amounts of supersymmetric
scalar particles can be stored in Q-balls. Final Q-ball decay
products will include LSPs, assuming R-parity is conserved. If
Q-balls decay after the LSP freezes out and before
nucleosynthesis, the LSP could be the CDM candidate and explain
the observed $\Omega_{CDM}$. The relic density of $\wtd N_1$ can
be estimated as\cite{Fujii:2001xp} \be \Omega_{\wtd N_1}\approx
0.5\times\left(\frac{0.7}{h}\right)^2 \times\left(\frac{m_{\wtd
N_1}}{100\mbox{GeV}}\right)
\left(\frac{10^{-7}\mbox{GeV}^2}{\langle\sigma v\rangle }\right)
\left(\frac{100\mbox{MeV}}{T_d}\right)
\left(\frac{10}{g_*(T_d)}\right)^{1/2} \ee For our first two
models, $\langle\sigma v\rangle\sim 10^{-7}\mbox{GeV}^2$. Thus if
the temperature $(T_d)$ when Q-balls decay is about $100$ MeV, we
will have $\Omega_{\wtd N_1}\approx 0.22$. One may attempt to
relate this number to the baryon number of the universe since
Q-balls may also carry baryon number. But probably, this wouldn't
happen in these models at least at the perturbative level, because
then baryon number is a global symmetry. Therefore Q-balls
wouldn't carry net baryon number. They may carry lepton number
because of lepton number violation. But since $T_d$ is well below
the temperature of the electroweak (EW) phase transition,
sphaleron effects are well suppressed so that the net lepton
number cannot be transferred to baryon number. Hence, in the
Q-ball scenario, baryon number probably is not directly related to
$\Omega_{\wtd N_1}$.

\begin{figure}[h!]
\center \epsfig{file=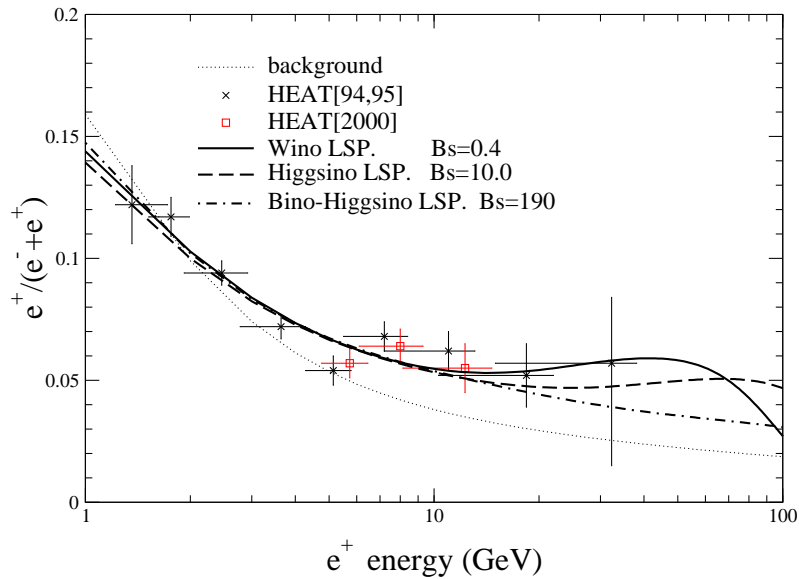,height=12cm,angle=-90}
\caption{Cosmic ray positron excess due to LSP annihilations.
Crossed points are HEAT experimental data. The dotted line is from
the standard cosmological model without taking into account the
LSP annihilation contributions. The solid line is for the $\wtd W$
LSP model, the dashed line is for the $\wtd H$ LSP model and the
dashed-dotted line is for the mixed $\wtd B$-$\wtd H$ LSP model.
Bs is the boost factor.} \label{idb:fig:posi}
\end{figure}

If we assume $\Omega_{\wtd N_1}\approx 0.23$ by either thermal
(for the third model) or non-thermal production (for the first two
models), then there will be interesting experimental signatures.
%The spin independent
%cross sections for proton and LSP scattering are $2.7\times
%10^{-11}$pb for the $\wtd W$ LSP model and $2.8\times 10^{-10}$pb
%for the $\wtd H$ LSP model\cite{Gondolo:2000ee}. Thus both of them
%are testable in future underground WIMP detection
%experiments.({\bf need to check}).
One of them is a positron excess in cosmic rays. Such an excess
has been reported by the HEAT experiment\cite{Barwick:1995gv} and
two other experiments (AMS and PAMELA)
\cite{Lechanoine-Leluc:2004dm} have been planned which will give
improved results in the future. In fact the HEAT excess is
currently the only reported dark matter signal that is consistent
with other data. In both the wino LSP and the higgsino LSP models,
LSPs annihilate to $W$ bosons quite efficiently. In particular,
$W^+$ decays to a positron(and a neutrino). We used
DARKSUSY\cite{Gondolo:2000ee} to fit the HEAT experiment
measurements\cite{Kane:2002nm}. The fitted curves are shown in
Figure \ref{idb:fig:posi}. The ``boost'' factors, which
characterize the local CDM density fluctuation against the
averaged halo CDM density,  in both models are not large. Small
changes in other parameters could give boost factors very near
unity. Thus both models give a nice explanation for the measured
positron excess in the cosmic ray data. For the mixed LSP model,
because the LSPs do not annihilate efficiently, one needs a
``boost'' factor of $\mathcal{O}$(100) to get a good fit, which
may be a bit too high.

\subsection{Summary of Results}\label{idb:sec:conc}
%%% ----------------------------------------------------------------------
We have investigated some phenomenological implications of
intersecting D-brane models, with emphasis on dark matter. We
calculated the soft SUSY breaking terms in these models focussing
on the $u$ moduli dominated SUSY breaking scenario in type IIA
string theory, in which case the results do not depend on the
Yukawa couplings and Wilson lines. The results depend on the brane
wrapping numbers as well as SUSY breaking parameters. Our main
result is providing in detail the soft-breaking Lagrangian for
intersecting brane models, which provides a new set of soft
parameters to study phenomenologically. In the absence of a
satisfactory mechanism of supersymmetry breaking and moduli
stabilization in these compactifications, we use a general self
consistent parameterization of F-term vevs, based on
\cite{Brignole:1997dp}. We applied our results to a particular
intersecting brane model\cite{Cremades:2003qj} which gives an
MSSM-like particle spectrum, and then selected three
representative points in the parameter space with relatively light
gluinos, in order to reduce fine-tuning, and calculated the weak
scale spectrum for them. The phenomenology of the three models
corresponding to the three points is very interesting. The LSPs
have different properties. They can be either wino-like,
higgsino-like, or a mixture of bino and higgsino. All of them can
be good candidates for the CDM.

\section{$M$ Theory compactifications on singular $G_2$ Holonomy Manifolds}

In the previous section, we saw that it is possible to construct
type IIA orientifold models giving rise to MSSM-like chiral
spectra in a natural way. However, the issues of moduli
stablization and supersymmetry breaking such as to generate the
hierarchy have not been well understood in this picture, which led
us to parametrize the effects of supersymmetry breaking in a
self-consistent manner and then examine their phenomenological
consequences.

In this section, we will focus on another corner of the entire $M$
theory moduli space, that of $M$ theory compactifications on
singular $G_2$ holonomy manifolds, where the issues of moduli
stabilization and supersymmetry breaking can be solved in a
natural manner such as to give rise to a stable hierarchy between
the Planck and electroweak scales. Thus far, there has been
essentially one good idea proposed to explain the relatively small
value of the weak scale. This is that the weak scale might be
identified with, or related to, the strong coupling scale of an
asymptotically free theory which becomes strongly coupled at low
energies and exhibits a mass gap at that strong coupling scale.
%\cite{Witten:1981kv} (I removed this because we would also have to refer to the
%origin of technicolor as well and probably others).
Holographically dual to this is the idea of warped extra
dimensions \cite{Randall:1999ee}. Strong dynamics (or its dual)
can certainly generate a small scale in a natural manner, but can
it also be compatible with the stabilization of all the moduli
fields?

One context for this question, which we will see is particularly
natural, is $M$ theory compactification on manifolds $X$ of
$G_2$-holonomy without fluxes. In these vacua, the only moduli one
has are zero modes of the metric on $X$, whose bosonic
superpartners are axions. Thus each moduli supermultiplet has a
Peccei-Quinn shift symmetry (which originates from 3-form gauge
transformations in the bulk 11d supergravity). Since such
symmetries can only be broken by non-perturbative effects, the
entire moduli superpotential $W$ is non-perturbative. In general
$W$ can depend on all the moduli. Therefore, in addition to the
small scale generated by the strong dynamics we might expect that
all the moduli are actually stabilized. This work will demonstrate
in detail that this is indeed the case.

Having established that the basic idea works well, the next
question we address is ``what are the phenomenological
implications?'' Since string/$M$ theory has many vacua, it would
be extremely useful if we could obtain a general prediction from
{\it all} vacua or at least some well-defined subset of vacua.
Remarkably, we are able to give such a prediction for all the
fluxless $M$ theory vacua (at least within the supergravity
approximation to which we are restricted for calculability): {\it
gaugino masses are generically suppressed relative to the
gravitino mass}.

A slightly more detailed elucidation of this result is that in all
de Sitter vacua within this class, gaugino masses are always
suppressed. In AdS vacua - which are obviously less interesting
phenomenologically - the gaugino masses are suppressed in `most'
of the vacua. This will be explained in more detail later.

The reason why we are able to draw such a generic conclusion is
the following: any given non-perturbative contribution to the
superpotential depends on various constants which are determined
by a specific choice of $G_2$-manifold $X$. {\it These constants
determine entirely the moduli potential}. They are given by the
one-loop beta-function coefficients $b_k$, the normalization of
each term $A_k$ and the constants $a_i$ (see (\ref{vol})) which
characterize the K\"{a}hler potential for the moduli. Finally,
there is a dependence on the gauge kinetic function, and in $M$
theory this is determined by a set of integers $N_i$ which specify
the homology class of the 3-cycle on which the non-Abelian gauge
group is localized. Rather than study a particular $X$ which fixes
a particular choice for these constants we have studied the
effective potential as a function of the $(A_k, b_k , a_i, N_i )$
. The result of gaugino mass suppression holds essentially for
arbitrary values of the $(A_k, b_k , a_i , N_i )$, at least in the
supergravity regime where we have been able to calculate. Thus,
any $G_2$-manifold which has hidden sectors with strong gauge
dynamics will lead to suppressed gaugino masses.

At a deeper level, however, the reason that this works is that the
idea of strong gauge dynamics to solve the hierarchy problem is a
good and simple idea which guides us to the answers directly. If
one's theory does not provide a simple mechanism for how the
hierarchy is generated, then it is difficult to see how one could
obtain a reliable prediction for, say, the spectrum of beyond the
Standard Model particles. In a particular subset of Type IIB
compactifications, Conlon and Quevedo have also discovered some
general results \cite{Conlon:2006us}. In fact, they remarkably
also find that gaugino masses are suppressed at tree level, though
the nature of the suppression is not quite the same. Some
heterotic compactifications also exhibit a suppression of
tree-level gaugino masses \cite{Binetruy:2000md}.

The suppression of gaugino masses relative to $m_{3/2}$ applies
for all the vacua, independent of the value of $m_{3/2}$. However,
in a generic vacuum the cosmological constant is too large. If we
therefore consider only those vacua in which the cosmological
constant is acceptable at leading order, this constrains the scale
of $m_{3/2}$ further. Remarkably, we find evidence that for such
vacua, $m_{3/2}$ is of order $1-100$ TeV. This result certainly
deserves much further investigation. The fact that such general
results emerge from these studies makes the task of predicting
implications for various collider observables as well as
distinguishing among different vacua with data from the LHC (or
any other experiment) easier. A more detailed study of the
collider physics and other phenomenology is currently being worked
out and will appear in the future.

\subsection{The Moduli Potential} \label{secII}

In this section we quickly summarize the basic relevant features
of $G_2$-compactifications, setup the notation and calculate the
potential for the moduli generated by strong hidden sector gauge
dynamics.

In $M$ theory compactifications on a manifold $X$ of
$G_2$-holonomy the moduli are in correspondence with the harmonic
3-forms. Since there are $N \equiv {b_3}(X)$ such independent
3-forms there are $N$ moduli $z_i=t_i+is_i$. The real parts of
these moduli $t_i$ are axion fields which originate from the
3-form field $C$ in $M$ theory and the imaginary parts $s_i$ are
zero modes of the metric on $X$ and characterize the size and
shape of $X$. Roughly speaking, one can think of the $s_i$'s as
measuring the volumes of a basis of the $N$ independent three
dimensional cycles in $X$.

Non-Abelian gauge fields are localized on three dimensional
submanifolds $Q$ of $X$ along which there is an orbifold
singularity \cite{Acharya:1998pm} while chiral fermions are
localized at point-like conical singularities
\cite{Atiyah:2001qf,Acharya:2004qe,Acharya:2001gy}. Thus these
provide $M$ theory realizations of theories with localized matter.
A particle localized at a point $p$ will be charged under a gauge
field supported on $Q$ if $p$ $\in Q$. Since generically, two
three dimensional submanifolds do not intersect in a seven
dimensional space, there will be no light matter fields charged
under both the standard model gauge group and any hidden sector
gauge group.  {\it Supersymmetry breaking is therefore gravity
mediated in these vacua.}\footnote{This is an example of the sort
of general result one is aiming for in string/$M$ theory. We can
contrast this result with Type IIA vacua. Here the non-Abelian
gauge fields are again localized on 3-cycles, but since
generically a pair of three cycles intersect at points in six
extra dimensions, {\it In Type IIA supersymmetry breaking will
generically be gauge mediated.}}

In general the K\"{a}hler potentials for the moduli are difficult
to determine in these vacua. However a set of K\"{a}hler
potentials, consistent with $G_2$-holonomy and known to describe
accurately some explicit examples of $G_2$ moduli dynamics were
given in \cite{Acharya:2005ez}. These models are given by

\be \label{kahler} K = -3 \ln(4\pi^{1/3}\,V_X) \ee where the
volume as a function of $s_i$ is \be \label{vol} V_X =
\prod_{i=1}^{N} s_i^{a_i},\;\; \mathrm{with} \;\; \sum_{i=1}^{N}
a_i = 7/3 \ee

We will assume that this $N$-parameter family of K\"{a}hler
potentials represents well the moduli dynamics. More general
K\"{a}hler potentials outside this class have the volume
functional multiplied by a function invariant under rescaling of
the metric. It would be extremely interesting to investigate the
extension of our results to these cases.

As motivated in the introduction, we are interested in studying
moduli stabilization induced via strong gauge dynamics. We will
begin by considering hidden sector gauge groups with no chiral
matter. Later sections will describe the cases with hidden sector
chiral matter.

In this `no matter' case, a superpotential of the following form
is generated \be \label{generalsuper} W = \sum_{k=1}^{M}A_ke^{ib_k
f_{k}}\, \ee \noindent where $M$ is the number of hidden sectors
undergoing gaugino condensation, $b_k=\frac {2\pi}{c_k}$ with
$c_k$ being the dual coxeter numbers of the hidden sector gauge
groups, and $A_k$ are numerical constants. The $A_k$ are RG-scheme
dependent and also depend upon the threshold corrections to the
gauge couplings; the work of \cite{Friedmann:2002ty} shows that
their ratios (which should be scheme independent) can in fact take
a reasonably wide range of values in the space of $M$ theory
vacua. We will only consider the ratios to vary from
$\mathcal{O}$(0.1-10) in what follows.

The gauge coupling functions $f_{k}$ for these singularities are
integer linear combinations of the $z_i$, because a 3-cycle $Q$
along which a given non-Abelian gauge field is localized is a
supersymmetric cycle, whose volume is linear in the moduli. \be
\label{generalgaugecoupl} f_k=\sum_{i=1}^{N} N^k_i\,z_i\,. \ee
Notice that, given a particular $G_2$-manifold $X$ for the extra
dimensions, the constants $( a_i, b_k , A_k , N^k_i )$ are
determined. Then, the K\"{a}hler potential and superpotential for
that particular $X$ are {\it completely determined} by the
constants $( a_i, b_k , A_k , N^k_i )$. This is as it should be,
since $M$ theory has no free dimensionless parameters.

We are ultimately aiming for an answer to the question, ``do $M$
theory vacua in general make a prediction for the beyond the
standard model spectrum?''. For this reason, since a fluxless $M$
theory vacuum is completely specified by the constants $(a_i, b_k
, A_k , N^k_i )$ we will try as much as possible ${\it not}$ to
pick a particular value for the constants and try to first
evaluate whether or not there is a prediction for {\it general
values of the constants}. Our results will show that at least
within the supergravity approximation there is indeed a general
prediction: the suppression of gaugino masses relative to the
gravitino mass.

At this point the simplest possibility would be to consider a
single hidden sector gauge group. Whilst this does in fact
stabilize all the moduli, it is a) non-generic and b) fixes the
moduli in a place which is strictly beyond the supergravity
approximation. Therefore we will begin, for simplicity, by
considering two such hidden sectors, which is more representative
of a typical $G_2$ compactification as well as being tractable
enough to analyze. The superpotential therefore has the following
form \be \label{super} W^{np} = A_1e^{ib_1 f_1}+ A_2e^{ib_2
f_2}\,. \ee The metric corresponding to the K\"{a}hler potential
(\ref{kahler}) is given by \be \label{metric}
K_{i\,\bar{j}}=\frac{3a_i}{4s_i^2}\delta_{i\,\bar{j}}\,. \ee The
${\cal N}=1$ supergravity scalar potential given by \be V=e^K
(K^{i\,\bar{j}} F_i\bar F_{\bar{j}} -3|W|^2)\,, \ee where \be
F_i=\partial_iW+(\partial_iK)W \,, \ee can now be computed. The
full expression for the scalar potential is given by \ba
\label{fullpotential}
V&=&\frac{1}{48\pi V_X^3}\,[\sum_{k=1}^{2}\sum_{i=1}^{N}a_i{\nu_i^k}\left({\nu_i^k}b_k+3\right)b_kA_k^2e^{-2b_k
\vec\nu^{\,k}\cdot\,\vec a}+3\sum_{k=1}^{2}A_k^2e^{-2b_k\vec\nu^{\,k}\cdot\,\vec a}\nonumber\\
& &+2{\rm cos}[(b_1\vec N^1-b_2\vec N^2)\cdot \vec t\,\,]\sum_{i=1}^{N}a_i
\prod_{k=1}^2\nu_i^kb_kA_ke^{-b_k\vec\nu^{\,k}\cdot\,\vec a}\\
& &+3{\rm cos}[(b_1\vec N^1-b_2\vec N^2)\cdot \vec
t\,\,]\left(2+\sum_{k=1}^{2}b_k\,\vec\nu^{\,k}\cdot\vec a
\right)\prod_{j=1}^2A_je^{-b_j\vec\nu^{\,j}\cdot\,\vec
a}]\,\nonumber \ea where we introduced a variable \be \label{nu}
\nu_i^k\equiv\frac{N_i^k s_i}{a_i} \;\;(\mathrm{no\;\,sum}) \ee
such that \be {\rm Im}f_k=\vec\nu^{\,k}\cdot\,\vec a\,. \ee By
extremizing (\ref{fullpotential}) with respect to the axions $t_i$
we obtain an equation \be {\rm sin}[(b_1\vec N^1-b_2\vec N^2)\cdot
\vec t\,\,]=0\,, \ee which fixes only one linear combination of
the axions. In this case \be {\rm cos}[(b_1\vec N^1-b_2\vec
N^2)\cdot \vec t\,\,]=\pm 1\,. \ee It turns out that in order for
the potential (\ref{fullpotential}) to have minima, the axions
must take on the values such that ${\rm cos}[(b_1\vec N^1-b_2\vec
N^2)\cdot \vec t\,\,]=-1$ for $A_1, A_2 > 0$. Otherwise the
potential has a runaway behavior. After choosing the minus sign,
the potential takes the form \ba \label{potential}
V=\frac{1}{48\pi V_X^3}\,[\sum_{k=1}^{2}\sum_{i=1}^{N}a_i{\nu_i^k}\left({\nu_i^k}b_k+3\right)b_kA_k^2e^{-2b_k\vec\nu^{\,k}\cdot\,\vec a}+3\sum_{k=1}^{2}A_k^2e^{-2b_k\vec\nu^{\,k}\cdot\,\vec a}&&\\
-2\sum_{i=1}^{N}a_i\prod_{k=1}^2\nu_i^kb_kA_ke^{-b_k\vec\nu^{\,k}\cdot\,\vec
a} -3\left(2+\sum_{k=1}^{2}b_k\,\vec\nu^{\,k}\cdot\vec a
\right)\prod_{j=1}^2A_je^{-b_j\vec\nu^{\,j}\cdot\,\vec
a}]&&\,\nonumber \ea

In the next section we will go on to analyze the vacua of this
potential with unbroken supersymmetry. The vacua in which
supersymmetry is spontaneously broken are described in sections
\ref{nonsusyadsvac} and \ref{chargedmattervac}.
%%%%%%%%%%%%%%%%%%%%%%%%%%%%%%%%%%%%%%%%%%%%%%%
\subsection{Supersymmetric Vacua} \label{susy}

In this section we will discuss the existence and properties of
the supersymmetric vacua in our theory. This is comparatively easy
to do since such vacua can be obtained by imposing the
supersymmetry conditions instead of extremizing the full scalar
potential (\ref{potential}). Therefore, we will study this case
with the most detail. Experience has also taught us that
potentials possessing rigid isolated supersymmetric vacua, also
typically have other non-supersymmetric vacua with many
qualitatively similar features.

The conditions for a supersymmetric vacuum are: \be \label{Fterm}
F_i = 0\,, \nonumber \ee which implies \ba
(b_1N^1_i+\frac{3a_i}{2s_i})A_1+(b_2N^2_i+\frac{3a_i}{2s_i})A_2\,
[\cos[(b_1\vec N^1-b_2\vec N^2)\cdot \vec t\,]\,
e^{(b_1\vec {N}^1-b_2\vec {N}^2)\cdot \vec s}\nonumber \\
+\; i\sin[(b_1\vec {N}^1-b_2\vec {N}^2)\cdot \vec t\,]\,
e^{(b_1\vec {N}^1-b_2\vec {N}^2)\cdot \vec s}]\,\, = \,\, 0 \ea
Equating the imaginary part of (\ref{Fterm}) to zero, one finds
that \ba
{\sin}[(b_1\vec N^1-b_2\vec N^2)\cdot \vec t\,\,] &=& 0\,, \nonumber\\
\mathrm{which} \;\; \mathrm{implies}\;\; \cos[(b_1\vec N^1-b_2\vec
N^2)\cdot \vec t\,\,] &=& \pm 1\,. \ea

For $A_1,A_2>0$, a solution with positive values for the moduli
($s_i$) exists when the axions take on the values such that
$\cos[(b_1\vec N^1-b_2\vec N^2)\cdot \vec t\,]=-1\,$. Now,
equating the real part of (\ref{Fterm}) to zero, one obtains \ba
\label{systemeqns}
(b_1{N}^1_i+\frac{3a_i}{2s_i})A_1+(b_2{N}^2_i+\frac{3a_i}{2s_i})A_2
\,e^{(b_1\vec{N}^1-b_2\vec {N}^2)\cdot \vec s} = 0\,. \ea This is
a system of $N$ transcendental equations with $N$ unknowns. As
such, it can only be solved numerically, in which case, it is
harder to get a good understanding of the nature of solutions
obtained. Rather than doing a brute force numerical analysis of
the system (\ref{systemeqns}) it is very convenient to introduce a
new auxiliary variable $\alpha$ to recast (\ref{systemeqns}) into
a system of \emph{linear} equations with $N$ unknowns coupled to a
single transcendental constraint as follows: \ba
&&\alpha(b_1N^1_i+\frac{3a_i}{2s_i})-(b_2N^2_i+\frac{3a_i}{2s_i}) = 0 \label{linear}\\
&&\frac {A_2} {A_1}\, = \frac 1 \alpha e^{-(b_1\vec {N}^1-b_2\vec
{N}^2)\cdot \vec s}\,.\label{constraint} \ea The system of linear
equations (\ref{linear}) can then be formally solved for $s_i$ in
terms of $\{b_1,b_2,N^1_i,N^2_i,a_i\}$ and $\alpha$: \be
\label{si}
 s_i = -\frac{3\,a_i\,(\alpha-1)}{2\,(b_1N^1_i\alpha-b_2N^2_i)};\;\;\;\;i=1,2,..,N\,.
\ee One can then substitute the solutions for $s_i$ into the
constraint (\ref{constraint}) and self-consistently solve for the
parameter $\alpha$ in terms of the input quantities
$\{A_1,A_2,b_1,b_2,N^1_i,N^2_i,a_i\}$. This, of course, has to be
done numerically, but we have indeed verified that solutions
exist. Thus, we have shown explicitly that the moduli can be
stabilized. We now go on to discuss the solutions, in particular
those which lie within the supergravity approximation.

%%%%%%%%%%%%%%%%%%%%%%%%%%%%%%%%%%%%%%%%%%%%%%%%%%%%%%%%%%%
\subsubsection{Solutions and the Supergravity Approximation}

Not all choices of the constants
$\{A_1,A_2,b_1,b_2,N^1_i,N^2_i,a_i\}$ lead to solutions consistent
with the approximation that in the bulk of spacetime, eleven
dimensional supergravity is valid. Although this is not a
precisely (in the numerical sense) defined approximation, a
reasonable requirement would seem to be that the values of the
stabilized moduli ($s_i$) obtained from (\ref{si}) are greater
than 1. It is an interesting question, certainly worthy of further
study, whether or not this is the correct criterion. In any case,
this is the criterion that we will use and discuss further.

\noindent From (\ref{si}) and (\ref{constraint}), and requiring
the $s_i$ to be greater than 1, we get the following two branches
of conditions on parameter $\alpha$ : \ba a) \;\;\;
\frac{A_2}{A_1} > 1;\;\;\;
min\,\{\frac{b_2N^2_i}{b_1N^1_i}\,;i=\overline{1,N}\,\} > \alpha >
\;
max\,\{\frac{b_2N^2_i+3a_i/2}{b_1N^1_i+3a_i/2}\,;i=\overline{1,N}\,\}
\nonumber \\ b) \;\;\; \frac{A_2}{A_1} < 1;\;\;\;
max\,\{\frac{b_2N^2_i}{b_1N^1_i}\,;i=\overline{1,N}\,\} < \alpha <
\;
min\,\{\frac{b_2N^2_i+3a_i/2}{b_1N^1_i+3a_i/2}\,;i=\overline{1,N}\,\}
 \label{inequality}\ea

\begin{figure}[hbtp]
  \centerline{\hbox{ \hspace{0.0in}
    \epsfxsize=4.5in
    \epsfbox{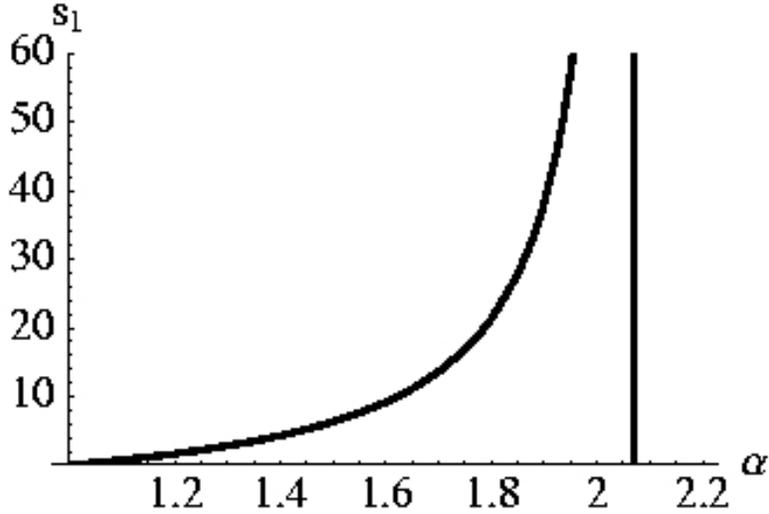}
    }
  }
\caption{Positive values of $s_1$ plotted as a function of
$\alpha$ for a case with two condensates and three bulk moduli for
the following choice of constants
$b_1=\frac{2\pi}{30},b_2=\frac{2\pi}{29},N^1_i=\{1,2,2\},N^2_i=\{2,3,5\},a_i=\{1,1/7,25/21\}$.
The qualitative feature of this plot remains the same for
different choices of constants as well as for different $i$. The
vertical line is the locus for $\alpha=\frac{b_2N^2_i}{b_1N^1_i}$,
where the denominator of (\ref{si}) vanishes.} \label{sA2tilde}
\end{figure}

Notice that the solution for $s_i$ (\ref{si}) has a singularity at
$\alpha=\frac{b_2N^2_i}{b_1N^1_i}$. This can be seen clearly from
Figure (\ref{sA2tilde}). We see that the modulus $s_1 \,(> 0) $
falls very rapidly as one moves away from the vertical asymptote
representing the singularity and can become smaller than one very
quickly, where the supergravity approximation fails to be valid.
The relative location of the singularities for different moduli
will turn out to be very important as we will see shortly.  From
 (\ref{inequality}), we know that there are two branches for
 allowed values of $\alpha$. Here we consider branch a) for
 concreteness, branch b) can be analyzed similarly.
\begin{figure}[hbtp]
  \centerline{\hbox{ \hspace{0.0in}
    \epsfxsize=3.3in
    \epsfbox{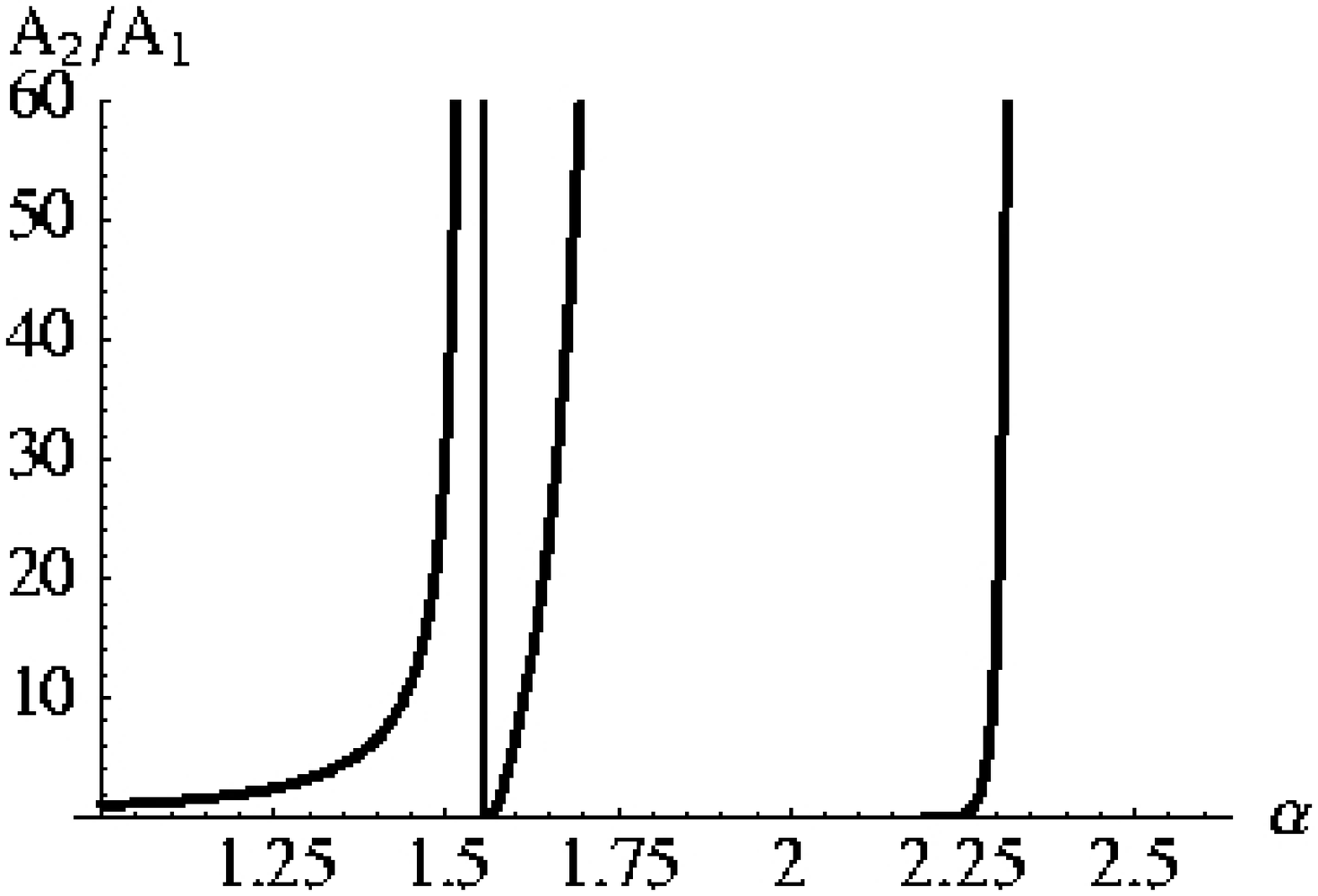}
    \hspace{0.25in}
    \epsfxsize=3.3in
    \epsfbox{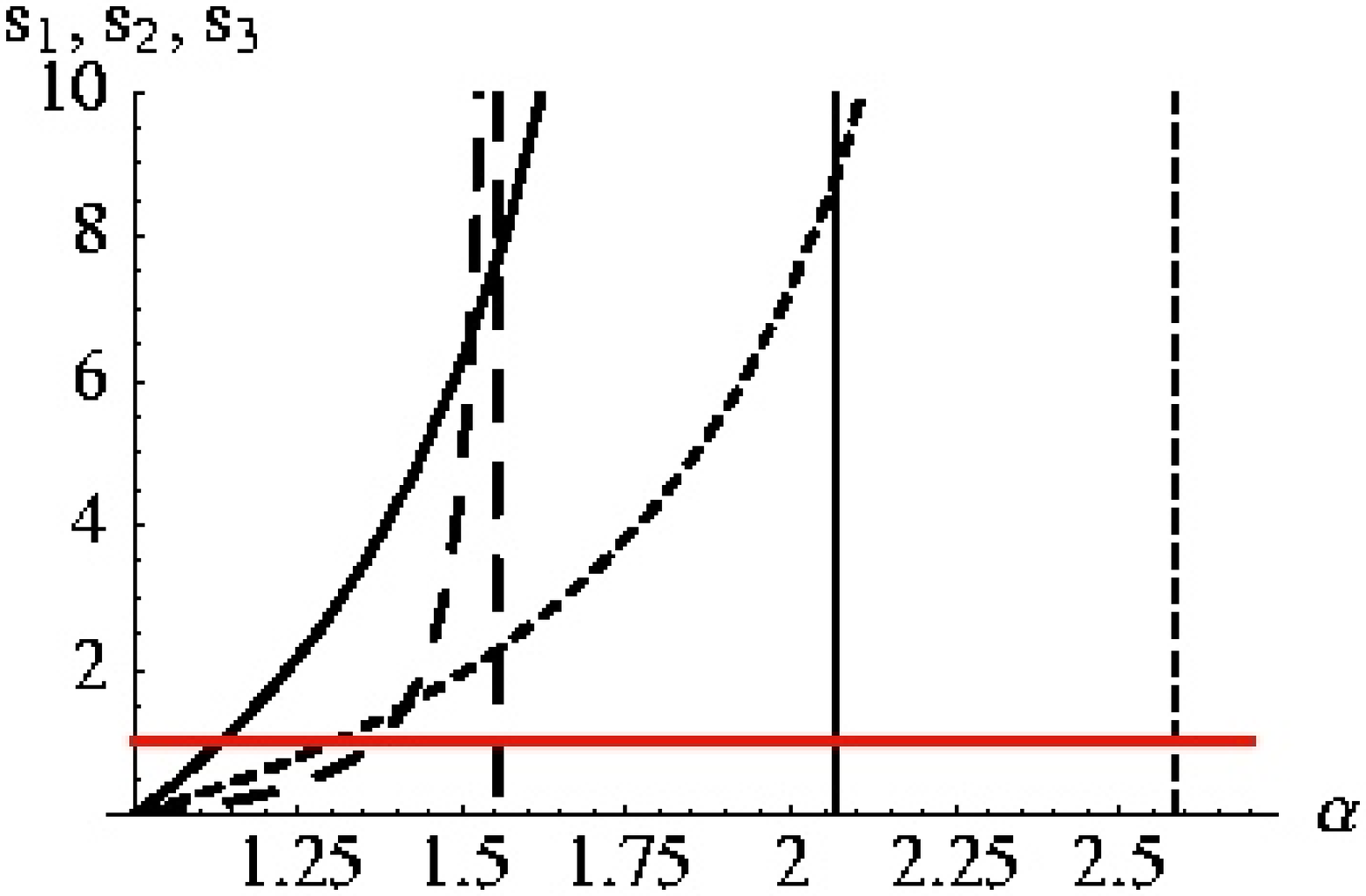}
    }
  }
   \caption{Left - $A_2/A_1$ plotted as a function of $\alpha$ for a case with two
    condensates and three bulk moduli. The function diverges as it approaches the loci of
    singularities of (\ref{si}), \textit{viz.} $\alpha=\frac{b_2N^2_i}{b_1N^1_i}$.
   \newline
    Right - Positive $s_i,\,i=1,2,3$ for the same case plotted as functions
    of $\alpha$. $s_1$ is represented by the solid curve, $s_2$ by the long dashed curve and $s_3$ by the short dashed
    curve.The vertical lines again represent the loci of singularities of (\ref{si}) which the respective moduli $s_i$ asymptote
    to. The horizontal solid (red) line shows the value unity for the
    moduli, below which the supergravity approximation is not valid.
    \newline
    Both plots are for $b_1=\frac{2\pi}{30},b_2=\frac{2\pi}{29},N^1_i=\{1,2,2\},N^2_i=\{2,3,5\},a_i=\{1,1/7,25/21\}$.}
   \label{Plots2-3}
\end{figure}
Figure (\ref{Plots2-3}) shows plots for $A_2/A_1$ and $s_i$ as
functions of $\alpha$ for a case with two condensates and three
bulk moduli. The plots are for a given choice of the constants
$\{b_1,b_2,N^1_i,N^2_i,a_i,\,i=1,2,3\}$. The qualitative feature
of the plots remains the same even if one has a different value
for the constants.

Since the $s_i$ fall very rapidly as one goes to the left of the
vertical asymptotes, there is a small region of $\alpha$ between
the origin and the leftmost vertical asymptote which yields
allowed values for all $s_i>1$. Thus, for a solution in the
supergravity regime all (three) vertical lines representing the
loci of singularities of the (three) moduli $s_i$ should be
(sufficiently) close to each other. This means that the positions
of the vertical line for the $i$th modulus
($\alpha=\frac{b_2N^2_i}{b_1N^1_i}$) and the $j$th modulus
($\alpha=\frac{b_2N^2_j}{b_1N^1_j}$) can not be too far apart.
This in turn implies that the ratio of integer coefficients
$(N^1_i/N^2_i)$ and $(N^1_j/N^2_j)$ for the $i$th and $j$th
modulus cannot be too different from each other in order to remain
within the approximation. Effectively, this means that the integer
combinations in the gauge kinetic functions
(\ref{generalgaugecoupl}) of the two hidden sector gauge groups in
(\ref{super}) can not be too linearly independent. We will give
explicit examples of $G_2$ manifolds in which $(N^1_i/N^2_i)$ and
$(N^1_j/N^2_j)$ are the same for all $i$ and $j$, so the
constraint of being within the supergravity approximation is
satisfied.
\begin{figure}[hbtp]
  \centerline{\hbox{ \hspace{0.0in}
    \epsfxsize=3.0in
    \epsfbox{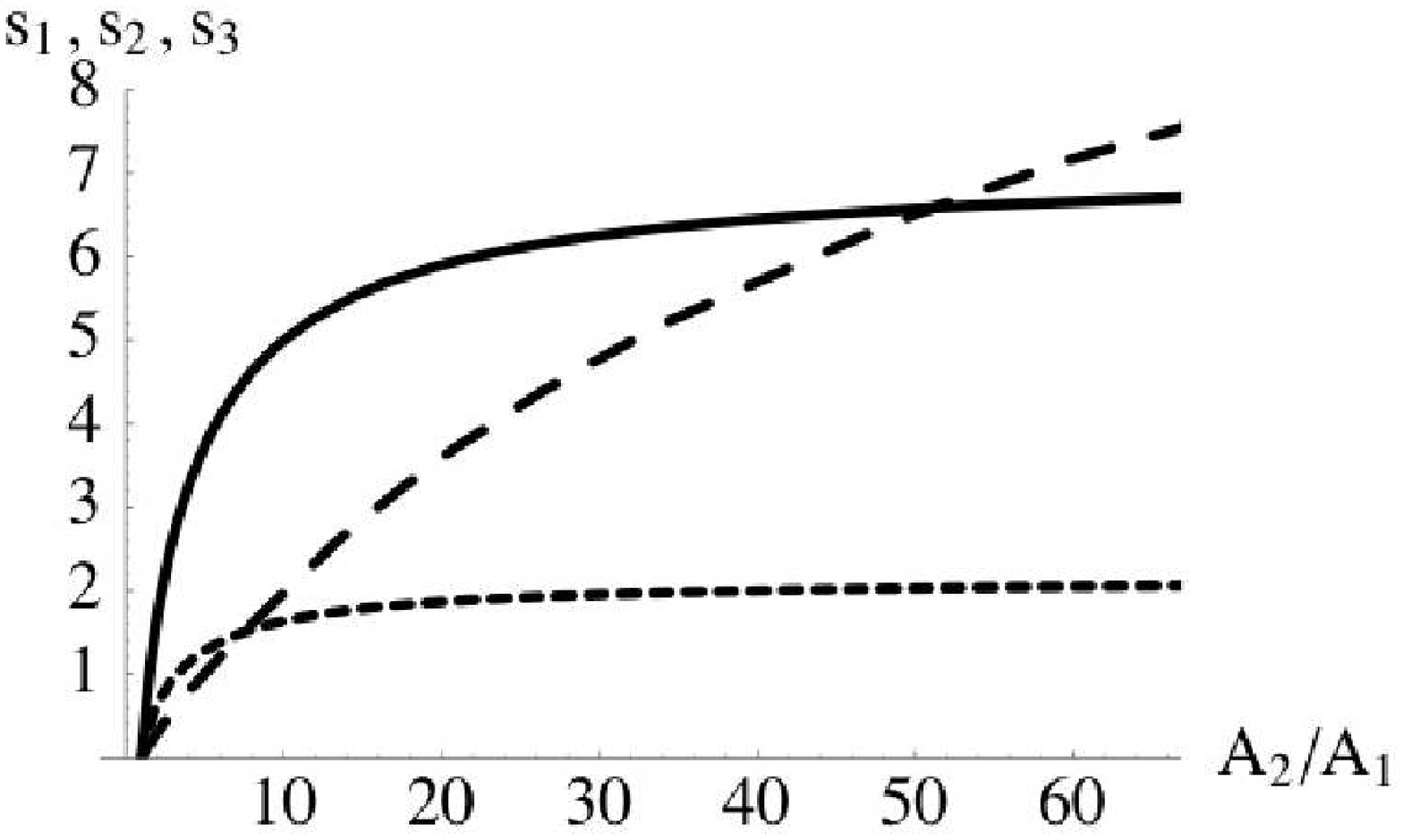}
    \hspace{0.25in}
    \epsfxsize=3.0in
    \epsfbox{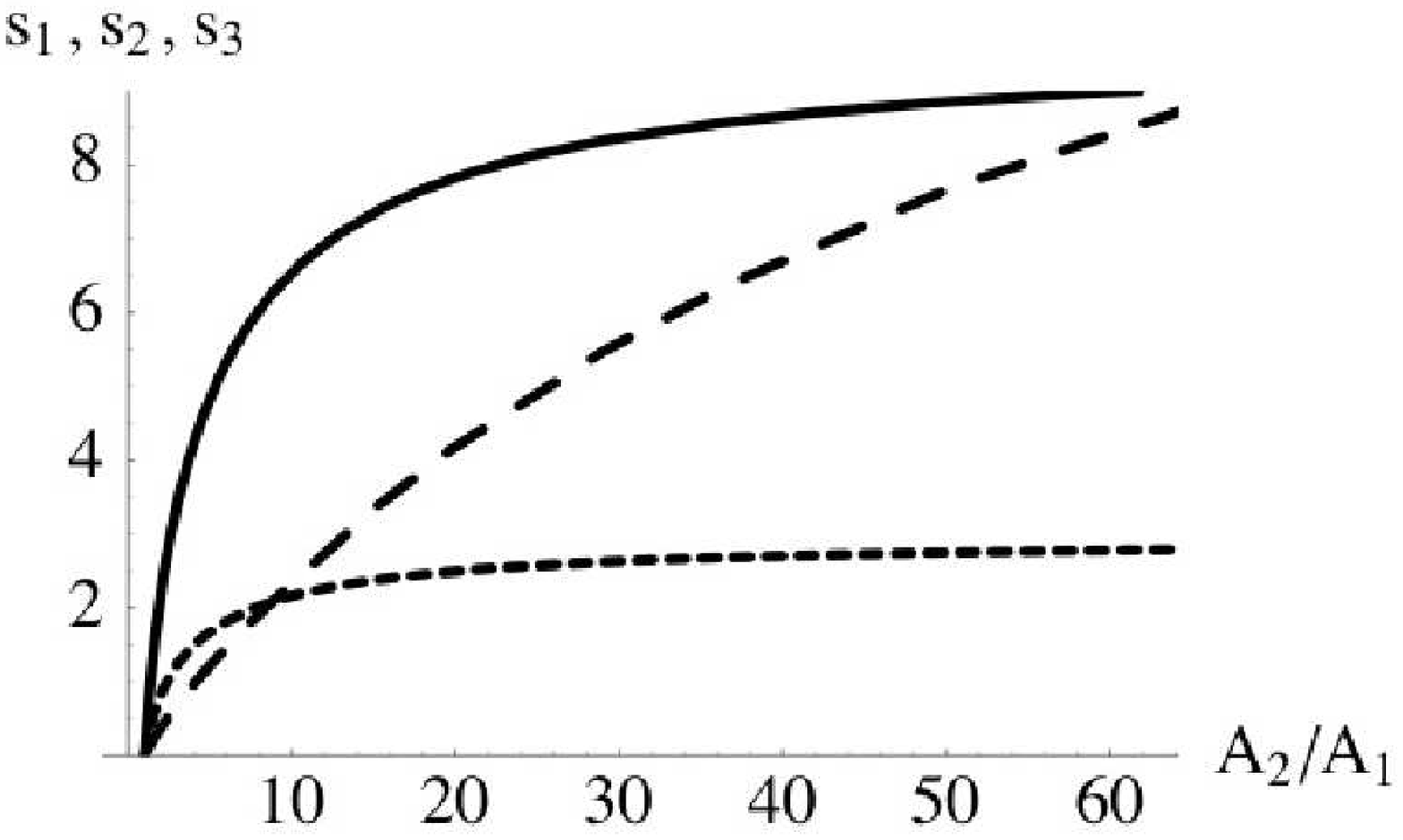}
    }
  }
  \centerline{\hbox{ \hspace{0.0in}
    \epsfxsize=3.0in
    \epsfbox{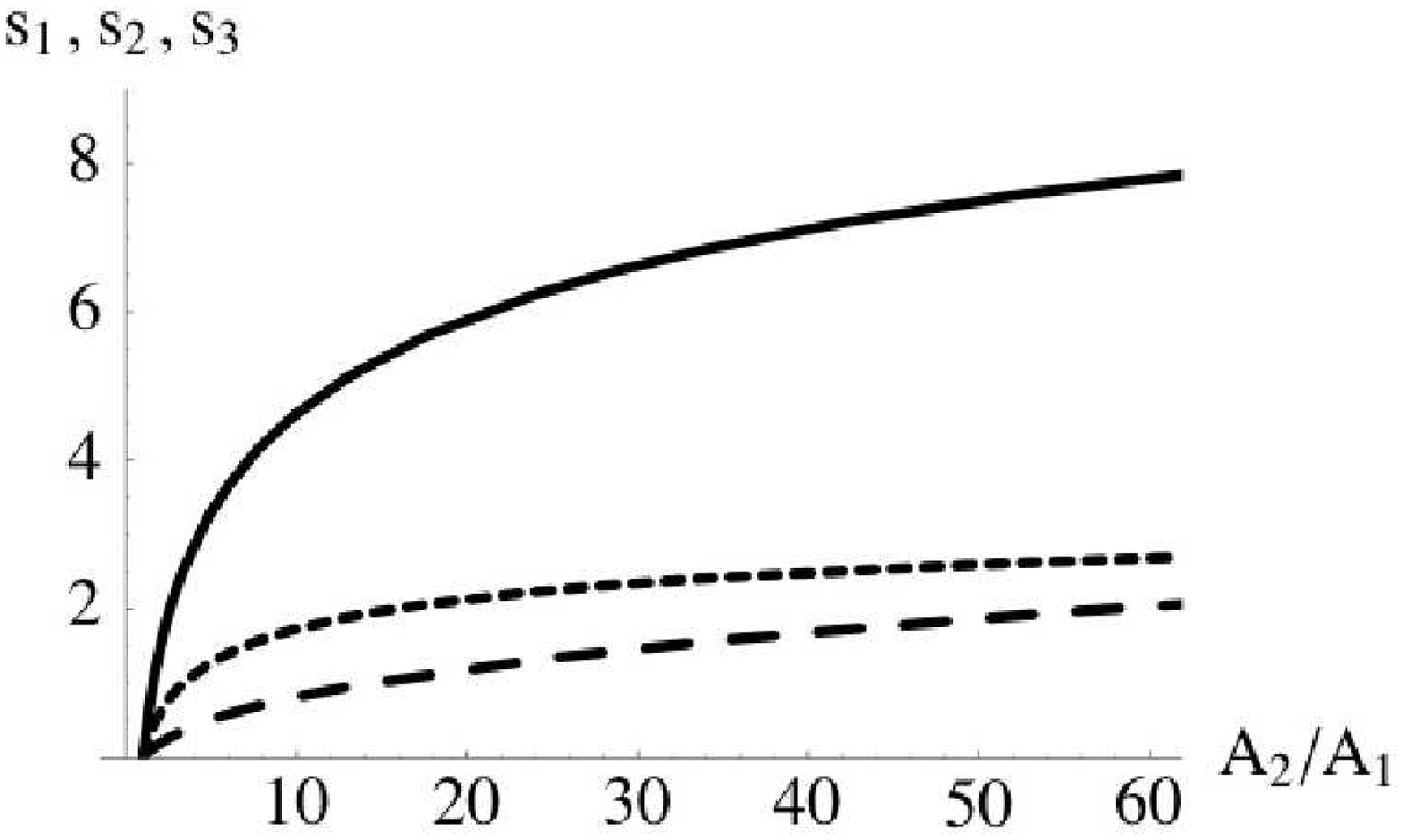}
    \hspace{0.25in}
    \epsfxsize=3.0in
    \epsfbox{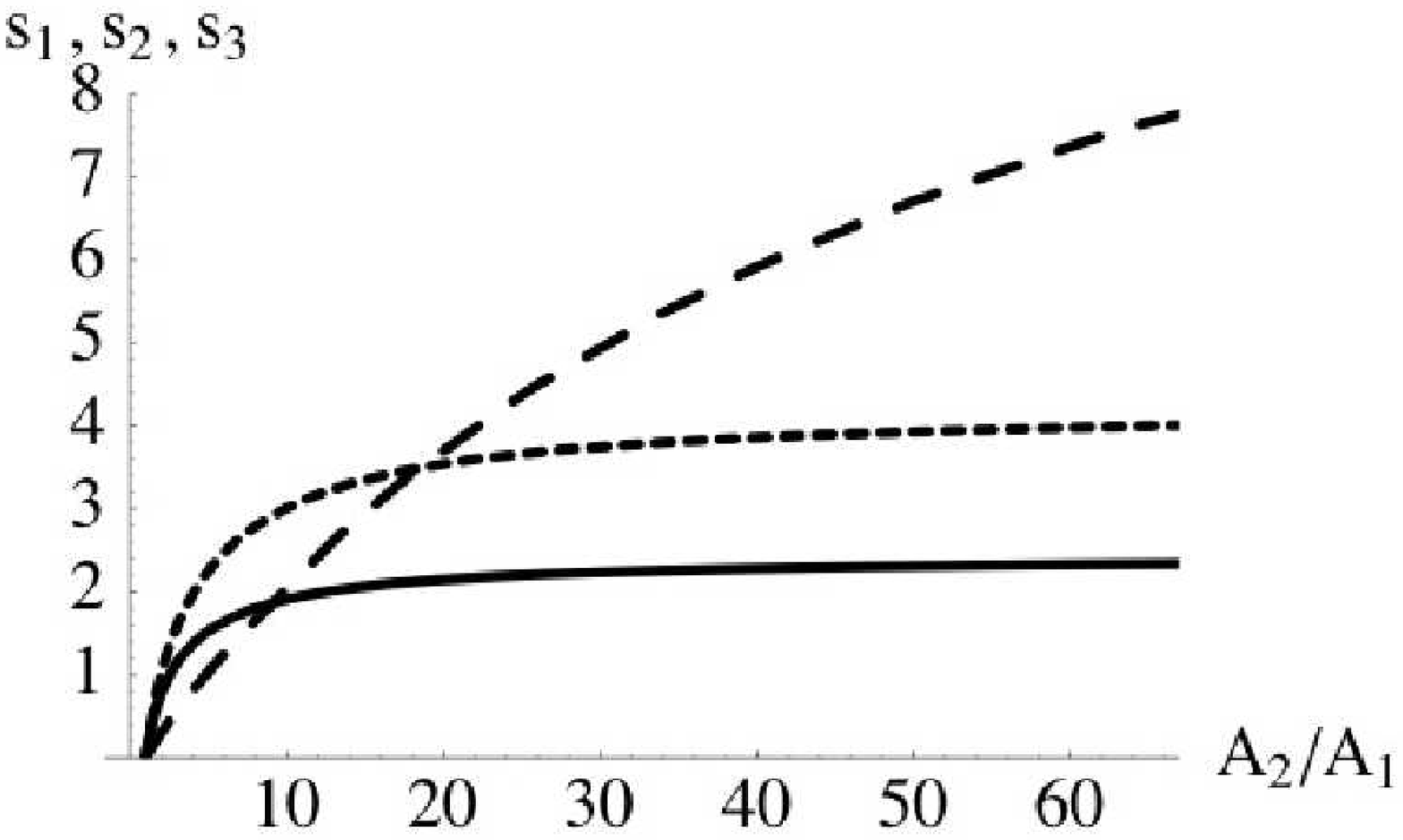}
    }
  }
\caption{Plots of positive $s_i,i=1,2,3$ as functions of
$A_2/A_1$.\newline
    Top Left: Same choice of constants as in Figure(\ref{Plots2-3}), i.e.
    $b_1=\frac{2\pi}{30},\,b_2=\frac{2\pi}{29},\,N^1_i=\{1,2,2\},\,N^2_i=\{2,3,5\},\,a_i=\{1,1/7,25/21\}.$\newline
    Top Right: We increase the ranks of the gauge groups but keep them close (keeping everything else same) -
    $b_1=\frac{2\pi}{40},\,b_2=\frac{2\pi}{38}$.\newline
    Bottom Left: We introduce a large difference in the ranks of
    the gauge groups (with everything else same) - $b_1=\frac{2\pi}{40},\,b_2=\frac{2\pi}{30}$.\newline
    Bottom Right: We keep the ranks of the gauge groups as in Top
    Left but change the integer coefficients to
    $N^1_i=\{1,2,2\},\,N^1_i=\{3,3,4\}$.}
    \label{Plots4-7}
\end{figure}
We now turn to the effect of the other constants on the nature of
solutions obtained. From the top right plot in Figure
(\ref{Plots4-7}), we see that increasing the ranks of the gauge
groups while keeping them close to each other (with all other
constants fixed) increases the size of the moduli in general. On
the other hand, from the bottom left plot we see that introducing
a large difference in the ranks leads to a decrease in the size of
the moduli in general. Hence, typically it is easier to find
solutions with comparatively large rank gauge groups which are
close to each other. The bottom right plot shows the sizes of the
moduli as functions of $A_2/A_1$ while keeping the ranks of the
gauge groups same as in the top left plot but changing the integer
coefficients. We typically find that if the integer coefficients
are such that the two gauge kinetic functions are almost
dependent, then it is easier to find solutions with values of
moduli in the supergravity regime.

The above analysis performed for three moduli can be easily
extended to include many more moduli. Typically, as the number of
moduli grows, the values of $a_i$ in (\ref{si}) decrease because
of (\ref{vol}). Therefore the ranks of the gauge groups should be
increased in order to remain in the supergravity regime as one can
see from the structure of (\ref{si}). At the same time, for
reasons described above, the integer combinations for the two
gauge kinetic functions should not be too linearly independent. In
addition, the integers $N^k_i$ should not be too large as they
also decrease the moduli sizes in (\ref{si}).

\begin{figure}[hbtp]
  \vspace{9pt}

  \centerline{\hbox{ \hspace{0.0in}
    \epsfxsize=3.3in
    \epsffile{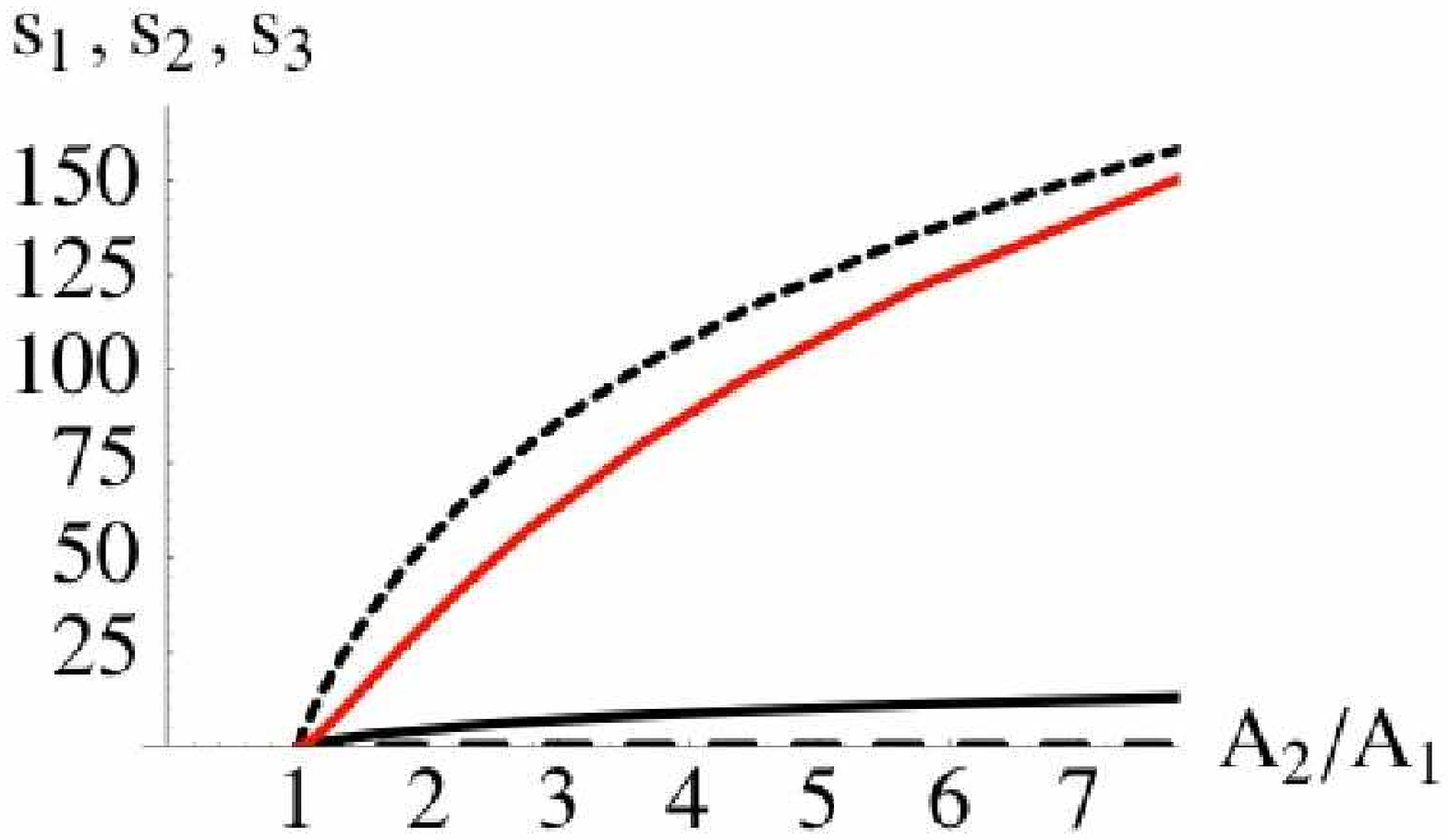}
    \hspace{0.25in}
    \epsfxsize=3.3in
    \epsffile{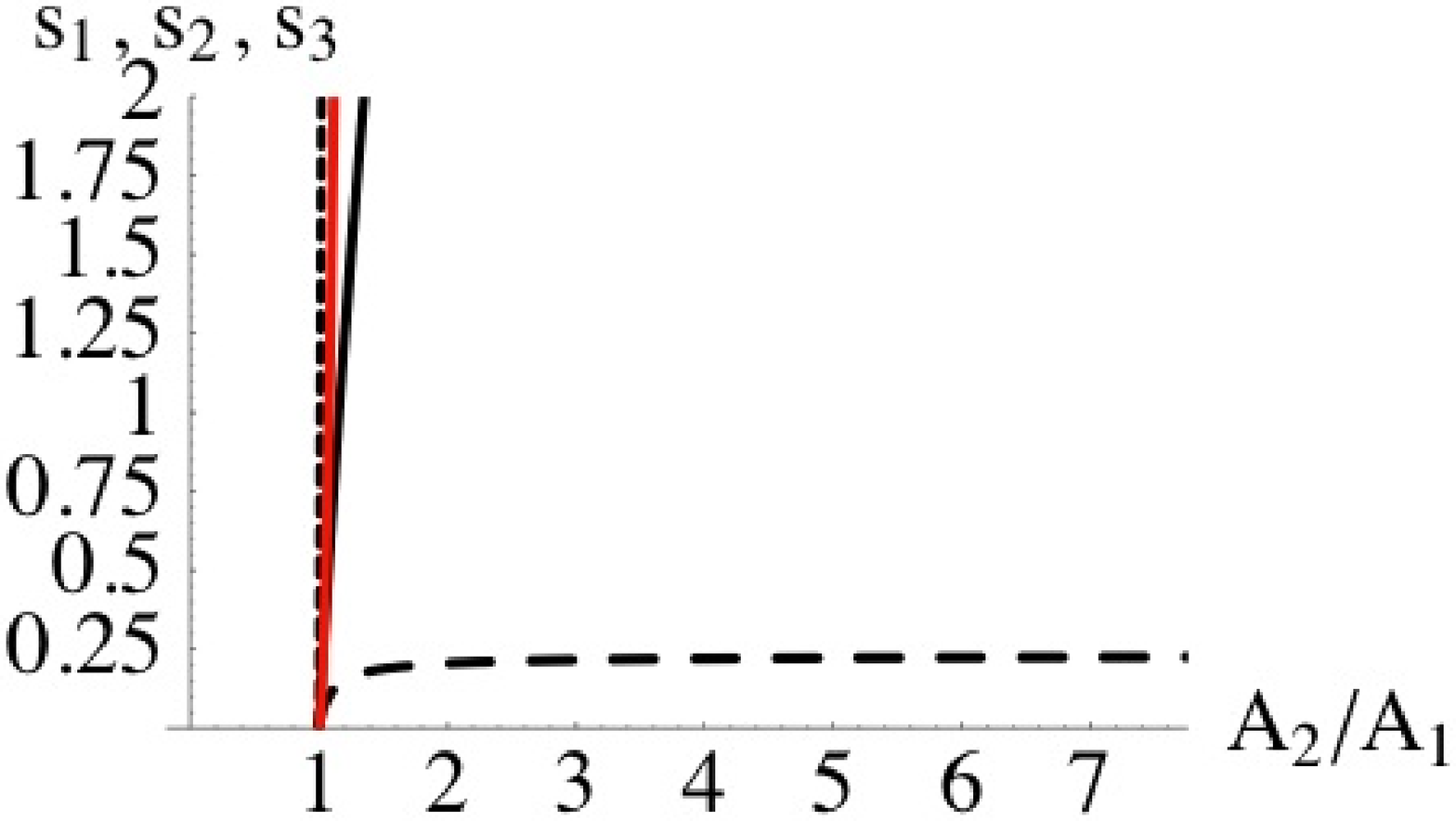}
    }
  }
\caption{Plots of positive $s_i,i=1,2,3$ as functions of
$A_2/A_1$. The constants are
$b_1=\frac{2\pi}{30},b_2=\frac{2\pi}{29},N^1_i=\{1,0,1\},N^2_i=\{1,1,1\},a_i=\{1/10,1,37/30\}$.
$s_1$ is represented by the solid curve, $s_2$ by the long dashed
curve and $s_3$ by the short dashed curve. The red curve
represents the volume of the internal manifold as a function of
$A_2/A_1$.
\newline
Right - the same plot with the vertical plot range decreased.}
\label{Plot15}
\end{figure}

What happens if some of the integers $N_i^1$ or $N_i^2$ are zero.
Figure \ref{Plot15} corresponds to this type of a situation when
the integer combinations  are given by
$N^1_i=\{1,0,1\},\,N^1_i=\{1,1,1\}$.
As we can see from the plots, all the moduli can still be
stabilized although one of the moduli, namely $s_2$ is stabilized
at values less than one in 11-dim Planck units. This gets us back
to the previous discussion as to when the supergravity
approximation can be valid. We will not have too much to say about
this point, except to note that a) the volume of $X$ can still be
large ((\ref{vol}) is large, greater than one in 11-dim Plank
units), b) the volumes of the associative three-cycles $Q_k$ which
appear in the gauge kinetic function (\ref{generalgaugecoupl}),
i.e. $Vol(Q_k)=\sum_{i=1}^nN_i^k\,s_i$  can also be large and c)
that the top Yukawa in these models comes from a small modulus vev
\cite{Atiyah:2001qf}. From Figure \ref{Plot15} we see that
although the modulus $s_2$ is always much smaller than one, the
overall volume of the manifold $V_X$ represented by the solid red
curve is much greater than one. Likewise, the volumes of the
associative three cycles $Vol(Q_1)=s_1+s_3$ and
$Vol(Q_2)=s_1+s_2+s_3$ are also large. Therefore if one interprets
the SUGRA approximation in this way, it seems possible to have
zero entries in the gauge kinetic functions for some of the moduli
and still stabilize all the moduli, as demonstrated by the
explicit example given above. In general, however, there is no
reason why any of the integers should vanish in the basis in which
the K\"{a}hler metric is given by (\ref{metric}).

%%%%%%%%%%%%%%%%%%%%%%%%%%%%%%%%%%%%%%%%%%%%%%%%%%%%%%%%%%%%%%%%%%%%%%%%%%%%%
\subsubsection{Special Case} \label{spl}
A very interesting special case arises when the gauge kinetic
functions $f_1$ and $f_2$ in (\ref{super}) are equal. Recall, that
since, in this case $N^1_i = N^2_i$, the moduli vevs are larger in
the supersymmetric vacuum and hence this case is representative of
the vacua to be found within the supergravity approximation.

Even though this is a special case, in section \ref{exampleG2}, we
will describe explicit examples of $G_2$ manifolds in which $N^1_i
= N^2_i$.

\begin{figure}[hbtp]
  \centerline{\hbox{ \hspace{0.0in}
    \epsfxsize=4.5in
    \epsfbox{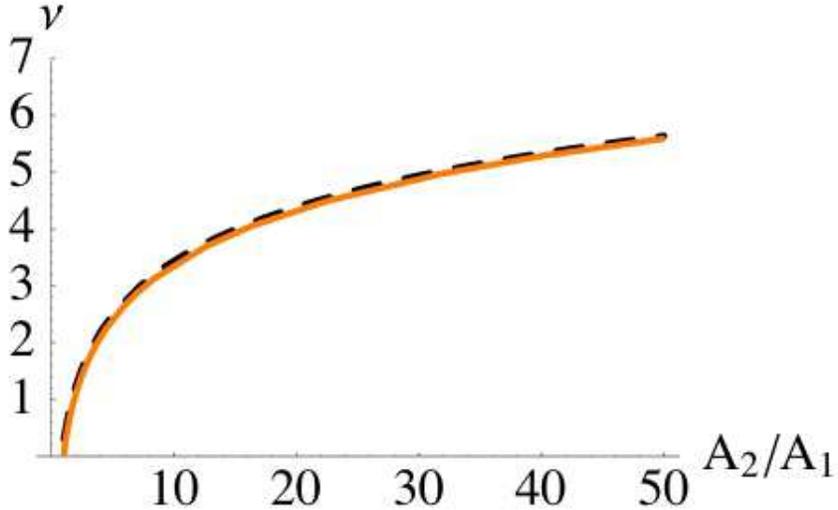}
    }
  }
\caption{Plot of $\nu$ as a function of $A_2/A_1$ for the choice
$b_1=\frac{2\pi}{5},\,b_2=\frac{2\pi}{4}$. The red solid curve
represents the exact numerical solution whereas the black dashed
curve is the leading order approximation given by (\ref{nu10}).}
    \label{Plot8}
\end{figure}

In the special case, we have \be N^1_i=N^2_i \equiv N_i\,, \ee and
therefore \be \label{nu1} \nu_i^1=\nu_i^2=\nu_i\equiv\frac{N_i
s_i}{a_i}\, \ee For this special case, the system of equations
(\ref{systemeqns}) can be simplified even further. We have
\be\label{constraint2} (b_1{\nu}_i+\frac{3}{2})A_1 -
(b_2{\nu}_i+\frac{3}{2})A_2e^{(b_1-b_2)\vec {\nu} \cdot \vec a}= 0
\ee with $\nu_i$ actually \emph{independent} of $i$. Thus, we are
left with just \emph{one} simple algebraic equation and one
transcendental constraint. The solution for $\nu_i$ is given by :
\ba\label{nu5} \nu_i \equiv \nu = -\frac{3(\alpha-1)}{2(\alpha
b_1-b_2)}\,, \ea with \ba\label{con}
 \frac{A_2}{A_1}\, =\frac 1 \alpha e^{-\frac 7 3(b_1-b_2)\nu}\,.
\ea Since $\nu_i$ is independent of $i$, it is also independent of
the number of moduli $N$. In Figure (\ref{Plot8}) we plotted $\nu$
as a function of $A_2/A_1$ when the hidden sector gauge groups are
$SU(5)$ and $SU(4)$. Notice that here the ranks of the gauge
groups don't have to be large for the moduli to be greater than
one. This is in contrast with the linearly independent cases
plotted in Figure (\ref{Plots4-7}). Once $\nu$ is determined in
terms of $A_2/A_1$, the moduli are given by: \be s_i=\frac
{a_i\nu}{N_i}\,. \ee Therefore, the hierarchy between the moduli
sizes is completely determined by the ratios $a_i/N_i$ for
different values of $i$.
In addition, from Figure \ref{Plot8} it can be seen that $\nu$
keeps increasing indefinitely if we keep increasing $A_2/A_1$
(though theoretically there may be a reasonable upper limit for
$A_2/A_1$), which is not possible for the general case as there
are $N$ $\nu_i$'s. This implies that it is possible to have a wide
range of the constants which yield a solution in the supergravity
regime.

Although the numerical solutions to the system
(\ref{nu5}-\ref{con}) described above are easy to generate, having
an explicit analytic solution, even an approximate one, which
could capture the dependence of $\nu$ on the constants $A_2/A_1$,
$b_1$ and $b_2$ would be very useful.

Fortunately there exists a good approximation, namely a large
$\nu$ limit, which allows us to find an analytical solution for
$\nu$ in a straightforward way. Expressing $\alpha$ from
(\ref{nu5}), in the leading order approximation when $\nu$ is
large we obtain \be\label{alpha} \alpha^{(0)}=\frac{b_2}{b_1}\,.
\ee

After substituting (\ref{alpha}) into (\ref{con}) we obtain the
approximate solution for $\nu$ in the leading order:
\be\label{nu10} \nu^{(0)}=\frac 3 7 \frac
1{b_2-b_1}\ln\left(\frac{A_2\,b_2}{A_1\,b_1}\right)=\frac 3
{14\pi} \frac {P\,Q}{P-Q} \ln\left(\frac{A_2\,P}{A_1\,Q}\right)\,,
\ee where the last expression corresponds to $SU(P)$ and $SU(Q)$
hidden sector gauge groups. For the moduli to be positive either
of the two following conditions have to be satisfied
\ba\label{branches}
&&a) \;\;\; A_1 Q < A_2 P;\;\;\;P>Q \nonumber \\
&&b) \;\;\; A_1 Q > A_2 P;\;\;\;P<Q\,. \ea From the plots in
Figure \ref{Plot8} we notice that the above approximation is
fairly accurate even when $\nu$ is $O(1)$. This is very helpful
and can be seen once we compute the first subleading contribution.
By substituting (\ref{nu10}) back into (\ref{nu5}) and solving for
$\alpha$ we now have up to the first subleading order:
\be\label{alpha4} \alpha=\alpha^{(0)}+\alpha^{(1)}=\frac P Q-\frac
7
{2\ln\left(\frac{A_2\,P}{A_1\,Q}\right)}\left(\frac{P-Q}{Q}\right)^2\,.
\ee It is then straightforward to compute $\nu$ which includes the
first subleading order contribution \be\label{nu40}
\nu=\nu^{(0)}+\nu^{(1)}=\frac 3 {14\pi} \frac
{P\,Q}{P-Q}\ln\left(\frac{A_2\,P}{A_1\,Q}\right)-\frac 3
{4\pi}\frac{P-Q} {\ln\left(\frac{A_2\,P}{A_1\,Q}\right)}\,. \ee We
can now examine the accuracy of the leading order approximation
when $\nu$ is $O(1)$ by considering the region where the ratio
$A_2/A_1$ is small. A quick check for the $SU(5)$ and $SU(4)$
hidden sector gauge groups chosen in the case presented in Figure
\ref {Plot8} yields for $A_2/A_1=4$: \ba
&&\alpha=\alpha^{(0)}+\alpha^{(1)}=1.25-0.136\,,
\\
&&\nu=\nu^{(0)}+\nu^{(1)}=2.195-0.148\,. \ea which results in a
12\% and 7\% error for $\alpha^{(0)}$ and $\nu^{(0)}$
respectively. The errors get highly suppressed when $\nu$ becomes
$O(10)$ and larger. Also, when the ranks of the gauge groups
$SU(P)$ and $SU(Q)$ are $O(10)$ and $P-Q$ is small, the ratio
$A_2/A_1$ can be $O(1)$ and still yield a large $\nu$. The
dependence of $\nu$ on the constants in (\ref{nu10}) is very
similar to the moduli dependence obtained for SUSY Minkowski vacua
in the Type IIB racetrack models \cite{Krefl:2006vu}.

We have demonstrated that there exist isolated supersymmetric
vacua in $M$ theory compactifications on $G_2$-manifolds with two
strongly coupled hidden sectors which give non-perturbative
contributions to the superpotential. Given the existence of
supersymmetric vacua, it is very likely that the potential also
contains non-supersymmetric critical points. Previous examples
have certainly illustrated this \cite{Acharya:2005ez}. Before
analyzing the non-supersymmetric critical points, however, we will
now present some examples of vacua which give rise to two strongly
coupled hidden sectors.
%%%%%%%%%%%%%%%%%%%%%%%%%%%%%%%%%%%%%%%%%%%%%%%%%%%%%%%%%

\subsection{Examples of $G_2$ Manifolds}\label{exampleG2}

Having shown that the potential stabilizes all the moduli, it is
of interest to construct explicit examples of $G_2$-manifolds
realizing these vacua. To demonstrate the existence of a
$G_2$-holonomy metric on a compact 7-manifold is a difficult
problem in solving non-linear equations \cite{Kovalev:2001zr}.
There is no analogue of Yau's theorem for Calabi-Yau manifolds
which allows an ``algebraic'' construction. However, Joyce and
Kovalev have successfully constructed many smooth examples
\cite{Kovalev:2001zr}. Furthermore, dualities with heterotic and
Type IIA string vacua also imply the existence of many singular
examples. The vacua of interest to us here are those with two or
more hidden sector gauge groups These correspond to
$G_2$-manifolds which have two three dimensional submanifolds
$Q_1$ and $Q_2$ along which there are orbifold singularities. In
order to describe such examples we will a) outline an extension of
Kovalev's construction to include orbifold singularities and b)
use duality with the heterotic string.

Kovalev constructs $G_2$ manifolds which can be described as the
total space of a fibration. The fibres are four dimensional $K3$
surfaces, which vary over a three dimensional sphere. Kovalev
considers the case in which the $K3$ fibers are generically
smooth, but it is reasonably straightforward to also consider
cases in which the (generic) $K3$ fiber has orbifold
singularities. This gives $G_2$-manifolds which also have orbifold
singularities along the sphere and give rise to Yang-Mills fields
in $M$ theory. For example if the generic fibre has both an
$SU(4)$ and an $SU(5)$ singularity, then the $G_2$ manifold will
have two such singularities, both parameterized by disjoint copies
of the sphere. In this case $N^1_i$ and $N^2_i$ are equal because
$Q_1$ and $Q_2$ are in the same homology class, which is precisely
the special case that we consider both above and below.

We arrive at a very similar picture by considering the $M$ theory
dual of the heterotic string on a Calabi-Yau manifold at large
complex structure. In this limit, the Calabi-Yau is $T^3$ fibered
and the $M$ theory dual is $K3$-fibered, again over a three-sphere
(or a discrete quotient thereof). Then, if the hidden sector $E_8$
is broken by the background gauge field to, say, $SU(5) \times
SU(2)$ the $K3$-fibers of the $G_2$-manifold generically have
$SU(5)$ and $SU(2)$ singularities, again with $N^1_i$ = $N^2_i$.
More generally, in $K3$ fibered examples, the homology class of
$Q_1$ could be $k$ times that of $Q_2$ and in this case $N^1_i = k
N^2_i$. As a particularly interesting example, the $M$ theory dual
of the heterotic vacua described in \cite{Braun:2005nv} include a
$G_2$ manifold whose singularities are such that they give rise to
an observable sector with precisely the matter content of the MSSM
whilst the hidden sector has gauge group $G = E_8$.

Finally, we also note that Joyce's examples typically can have
several sets of orbifold singularities which often fall into the
special class \cite{Kovalev:2001zr}. We now go on to describe the
vacua in which supersymmetry is spontaneously broken.

\subsection{Vacua with spontaneously broken Supersymmetry}\label{nonsusyadsvac}

The potential (\ref{fullpotential}) also possesses vacua in which
supersymmetry is spontaneously broken. Again these are isolated,
so the moduli are all fixed. These all turn out to have negative
cosmological constant. We will see in section
\ref{chargedmattervac} that adding matter in the hidden sector can
give a potential with de Sitter vacua.

Since the scalar potential (\ref{fullpotential}) is extremely
complicated, finding solutions is quite a non-trivial task. As for
the supersymmetric solution, it is possible to simplify the system
of $N$ transcendental equations obtained. However, unlike the
supersymmetric solution, we have only been able to do this so far
for the special case as in \ref{spl}. Therefore, for simplicity we
analyze the special case in detail. As we described above, there
are examples of vacua which fall into this special class.
Moreover, as explained previously, we expect that typically vacua
not in the special class are beyond the supergravity
approximation.

\noindent By extremizing (\ref{potential}) with respect to $s_k$
we obtain the following system of equations \ba \label{e1}
&&2\nu^2_k(b_1\alpha-b_2)^2-\nu_k[2(b_1\alpha-b_2)(b_1^2\alpha-b_2^2)\sum_{i=1}^{N}a_i\nu^2_i+3(\alpha-1)(b_1^2\alpha-b_2^2)
\vec\nu\cdot\vec a \nonumber\\
&&+3(b_1\alpha-b_2)^2\vec\nu\cdot\vec a+3(\alpha-1)(b_1\alpha-b_2)]-3[(b_1\alpha-b_2)^2\sum_{i=1}^{N}a_i\nu^2_i\nonumber\\
&&+3(\alpha-1)(b_1\alpha-b_2)\vec\nu\cdot\vec a+3(\alpha-1)^2]
\;=\; 0\,, \ea

\noindent where we have again introduced an auxiliary variable
$\alpha$ defined by \be \label{aux} \frac {A_2}{A_1}\equiv \frac 1
{\alpha}\, e^{-(b_1-b_2)\vec\nu\cdot\vec a}\,. \ee

\noindent similar to that in section \ref{spl}. The definition
(\ref{aux}) together with the system of polynomial equations
(\ref{e1}) can be regarded as a coupled system of equations for
$\alpha$ and $\nu_k$. We introduce the following notation:
\be\label{xyzw}
x\equiv(\alpha-1)\,,\,\,y\equiv(b_1\alpha-b_2)\,,\,\,z\equiv(b_1^2\alpha-b_2^2)\,,\,\,w\equiv
\frac{xz}{y^2}\,. \ee In this notation, from (\ref{e1}) (divided
by $x^2$) we obtain the following system of coupled equations \be
\label{u2} 2\frac {y^2} {x^2} \nu_k^2-\left(2\,\frac {y^2}
{x^2}w\sum_{i=1}^{N}a_i\nu^2_i+3\frac y x(w+1)\vec\nu\cdot\vec
a+3\right)\frac y x\nu_k-3\left(\frac {y^2}
{x^2}\sum_{i=1}^{N}a_i\nu_i^2+3\frac y x\vec\nu\cdot\vec
a+3\right)=0\,. \ee It is convenient to recast this system of $N$
{\it cubic} equations into a system of $N$ {\it quadratic}
equations plus a constraint. Namely, by introducing a new variable
$T$ as\be \label{T} 4\,T\equiv2\frac {y^2}
{x^2}w\sum_{i=1}^{N}a_i\nu^2_i+3\frac y x(w+1)\vec\nu\cdot\vec
a+3\, , \ee where the factor of four has been introduced for
future convenience, the system in (\ref{u2}) can be expressed as
\be \label{e2} 2\frac {y^2} {x^2} \nu_k^2-4\,T\frac y
x\nu_k-3\left(\frac {y^2} {x^2}\sum_{i=1}^{N}a_i\nu_i^2+3\frac y
x\vec\nu\cdot\vec a+3\right)=0\,. \ee

An important property of the system (\ref{e2}) is that {\it all of
its equations are the same  independent of the index k}. However,
since the combination in the round brackets in (\ref{e2}) is not a
constant with respect to $\vec\nu$ this system of quadratic
equations does not decouple. Nevertheless, because both the first
and the second monomials in (\ref{e2}) with respect to $\nu_k$ are
{\it independent of $\vec\nu$}, the standard solution of a
quadratic equation dictates that the solutions for $\nu_k$ of
(\ref{e2}) have the form \be \label{sol14} \nu_k=\frac x y
\left({T}+m_k H\right)\,,\,\, {\rm with}\,\, m_k=\pm
1\,,\,\,k=\overline{1,N}\,, \ee where we introduced another
variable $H$ and pulled out the factor of $x/y$ for future
convenience.

We have now reduced the task of determining $\nu_k$ for each
$k=\overline{1,N}$ to finding {\it only two} quantities - $T$ and
$H$. By substituting (\ref{sol14}) into equations
(\ref{T}-\ref{e2}) and using (\ref{vol}), we obtain a system of
two coupled quadratic equations
\ba \label{e3} \frac{14w}3\left({T_{A}^2}+2 A{T_{A}} H_{A}+H_{A}^2\right)+7(w+1) \left({T_{A}}+A H_{A}\right)+3-4T_{A}=0\,\,\,&& \\
9\left({T_{A}^2}+2 A {T_{A}}
H_{A}+H_{A}^2\right)-4H_{A}\left(H_{A}+A{T_{A}}
\right)+21\left({T_{A}}+AH_{A}\right)+9=0\,,&& \nonumber \ea where
parameter $A$ defined by \be \label{parA} A\equiv\frac 3 7\vec
m\cdot\vec a\,, \ee is now labelling each solution. Note that by
factoring out $x/y$ in (\ref{sol14}), the system obtained in
(\ref{e3}) is independent of either $x$ or $y$. However, it does
couple to the constraint (\ref{aux}) via $w$. In subsection
\ref{approxsol} we will see that there exists a natural limit when
the system (\ref{e3}) completely decouples from the constraint
(\ref{aux}). Since both $T_A$ and $H_A$ both depend on the
parameter A, the solution in (\ref{sol14}) is now written as \be
\label{sol1} \nu_k^{A}=\frac x y \left(T_{A}+m_k H_{A}\right)\,.
\ee Since $k=\overline{1,N}$ and $m_k=\pm 1$, vector $\vec m$
represents one of $2^{N}$ possible combinations. Thus, parameter
$\vec m\cdot\vec a$ can take on $2^{N}$ possible rational values
within the range: \be -\frac 7 3\leq \vec m\cdot\vec a\leq\frac 7
3\,, \ee so that parameter $A$ defined in (\ref{parA}) labelling
each solution can take on $2^{N}$ rational values in the range:
\be -1\leq A \leq 1\,. \ee
 For example, when $N=2$, there are four possible
combinations for $\vec m=(m_1,m_2)$, namely \be
(m_1,m_2)=\{(-1,-1),\,(1,-1),\,(-1,1),\,(1,1)\}\,. \ee These
combinations result in the following four possible values for $A$:
\be A=\{-1,\,\frac3 7(a_1-a_2)\,,\frac3 7(-a_1+a_2)\,,1\}\,, \ee
where we used (\ref{vol}) for the first and last combinations.

In general, for an arbitrary value of $A$, system (\ref{e3}) has
four solutions. However, with the exception of the case when
$A=1$, out of the four solutions only two are actually real, as we
will see later in subsection \ref{approxsol}. The way to find
those solutions is the following:

Having found $\nu_k^{A}$ analytically in terms of $\alpha$ and the
other constants, we can substitute it into the transcendental
constraint (\ref{aux}) to determine $\alpha$ numerically for
particular values of  $\{A_1, \,A_2,\,b_1,\,b_2,\,N_k,\,a_k\}$.
Again, in general there will be more than one solution for
$\alpha$. We can then substitute those values back into the
analytical solution for $\nu_k^{A}$ to find the corresponding
extrema, having chosen only those $\alpha$, obtained numerically
from (\ref{aux}), which result in real values of $\nu_k^{A}$. We
thus have $2^{N +1}$ real extrema. However, after a closer look at
the system of equations (\ref{e3}) we notice that when
$A\rightarrow -A$, equations remain invariant if
$H_{-A}\rightarrow - H_{A}$, and $T_{-A}\rightarrow T_{A}$, thus
simply exchanging the solutions $\nu_{(k,+)}^{A}$ where $m_k=1$
with $\nu_{(k,-)}^{A}$ where $m_k=-1$, i.e. \be\label{symmet}
\nu_{(k,+)}^{-A}=\nu_{(k,-)}^{A}\,, \ee which implies that the
scalar potential (\ref{potential}) in general has a total number
of $2^{N}$ real independent extrema. However, as we will see later
in section \ref{approxsol}, many of those vacua will be
incompatible with the supergravity approximation.

For general values of $A$, equations (\ref{e3}) have analytical
solutions that are too complicated to be presented here. In
addition to restricting to the situation with the same gauge
kinetic function $f$ in both hidden sectors, we now further
restrict to special situations where $A$ takes special values, so
that the expressions are simple. However, it is important to
understand that they still capture the main features of the
general solution. In the following, we provide explicit solutions
(in the restricted situation as mentioned above) for $M$ theory
compactifcations on $G_2$ manifolds with one and two moduli
respectively. In subsection \ref{approxsol} we will generalize our
results to the case with many moduli and give a complete
classification of all possible solutions. We will then consider
the limit when the volume of the associative cycle
$Vol(Q)=\vec\nu\cdot\vec a$ is large and obtain explicit analytic
solutions for the moduli.
%%%%%%%%%%%%%%%%%%%%%%%%%%%%%%%%%%%%%%%%%%%%%%%%%%%%%%%%%%%%%%%%
\subsubsection{One modulus case}
The first, and the simplest case is to consider a manifold with
only one modulus, i.e. $N=1$, $a=\frac 7 3$. In this case, $A=\pm
1$. From the previous discussion we only need to consider the case
$A=1$. It turns out that this is a special case for which the
system (\ref{e3}) degenerates to yield three solutions instead of
four. All three are real, however, only two of them result in
positive values of the modulus: \be \label{NH1}
T_{1}^{(1)}=-\frac{15} 8\,,\,\,\,\,H_{1}^{(1)}=\frac 3 8 \ee and
\ba \label{NH2} T_{1}^{(2)}&=&\frac{3}{28 \left(243-441 w+196
w^2\right)}
(-13419+\frac{3645}{w}+15288 w-5488 w^2\nonumber\\
&&-329 \sqrt{729-1701 w+1323 w^2-343 w^3}\\
&&+\frac{135} w \sqrt{729-1701 w+1323 w^2-343 w^3}\nonumber\\
&&+196 w \sqrt{729-1701 w+1323 w^2-343 w^3})\nonumber\\
H_{1}^{(2)}&=&\frac{3} {28 w}\left(-27+28 w-\sqrt{729-1701 w+1323
w^2-343 w^3}\right)\,,\nonumber \ea which give the following two
values for the modulus \ba \label{sol2}
&&s^{(1)}=\frac a{N_1}\frac x y\left(T_{1}^{(1)}+H_{1}^{(1)}\right)=-\frac {7x}{2Ny}\\
&&s^{(2)}=\frac a{N_1}\frac x
y\left(T_{1}^{(2)}+H_{1}^{(2)}\right)=-\frac
{x}{Ny}\left(\frac{3+\sqrt{9-7w}}w\right)\,.\nonumber \ea In
addition, each solution in (\ref{sol2}) is a function of the
auxiliary variable $\alpha$ defined in (\ref{aux}). By
substituting (\ref{sol2}) into (\ref{aux}) we obtain two equations
for $\alpha^{(1)}$ and $\alpha^{(2)}$ \be \label{e4} \frac
{A_2}{A_1}= \frac 1
{\alpha^{(1)}}\,e^{-(b_1-b_2)s^{(1)}N_1}\,,\,\,\,\, \frac
{A_2}{A_1}= \frac 1 {\alpha^{(2)}}\,e^{-(b_1-b_2)s^{(2)}N_1} \,.
\ee The transcendental equations (\ref{e4}) can only be solved
numerically. Here we will choose the following values for this
simple toy model \be \label{choice}
A_1=0.12\,,\,\,\,A_2=2\,,\,\,\,b_1=\frac
{2\pi}{8}\,,\,\,\,b_2=\frac {2\pi}{7}\,,\,\,\,N_1=1\,. \ee By
solving (\ref{e4}) numerically and keeping only those solutions
that result in real positive values for the modulus $s$ in
(\ref{sol2}) we get \be \label{sol3}
s^{(1)}=26.101\,,\,\,\,s^{(2)}=27.185\,. \ee with \be
\alpha^{(1)}=1.122\,,\,\,\,\alpha^{(2)}=1.267\,. \ee
\begin{figure}[hbtp]
  \centerline{\hbox{ \hspace{0.0in}
    \epsfxsize=4.0in
    \epsfbox{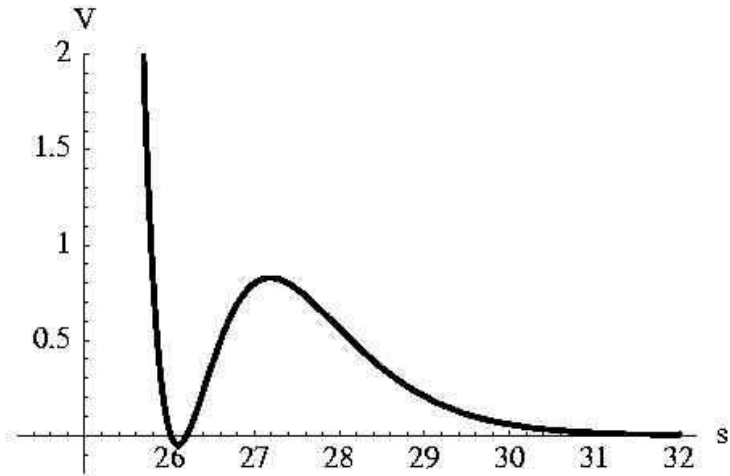}
    }
  }
\caption{Potential multiplied by $10^{32}$ plotted as a function
of one modulus $s$. For our particular choice of constants in
(\ref{choice}), the modulus is stabilized at the supersymmetric
AdS minimum $s^{(1)}=26.101$. The maximum is de Sitter, given by
$s^{(2)}=27.185$.} \label{onemodulusplot}
\end{figure}
In figure (\ref{onemodulusplot}) we see that the two solutions in
(\ref{sol3}) correspond to an AdS minimum and a de Sitter maximum.
In fact, the AdS minimum at $s^{(1)}$ is supersymmetric. The
general solution for $s^{(1)}$ given in (\ref{sol2}) can also be
obtained by methods of section \ref{susy}, imposing the SUSY
condition on the corresponding $F$-term by setting it to zero,
while introducing the same auxiliary constraint as in (\ref{aux}).
%%%%%%%%%%%%%%%%%%%%%%%%%%%%%%%%%%%%%%%%%%%%%%%%%%%%%%%%%%%%%%%%
\subsubsection{Two moduli case}
While the previous example with one modulus is interesting, it
does not capture some very important properties of the vacua which
arise when two or more moduli are considered. In particular, in
this subsection we will see that the supersymmetric AdS minimum,
obtained in the one-dimensional case, actually turns into a saddle
point whereas the stable minima are AdS with spontaneously broken
supersymmetry. Let us now consider a particularly simple example
with two moduli. Here we will choose both moduli to appear on an
equal footing in the K\"{a}hler potential (\ref{kahler}) by
choosing \be a_1=\frac 7 6\,,\,\,\,a_2=\frac 7 6\,. \ee We now
have four possible combinations for $\vec m=(m_1,m_2)$: \be
(1,1)\,,\,\,\,(1,-1)\,,\,\,\,(-1,1)\,,\,\,\,(-1,-1)\,, \ee
corresponding to the following possible values of $A$: \be
1\,,\,\,\,0\,,\,\,\,0\,,\,\,\,-1\,, \ee where only two of the four
actually produce independent solutions. The case when $A=1$ has
been solved in the previous subsection with $T_{1}^{(1)}$,
$H_{1}^{(1)}$ and  $T_{1}^{(2)}$, $H_{1}^{(2)}$ given by
(\ref{NH1}-\ref{NH2}) with the moduli taking on the following
values for the supersymmetric AdS extremum \ba \label{sol5}
&&s^{(1)}_1=\frac {a_1x}{N_1y}\left(-\frac 3 2\right)=-\frac {7x}{4N_1y}\,,\\
&&s^{(1)}_2=\frac {a_2x}{N_2y}\left(-\frac 3 2\right)=-\frac
{7x}{4N_2y}\,,\nonumber \ea and the de Sitter extremum \ba
\label{sol6}
&&s^{(2)}_1=\frac {a_1x}{N_1y}\left(-\frac 3{7w}(3+\sqrt{9-7w})\right)=-\frac {x}{2N_1y}{\left(\frac{3+\sqrt{9-7w}}w\right)}\,,\\
&&s^{(2)}_2=\frac {a_2x}{N_2y}\left(-\frac
3{7w}(3+\sqrt{9-7w})\right)=-\frac
{x}{2N_2y}{\left(\frac{3+\sqrt{9-7w}}w\right)}\,.\nonumber \ea As
mentioned earlier, the supersymmetric solution can also be
obtained by the methods of section \ref{susy}. Now, we also have a
new case when $A=0$. The corresponding two real solutions for
$T_{0}$ and $H_{0}$ are \ba \label{NH3}
&&T_{0}^{(1)}=\frac 3{112w}(15 - 63w - D)\,,\\
&&H_{0}^{(1)}=\frac 1{4\sqrt{5}}\sqrt{-\frac{585}8 -
\frac{18225}{392w^2} + \frac{3915}{28w} + \frac{1215}{392w^2}D-
\frac{225}{56 w}D}\,,\nonumber \ea and \ba \label{NH4}
&&T_{0}^{(2)}=\frac 3{112w}(15 - 63w - D)\,,\\
&&H_{0}^{(2)}=-\frac 1{4\sqrt{5}}\sqrt{-\frac{585}8 -
\frac{18225}{392w^2} + \frac{3915}{28w} + \frac{1215}{392w^2}D-
\frac{225}{56 w}D}\,,\nonumber \ea where we defined \be
D\equiv\sqrt{225 - 770w + 833w^2}\,. \ee The moduli are then
extremized at the values given by \ba
&&s^{(3)}_1=\frac {a_1x}{N_1y}\left({T_{0}^{(1)}}+H_{0}^{(1)}\right)\,,\\
&&s^{(3)}_2=\frac
{a_2x}{N_2y}\left({T_{0}^{(1)}}-H_{0}^{(1)}\right)\,, \nonumber
\ea and \ba
&&s^{(4)}_1=\frac {a_1x}{N_1y}\left({T_{0}^{(2)}}+H_{0}^{(2)}\right)\,,\\
&&s^{(4)}_2=\frac
{a_2x}{N_2y}\left({T_{0}^{(2)}}-H_{0}^{(2)}\right)\,.\nonumber \ea
To completely determine the extrema we again need to substitute
the solutions given above into the constraint equation (\ref{aux})
and choose a particular set of values for $A_1$, $A_2$, $b_1$, and
$b_2$ to find numerical solutions that result in real positive
values for the moduli $s_1$ and $s_2$. Here we again use the same
values as we chose in the previous case given by \be
\label{choice1} A_1=0.12\,,\,\,\,A_2=2\,,\,\,\,b_1=\frac
{2\pi}{8}\,,\,\,\,b_2=\frac
{2\pi}{7}\,,\,\,\,N_1=1\,,\,\,\,N_2=1\,. \ee For the SUSY extremum
we have \be \label{sol7}
s^{(1)}_1=13.05\,,\,\,\,s^{(1)}_2=13.05\,. \ee The de Sitter
extremum is given by \be \label{sol8}
s^{(2)}_2=13.59\,,\,\,\,s^{(2)}_2=13.59\,. \ee The other two
extrema are at the values \be \label{sol9}
s^{(3)}_1=2.61\,,\,\,\,s^{(3)}_2=23.55\,,\,{\rm
and}\,\,s^{(4)}_1=23.55\,,\,\,\,s^{(4)}_2=2.61 \ee
\begin{figure}[hbtp]
  \centerline{\hbox{ \hspace{0.0in}
    \epsfxsize=4.0in
    \epsfbox{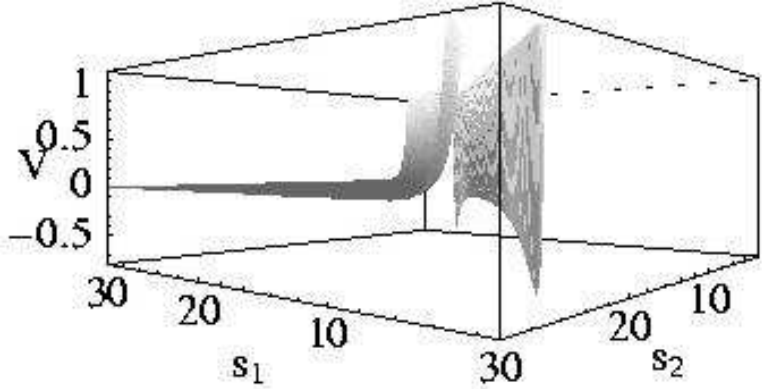}
    }
  }
\caption{Potential multiplied by $10^{32}$ plotted as a function
of two moduli $s_1$ and $s_2$ for the values in (\ref{choice1}).
The SUSY AdS extremum given by (\ref{sol7}) is a saddle point,
located between the non-supersymmetric AdS minima given by
(\ref{sol9}).} \label{twomoduliplot1}
\end{figure}

It is interesting to note that the supersymmetric extremum in
(\ref{sol7}) is no longer a stable minimum but instead, a saddle
point. The two symmetrically located stable minima seen in figure
(\ref{twomoduliplot1}) are non-supersymmetric. Thus we have an
explicit illustration of a potential where spontaneous breaking of
supersymmetry can be realized. The stable minima appear
symmetrically since both moduli were chosen to be on an equal
footing in the scalar potential. With a slight deviation where
$a_1\not=a_2$ and/or  $N_1\not =N_2$ one of the minima will be
deeper that the other. It is important to note that at both
minima, the volume given by (\ref{vol}) is stabilized at the value
$V_X=122.28$ which is large enough for the supergravity analysis
presented here to be valid.
%%%%%%%%%%%%%%%%%%%%%%%%%%%%%%%%%%%%%%%%%%%%%%%%%%%%%%%%%%%%%%%%%
\subsubsection{Generalization to many moduli}\label{approxsol}

\begin{figure}[hbtp]
  \centerline{\hbox{ \hspace{0.0in}
    \epsfxsize=3.3in
    \epsffile{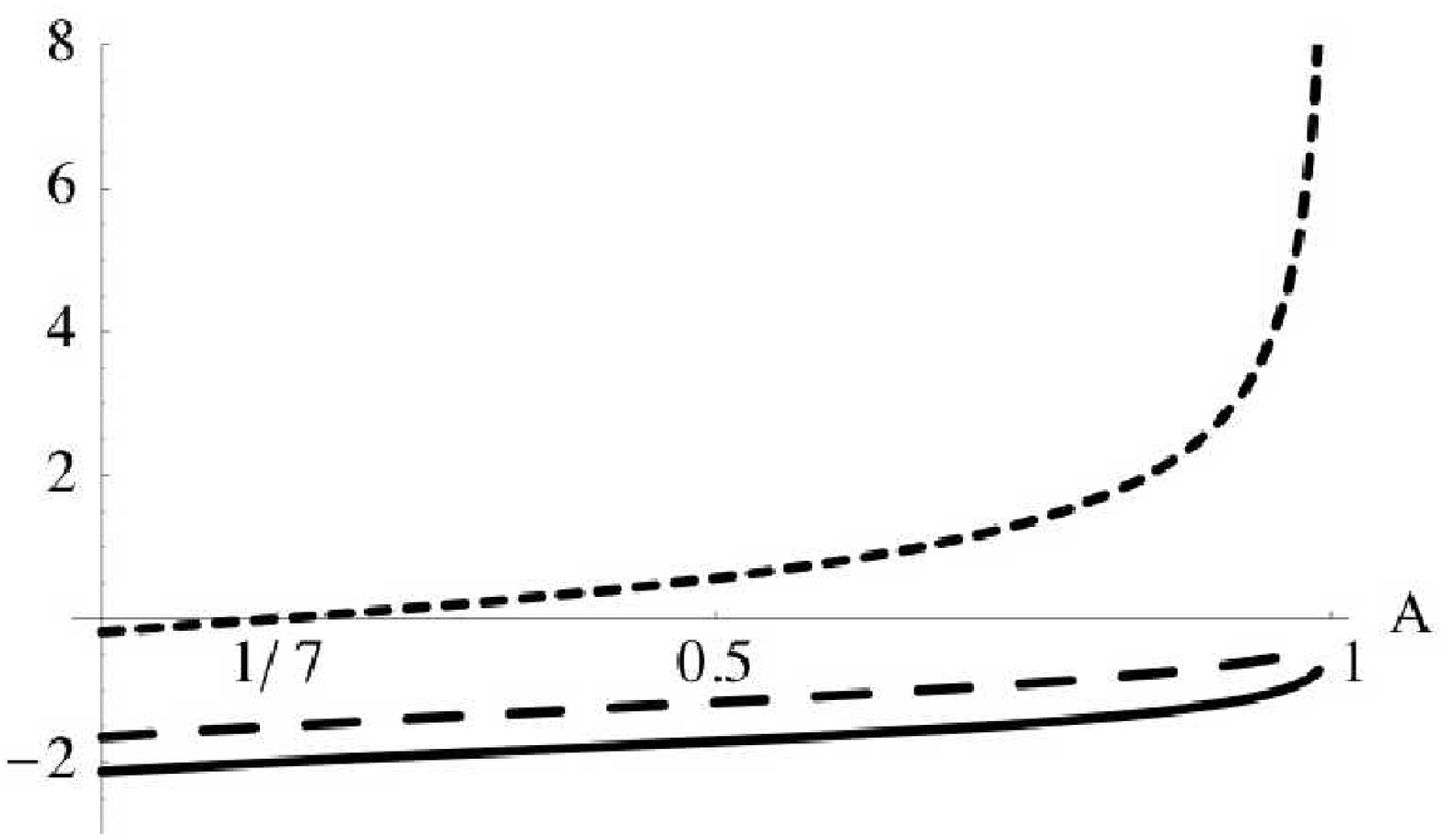}
    \hspace{0.25in}
    \epsfxsize=3.3in
    \epsffile{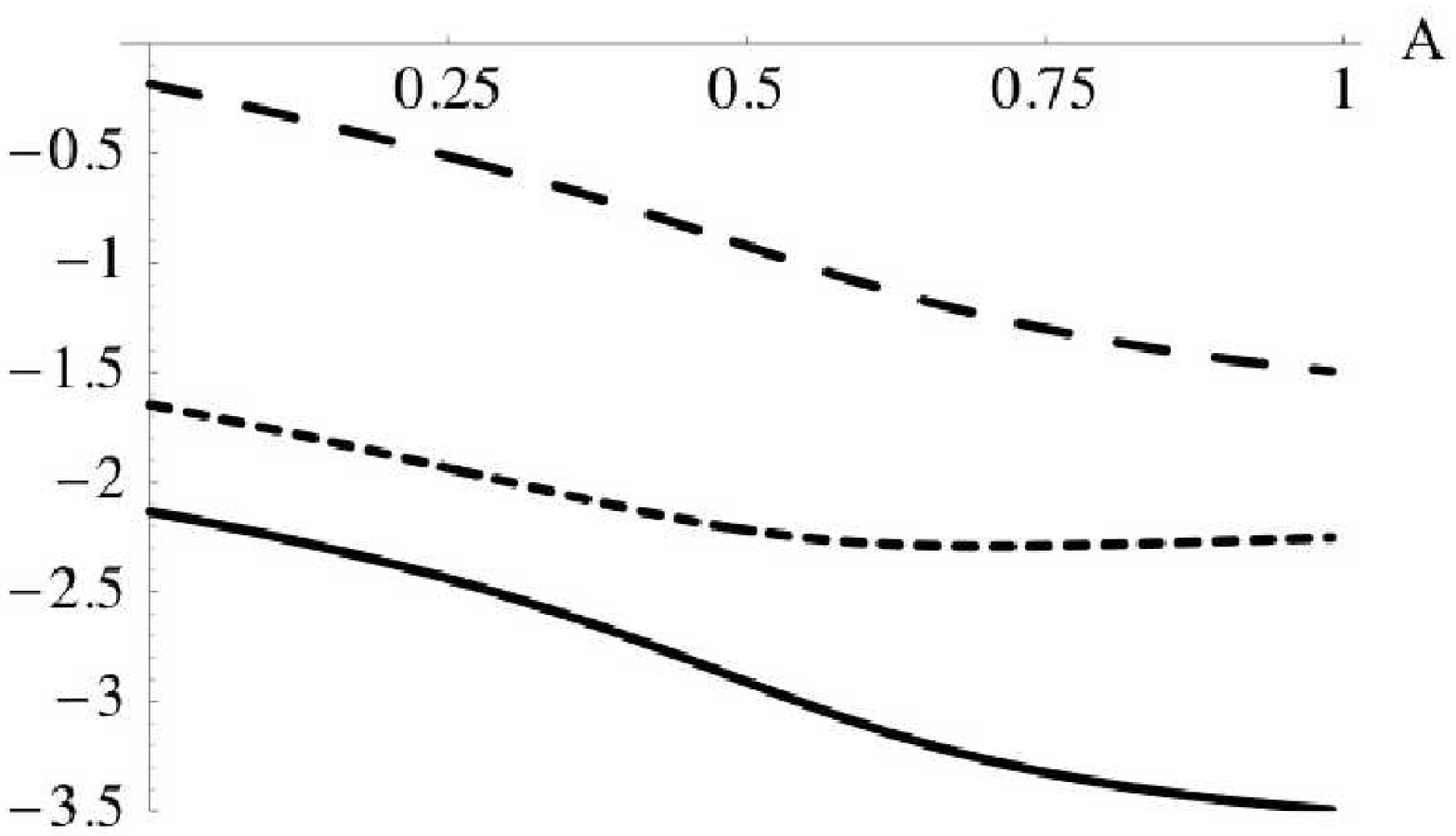}
    }
  }
\caption{Plots of $L_{A,\,+}^{(c)}$, $L_{A,\,-}^{(c)}$ and
${B_A^{(c)}}$, where $c=\overline{1,2}$, corresponding to the two
real solutions of the system (\ref{e20}) as functions of parameter
$A$ in the range $0\leq A<1$. Both left and right graphs have
$L_{A,\,+}^{(c)}$ - long dashed line, $L_{A,\,-}^{(c)}$ -short
dashed line, ${B_A^{(c)}}$ - solid line.
\newline
Left: Plots of $L_{A,\,+}^{(1)}$,  $L_{A,\,-}^{(1)}$ and
${B_A^{(1)}}$ corresponding to the first real solution at each
$A$. There is a critical value $A=1/7$ where $L_{A,\,-}^{(1)}=0$
and becomes positive for $A>1/7$.
    \newline
Right: Plots of $L_{A,\,+}^{(2)}$,  $L_{A,\,-}^{(2)}$ and
${B_A^{(2)}}$ corresponding to the second real solution at each
$A$. }\label{Plot24}
\end{figure}

In the previous section we demonstrated the existence of stable
vacua with broken SUSY for the special case with two moduli. Here
we will extend the analysis to include cases with an arbitrary
number of moduli for any value of the parameter $A$. It was
demonstrated in section \ref{spl} that the SUSY extremum has an
approximate analytical solution given by (\ref{nu10}). Therefore,
it would be highly desirable to obtain approximate analytical
solutions for the other extrema in a similar way. We will start
with the observation that for the SUSY extremum (\ref{sol5})
obtained for the special case when $A=1$, both $T_1^{(1)}$ and
$H_1^{(1)}$ given by (\ref{NH1}) are independent of $w$. On the
other hand, if in the leading order parameter $\alpha$ is given by
(\ref{alpha}), from the definitions in (\ref{xyzw}) it follows
that in this case \be\label{lim} y\rightarrow 0\,,\,\,{\rm
and}\,\,\,w\rightarrow -\infty\,. \ee Thus, if we consider the
system (\ref{e3}) in the limit when $w\rightarrow -\infty$, we
should be able to still obtain the SUSY extremum exactly. In
addition, one might also expect that the solutions for the vacua
with broken SUSY may also be located near the loci where
$y\rightarrow 0$. With this in mind we will take the limit
(\ref{lim}) which results in the following somewhat simplified
system of equations for $T_A$ and $H_A$:
\ba \label{e20} &&2\left({T_{A}^2}+2 A{T_{A}} H_{A}+H_{A}^2\right)+ 3\left({T_{A}}+A H_{A}\right)=0 \\
&&9\left({T_{A}^2}+2 A {T_{A}}
H_{A}+H_{A}^2\right)-4H_{A}\left(H_{A}+A{T_{A}}
\right)+21\left({T_{A}}+AH_{A}\right)+9=0\,. \nonumber \ea It is
straightforward to see that (\ref{NH1}) is an exact solution to
the above system when $A=1$. Moreover, unlike the general case
when $w$ is finite, where the system had three real solutions two
of which resulted in positive moduli, system (\ref{e20}) above
completely degenerates when $A=1$ yielding only one solution
corresponding to the SUSY extremum. On the other hand, for an
arbitrary $0\leq A < 1$ the system has four solutions. One can
check that at every point $A$ in the range $0\leq A < 1$ exactly
two out of these four solutions are real. The corresponding plots
are presented in Figure \ref{Plot24}. Before we discuss the plots
we would like to introduce some new notation: \be\label{lk}
L_{A,\,k}^{(c)}=T_A^{(c)}+m_kH_A^{(c)}\,, \ee where $c=\overline
{1,2}$ corresponding to the two real solutions. In this notation
(\ref{sol1}) can be reexpressed as: \be\label{sol32}
\nu_{A,\,k}^{(c)}=\frac x yL_{A,\,k}^{(c)}\,. \ee The volume of
the associative three cycle $Q$ for these vacua is then:
\be\label{deft} {\cal T}^{\,(c)}_A\equiv Vol(Q)^{(c)}_A=
Im(f^{(c)}_A)=\sum_{i=1}^{N}N_i\,
s^{(c)}_{A,\,i}=\sum_{i=1}^{N}a_i\,\nu^{(c)}_{A,\,i}=\frac x
y\,\vec a\cdot\vec L_{A}^{\,(c)}\,. \ee For future convenience we
will also introduce \be\label{BAc} B_A^{(c)}\equiv\vec a\cdot\vec
L_{A}^{\,(c)}=\frac 7 3\left(T_A^{(c)}+AH_A^{(c)}\right)\,. \ee
Constraint (\ref{aux}) is then given by: \be
\label{aux2}{\alpha^{(c\,)}_A}= \frac {A_1}{A_2}\,
e^{-(b_1-b_2){\cal T}^{\,(c)}_A}\,, \ee which is coupled to
\be\label{eq32} \frac x
y=\frac{{\alpha^{(c\,)}_A}-1}{b_1\,{\alpha^{(c\,)}_A}-b_2}=\frac
{{\cal T}^{\,(c)}_A}{B_A^{(c)}}\,, \ee where definitions
(\ref{xyzw}) were used to substitute for $x$ and $y$. Both
$L_{A,\,k}^{(c)}$ and $B_A^{(c)}$ are completely determined by the
system (\ref{e20}), whereas ${\cal T}^{\,(c)}_A$ is determined
from (\ref{aux2}-\ref{eq32}). Then solution (\ref{sol32}) can be
conveniently expressed as \be\label{sol34} \nu_{A,\,k}^{(c)}=\frac
{{\cal T}^{\,(c)}_A}{B_A^{(c)}}\,L_{A,\,k}^{(c)}\,. \ee
 Recall that $m_k=\pm 1$. Thus the only two
possibilities for $L_{A,\,k}^{(c)}$ for any $k=\overline{1,N}$ are
\be\label{lpm} L_{A,\,\pm}^{(c)}=T_A^{(c)}\pm H_A^{(c)}\,. \ee As
we vary parameter $A$ over the range $0\leq A<1$ point by point,
system (\ref{e20}) always has exactly two real solutions. In
Figure \ref{Plot24} we present plots of $L_{A,\,+}^{(c)}$,
$L_{A,\,-}^{(c)}$ and ${B_A^{(c)}}$, where $c=\overline{1,2}$ as
functions of $A$. We only need to consider the positive range
$0\leq A<1$ because of the symmetry (\ref{symmet}).
What happens to these solutions when $A=1$? We already know from
the previous discussion that the system (\ref{e20}) obtained in
the limit $w\rightarrow -\infty$ degenerates for $A=1$ and one
obtains the solution that corresponds to the SUSY extremum
explicitly. The solutions plotted in Figure \ref{Plot24} were
obtained assuming $A\not =1$ and therefore have an apparent
singularity when $A=1$. Thus they cannot capture either the SUSY
or the de Sitter extrema that arise in this special case. To
explain what happens to the de Sitter extremum we need to examine
the exact solution in (\ref{NH2}), in the same limit. Indeed,
bearing in mind that $w$ is negative, from (\ref{NH2}) we have \be
L_{A,+}^{(2)}=T_1^{(2)}+H_1^{(2)}=-\frac 3
7\left(\frac{3+\sqrt{9-7w}}w\right)\,. \ee Here we see immediately
that in the limit $w\rightarrow -\infty$ for the solution above
$L_{A,+}^{(2)}\rightarrow 0$. Therefore we conclude that the de
Sitter extremum cannot be obtained from (\ref{e20}) which
correlates with the previous observation that for $A=1$
(\ref{e20}) has only one solution - the SUSY extremum.
Nevertheless, as we will see in the next subsection the real
solutions plotted in Figure \ref{Plot24} are a very good
approximation to the exact numerical solutions corresponding to
the AdS vacua with spontaneously broken supersymmetry.

Now we would like to classify which of these AdS vacua have all
the moduli stabilized at positive values. Indeed if some of the
moduli are fixed at negative values we can automatically exclude
such vacua from further consideration since the supergravity
approximation assumes that all the moduli are positive. Since the
volume ${{\cal T}^{\,(c)}_A}$ is always positive by definition,
from (\ref{sol34}) we see immediately that for all moduli to be
stabilized in the positive range, all three quantities
$L_{A,\,+}^{(c)}$,  $L_{A,\,-}^{(c)}$ and ${B_A^{(c)}}$ must have
the same sign. In Figure {\ref{Plot24}} the plots on the right
satisfy this requirement for the entire range $0\leq A<1$. On the
other hand, the short-dashed curve corresponding to
$L_{A,-}^{(1)}$ on the left plot is negative when $0\leq A<1/7$,
features a zero at $A=1/7$ and becomes positive for $1/7< A<1$.
Yet, both $L_{A,\,+}^{(1)}$ and ${B_A^{(1)}}$ remain negative
throughout the entire range. Moreover, it is easy to verify that
the solution with $T_{1/7}^{(1)}=-3/4$ and $H_{1/7}^{(1)}=-3/4$,
such that $L_{1/7,-}^{(1)}=0$ is also an exact solution for the
general case (\ref{e3}) when $w$ is finite. Therefore, all
solutions compatible with the SUGRA approximation can be
classified as follows:

\vspace{0.5cm}

Given a set of $\{a_i\}$ with $i=\overline{1,N}$, there are
$2^{N}$ possible values of $A$, including the negative ones. From
the symmetry in (\ref{symmet}), only half of those give
independent solutions. This narrows the possibilities to $2^{N-1}$
positive combinations that fall in the range $0\leq A\leq 1$. For
each $A$ in the range $0\leq A<1/7$ there exist exactly two
solutions describing AdS vacua with broken SUSY with all the
moduli fixed at positive values.

For each  $A$ in the range $1/7\leq A<1$ there exists exactly one
solution describing an AdS vacuum with broken SUSY with all the
moduli stabilized in the positive range of values. For $A=1$ there
are exactly two solutions with all the moduli stabilized in the
positive range - de Sitter extremum in (\ref{NH2}) and the SUSY
AdS extremum in (\ref{NH1}). These two solutions are always
present for any set of $\{a_i\}$.

%%%%%%%%%%%%%%%%%%%%%%%%%%%%%%

\subsubsection{Explicit approximate solutions}\label{approxmod}
In this section we will complete our analysis of the AdS vacua and
obtain explicit analytic solutions for the moduli. We will take an
approach similar to the one we employed in section \ref{spl} when
we obtained an approximate formula (\ref{nu10}). Expressing
$\alpha^{(c\,)}$ from (\ref{eq32}) we obtain \be \label{eq37}
\alpha^{(c\,)}_A=\frac{b_2{{{\cal T}^{\,(c)}_A}}-{B_A^{(c)}}}
{b_1{{{\cal T}^{\,(c)}_A}}-{B_A^{(c)}}}\,. \ee There exists a
natural limit when the volume of the associative cycle ${\cal
T}^{\,(c)}_A$ is large. Just like in the approximate SUSY case in
(\ref{alpha}), the leading order solution to (\ref{eq37}) in this
limit is given by \be \label{eq38}
\alpha^{(c\,)}_A=\frac{b_2}{b_1}\,, \ee independent of $A$ and
$c$. Plugging this into (\ref{aux2}) and solving for ${\cal
T}^{\,(c)}_A$ we have in the leading order \be\label{eq39} {\cal
T}^{\,(c)}_A=\frac
1{b_2-b_1}\ln\left(\frac{A_2b_2}{A_1b_1}\right)=\frac
1{2\pi}\frac{P\,Q}{P-Q}\ln\left(\frac{A_2P}{A_1Q}\right)\,, \ee
where we again assumed the hidden sector gauge groups to be
$SU(P)$ and $SU(Q)$. Notice that this approximation automatically
results in the limit $w\rightarrow -\infty$ and therefore,
$L_{A,\,+}^{(c)}$,  $L_{A,\,-}^{(c)}$ and ${B_A^{(c)}}$ computed
by solving (\ref{e20}) and plotted in Figure \ref{Plot24} are
consistent with this approximation. Thus, combining (\ref{eq39})
with (\ref{sol34}) and (\ref{nu1}) we have the following
approximate analytic solution for the moduli in the leading order:
\be\label{eq43} s^{\,(c)}_{A,\,k}=\frac
1{2\pi}\left(\frac{a_k}{N_k}\right)\left(\frac
{L_{A,\,k}^{(c)}}{B_A^{(c)}}\right)\frac{P\,Q}{P-Q}\ln\left(\frac{A_2P}{A_1Q}\right)\,.
\ee To verify the approximation we can check it for the previously
considered special case with two moduli when $a_1=a_2=7/6$, i.e.
the case when $A=0$. By solving (\ref{e20}) we obtain: \ba
&&L_{0,\,+}^{(1)}=L_{0,\,-}^{(2)}=\frac
3{16}\left(-9+\sqrt{17}+\sqrt{-26+10\sqrt{17}}\right)\,,\\\nonumber
&&L_{0,\,-}^{(1)}=L_{0,\,+}^{(2)}=\frac
3{16}\left(-9+\sqrt{17}-\sqrt{-26+10\sqrt{17}}\right)\,,\\\nonumber
&&B_{0}^{(1)}=B_{0}^{(2)}=\frac 7{16}\left(-9+\sqrt{17}\right)\,.
\ea Thus, we have the following two solutions for the moduli for
the AdS vacua with broken SUSY: \ba\label{mod43}
&&s^{\,(1)}_{0,\,1}=\left(1-\frac{\sqrt{-26+10\sqrt{17}}}{9-\sqrt{17}}\right)\frac
1{4\pi{N_1}}\frac{P\,Q}{P-Q}\ln\left(\frac{A_2P}{A_1Q}\right)\sim
\frac
{0.016}{N_1}\frac{P\,Q}{P-Q}\ln\left(\frac{A_2\,P}{A_1\,Q}\right)\,,\\\nonumber
&&s^{\,(1)}_{0,\,2}=\left(1+\frac{\sqrt{-26+10\sqrt{17}}}{9-\sqrt{17}}\right)\frac
1{4\pi{N_2}}\frac{P\,Q}{P-Q}\ln\left(\frac{A_2P}{A_1Q}\right)\sim
\frac
{0.143}{N_2}\frac{P\,Q}{P-Q}\ln\left(\frac{A_2\,P}{A_1\,Q}\right)\,.\nonumber
\ea and \ba\label{mod44}
&&s^{\,(2)}_{0,\,1}=\left(1+\frac{\sqrt{-26+10\sqrt{17}}}{9-\sqrt{17}}\right)\frac
1{4\pi{N_1}}\frac{P\,Q}{P-Q}\ln\left(\frac{A_2P}{A_1Q}\right)\sim
\frac
{0.143}{N_2}\frac{P\,Q}{P-Q}\ln\left(\frac{A_2\,P}{A_1\,Q}\right)\,,\\\nonumber
&&s^{\,(2)}_{0,\,2}=\left(1-\frac{\sqrt{-26+10\sqrt{17}}}{9-\sqrt{17}}\right)\frac
1{4\pi{N_2}}\frac{P\,Q}{P-Q}\ln\left(\frac{A_2P}{A_1Q}\right)\sim
\frac
{0.016}{N_1}\frac{P\,Q}{P-Q}\ln\left(\frac{A_2\,P}{A_1\,Q}\right)\,.\nonumber
\ea The choice of the constants given in (\ref{choice1}) results
the following values: \be
s^{\,(1)}_{0,\,1}=s^{\,(2)}_{0,\,2}=2.62\,,\,\,\,\,\,s^{\,(1)}_{0,\,2}=s^{\,(2)}_{0,\,1}=23.64\,.
\ee A quick comparison with the exact values in (\ref{sol9})
obtained numerically leads us to believe that the approximate
analytical solutions presented here are highly accurate. This is
especially true when the volume of the associative cycle ${\cal
T}^{\,(c)}_A$ is large. For the particular choice above the
approximate value is: \be {\cal T}^{\,(1)}_0={\cal T}^{\,(2)}_0=
26.16\,, \ee which is indeed fairly large. To complete the
picture, we also would like to include the first subleading order
contributions to the approximate solutions presented here. After a
straightforward computation we have the following: \be
\label{eq45} \alpha^{(c\,)}_A=\frac P
Q+\frac{B_A^{(c)}}{\ln\left(\frac{A_2 P}{A_1 Q}\right)}
\left(\frac{P-Q}Q\right)^2\,, \ee and \be \label{eq46} {\cal
T}^{\,(c)}_A=\frac
1{2\pi}\frac{P\,Q}{P-Q}\ln\left(\frac{A_2P}{A_1Q}\right) +\frac
{B_A^{(c)}}{2\pi}\left(\frac{P-Q}{\log\left(\frac{A_2 P}{A_1
Q}\right)}\right)\,. \ee By combining (\ref{eq46}) with
(\ref{sol34}) and (\ref{nu1}) it is easy to obtain the
corresponding expressions for the moduli that include the first
subleading order correction: \be\label{eq47}
s^{\,(c)}_{A,\,k}=\frac
1{2\pi}\left(\frac{a_k}{N_k}\right)\left(\frac
{L_{A,\,k}^{(c)}}{B_A^{(c)}}\right)\frac{P\,Q}{P-Q}\ln\left(\frac{A_2P}{A_1Q}\right)
+\frac
{L_{A,\,k}^{(c)}}{2\pi}\left(\frac{a_k}{N_k}\right)\left(\frac{P-Q}{\ln\left(\frac{A_2
P}{A_1 Q}\right)}\right)\,. \ee

%%%%%%%%%%%%%%%%%%%%%%%%%%%%%%%%%%%%%%%%%%%%%%%%%%%%%%%%%

\subsection{Vacua with charged matter in the Hidden Sector}\label{chargedmattervac}
%Susy QCD is non-chiral: it has left handed fermions in the fundamental and
%antifundamental representations!!
Thus far, we have studied in reasonable detail, the vacuum
structure in the cases when the hidden sector has two strongly
coupled gauge groups without any charged matter. It is of interest
to study how the addition of matter charged under the hidden
sector gauge group changes the conclusions, as done for type IIB
compactifications in \cite{Lebedev:2006qq}. We argue that the
addition of charged matter can give rise to Minkowski or
metastable de Sitter (dS) vacua. This is due to the additional
$F$-term's of the matter fields. Moreover, we explain why it is
reasonable to expect that for a given choice of $G_2$-manifold,
the dS vacuum obtained is \emph{unique}.

%%%%%%%%%%%%%%%%%%%%%%%%%%%%%%%%%
\subsubsection{Scalar Potential}

Generically we would expect that a hidden sector gauge theory can
possess a fairly rich particle spectrum which, like the visible
sector, may include chiral matter. For example, an $SU(N_c)$ gauge
theory apart from the ``pure glue'' may also include massless
quark states $Q$ and $\tilde Q$ transforming in $N_c$ and $\bar
N_c$ of $SU(N_c)$. When embedded into $M$ theory the effective
superpotential due to gaugino condensation for such a hidden
sector with $N_f$ quark flavors has the following form
\cite{Seiberg:1994bz}: \be \label{mattersup}
W=A_1\,e^{i\frac{2\pi}{N_c-N_f}\sum_{i=1}^{N} N_i^{(1)} z_i}
\det(Q\tilde Q)^{-\frac{1}{N_c-N_f}}\,. \ee We can introduce an
effective meson field $\phi$ to replace the quark bilinear \be
\label{meson} \phi\equiv\left(Q\tilde
Q\right)^{1/2}=\phi_0e^{i\theta}\,, \ee and for notational brevity
we define \be \label{b1anda}
b_1\equiv\frac{2\pi}{N_c-N_f}\,,\,\,\,\,\,\,\,a\equiv-\frac
2{N_c-N_f}\,. \ee Here we will consider the case when the hidden
sector gauge groups are $SU(N_c)$ and  $SU(Q)$ with $N_f$ flavors
of the quarks $Q$ ($\tilde Q$) transforming as $N_c$ ($\bar N_c$)
under $SU(N_c)$ and as singlets under $SU(Q)$. In this case, when
$N_f=1$, the effective nonperturbative superpotential has the
following form: \ba \label{mattersuptwosectors}
W=A_1{\phi}^a\,e^{ib_1\,f}+A_2e^{ib_2\,f}\,. \ea

One serious drawback of considering hidden sector matter is that
we cannot explicitly calculate the moduli dependence of the matter
K\"{a}hler potential. Therefore we will have to make some (albeit
reasonable) assumptions, unlike the cases studied in the previous
sections. In what follows we will assume that we work in a
particular region of the moduli space where the K\"{a}hler metric
for the matter fields in the hidden sector is a very slowly
varying function of the moduli, essentially a constant. This
assumption is based on the fact that the chiral fermions are
localized at point-like conical singularities so that the bulk
moduli $s_i$ should have very little effect on the local physics.
In general, a singularity supporting a chiral fermion has no local
moduli, since there are no flat directions constructed from a
single chiral matter representation. Our assumption is further
justified by the $M$ theory lift of some calculable Type IIA
matter metrics as described in the Appendix \ref{kahlermetricM}.
It is an interesting and extremely important problem to properly
derive the matter K\"{a}hler potential in $M$ theory and test our
assumptions.

Thus we will consider the case when the hidden sector chiral
fermions have ``modular weight zero'' and assume a canonically
normalized K\"{a}hler potential. Furthermore, for the sake of
simplicity,  we will only study the case $N_f=1$. Since the
potential is invariant under $Q \leftrightarrow {\tilde Q}$ we
will focus only on vacua with $Q = {\tilde Q}$. With these
assumptions, the meson field $ \phi\equiv(Q\tilde Q)^{1/2}$ is
such that the corresponding K\"{a}hler potential for $\phi$ is
canonical. Therefore, the total K\"{a}hler potential, i.e. moduli
plus matter takes the form: \be \label{kahlerwithmatter} K = -3
\ln(4\pi^{1/3}\,V_X)+\phi\bar\phi\,. \ee The moduli $F$-terms are
then given by \ba\label{modf}
F_k&=&ie^{i b_2\vec N\cdot \vec t}[N_k(b_1A_1\phi_0^a e^{-b_1\vec N\cdot\vec s+i(b_1-b_2)\vec N\cdot \vec t+ia\theta}+b_2A_2 e^{-b_2\vec N\cdot\vec s})\nonumber\\
&&+\frac{3a_k}{2s_k}(A_1\phi_0^a e^{-b_1\vec N\cdot\vec
s+i(b_1-b_2)\vec N\cdot \vec t+ia\theta}+A_2 e^{-b_2\vec
N\cdot\vec s})]\,. \ea

In addition, an $F$-term due to the meson field is also generated
\be\label{mesf} F_\phi=\phi_0e^{-i\theta+i b_2\vec N\cdot \vec
t}\left[\left(\frac a{\phi_0^2}+1\right)A_1\phi_0^a e^{-b_1\vec
N\cdot\vec s+i(b_1-b_2)\vec N\cdot \vec t+ia\theta}+A_2
e^{-b_2\vec N\cdot\vec s}\right]\,. \ee The supergravity scalar
potential is then given by: \ba \label{potential_with_matter}
V&=&\frac{e^{\phi_0^2}}{48\pi
V_X^3}\,[(b_1^2A_1^2\phi_0^{2a}e^{-2b_1\vec\nu\cdot\,\vec
a}+b_2^2A_2^2e^{-2b_2\vec\nu\cdot\,\vec a}
+2b_1b_2A_1A_2\phi_0^{a}e^{-(b_1+b_2)\vec\nu\cdot\,\vec a}{\rm cos}((b_1-b_2)\vec N\cdot \vec t+a\theta))\nonumber\\\nonumber\\
&&\times\sum_{i=1}^{N}a_i({\nu_i})^2+3(\vec\nu\cdot\,\vec a)(b_1A_1^2\phi_0^{2\alpha}e^{-2b_1\vec\nu\cdot\,\vec a}+b_2A_2^2e^{-2b_2\vec\nu\cdot\,\vec a}+(b_1+b_2)A_1A_2\phi_0^{a}e^{-(b_1+b_2)\vec\nu\cdot\,\vec a}\,\nonumber\\\nonumber\\
&&\times\,{\rm cos}((b_1-b_2)\vec N\cdot \vec t+a\theta))
+3(A_1^2\phi_0^{2a}e^{-2b_1\vec\nu\cdot\,\vec a}+A_2^2e^{-2b_2\vec\nu\cdot\,\vec a}+2A_1A_2\phi_0^{a}e^{-(b_1+b_2)\vec\nu\cdot\,\vec a}\,\\\nonumber\\
&&\times{\rm cos}((b_1-b_2)\vec N\cdot \vec t+a\theta))
+\frac 3 4{\phi_0^2}\,(A_1^2\phi_0^{2\alpha}\left(\frac a{\phi_0^2}+1\right)^2e^{-2b_1\vec\nu\cdot\,\vec a}+A_2^2e^{-2b_2\vec\nu\cdot\,\vec a}\,\nonumber\\\nonumber\\
&&+2A_1A_2\phi_0^{a}\left(\frac
a{\phi_0^{2}}+1\right)e^{-(b_1+b_2)\vec\nu\cdot\,\vec a} {\rm
cos}((b_1-b_2)\vec N\cdot \vec t+a\theta))]\,.\nonumber \ea
Minimizing this potential with respect to the axions and $\theta$
we obtain the following condition: \be {\rm sin}((b_1-b_2)\vec
N\cdot \vec t+a\theta)=0\,. \ee The potential has local minima
with respect to the moduli $s_i$ when \be\label{axcond} {\rm
cos}((b_1-b_2)\vec N\cdot \vec t+a\theta)=-1\,. \ee In this case
(\ref{potential_with_matter}) reduces to \ba
\label{potential_noaxions}
V&=&\frac{e^{\phi_0^2}}{48\pi V_X^3}\,[(b_1A_1\phi_0^{a}e^{-b_1\vec\nu\cdot\,\vec a}-b_2A_2e^{-b_2\vec\nu\cdot\,\vec a})^2\sum_{i=1}^{N}a_i({\nu_i})^2\nonumber\\\nonumber\\
&&+3(\vec\nu\cdot\,\vec a)(A_1\phi_0^{a}e^{-b_1\vec\nu\cdot\,\vec a}-A_2e^{-b_2\vec\nu\cdot\,\vec a})(b_1A_1\phi_0^{a}e^{-b_1\vec\nu\cdot\,\vec a}-b_2A_2e^{-b_2\vec\nu\cdot\,\vec a})\\\nonumber\\
&&+3(A_1\phi_0^{a}e^{-b_1\vec\nu\cdot\,\vec
a}-A_2e^{-b_2\vec\nu\cdot\,\vec a})^2+\frac 3
4(A_1\phi_0^{a}\left(\frac
a\phi_0+\phi_0\right)e^{-b_1\vec\nu\cdot\,\vec
a}-A_2\phi_0e^{-b_2\vec\nu\cdot\,\vec a})^2]\,.\nonumber \ea

%%%%%%%%%%%%%%%%%%%%%%%%%%%%%%%%%
\subsubsection{Supersymmetric extrema}
Here we consider a case when the scalar potential
(\ref{potential_noaxions}) possess SUSY extrema and find
approximate solutions for the moduli and the meson field vevs.
Taking into account (\ref{axcond}) and setting the moduli
$F$-terms (\ref{modf}) to zero we obtain \be\label{nueq}
\nu_k=\nu=-\frac 3 2\frac{\tilde\alpha-1}{b_1\tilde\alpha-b_2}\,,
\ee together with the constraint \be\label{trcons}
\tilde\alpha\equiv\frac {A_1}{A_2}\phi_0^ae^{-\frac 7
3(b_1-b_2)\nu}\,. \ee At the same time, setting the matter
$F$-term (\ref{mesf}) to zero results in the following condition:
\be\label{mescon} \left(\frac
a{\phi_0^2}+1\right)\tilde\alpha-1=0\,. \ee Expressing
$\tilde\alpha$ from (\ref{nueq}) and substituting it into
(\ref{mescon}) we obtain the following solution for the meson vev
at the SUSY extremum: \be\label{phisol} \phi_0^2=a\frac{b_2+
3/(2\nu)}{b_1-b_2}\,. \ee Recall that in our analysis we are
considering the case when $P\equiv N_c-N_f>0$, which implies that
parameter $a$ defined in (\ref{b1anda}) is negative. Thus, since
the left hand side of (\ref{phisol}) is positive, for the SUSY
solution to exist, it is necessary to satisfy \be\label{con21}
b_2>b_1\,\,\,\,\,\,=>\,\,\,\,\,\,\,P>Q\,. \ee Recall that for the
moduli to be positive, the constants have to satisfy certain
conditions resulting in two possible branches (\ref{branches}).
Therefore, condition (\ref{con21}) implies that the SUSY AdS
extremum exists only for branch a) in (\ref{branches}). In the
limit, when $\nu$ is large, the approximate solution is given by:
\ba\label{app21} &&\tilde\alpha=\frac P Q\,,\\\nonumber
&&s_i=\frac{a_i\nu}{N_i}\,,\,\,\,\,\,\,{\rm with}\,\,\,\,\,\,\,
\nu=\frac
3{14\pi}\frac{P\,Q}{P-Q}\ln\left(\frac{A_2P}{A_1Q}\right)\,,\\\nonumber
&&\phi_0^2=\frac 2{P-Q}+\frac
7{P\,\ln\left(\frac{A_2P}{A_1Q}\right)}\,,\nonumber \ea where we
also assumed that $P\sim {\mathcal O}(10)$, such that
$\phi_0^a\approx 1$. For the case with two moduli where
$a_1=a_2=7/6$ and the choice \be \label{choice3}
A_1=4.1\,,\,\,\,A_2=30\,,\,\,\,b_1=\frac
{2\pi}{30}\,,\,\,\,b_2=\frac
{2\pi}{27}\,,\,\,\,N_1=1\,,\,\,\,N_2=1\,, \ee the numerical
solution for the SUSY extremum obtained by minimizing the scalar
potential (\ref{potential_noaxions}) gives \be \label{sol17}
s_1\approx44.5\,,\,\,\,s_2\approx44.5\,,\,\,\,
\phi_0\approx0.883\,, \ee whereas the approximate analytic
solution obtained in (\ref{app21}) yields \be \label{sol18}
s_1\approx45.0\,,\,\,\,s_2\approx45.0\,,\,\,\,
\phi_0\approx0.882\,. \ee This vacuum is very similar to the SUSY
AdS extremum obtained previously for the potential arising from
the ``pure glue'' Super Yang-Mills (SYM) hidden sector gauge
theory. Thus, we will not discuss it any further and instead move
to the more interesting case, for which condition (\ref{con21}) is
not satisfied.
%%%%%%%%%%%%%%%%%%%%%%%%%%%%%%%%%
\subsubsection{Metastable de Sitter (dS) minima}
Below we will use the same approach and notation we used in
section \ref{nonsusyadsvac}, to describe AdS vacua with broken
SUSY. Again, for brevity we denote \be\label{xyzw1} \tilde
x\equiv(\tilde\alpha-1)\,,\,\,\tilde
y\equiv(b_1\tilde\alpha-b_2)\,,\,\,\tilde
z\equiv(b_1^2\tilde\alpha-b_2^2)\,,\,\, \tilde w\equiv\frac{\tilde
x\tilde z}{{\tilde y}^2}\,. \ee Extremizing
(\ref{potential_noaxions}) with respect to the moduli $s_i$ and
dividing by ${\tilde x}^2$ we obtain the following system of
coupled equations \ba \label{u24} 2\frac {{\tilde y}^2} {{\tilde
x}^2} \nu_k^2&-&\left(2\,\frac {{\tilde y}^2} {{\tilde x}^2}\tilde
w\sum_{i=1}^{N}a_i\nu^2_i+3\frac {\tilde y} {\tilde x}\left(\tilde
w+1\right)\vec\nu\cdot\vec a+3+\frac 3 2
\phi_0^2\left(\frac{a\tilde\alpha}{\phi_0^2\,\tilde x}+1\right)
\left(\frac{a\tilde\alpha b_1}{\phi_0^2\,\tilde y}+1\right)\right)\frac {\tilde y}{\tilde x}\nu_k\,\\
&-&3\left(\frac {{\tilde y}^2} {{\tilde
x}^2}\sum_{i=1}^{N}a_i\nu_i^2+3\frac {\tilde y}{\tilde x}
\vec\nu\cdot\vec a+3+\frac 3 4
\phi_0^2\left(\frac{a\tilde\alpha}{\phi_0^2\,\tilde
x}+1\right)^2\right)=0\,,\nonumber \ea plus the constraint
(\ref{trcons}). Next, we extremize (\ref{potential_noaxions}) with
respect to $\phi_0$ and divide it by $2\phi_0{\tilde x}^2$ to
obtain: \ba \label{u25} \frac {{\tilde y}^2} {{\tilde
x}^2}\sum_{i=1}^{N}a_i\nu_i^2+\frac 32\frac {\tilde y}{\tilde
x}\vec\nu\cdot\vec a+\frac 34 \left(2\frac {\tilde y}{\tilde
x}\vec\nu\cdot\vec a+\frac{a\tilde\alpha}{\tilde
x}\left(\frac{a-1}{\phi_0^2}+2\right)+
5+\phi_0^2\right)\left(\frac{a\tilde\alpha}{\phi_0^2\,\tilde x}+1\right)&&\\
+\frac{a\tilde\alpha b_1}{\phi_0^2\,\tilde x}\left(\frac {{\tilde
y}} {{\tilde x}}\sum_{i=1}^{N}a_i\nu_i^2+ \frac 3
2\vec\nu\cdot\vec a\right)=0\,.&&\,\nonumber \ea To solve the
system of $N$ cubic equations (\ref{u24}), we introduce a
quadratic constraint \be \label{Nt} 4\,\tilde T\equiv2\,\frac
{{\tilde y}^2} {{\tilde x}^2}\tilde
w\sum_{i=1}^{N}a_i\nu^2_i+3\frac {\tilde y} {\tilde x}\left(\tilde
w+1\right)\vec\nu\cdot\vec a+3+\frac 3 2
\phi_0^2\left(\frac{a\tilde\alpha}{\phi_0^2\,\tilde
x}+1\right)\left(\frac{a\tilde\alpha b_1}{\phi_0^2\,\tilde
y}+1\right)\,, \ee such that (\ref{u24}) turns into a system of
$N$ coupled quadratic equations: \be \label{u26} 2\frac {{\tilde
y}^2} {{\tilde x}^2} \nu_k^2-4\tilde T\frac {\tilde y}{\tilde
x}\nu_k -3\left(\frac {{\tilde y}^2} {{\tilde
x}^2}\sum_{i=1}^{N}a_i\nu_i^2+3\frac {\tilde y}{\tilde
x}\vec\nu\cdot\vec a+3+\frac 3 4
\phi_0^2\left(\frac{a\tilde\alpha}{\phi_0^2\,\tilde
x}+1\right)^2\right)=0\,. \ee Again, the standard solution of a
quadratic equation dictates that the solutions for $\nu_k$ of
(\ref{u26}) have the form \be \label{sol24} \nu_k=\frac {\tilde x}
{\tilde y} \left({\tilde T}+m_k \tilde H\right)\,,\,\, {\rm
with}\,\, m_k=\pm 1\,,\,\,k=\overline{1,N}\,. \ee We have now
reduced the task of determining $\nu_k$ for each
$k=\overline{1,N}$ to finding {\it only two} quantities - $\tilde
T$ and $\tilde H$. By substituting (\ref{sol24}) into equations
(\ref{u25}-\ref{u26}) and using (\ref{vol}), we obtain a system of
three coupled equations \ba \label{e43} &&\frac 7 3\left({\tilde
T_{A}^2}+2 A{\tilde T_{A}} \tilde H_{A}+\tilde
H_{A}^2\right)+\frac 72\left({\tilde T_{A}}+A \tilde H_{A}\right)
+\frac 3 4[\frac{14}3\left({\tilde T_{A}}+A \tilde H_{A}\right)+
\\ && \frac{a\tilde\alpha}x\left(\frac{a-1}{\phi_0^2}+2\right)
+5+\phi_0^2]\times\left(\frac{a\tilde\alpha}{\phi_0^2\,\tilde
x}+1\right)+ \nonumber \\ && \frac{a\tilde\alpha
b_1}{\phi_0^2}\frac {7} {3{\tilde y}} \left(\left({\tilde
T_{A}^2}+2 A{\tilde T_{A}} \tilde H_{A}+\tilde
H_{A}^2\right)+\frac 3 2\left({\tilde T_{A}}+A \tilde H_{A}
\right)\right)=0 \nonumber\\ && \frac{14w}3\left({\tilde
T_{A}^2}+2 A{\tilde T_{A}} \tilde H_{A}+\tilde
H_{A}^2\right)+7(w+1) \left({\tilde T_{A}}+A \tilde
H_{A}\right)+3+ \nonumber \\ && \frac 3 2
\phi_0^2\left(\frac{a\tilde\alpha}{\phi_0^2\,\tilde x}+1\right)
\left(\frac{a\tilde\alpha b_1}{\phi_0^2\,\tilde y}+1\right)-4\tilde T_{A}=0\,\nonumber \\
&& 9\left({\tilde T_{A}^2}+2 A {\tilde T_{A}} \tilde H_{A}+\tilde
H_{A}^2\right)-4\tilde H_{A}\left(\tilde H_{A}+A{\tilde T_{A}}
\right)+21\left({\tilde T_{A}}+A\tilde H_{A}\right)+9+ \nonumber
\\ && \frac 9 4 \phi_0^2\left(\frac{a\tilde\alpha}{\phi_0^2\,\tilde
x}+1\right)^2=0\,,\nonumber \ea plus the constraint
(\ref{trcons}). Note that each solution is again labelled by
parameter $A$ so that (\ref{sol24}) becomes \be \label{sol36}
\nu_k^{A}=\frac {\tilde x} {\tilde y} \left(\tilde T_{A}+m_k
\tilde H_{A}\right)\,. \ee Let us consider the case when $A=1$. In
this case, the solution is given by \be \label{sol37}
\nu_k^{1}=\nu=\frac {\tilde x} {\tilde y} \left(\tilde
T_{1}+\tilde H_{1}\right)=\frac {\tilde x} {\tilde y}\tilde
L_{1,+}\,. \ee and (\ref{e43}) is reduced to \ba \label{e46} \frac
7 3\left(\tilde T_1+\tilde H_1\right)^2+\frac 72\left({\tilde
T_1}+\tilde H_1\right)+ \frac 3 4\left(\frac {14}3\left(\tilde
T_1+\tilde
H_1\right)+\frac{a\tilde\alpha}x\left(\frac{a-1}{\phi_0^2}+2\right)
+\phi_0^2+5\right) && \\
\left(\frac{a\tilde\alpha}{\phi_0^2\,\tilde
x}+1\right)+\frac{a\tilde\alpha b_1}{\phi_0^2}\frac {7} {3{\tilde
y}}\left(\left(\tilde T_1+\tilde H_1\right)^2+
\frac 3 2\left({\tilde T_1}+\tilde H_1\right)\right)=0&&\,\nonumber\\
\frac{14w}3\left(\tilde T_1+\tilde H_1\right)^2+7(w+1) \left({\tilde T_1}+
\tilde H_1\right)+3+\frac 3 2 \phi_0^2\left(\frac{a\tilde\alpha}{\phi_0^2\,\tilde x}+
1\right)\left(\frac{a\tilde\alpha b_1}{\phi_0^2\,\tilde y}+1\right)-4\tilde T_1=0\,\nonumber&& \\
9\left(\tilde T_1+\tilde H_1\right)^2-4\tilde H_1\left(\tilde
H_1+{\tilde T_1} \right)+21\left({\tilde T_1}+\tilde
H_1\right)+9+\frac 9 4
\phi_0^2\left(\frac{a\tilde\alpha}{\phi_0^2\,\tilde
x}+1\right)^2=0\,.&& \nonumber \ea In the notation introduced in
(\ref{xyzw1}), the SUSY condition (\ref{mescon}) can be written as
\be\label{sucon} \frac{a\tilde\alpha}{\phi_0^2}+\tilde x=0\,. \ee
It is then straightforward to check that in the SUSY case, the
system (\ref{e46}) yields \be \tilde T_1=-\frac {15}
8\,,\,\,\,\,\,\,\tilde H_1=\frac 3 8\,,\,\,\,\,\,\,\tilde
L_{1,+}=-\frac 3 2\,, \ee as expected. We will now consider branch
b) in (\ref{branches}) for which (\ref{sucon}) is not satisfied.
Moreover, in order to obtain analytical solutions for the moduli
and the meson vev $\phi_0$ we will again consider the large three
cycle volume approximation. Recall that in this case we take
$\tilde y \rightarrow 0$ and $\tilde w\rightarrow -\infty$ limit
to obtain the following reduced system of equations when $A=1$ for
$\tilde L_{1,+}$ and $\phi_0$: \ba \label{e47}
\frac 7 3\left(\tilde L_{1,+}\right)^2+\frac 72\tilde L_{1,+}+\frac 3 4\left(\frac{14}3\tilde L_{1,+}+\frac{a\tilde\alpha}x\left(\frac{a-1}{\phi_0^2}+2\right)+\phi_0^2+5\right)\left(\frac{a\tilde\alpha}{\phi_0^2\,\tilde x}+1\right)&&\,\\
+\frac{a\tilde\alpha b_1}{\phi_0^2}\frac 7 {3{\tilde y}}
\left(\left(\tilde L_{1,+}\right)^2+\frac 3 2\tilde L_{1,+}\right)=0&&\,\nonumber\\
\frac 23\left(\tilde L_{1,+}\right)^2+\tilde L_{1,+}+\frac
{3{a\tilde\alpha b_1}\tilde y} {14 \tilde x\tilde z}
\left(\frac{a\tilde\alpha} {\phi_0^2\,\tilde x}+1\right)=0\,,&&
\nonumber \ea Note that in (\ref{e47}), we have dropped the third
equation since for $A=1$ we only need to know $\tilde{L}_{1,+}$
and the third equation in (\ref{e46}) determines $\tilde{H}_{1,+}$
in terms of $\tilde{L}_{1,+}$. We also kept the first subleading
term in the second equation. Note that the term in the second line
of the first equation proportional to $\sim 1/{\tilde y}$ appears
to blow up as $\tilde y \rightarrow 0$. However, from the second
equation one can see that the combination $\left(\tilde
L_{1,+}\right)^2+\frac 3 2\tilde L_{1,+}$ is proportional to
$\tilde y$ which makes the corresponding term finite. By keeping
the subleading term in the second equation, we can express
\be\label{e51} \left(\tilde L_{1,+}\right)^2=-\frac 3 2\tilde
L_{1,+}-\frac {9{a\tilde\alpha b_1}\tilde y} {28 \tilde x\tilde z}
\left(\frac{a\tilde\alpha}{\phi_0^2\,\tilde x}+1\right) \ee from
the second equation to substitute into the first equation to
obtain in the leading order \be \label{e49} \left(\frac {14}
3\tilde
L_{1,+}+5+\phi_0^2+\frac{a\tilde\alpha}x\left(\frac{a-1}{\phi_0^2}+2\right)-\frac
1 {\tilde z\tilde x}\left(\frac{a\tilde\alpha
b_1}{\phi_0}\right)^2\right)\left(\frac{a\tilde\alpha}{\phi_0^2\,\tilde
x}+1\right)=0\,. \ee Since we are now considering branch b) in
(\ref{branches}), the second factor in (\ref{e49}) is
automatically non-zero. Therefore, the first factor in (\ref{e49})
must be zero. Thus, after substituting \be\label{e61} \tilde
L_{1,+}\approx-\frac 3 2+\frac {3{a\tilde\alpha b_1}\tilde y} {14
\tilde x\tilde z} \left(\frac{a\tilde\alpha}{\phi_0^2\,\tilde
x}+1\right)\,, \ee obtained from (\ref{e51}), we have the
following equation for $\phi_0$ \be \label{e50} \phi_0^2-2+\frac
{{a\tilde\alpha b_1}\tilde y} {\tilde x\tilde z}
\left(\frac{a\tilde\alpha}{\phi_0^2\,\tilde x}+1\right)+
\frac{a\tilde\alpha}x\left(\frac{a-1}{\phi_0^2}+2\right)-\frac 1
{\tilde z\tilde x}\left(\frac{a\tilde\alpha
b_1}{\phi_0}\right)^2=0\,. \ee

Also, since in the leading order $\tilde L_{1,+}=-3/2$, using the
definitions in (\ref{xyzw1}) we can express $\tilde\alpha$ from
(\ref{sol37}) in the limit when $\nu$ is large, including the
first subleading term \be\label{e54}
\tilde\alpha\approx\frac{b_2}{b_1}+\frac{3(b_1-b_2)}{2b_1^2\,\nu}\,.
\ee By combining (\ref{trcons}) with the leading term in
(\ref{e54}) and taking into account that $\phi_0^a\sim 1$ we again
obtain \be\label{app24} s_i=\frac{a_i\nu}{N_i}\,,\,\,\,\,\,\,{\rm
with}\,\,\,\,\,\,\, \nu\approx\frac
3{14\pi}\frac{P\,Q}{Q-P}\ln\left(\frac{A_1Q}{A_2P}\right)\,. \ee
Thus, from (\ref{e54}) we have \be\label{e57}
\tilde\alpha\approx\frac{P}{Q}+\frac{7\,(Q-P)^2}{2\,Q^2\,\ln\left(\frac
{A_1Q}{A_2P}\right)}\,. \ee Finally, using (\ref{e57}) along with
the definitions of $\tilde x$, $\tilde y$ and $\tilde z$ in
(\ref{xyzw1}) in terms of $\tilde\alpha$ we can solve for
$\phi_0^2$ from (\ref{e50}) and assuming that $Q-P\sim {\mathcal
O}(1)$, in the limit when $P$ is large we obtain \be\label{e58}
\phi_0^2\approx1-\frac 2{Q-P}+\sqrt{1-\frac 2{Q-P}}-\frac
7{P\,\ln\left(\frac{A_1\,Q}{A_2 P}\right)}\left(\frac 3
2+\sqrt{1-\frac 2{Q-P}}\right)\,. \ee We notice immediately that
since $\phi_0^2$ is real and positive it is necessary that
\be\label{e74} Q-P>2\,. \ee We will show shortly that the extremum
we found above corresponds to a metastable minimum. Also, for a
simple case with two moduli, via an explicit numerical check we
have confirmed that if $Q-P\leq 2$ the local minimum is completely
destabilized yielding a runaway potential. Also note that for
(\ref{e58}) to be accurate, it is not only $P$ which has to be
large but also the product $P\,\ln\left(\frac{A_1\,Q}{A_2
P}\right)$ has to stay large to keep the subleading terms
suppressed. To check the accuracy of the solution we again
consider a manifold with two moduli where $a_1=a_2=7/6$ and
$P=27$, $Q=30$, $A_1=27$, $A_2=4$, $N_1=N_2=1$. The exact values
obtained numerically are:
\begin{equation}\label{exact23}
s_1\approx 43.2917\,,\,\,\,s_2\approx 43.2917\,,\,\,\phi_0\approx
0.809\,.
\end{equation}
The approximate equations above yield the following values:
\begin{equation}\label{approxim23}
s_1\approx 43.292\,,\,\,\,s_2\approx 43.292\,,\,\,\phi_0\approx
0.802\,.
\end{equation}
Note the high accuracy of the leading order approximation for the
moduli $s_i$.

It is now straightforward to compute the vacuum energy using the
approximate solution obtained above. First, we compute
\be\label{e63} K^{i\,\bar j}F_i{\bar F_{\bar
j}}-3|W|^2=4\left(A_2\tilde x\right)^2\left(\frac 7
9\left(L_{1,+}\right)^2+\frac 7 3L_{1,+}+1\right) \left(\frac
{A_1Q}{A_2P}\right)^{-\frac{2P}{Q-P}} \ee and \be\label{e62}
K^{\phi\bar{\phi}}F_{\phi}{\bar F_{\bar{\phi}}}=\left(A_2\tilde
x\phi_0\right)^2\left(\frac{a\tilde\alpha}{\phi_0^2\,\tilde
x}+1\right)^2 \left(\frac {A_1Q}{A_2P}\right)^{-\frac{2P}{Q-P}}
\ee

Using (\ref{e61}), (\ref{e63}) and (\ref{e62}) we obtain the
following expression for the potential at the extremum with
respect to the moduli $s_i$ as a function of $\phi_0$
\be\label{po45} V_0=\frac {(A_2\tilde x)^2}{64\pi
V_X^3}\left[\phi_0^4+ \left(\frac {2\,a\tilde\alpha}{\tilde
x}-3\right)\phi_0^2+\left(\frac {a\tilde\alpha}{\tilde
x}\right)^2\right]\frac {e^{\phi_0^2}}{\phi_0^2}\left(\frac
{A_1Q}{A_2P}\right)^{-\frac{2P}{Q-P}}\,, \ee where the terms
linear in $\tilde y$ cancelled and the quadratic terms were
dropped. A quick look at the structure of the potential
(\ref{po45}) as a function of $\phi_0^2$, where $\phi_0^2>0$, is
enough to see that there is a single extremum with respect to
$\phi_0^2$ which is indeed, a minimum. The polynomial in the
square brackets is quadratic with respect to $\phi_0^2$. Moreover,
the coefficient of the $\phi_0^4$ monomial is equal to unity and
therefore is always positive. This implies that for the minimum of
such a biquadratic polynomial to be positive, it is necessary for
the corresponding discriminant to be negative, which results in
the following condition:
\begin{equation}\label{e70}
3-4\frac {a\tilde\alpha}{\tilde x}<0\,.
\end{equation}
Again, since in the leading order $\tilde L_{1,+}=-3/2$, using the
definitions in (\ref{app24}), we can express $\tilde\alpha$ from
(\ref{sol37}) in terms of $\nu$ to get
\begin{equation}\label{eqab}
\frac{\tilde\alpha}{\tilde
x}=\frac{\tilde\alpha}{\tilde\alpha-1}=\frac{P}{P-Q}+\frac{3PQ}{4\pi\nu(P-Q)}\,.
\end{equation}
We then substitute $\nu$ from (\ref{app24}) into (\ref{eqab}) and
use it together with $a=-2/P$ we obtain from (\ref{e70}) the
following condition
\begin{equation}\label{e80}
3-\frac 8{Q-P}-\frac{28}{P\ln\left(\frac{A_1Q}{A_2P}\right)}<0\,.
\end{equation}
The above equation is the leading order requirement for the energy
density at the minimum to be positive. It is also clear that the
minimum is {\it metastable}, as in the decompactification limit
($V_X \rightarrow \infty$), the scalar potential vanishes from
above, leading to an absolute Minkowski minimum.

%%%%%%%%%%%%%%%%%%%%%%%%%%%%%%%%%

\subsubsection{The uniqueness of the dS vacuum}\label{unique}
In the previous subsection we found a particular solution of the
system in (\ref{e43}) corresponding to $A=1$. Here we would like
to investigate if solutions for $0\leq A<1$ are possible when the
vacuum for $A=1$ is de Sitter. Just like for the pure Super
Yang-Mills (SYM) case, we can recast (\ref{sol36}) as
\be\label{eq786} \nu^A_{k}=\frac{{\cal T}_A}{\tilde B_A}\,\tilde
L_{A,\,k}\,, \ee where the volume of the associative three cycle
$Q$ is again \be\label{def456} {\cal T}_A\equiv Vol(Q)_A=\vec
a\cdot\vec\nu^{A}=\frac {\tilde x}{\tilde y}\,\tilde B_{A}\,, \ee
and we have introduced \be\label{BAcbt} \tilde B_A\equiv\vec
a\cdot\vec{\tilde L}_{A}=\frac 7 3\left(\tilde T_A+A\tilde
H_A\right)\,. \ee Just like we did in equation (\ref{eq37}) for
the pure SYM case, we can also express $\tilde\alpha_A$ as \be
\label{eq370} \tilde\alpha_A=\frac{b_2{{{\cal T}_A}}-{\tilde
B_A}}{b_1{{{\cal T}_A}}-{\tilde B_A}}\,. \ee If we again consider
the large associative cycle volume limit and take $\tilde y
\rightarrow 0$ and $\tilde w\rightarrow -\infty$, the second and
third equations in (\ref{e43}) in the leading order reduce to
\ba \label{e202} &&2\left({{\tilde T}_{A}^2}+2 A{\tilde T_{A}} \tilde H_{A}+\tilde H_{A}^2\right)+ 3\left({\tilde T_{A}}+A \tilde H_{A}\right)=0 \\
&&9\left({\tilde T_{A}^2}+2 A {\tilde T_{A}} \tilde H_{A}+\tilde
H_{A}^2\right)-4\tilde H_{A}\left(\tilde H_{A}+A{\tilde T_{A}}
\right)+21\left({\tilde T_{A}}+A\tilde H_{A}\right)+9+\frac 9 4
\phi_0^2\left(\frac{a\tilde\alpha}{\phi_0^2\,\tilde
x}+1\right)^2=0\,. \nonumber \ea Note that the only difference
between (\ref{e20}) and (\ref{e202}) is the presence of the term
\be\label{exter} \delta\equiv\frac 9 4
\phi_0^2\left(\frac{a\tilde\alpha}{\phi_0^2\,\tilde
x}+1\right)^2\,, \ee which couples the system (\ref{e202}) to the
first equation in (\ref{e43}) which determines $\phi_0$. Instead
of solving the full system to determine $\tilde T_A$, $\tilde H_A$
and $\phi_0$ and analyzing the solutions we choose a quicker
strategy for our further analysis. Namely, we can solve the system
of two equations in (\ref{e202}) and regard $\delta$ as a
continuous deformation parameter. One may object to this
proposition because $\tilde\alpha$ and therefore $\tilde
x=\tilde\alpha-1$ are not independent of parameter $A$. However,
in the limit when ${\cal T}_A$ is large, we notice from
(\ref{eq370}) that in the leading order, $\tilde\alpha_A$ is
indeed {\em independent} of $A$.

Recall that in the pure SYM case the system (\ref{e20})
corresponding to the case when $\delta=0$ has two {\em real}
solutions for all $0\leq A\leq 1$. Thus, one may expect that as we
continuously dial $\delta$, the system may still yield real
solutions for $A<1$. Let us first determine the range of possible
values of parameter $\delta$. A quick calculation yields that the
combination in (\ref{exter}) is the smallest with respect to
$\phi_0$ when $\phi_0^2=\frac{a\tilde\alpha}{\tilde x}$. In this
case \be\label{e97} \delta=9 \frac{a\tilde\alpha}{\tilde x}\,. \ee
Now, recall from the previous subsection that for the solution
corresponding to $A=1$ to have a positive vacuum energy, condition
(\ref{e70}) must hold. Since $\tilde\alpha$ and $\tilde x$ are
independent of $A$ in the leading order, condition (\ref{e70})
implies that \be\label{e970} \delta>\frac {27}4. \ee Again, since
the volume ${\cal T}_A$ is always positive, from (\ref{eq786}) we
see that for all moduli to be stabilized in the positive range,
all three quantities $\tilde L_{A,\,+}$, $\tilde L_{A,\,-}$ and
$\tilde B_A$ must have the same sign. For $\delta=27/4$, the
system (\ref{e202}) has two real solutions when $0.877781<A<1$.
\begin{figure}[hbtp]
  \centerline{\hbox{ \hspace{0.0in}
    \epsfxsize=3.3in
    \epsfbox{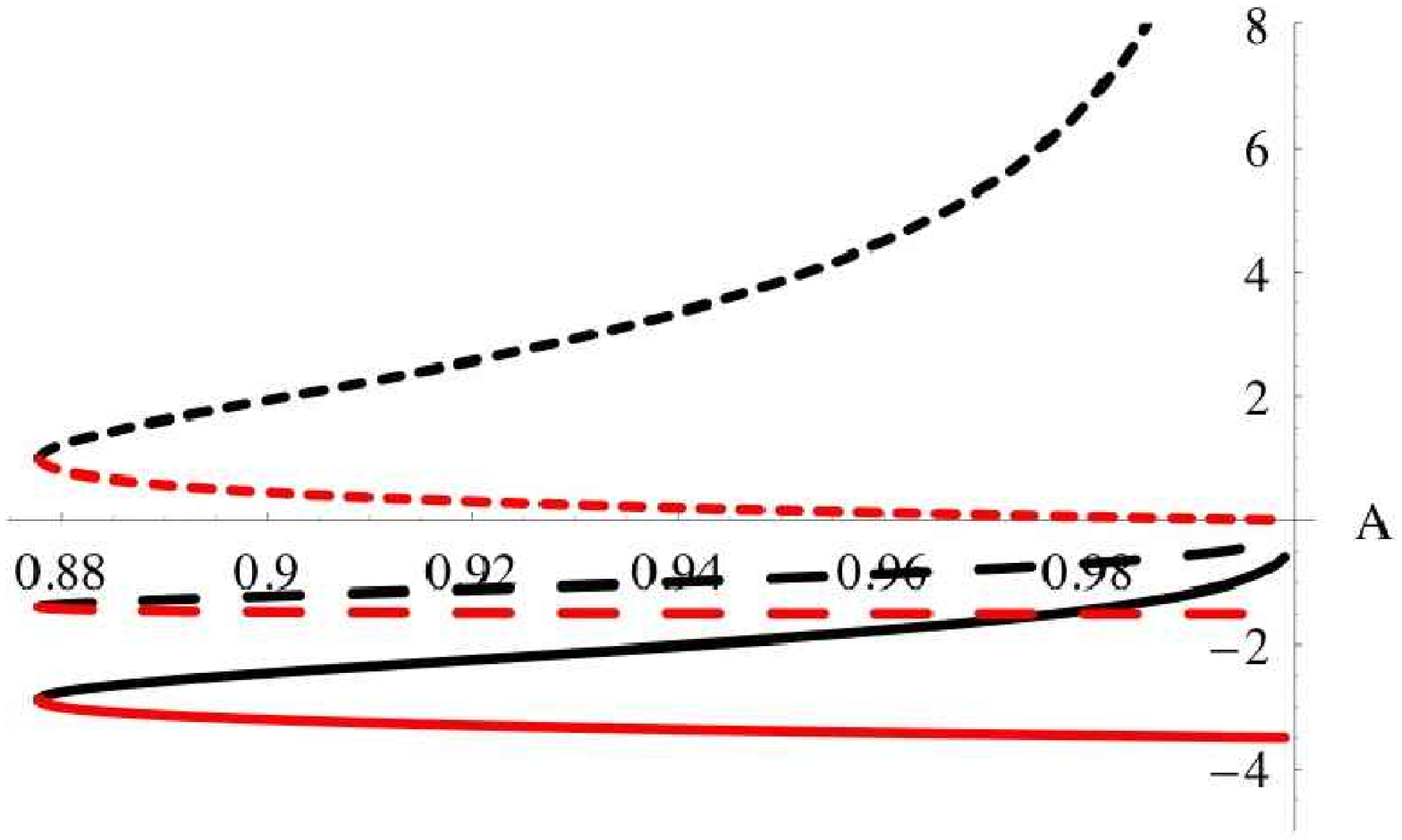}
    \hspace{0.25in}
    \epsfxsize=3.3in
    \epsfbox{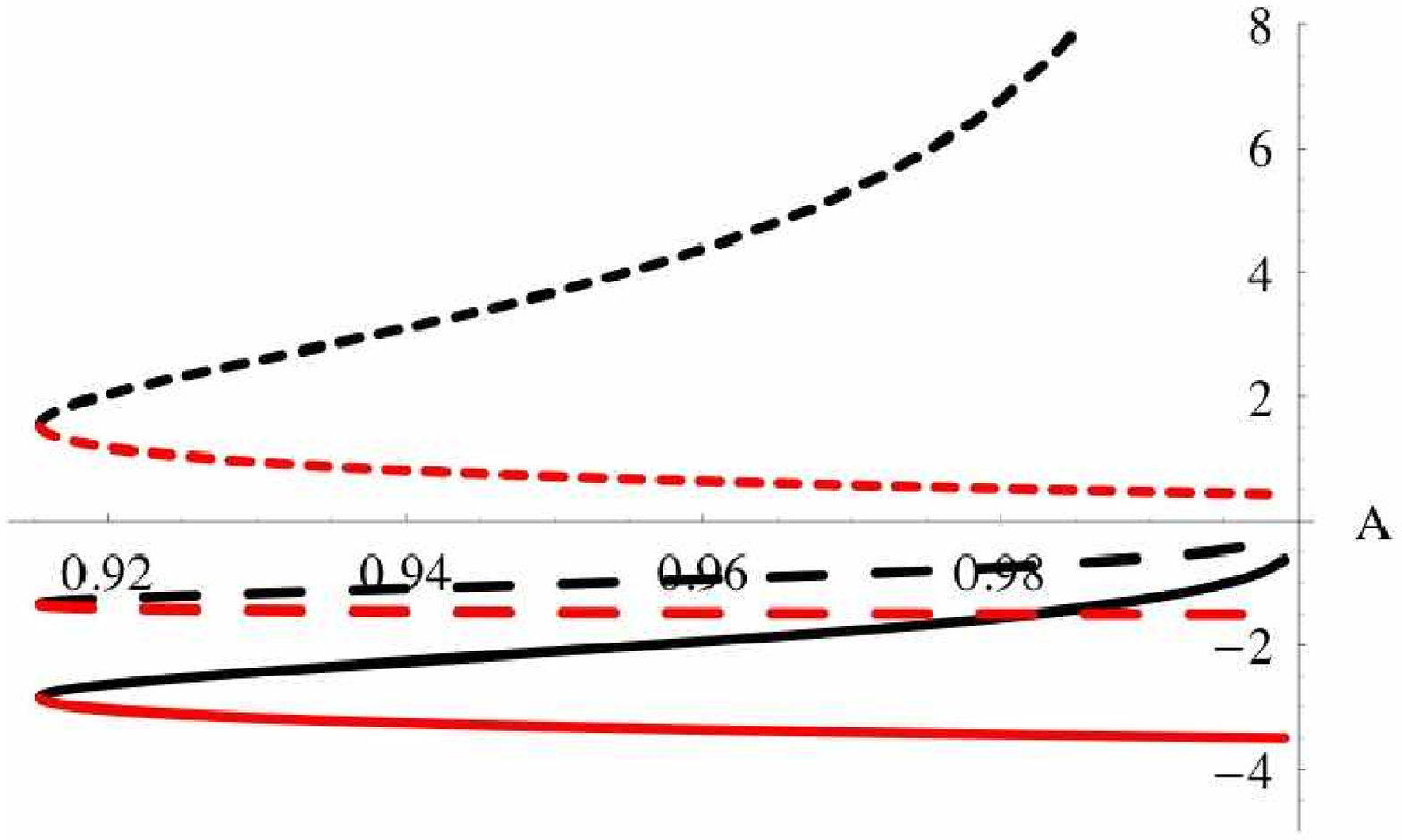}
    }
  }
\caption{Plots of $\tilde L_{A,\,+}^{(c)}$,  $\tilde
L_{A,\,-}^{(c)}$ and ${\tilde B_A^{(c)}}$, where
$c=\overline{1,2}$, corresponding to the two real solutions of the
system (\ref{e202})  as functions of parameter $A$. $\tilde
L_{A,\,+ }^{(c)}$ - long dashed curve, $\tilde L_{A,\,-}^{(c)}$
-short dashed curve, ${\tilde B_A^{(c)}}$ - solid curve. Black
color: $\tilde L_{A,\,+}^{(1)}$,  $\tilde L_{A,\,-}^{(1)}$ and
${\tilde B_A^{(1)}}$ corresponding to the first real solution. Red
color: $\tilde L_{A,\,+}^{(2)}$,  $\tilde L_{A,\,-}^{(2)}$ and
${\tilde B_A^{(2)}}$ corresponding to the second real solution.
Left plot: when $\delta=27/4$ the real solutions exist in the
range $0.877781< A<1$. Right plot: when $\delta=8$ the real
solutions exist st in the range $0.915342< A<1$.}
    \label{Plot240}
\end{figure}
However, from the left plot in Figure {\ref{Plot240}}
corresponding to the minimum value $\delta=27/4$ we see that
neither of the two solutions satisfy the above requirement since
both short-dashed curves corresponding to $\tilde L_{A,-}$ for the
two solutions are {\em always} positive for the entire range
$0.877781<A<1$, whereas both $\tilde L_{A,\,+}$ and $\tilde B_A$
remain negative. Therefore for $\delta=27/4$ and $A<1$ there are
no solutions for which all the moduli are stabilized at positive
values. Moreover, as parameter $\delta$ is further increased, the
range of possible values of $A$ for which the system has two real
solutions gets smaller and more importantly, the values of $\tilde
L_{A,-}$ remain positive and only increase while both $\tilde
L_{A,\,+}$ and $\tilde B_A$ remain negative, which can be seen
from the right plot in Figure {\ref{Plot240}}, where $\delta=8$.
This trend continues as we increase $\delta$.

Thus, we can make the following general claim: If the solution for
$A=1$ has a positive vacuum energy, condition (\ref{e970}) must
hold. When  this condition is satisfied the system (\ref{e202})
has no solutions in the range $0\leq A<1$ for which all the moduli
are stabilized at positive values. Therefore, if the vacuum found
for $A=1$ is de Sitter it is the only possible vacuum where all
the moduli are stabilized at positive values. Although the above
analysis was done in the limit when ${\cal T}_A$ is large, we have
run a number of explicit numerical checks for a manifold with two
moduli and various values of the constants confirming the above
claim. In addition, although we have not proved it, it seems
plausible from many numerical checks we carried out that it is
also not possible to have a metastable dS minimum for values of
$A$ different from unity, even if the dS condition on the $A=1$
vacuum is {\it not} imposed.

Finally, it should be noted that the situation with a ``unique" dS
vacuum is in sharp contrast to that when one obtains anti-de
Sitter vacua, where there are between $2^{N-1}$ and $2^N$
solutions for $N$ moduli depending on the value of $A$ (see
section \ref{nonsusyadsvac}). Let us explain this in a bit more
detail. Since the dS solution found for $A=1$ is located right in
the vicinity of the ``would be AdS SUSY extremum''\footnote{This
can be seen by comparing the leading order expression for the
moduli vevs in the dS case (\ref{app24}) with the corresponding
formula for the SUSY AdS extremum (\ref{alpha4}).} where the
moduli $F$-terms are nearly zero, it is the large contribution
from the matter $F$-term (\ref{e62}) which cancels the $-3|W|^2$
term in the scalar potential resulting in a positive vacuum
energy. Recall that in the leading order all the AdS vacua with
the moduli vevs $s_{A,i}^{(c)}$ are located within the hyperplane
\footnote{c=1,2 labels the two real solutions of the system
(\ref{e20}).} \be\label{hyp} \sum_{i=1}^{N}s_{A,i}^{(c)}N_i =\frac
1{2\pi}\frac{P\,Q}{P-Q}{\ln}\left(\frac{A_2P}{A_1Q}\right)=constant\,.
\ee The matter $F$-term contribution to the scalar potential
$K^{\phi\bar{\phi}}F_{\phi}{\bar F_{\bar{\phi}}}$ evaluated  at
the same $s_{A,i}^{(c)}$ but arbitrary $\phi_0$ is therefore also
{\em constant} along the hyperplane (\ref{hyp}). Thus, while the
matter $F$-term contribution stays constant, as we move along the
hyperplane (\ref{hyp}) away from the dS minimum, where the moduli
$F$-terms are the smallest, the moduli $F$-term contributions can
only get larger so that the scalar potential becomes even more
positive. This implies that the AdS minima with broken SUSY found
in Section \ref{nonsusyadsvac} completely disappear, as the AdS
SUSY extremum becomes a minimum.

%%%%%%%%%%%%%%%%%%%%%%%%%%%%%%%%%%%%%%%%%%%%%%%%%%%%%%%%%%%%%%%%%%%%%%%%%%%%%%%%%%%%%%%%%%%%
%%%%%%%%%%%%%%%%%%%%%%%%%%%%%%%%%%%%%%%%%%%%%%%%%%%%%%%

\subsection{Relevant Scales}\label{distribution}

We have demonstrated above that in fluxless $M$ theory vacua,
strong gauge dynamics can generate a potential which stabilizes
all the moduli. Since the entire potential is generated by this
dynamics, and the strong coupling scale is below the Planck scale,
we also have a hierarchy of scales. In this section we will
calculate some of the basic scales in detail. In particular, the
gravitino mass, which typically controls the scale of
supersymmetry breaking is calculated. By uniformly scanning over
the constants $( N, P, Q , A_k)$ with $N_i$ order one, we
demonstrate in \ref{samplingsection} that a reasonable fraction of
choices of constants have a TeV scale gravitino mass. We do not
know if the space of $G_2$-manifolds uniformly scans the $( P, Q ,
A_k , N_i )$ or not, and more importantly, the scale of variation
of the $A_k$'s in the space of manifolds is not clear. The
variation of the $A_k$'s is the most important issue here, since
one can certainly vary $P$ and $Q$ over an order of magnitude. We
begin with a discussion of the basic scales in the problem. We
will begin with the AdS vacua, then go on to discuss the de Sitter
case. In particular, in the dS case, requiring a small vacuum
energy seems to lead to superpartners.at around the TeV scale.

\subsubsection{Scales: AdS Vacua}

As an example, we consider one of the non-SUSY minima in our toy
model given by (\ref{sol9}) and compute some of the quantities
relevant for phenomenology. Namely, the vacuum energy \be
\label{lambda} \Lambda_0=-(5.1\times 10^{10}\,\mathrm{GeV})^4\,,
\ee the gravitino mass \be \label{mgrav} M_{3/2}=m_pe^{K/2}|W|
\approx2.081\,\mathrm{TeV}\,, \ee the 11-dimensional Planck scale
\be \label{m11} M_{11}=\frac{\sqrt{\pi}m_p}
{V_X^{1/2}}\approx3.9\times 10^{17}\, \mathrm{GeV}\,, \ee the
scale of gaugino condensation in the hidden sectors \ba
\Lambda^{(1)}_{g} &=& m_p\,e^{-\frac{b_1}{3}\Sigma_iN_is^i}\approx2.6\times 10^{15}\, \mathrm{GeV}\\
\Lambda^{(2)}_{g} &\approx& 9.7\times10^{14}\,\mathrm{GeV} \ea

where $m_p=(8\pi G_N)^{-1/2}=2.43\times 10^{18}$ GeV is the
reduced four-dimensional Planck mass.

\noindent From (\ref{mgrav}) and (\ref{m11}), we see that it is
possible to have a TeV scale gravitino mass together with $M_{11}
\geq M_{unif} (2\times 10^{16}$GeV). This feature survives in more
general cases as well, {\it implying that standard gauge
unification is compatible with low scale SUSY in these vacua.}

\subsubsection{Gravitino mass}
By definition, the gravitino mass is given by: \be\label{gr1}
m_{3/2}=m_p\,e^{K/2}|W|\,. \ee For the particular $M$ theory vacua
with K\"{a}hler potential given by (\ref{kahler}) and the
non-perturbative superpotential as in (\ref{super}) with $SU(P)$
and $SU(Q)$ hidden sector gauge groups we have: \be\label{gr2}
m_{3/2}=\,\frac
{m_p}{8\sqrt{\pi}{V_X}^{3/2}}\left|A_1e^{-{\frac{2\pi}{P}{\rm
Im}f}}-A_2e^{-{\frac{2\pi}{Q}{\rm Im}f}}\right|\,, \ee where the
relative minus sign inside the superpotential is due to the
axions. Before we get to the gravitino mass we first compute the
volume of the compactified manifold $V_X$ for the AdS vacua with
broken SUSY. By plugging the approximate leading order solution
for the moduli (\ref{eq43}) into the definition (\ref{vol}) of
$V_X$ we obtain: \be\label{volume} (V_X)^{(c)}_A=\left[\frac
1{2\pi}\frac{PQ}{P-Q} {\rm
ln}\left(\frac{A_2P}{A_1Q}\right)\right]^{7/3}\prod_{i=1}^{N}
\left(\frac{a_i\,L_{A,\,i}^{(c)}}{N_i\,B_A^{(c)}}\right)^{a_i}\,.
\ee Recalling the definition (\ref{deft}) of ${\cal T}^{\,(c)}_A$
and using (\ref{eq39}) together with (\ref{volume}) to plug into
(\ref{gr2}) the gravitino mass for these vacua in the leading
order approximation is given by: \be\label{gr3}
(m_{3/2})^{(c)}_A=\sqrt{2}\pi^3\,A_2P
\left|\frac{P-Q}{PQ}\right|\left[\frac{PQ}{P-Q}{\rm
ln}\left(\frac{A_2P}{A_1Q}\right) \right]^{-\frac
72}\left[\frac{A_2P}{A_1Q}\right]^{-\frac{P}{P-Q}}\prod_{i=1}^{N}
\left(\frac{N_i\,B_A^{(c)}}{a_i\,L_{A,\,i}^{(c)}}\right)^{\frac{3a_i}2}\,.
\ee For the special case with two moduli when $a_1=a_2=7/6$,
considered in the previous sections we obtain the following:
\begin{eqnarray}\label{gr4}
(m_{3/2})^{(1,2)}_0&=&m_p\,2^{1/2}\pi^3\left(7+\sqrt{17}\right)^{\frac 7 4}\left(N_1\,N_2\right)^{\frac 7 4}\,A_2\,P\left|\frac{P-Q}{P\,Q}\right|\left(\frac{A_2\,P}{A_1\,Q}\right)^{-\frac P{P-Q}}\left(\frac{PQ}{P-Q}{\rm ln}\frac{A_2\,P}{A_1\,Q}\right)^{-\frac 7 2}\nonumber\\
&\sim&m_p\,2.97\times 10^3\left(N_1\,N_2\right)^{\frac 7
4}\,A_2\,P\left|\frac{P-Q}{P\,Q}\right|\left(\frac{A_2\,P}{A_1\,Q}\right)^{-\frac
P{P-Q}}\left(\frac{PQ}{P-Q}{\rm
ln}\frac{A_2\,P}{A_1\,Q}\right)^{-\frac 7 2}
\end{eqnarray}
For the choice of constants as in (\ref{choice1}) the leading
order approximation (\ref{gr4}) yields: \be
(m_{3/2})^{(1,2)}_0=2061 {\rm GeV}\,, \ee whereas the exact value
computed numerically for the same choice of constants is: \be
m_{3/2}=2081 {\rm GeV}\,. \ee Again, we see a good agreement
between the leading order approximation and the exact values.

%%%%%%%%%%%%%%%%%%%%%%%%%%%%%%%%%%%%%%%%%%%%%%%%%%%%%%%%%%%%%%%%

\subsubsection{Scanning the Gravitino mass}\label{samplingsection}

In previous sections we found explicit solutions describing vacua
with spontaneously broken supersymmetry. Moreover, we also
demonstrated that for a particular set of the constants these
solutions can result in $m_{3/2}\sim O(1)\,{\rm TeV}$. It would be
extremely interesting and worthwhile to estimate (even roughly)
the fraction of all possible solutions which exhibit spontaneously
broken SUSY at the scales of $O(1)$-$O(10)$ TeV. We would first
like to do this for generic AdS/dS vacua with a large magnitude of
the cosmological constant ($\sim m^2_{3/2}m^2_{p}$). The analysis
for the AdS vacua is given below but as we will see, the results
obtained for the fraction of vacua are virtually the same for the
dS case as well. In the next subsection, we impose the requirement
of a small cosmological constant as a constraint and try to
understand its repercussions for the gravitino mass.

We do not yet know the range that the constants $(N, P , Q , A_1,
A_2 )$ take in the space of all $G_2$ manifolds. Nevertheless, we
do have a rough idea about some of them. For example, we expect
that the quantity given by the ratio \be\label{rat}
\rho\equiv\frac{A_2\,P}{A_1\,Q}\,, \ee which appears in several
equations, does deviate from unity. One reason for this may be due
the threshold corrections {\cite{Friedmann:2002ty}} which in turn
depend on the properties of a particular $G_2$-holonomy manifold.
On the other hand, since we do not expect large threshold
corrections, we might guess that the range of possible values of
$\rho$ should not be much greater than one. Thus, an upper limit
$\rho\leq 10$ is probably reasonable. Also, based on the duality
with the Heterotic String we can get some idea on the possible
range of integers $P$ and $Q$ corresponding to the dual coxeter
numbers of the hidden sector gauge groups. Namely, since for both
$SO(32)$ and $E_8$ gauge groups appearing in the Heterotic String
theories the dual coxeter numbers are $h^v=30$, we can tentatively
assume that both $P$ and $Q$ can be at least as large as $30$. Of
course, we do not rule out any values higher than $30$ but in this
section we will assume an upper bound $P,Q\leq 30$.

We now turn our attention to equation (\ref{gr3}) which will be
used to estimate the gravitino mass scale. It is clear from the
structure of the formula that $m_{3/2}$ is extremely sensitive to
$P$, $Q$ as well as the ratio $\rho$, given by (\ref{rat}). On the
other hand it is less sensitive to the other constants appearing
in the equation such as $N_i$, $a_i$ and the ratios
$B^{(c)}_A/L^{(c)}_{A,i}$. This is because the powers $3a_i/2$ for
each term under the product get much less than one as the number
of moduli increases because of the constraint on $a_i$ in
(\ref{vol}). This will smooth any differences between the
contributions coming from the individual factors inside the
product. Since for $0\leq A\leq 1$ ($A$ is defined in
(\ref{sol1})), the ratios $B^{(c)}_A/L^{(c)}_{A,i}$ vary only in
the range $O(1)$-$O(10)$, for our purposes it will be sufficient
to simply consider (\ref{gr3}) for the case when $A=1$
corresponding to the SUSY extremum so that
${B^{(1)}_1}/L^{(1)}_{1,i}=7/3$ for all $i$. This is certainly
good enough for the order of magnitude estimates we are interested
in. It also seems reasonable to assume that the integers $N_i$ are
all of $O(1)$. Yet, even if some $N_i$ are unnaturally large,
their individual contributions are generically washed out since
they are raised to the powers that are much less than one. Thus,
for simplicity we will take $N_i=1$ for all $i=\overline{1,N}$.
Finally, from field theory computations \cite{Finnell:1995dr},
$A_2 = Q$ (in a particular RG scheme) up to threshold corrections.
We therefore take $A_2 \sim Q$ for simplicity, allowing $A_1$ to
vary.

Thus, the gravitino mass in our analysis is given by
\be\label{gr7} m_{3/2}\sim\sqrt{2}\pi^3\,PQ
\left(\frac{P-Q}{PQ}\right)^{\frac 92}\left[{\rm ln}\rho
\right]^{-\frac 72}(\rho)^{-\frac{P}{P-Q}}\prod_{i=1}^{N}
\left(\frac 7{3 a_i}\right)^{\frac{3a_i}2}\,. \ee Finally, with
regard to the constants $a_i$ which are a subject to the
constraint \be\label{con2} \sum_{i=1}^{N}a_i=\frac 7 3\,, \ee we
will narrow our analysis to two opposite cases. For the first case
we make the following choice \be {\rm
1)}\,\,\,\,\,\,\,\,\,a_1=2\,,\,\,{\rm and}\,\,\, a_i=\frac 1
{3(N-1)},\,\,{\rm for}\,\,\,i=\overline{2,N}\,, \ee such that one
modulus is generically large and all the other moduli are much
smaller. This is a highly anisotropic $G_2$-manifold. The second
case is \be {\rm 2)}\,\,\,\,\,\,\,\,a_i=\frac7{3\,N}\,,\,\,\,{\rm
for\,all}\,\,\,i=\overline{1,N}\,,\,\,\,\,\,\,\,\,\,\,\,\,\,\,\,\,\,\,\,\,\,\,\,\,\,\,\,\,\,\,\,\,\,\,\,\,\,\,\,\,\,\,
\ee with all the moduli being on an equal footing. Therefore, by
considering these opposite cases we expect that most other
possible sets of $a_i$ will give similar results that are
somewhere in between. For each set of $a_i$ above, equation
(\ref{gr7}) gives \be\label{gr8} {\rm
1)}\,\,\,\,\,\,\,\,\,m_{3/2}^{(1)}\sim\frac{343\sqrt{14}\pi^3\,}{216}\,PQ
\left(\frac{P-Q}{PQ}\right)^{\frac 92}\left[{\rm ln}\rho
\right]^{-\frac 72}(\rho)^{-\frac{P}{P-Q}} \left(N-1\right)^{\frac
12}\,, \ee \be\label{gr9} {\rm
2)}\,\,\,\,\,\,\,\,\,m_{3/2}^{(2)}\sim\sqrt{2}\pi^3\,PQ
\left(\frac{P-Q}{PQ}\right)^{\frac 92}\left[{\rm ln}\rho
\right]^{-\frac 72}(\rho)^{-\frac{P}{P-Q}} \left(N\right)^{\frac
72}\,.\,\,\,\,\,\,\,\,\,\,\,\,\,\,\,\,\,\,\,\,\,\,\,\,\, \ee For a
typical compactification we expect $N\sim O(100)$, therefore the
variation of $m_{3/2}$ due to an $O(1)$ change in the number of
moduli for the first case is $O(1)$ whereas in the second case it
can be as large as $O(10)$. Thus, if we choose $N=100$, we expect
that our order of magnitude analysis will be fairly robust for
case $1)$. For case $2)$, however, we will perform the same
analysis for $N=100$ and $N=50$ to see how different the results
will be. Before we proceed further we need to impose a restriction
on the possible solutions to remain within the SUGRA framework.
Using (\ref{volume}), condition that $V_X$ must remain greater
than one for the two cases under consideration translates into the
following two conditions: \be\label{vol8} {\rm
1)}\,\,\,\,\,\,\,\,\,\frac 37\left(\frac
{64}{3(N-1)}\right)^{\frac 1 7}\frac 1{2\pi}\frac{P\,Q}{P-Q}{\rm
ln}\,\rho\,>\,1\,, \ee \be\label{vol9} {\rm
2)}\,\,\,\,\,\,\,\,\,\frac 1{2\pi\,N}\frac{P\,Q}{P-Q}{\rm
ln}\,\rho\,>\,1\,.\,\,\,\,\,\,\,\,\,\,\,\,\,\,\,\,\,\,\,\,\,\,\,\,\,\,\,\,\,\,\,\,\,\,\,\,
\ee Then, as long as conditions (\ref{vol8}-\ref{vol9}) hold,
volume of the associative cycle $\cal T$ is greater than one is
satisfied automatically - a necessary condition for the validity
of supergravity. This is obvious from comparing the right hand
side of (\ref{eq39}) with each condition above. From
(\ref{vol8}-\ref{vol9}) we can find a critical value of
$\rho=\rho_{crit}$ for both cases at which $V_X=1$:
\be\label{rho8} {\rm
1)}\,\,\,\,\,\,\,\,\,\rho_{crit}^{(1)}\equiv{\rm Exp}\left[\frac
{14\pi}3\left(\frac {3(N-1)}{64}\right)^{\frac 1
7}\frac{P-Q}{P\,Q}\right]\,, \ee \be\label{rho9} {\rm
2)}\,\,\,\,\,\,\,\,\,\rho_{crit}^{(2)}\equiv{\rm Exp}\left[{2\pi
N}\left(\frac{P-Q}{P\,Q}\right)\right]\,.\,\,\,\,\,\,\,\,\,\,\,\,\,\,\,\,\,\,\,\,\,\,\,\,\,\,\,\,
\ee By substituting (\ref{rho8}-\ref{rho9}) into
(\ref{gr8}-\ref{gr9}) we can find the corresponding upper limits
on $m_{3/2}$ as functions of $P$ and $Q$, below which our
solutions are going to be consistent with the SUGRA approximation.

\begin{figure}[hbtp]
  \centerline{\hbox{ \hspace{0.0in}
    \epsfxsize=3.3in
    \epsfbox{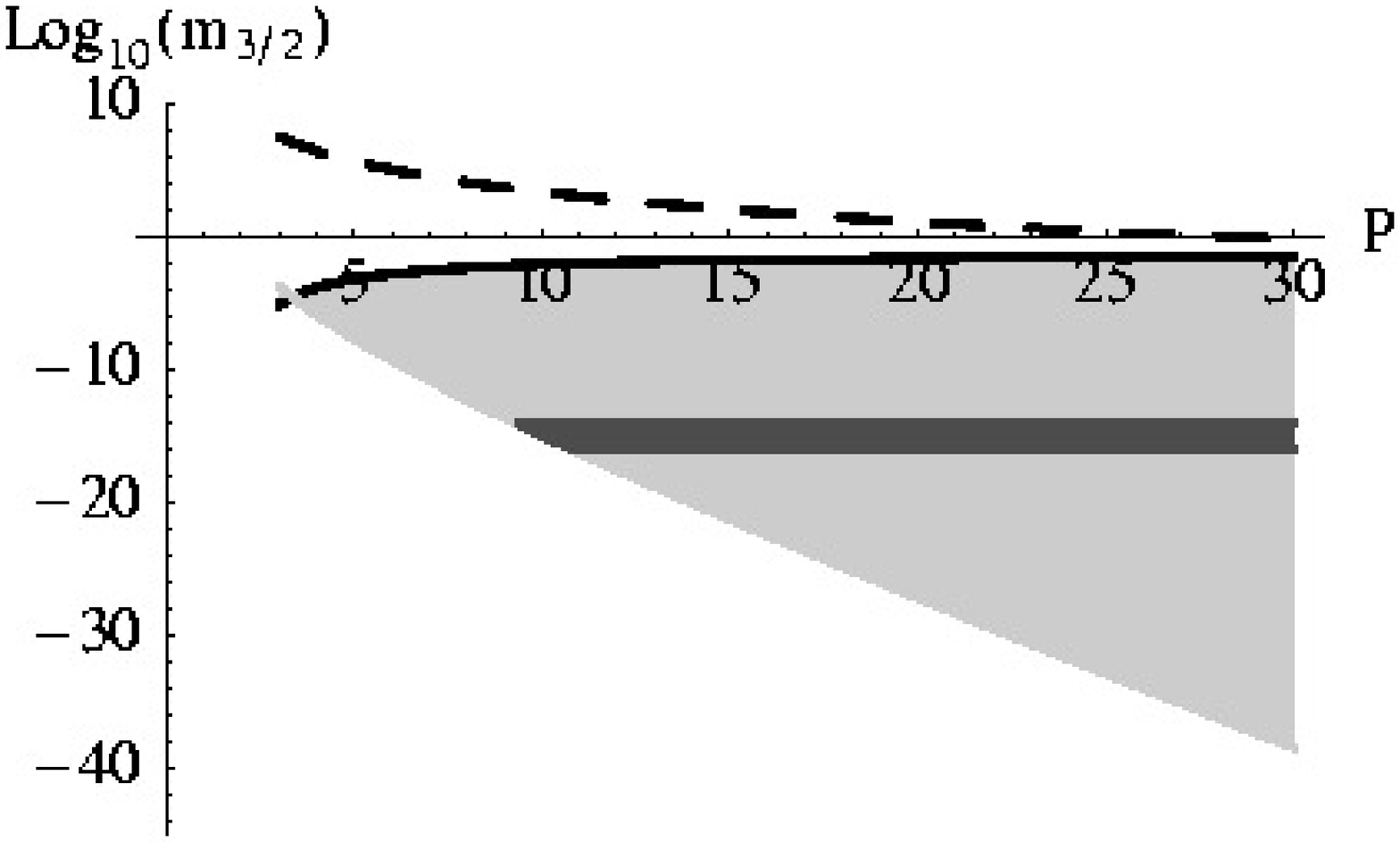}
    \hspace{0.25in}
    \epsfxsize=3.3in
    \epsfbox{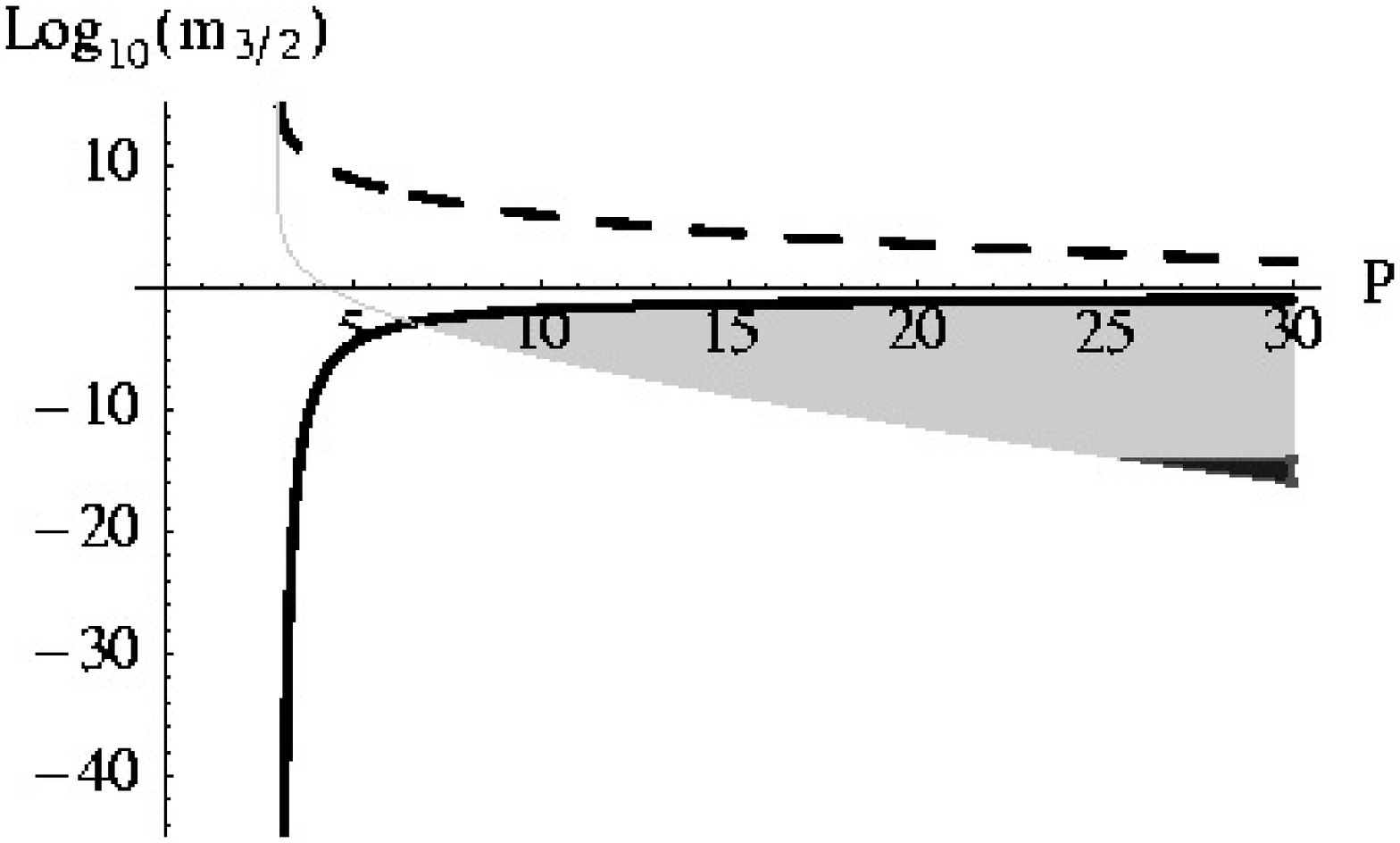}
     }
    }
  \centerline{\hbox{ \hspace{0.0in}
    \epsfxsize=3.3in
    \epsfbox{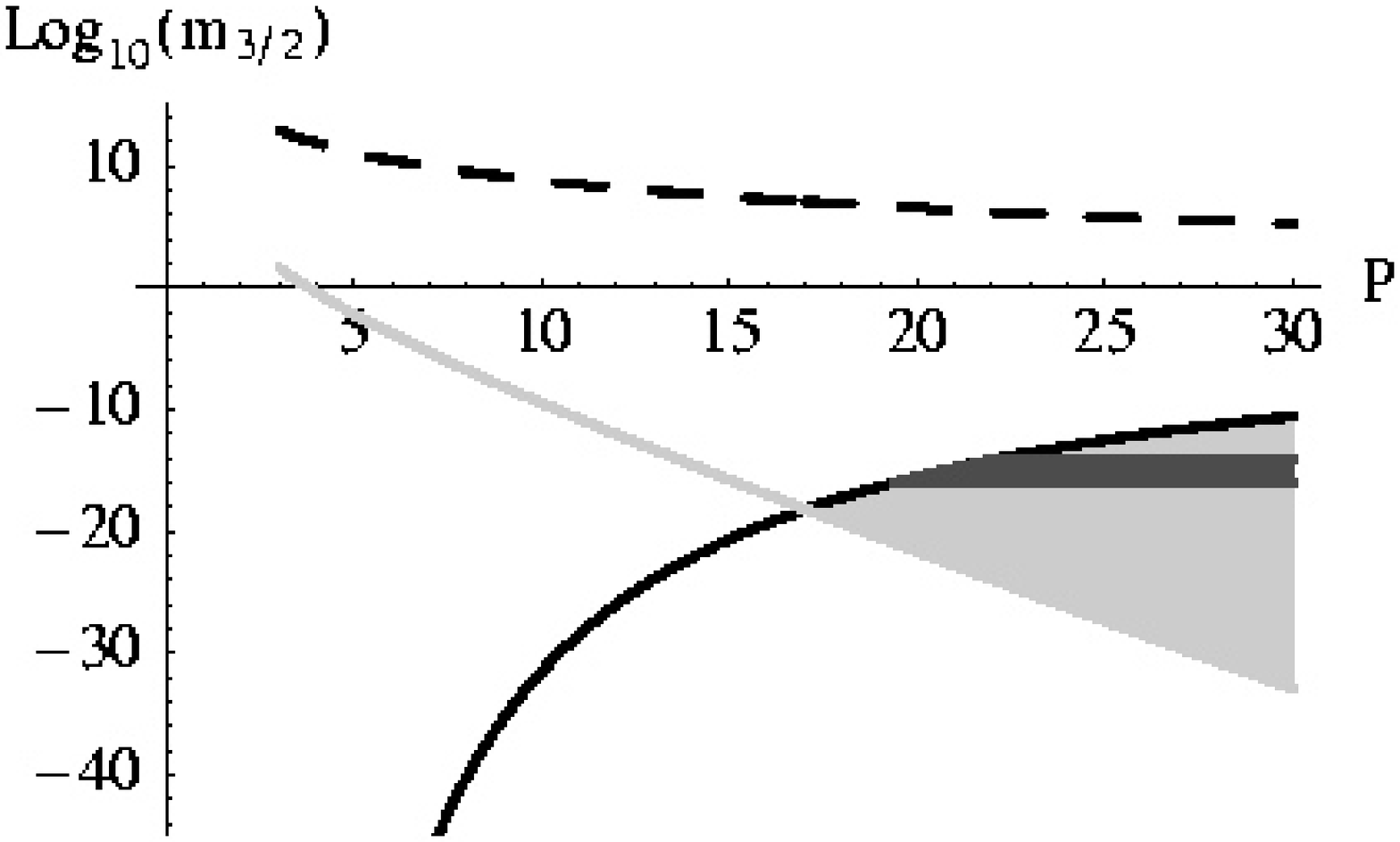}
    \hspace{0.25in}
    \epsfxsize=3.3in
    \epsfbox{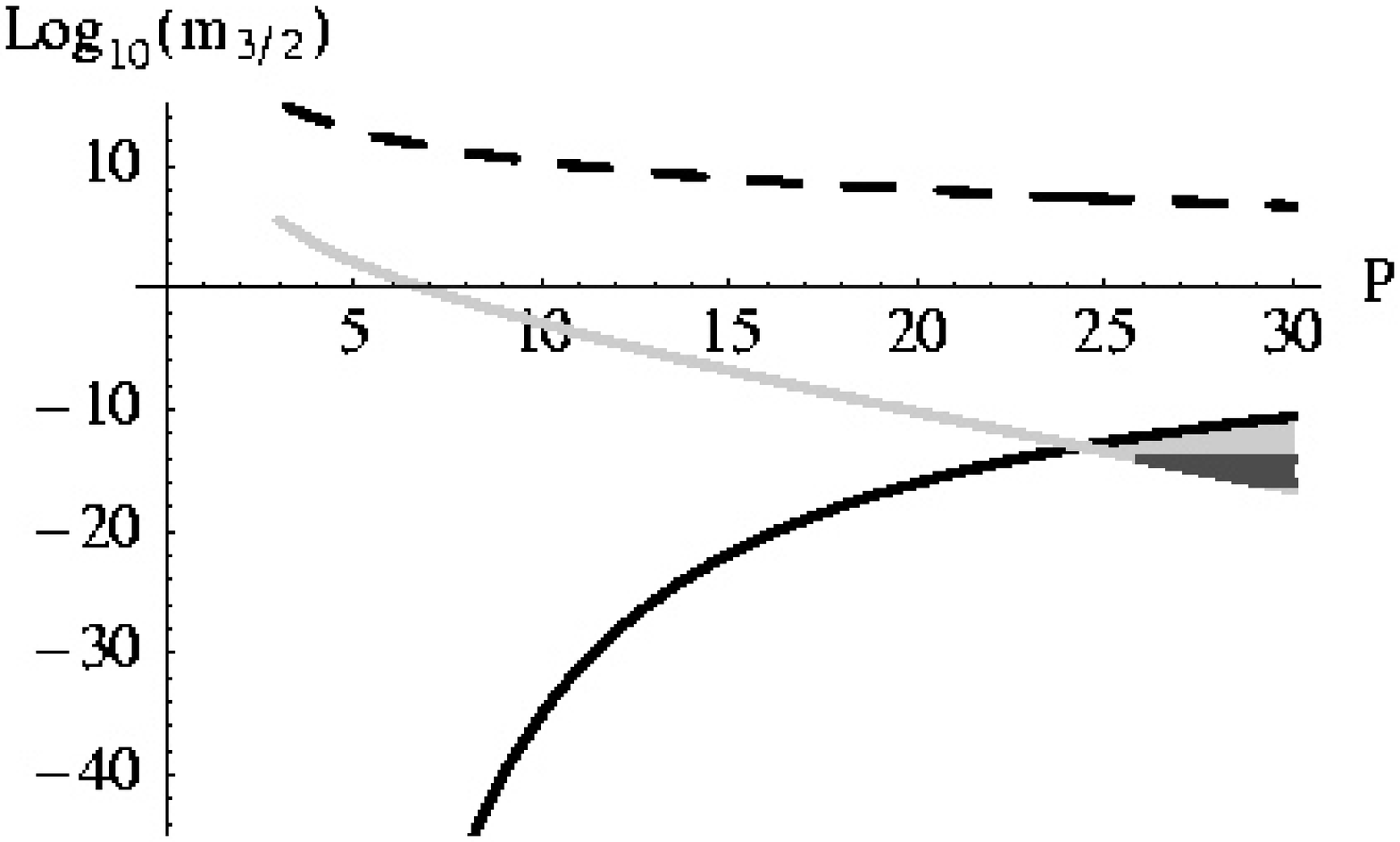}
    }
  }
\caption{${\rm log}_{10}(m_{3/2})$ as a function of $P$ for case
$1)$ - top plots and case $2)$ - bottom plots. The light grey area
represents possible values of ${\rm log}_{10}(m_{3/2})$ in the
range where $\rho_{crit}\leq\rho\leq 10$ consistent with the SUGRA
approximation. The dark area indicates the region of interest
where $-16\leq {\rm log}_{10}(m_{3/2})\leq -14$ such that
$240\,{\rm GeV}\leq m_{3/2}\leq 24\,{\rm TeV}$. The dark solid
curve corresponds to ${\rm log}_{10}(m_{3/2})$ when
$\rho=\rho_{crit}$. The lower boundary of the light grey area
represents the ${\rm log}_{10}(m_{3/2})$ curve when $\rho=10$. The
dashed curve corresponds to $\rho=1.01$. Top left: Case $1)$ when
$P-Q=1$. Top right: Case $1)$ when $P-Q=3$. Bottom left: Case $2)$
when $P-Q=1$. Bottom right: Case $2)$ when $P-Q=2$.}
\label{scans1}
\end{figure}

In Figure \ref{scans1} we present plots of ${\rm
log}_{10}(m_{3/2})$ for both cases as a function of $P$ in the
range where $\rho_{crit}\leq\rho\leq 10$ for different values of
$P-Q$.
On all the plots the light grey area represents possible values of
${\rm log}_{10}(m_{3/2})$ consistent with the supergravity
framework. For the sake of completeness we have also included the
formal plot of ${\rm log}_{10}(m_{3/2})$ corresponding to
$\rho=1.01$ represented by the dashed curve. From the plots it is
clear that as the difference $P-Q$ is increased from $1$ to $3$ -
top and from $1$ to $2$ - bottom, both the light grey area
representing all possible values of ${\rm log}_{10}(m_{3/2})$
consistent with the SUGRA approximation and the dark area
corresponding to $-16\leq {\rm log}_{10}(m_{3/2})\leq -14$ get
significantly smaller. If we further increase $P-Q$, the light
grey region shrinks even more for case $1)$, and does not exist
for case $2)$, while the dark region completely disappears in both
cases. Therefore, for case $2)$ the plots on the bottom of Figure
\ref{scans1} are the only possibilities where solutions for
$P\leq30$ and $N=100$ consistent with the SUGRA approximation are
possible, implying un upper bound $(P-Q)_{max}=2$. It turns out
that for case $1)$ the upper bound on $(P-Q)$ where such solutions
are possible is much higher $(P-Q)_{max}=23$.

Assuming that all values of the constants such as $P$, $Q$ and
$\rho$ are equally likely to occur in the ranges chosen above we
can perform a crude estimate of the number of solutions with
$-16\leq {\rm log}_{10}(m_{3/2})\leq -14$ relative to the total
number of possible solutions consistent with the SUGRA
approximation. In doing so we will use the following approach. For
each value of $(P-Q)$ in the range $1\leq (P-Q)\leq (P-Q)_{max}$
we compute the area of the grey region for each plot and then add
all of them to find the total volume corresponding to all possible
values of ${\rm log}_{10}(m_{3/2})$ consistent with the
supergravity approximation. \be
\Omega_{\,tot}=\sum_{\,\,\,\,\,\,\,\,(P-Q)\,=1}^{\,\,\,\,\,\,\,\,\,\,\,\,(P-Q)_{max}}\int_{P_{min}}^{30}dP\,{\rm
log}_{10}(m_{3/2})_{|_{\{\rho_{crit}\leq\rho\leq10\}}}\,. \ee
Likewise, we add all the dark areas for each plot to find the
volume corresponding to the region where $-16\leq {\rm
log}_{10}(m_{3/2})\leq -14$ \be
\Omega_{\,0}=\sum_{\,\,\,\,\,\,\,\,(P-Q)\,=1}^{\,\,\,\,\,\,\,\,\,\,\,\,(P-Q)^{*}_{max}}\int\,dP\,{\rm
log}_{10}(m_{3/2})_{|_{{\left\{{\rho_{crit}\leq\rho\leq10}\atop{-16\leq
{\rm log}_{10}(m_{3/2})\leq -14}\right\}}}}\,, \ee where
$(P-Q)^{*}_{max}$ is un upper bound before the dark region
completely disappears. From the previous discussion, for case
$1)$: $(P-Q)^{*}_{max}=3$; for case $2)$: $(P-Q)^{*}_{max}=2$ .
Then, the fraction of the volume where $240\,{\rm GeV}\leq
m_{3/2}\leq 24\,{\rm TeV}$ is given by the ratio \be \Delta=\frac
{\Omega_{\,0}}{\Omega_{\,tot}}\,. \ee Numerical computations yield
the following values for the two cases when $N=100$:
\be\label{delta1} {\rm 1)}\,\,\,\,\,\,\,\,\,\Delta_1=3.5\%\,, \ee
\be\label{delta2} {\rm 2)}\,\,\,\,\,\,\,\,\,\Delta_2=13.6\%\,. \ee
Because of the significant difference in the dependence of
$\rho_{crit}$ on the number of moduli $N$ in (\ref{rho8}) versus
(\ref{rho9}), the number of solutions consistent with the SUGRA
approximation is cut down dramatically in case $2)$ compared to
case $1)$. This also occurs because of the different dependence of
$m_{3/2}$ in (\ref{gr8}) and (\ref{gr9}) on $N$. Namely, for
$N\sim O(100)$, the values of $m_{3/2}$ for case $2)$ in
(\ref{gr9}) are $\sim O(10^6)$ greater than those for case $1)$.
Furthermore, for the same reasons, it turns out that if we keep
increasing the number of moduli $N$, for case $2)$ there is un
upper bound $N\leq 157$ for the solutions with $240\,{\rm GeV}\leq
m_{3/2}\leq 24\,{\rm TeV}$ compatible with the SUGRA approximation
to exist at all. Of course, this upper bound can be higher if we
allow $P$ to be greater than $30$.

By performing the same analysis for $N=50$ we get the following
estimates \be\label{delta3} {\rm
1)}\,\,\,\,\,\,\,\,\,\Delta_1=3.4\%\,, \ee \be\label{delta4}
\,\,\,{\rm 2)}\,\,\,\,\,\,\,\,\,\Delta_2=10.7\%\,. \ee As
expected, decreasing the number of moduli by a half has produced
little effect on $\Delta_1$ while decreasing $\Delta_2$ by a few
present. These numbers coming from our somewhat crude analysis
already demonstrate that a comparatively large fraction of vacua
in $M$ theory generate the desired hierarchy between the Planck
and the electroweak scale physics. Also, one can easily check that
all the solutions consistent with the SUGRA framework for which
$240\,{\rm GeV}\leq m_{3/2}\leq 24\,{\rm TeV}$ for any number of
moduli $N$ satisfy the following bound on the eleven dimensional
scale \be 3.6\times 10^{16}\,{\rm GeV}\leq m_{11}\leq 4.3\times
10^{18}\,{\rm GeV}\,, \ee which makes them compatible with the
standard unification at $M_{GUT}\sim 2\times 10^{16}$ GeV. This is
also a nice feature. Of course, apart from determining the upper
and lower bounds on the constants, it would be desirable to know
their distribution for all possible manifolds of $G_2$ holonomy.
In this case instead of using the flat statistical measure as we
did here, each solution would be assigned a certain weight making
the sampling analysis more accurate. However, this is an extremely
challenging task which goes beyond the scope of this work.

The simple analysis presented in this section clearly points to a
very restrictive nature of the solutions. Namely, the requirement
of consistency with the supergravity regime results in very strict
bounds on the properties of the compactification manifold. Further
requirements coming from the SUSY breaking scale to be in the
range required for supersymmetry to solve the hierarchy problem
narrows down the class of possible $G_2$ holonomy manifolds even
more. It would be extremely interesting to know to what extent
these results extend into the small volume, "stringy" regime,
about which we have nothing to say here.

\subsubsection{Results for dS Vacua}
We will now show that the results obtained in the previous
subsection also hold true for dS vacua with K\"{a}hler potential
given by (\ref{kahlerwithmatter}) and the non-perturbative
superpotential as in (\ref{mattersuptwosectors}) with $SU(N_c)$
and $SU(Q)$ hidden sector gauge groups. For this case, we have :
\be\label{gr22} m_{3/2}=m_p\,\frac
{e^{\phi_0^2/2}}{8\sqrt{\pi}{V_X}^{3/2}}\left|A_1\phi_0^a\,e^{-{\frac{2\pi}{P}{\rm
Im}f}}-A_2e^{-{\frac{2\pi}{Q}{\rm Im}f}}\right|\,, \ee where the
relative minus sign inside the superpotential is due to the axions
and $P\equiv N_c-1$. Before we get to the gravitino mass we first
compute the volume of the compactified manifold $V_X$ for the
metastable dS vacuum with broken SUSY. By substituting the
approximate leading order solution for the moduli (\ref{app24})
into the definition (\ref{vol}) of $V_X$ we obtain:
\be\label{volume2} V_X=\left(\frac
1{2\pi}\right)^{7/3}\left[\frac{PQ}{Q-P}\ln\left(\frac{A_1Q}{A_2P}\right)\right]^{7/3}\prod_{i=1}^{N}
\left(\frac{3a_i}{7N_i}\right)^{a_i}\,. \ee Recalling the
definition of Im($f$) in terms of $\nu$ and using the solution for
$\nu$ (eqn.(\ref{app24})) together with (\ref{volume2}), the
gravitino mass for the dS vacuum in the leading order
approximation is given by: \be\label{gravitino3}
m_{3/2}=m_p\sqrt{2}\pi^3\,A_2 \left|\frac P Q \,\phi_0^{-\frac
2P}-1\right|\left(\frac {P\,Q}{Q-P}\ln\left(\frac
{A_1Q}{A_2P}\right) \right)^{-\frac 72}\left(\frac
{A_1Q}{A_2P}\right)^{-\frac{P}{Q-P}}\prod_{i=1}^{N} \left(\frac {7
N_i}{3 a_i}\right)^{\frac{3a_i}2}e^{\phi_0^2/2}\,, \ee where
$\phi_0^2$ is given by (\ref{e58}). Since $\phi_0^{-2/P} \sim 1$
from section \ref{chargedmattervac} and $A_2 \sim Q$, we see that
the expression for the gravitino mass for dS vacua is almost the
same as that for the AdS vacua (eqn. \ref{gr7}) provided we
replace $\rho$ in (\ref{gr7}) by $\tilde{\rho}=A_1Q/A_2P$ in
(\ref{gravitino3}) and $P-Q$ in (\ref{gr7}) by $Q-P$ in
(\ref{gravitino3}). Therefore, the results obtained for the
fraction of vacua with a gravitino mass in the $O(1-10)$ TeV range
also hold for this case.

%%%%%%%%%%%%%%%%%%%%%%%%%%%%%%%%%%%%%%%%%%%%%%%%%%%%%%%%%%%%%%%%%%

\subsubsection{Small Cosmological Constant implies Low-scale Supersymmetry in dS Vacua}\label{CClowsusy}

In this subsection we will study the distribution of SUSY breaking
scales in the de Sitter vacua which as we showed earlier, can
arise when the hidden sector has chiral matter. In particular we
will see that the requirement of a small cosmological constant
leads to a scale of SUSY breaking of ${\cal O}(1-100)$ TeV.

In section VI, we saw that the minimum obtained is de Sitter if
the discriminant of the quadratic polynomial with respect to
$\phi_0^2$ in eqn. (\ref{po45}) is negative, while it is anti-de
Sitter if the discriminant in (\ref{po45}) is positive. For
$m_{3/2}\sim {\mathcal O}(1-10\,{\rm TeV})$ the magnitude of the
vacuum energy in both cases can be estimated to be \be |V_0| \sim
m_{p}^2\,m_{3/2}^2\sim (10^{10} {\rm GeV})^4 - (10^{11} {\rm
GeV})^4\,. \ee

On the other hand, if the discriminant in (\ref{po45}) vanishes,
one obtains a vanishing cosmological constant (to leading order in
the approximation). At present it is not known if there is a
physical principle which imposes this condition. However, one can
still use it as an observational constraint since the observed
value of the cosmological constant is known to be extremely small.
For instance, if the space of $G_2$ manifolds scans the constants
$(A_i, P, Q, N)$ finely enough there could simply exist vacua for
which the vacuum energy is acceptably low.

By setting the left hand side in (\ref{e80}) to zero we can then
express
\begin{equation}\label{er765}
{P\ln\left(\frac{A_1Q}{A_2P}\right)}=\frac{28(Q-P)}{3(Q-P)-8}\,,
\end{equation}

Of course, since the above constraint was obtained in the leading
order, the vacuum energy is only zero in the leading order in our
analytic expansion. The subleading contributions we neglected in
(225) although smaller than the leading contributions, are still
much larger than the observed value of the cosmological constant.
However, one can {\em in principle} take into account all the
subleading corrections and tune the ratio $A_1Q/A_2P$ inside the
logarithm to set the vacuum energy to a very small value
compatible with the observations. As will be seen later, since the
expression in the R.H.S. of (\ref{er765}) turns out to be large,
the subleading corrections which affect the value of the
cosmological constant will have little effect on the
phenomenological quantities calculated by imposing the constraint
to leading order.

We would now like to analyze in detail the phenomenological
implications of the solutions obtained by imposing (\ref{er765})
as a constraint. The most important phenomenological quantity in
this regard is the gravitino mass as it sets the scale of all soft
supersymmetry breaking parameters. We focus on the gravitino mass
in this section. The soft supersymmetry breaking parameters will
be discussed in section \ref{pheno}.

%%%%%%%%%%%%%%%%%%%%%%%%%%%%%%%%%%%%%%%%%%%%%%%%%%
\subsubsection{Gravitino mass with a small positive cosmological constant}

By substituting the constraint (\ref{er765}) into the gravitino
mass formula (\ref{gravitino3}), we
obtain:\begin{equation}\label{gravitino76}
m_{3/2}=m_p\sqrt{2}\pi^3\,A_2 \left|\frac P {Q} \,\phi_0^{-\frac
2P}-1\right|\left(\frac {28Q}{3(Q-P)-8} \right)^{-\frac
72}e^{-\frac{28}{3(Q-P)-8}}\prod_{i=1}^{N} \left(\frac {7 N_i}{3
a_i}\right)^{\frac{3a_i}2}e^{\phi_0^2/2}\,,
\end{equation} where the meson vev is now given by: \begin{equation}\label{mesonvevconstr}
\phi_0^2\approx-\frac 18+\frac 1{Q-P}+\frac 1 4\sqrt{1-\frac
2{Q-P}}+\frac 2{Q-P}\sqrt{1-\frac 2{Q-P}}\,.
\end{equation} In this case, the moduli vevs are given by
\begin{equation}\label{app247}
s_i=\frac{a_i\nu}{N_i}\,,\,\,\,\,\,\,{\rm with}\,\,\,\,\,\,\,
\nu\approx\frac{6\,Q}{{\pi}(3(Q-P)-8)}\,.
\end{equation}

From (\ref{gravitino76}), one notes that the gravitino mass (when
the cosmological constant is made tiny) is completely determined
by the dual coxeter numbers of the hidden sector gauge groups
$N_c\,{\rm and}\,Q$, the rational numbers $a_i$ (see (\ref{vol}))
characterizing the volume of the $G_2$ manifold and the integers
$N_i$.\footnote{From field theory computations
\cite{Finnell:1995dr}, $A_1 = N_c-N_f=P$ and $A_2=Q$ (in a
particular RG scheme), up to threshold corrections. We can
therefore express $A_1$, $A_2$ as $A_1=P\,C_1$ and $A_2=Q\,C_2$,
where coefficients $C_1$ and $C_2$ depend {\em only} on the
threshold corrections and are constant with respect to the moduli
\cite{Friedmann:2002ty}. In this case, the quantity
$\ln(A_1Q/A_2P)=\ln(C_1/C_2)$ is fixed by imposing (\ref{er765}).}

The rationals $a_i$ are subject to the constraint $\sum_{i=1}^N
a_i = 7/3$. It is reasonable to consider a ``democratic'' choice
for $a_i$, $a_i=7/(3N)$ for all $i=\overline{1,N}$ and also to
take for simplicity all the integers $N_i=1$. The integers $N_i$
will generically be of ${\mathcal O}(1)$; even if some of the
$N_i$ are unnaturally large, their individual contributions will
be typically washed out as they are raised to powers that are much
less than unity (see (\ref{gravitino3}) and the expression for
$a_i$ for the democratic choice). In this case, after setting
$A_2=Q\,C_2$, the gravitino mass formula is given by
\begin{equation}\label{gravitino44}
m_{3/2}=m_p\sqrt{2}\pi^3 C_2\left|P \,\phi_0^{-\frac
2P}-Q\right|\left(\frac{N(3(Q-P)-8)} {28Q} \right)^{\frac
72}e^{-\frac{28}{3(Q-P)-8}}e^{\phi_0^2/2}\,,
\end{equation}and the moduli vevs are
\begin{equation}\label{app243}
s_i=\frac{14\,Q}{\pi\,N(3(Q-P)-8)}\,.
\end{equation} From (\ref{gravitino44}), the gravitino mass depends on just four
constants - $C_2$, $P$, $Q$ and $N$ (the total number of moduli),
determined by the topology of the manifold. It should be kept in
mind that for the solution to exist, it is necessary that $Q-P>2$
(see (\ref{e74})). For the smallest possible value
$(Q-P)_{min}=3$, the expression for $m_{3/2}$ simplifies even
further\begin{equation}\label{gravitino49}
m_{3/2}=m_p\sqrt{2}\pi^3 C_2\left|P(\phi_0^{-\frac
2P}-1)-3\right|\left(\frac N{28(P+3)} \right)^{\frac
72}e^{-{28}}e^{\phi_0^2/2}\approx m_p\,3\sqrt{2}\pi^3
C_2\left(\frac N{28(P+3)} \right)^{\frac
72}e^{-{28}}e^{\phi_0^2/2}\,.
\end{equation}together with\begin{equation}\label{e772}
\phi_0^2\approx\frac 1{72}\left(15+22\sqrt{3}\right)\approx
0.7376\,,\,\,\,\,\,\,s_i=\frac{14\,(P+3)}{{\pi\,N}}\,.\nonumber
\end{equation}
Note that the dependence on $N$ and $P$ in (\ref{gravitino49}) is
due solely to the volume $V_X$ dependence on those parameters. The
expression for the gravitino mass has a more transparent form if
we don't substitute the expression for the volume (\ref{volume2})
into (\ref{gr22}). For $Q-P=3$ we obtain \be\label{gr780}
m_{3/2}\approx m_p\,\frac
{3\,e^{\phi_0^2/2}}{8\sqrt{\pi}{V_X}^{3/2}}e^{-{28}}C_2 \approx
514\,{\rm TeV}\frac{C_2}{V_X^{3/2}}\,, \ee where the detailed
dependence on $a_i$, $N_i$, $P$ and the number of moduli $N$ is
completely encoded inside the seven-dimensional volume $V_X$ which
appears to be the more relevant physical quantity. Furthermore, in
the supergravity approximation the volume $V_X>1$, which
translates into an upper bound on the gravitino mass when $Q-P=3$
\be m_{3/2}< {\cal O}(100\,{\rm TeV})\,. \ee
%%%%%%%%%%%%%%%%%%%%%%%%%%%%%%%%%%%%%%%%%%%%%%%%%%%%%%%%%%%%%%%%%%%%%

\subsubsection{The Gravitino mass Distribution and its consequences}

The gravitino mass (\ref{gravitino44}) depends on three integers:
the two gauge group dual coxeter numbers $N_c$ , $Q$ and the
number of moduli $N$.\footnote{We can set $C_2=1$ for the order of
magnitude estimates we are doing here.} This will give us an idea
about the distribution of the gravitino mass (which sets the
superpartner masses) obtained after imposing the constraint
(\ref{er765}) that the vacuum energy is acceptable. Asides from
only considering the vacua within the supergravity approximation
(ie $s_i >1$) we expect an upper bound on the dual coxeter numbers
of the hidden sector gauge groups $P$ and $Q$. Based on duality
with the heterotic string, it seems reasonable to assume that they
can be at least as large as 30 - the dual coxeter number of $E_8$.
Of course, values of $P,Q$ larger than 30 cannot be ruled out, and
here we assume an upper bound $P \leq 100$.

The distribution can be constructed as follows. The three
integers: $P$, $Q-P$ and $N$ are varied subject to (\ref{e74}) and
the supergravity constraint $s_i>1$. For each point in the
resulting two dimensional subspace, $\log_{10}(m_{3/2})$ can be
computed and rounded off to the closest integer value. One can
then count how many times each integer value is encountered in the
entire scan and plot the corresponding distribution.
\begin{figure}[hbtp]
  \centerline{\hbox{ \hspace{0.0in}
    \epsfxsize=2.1in
    \epsffile{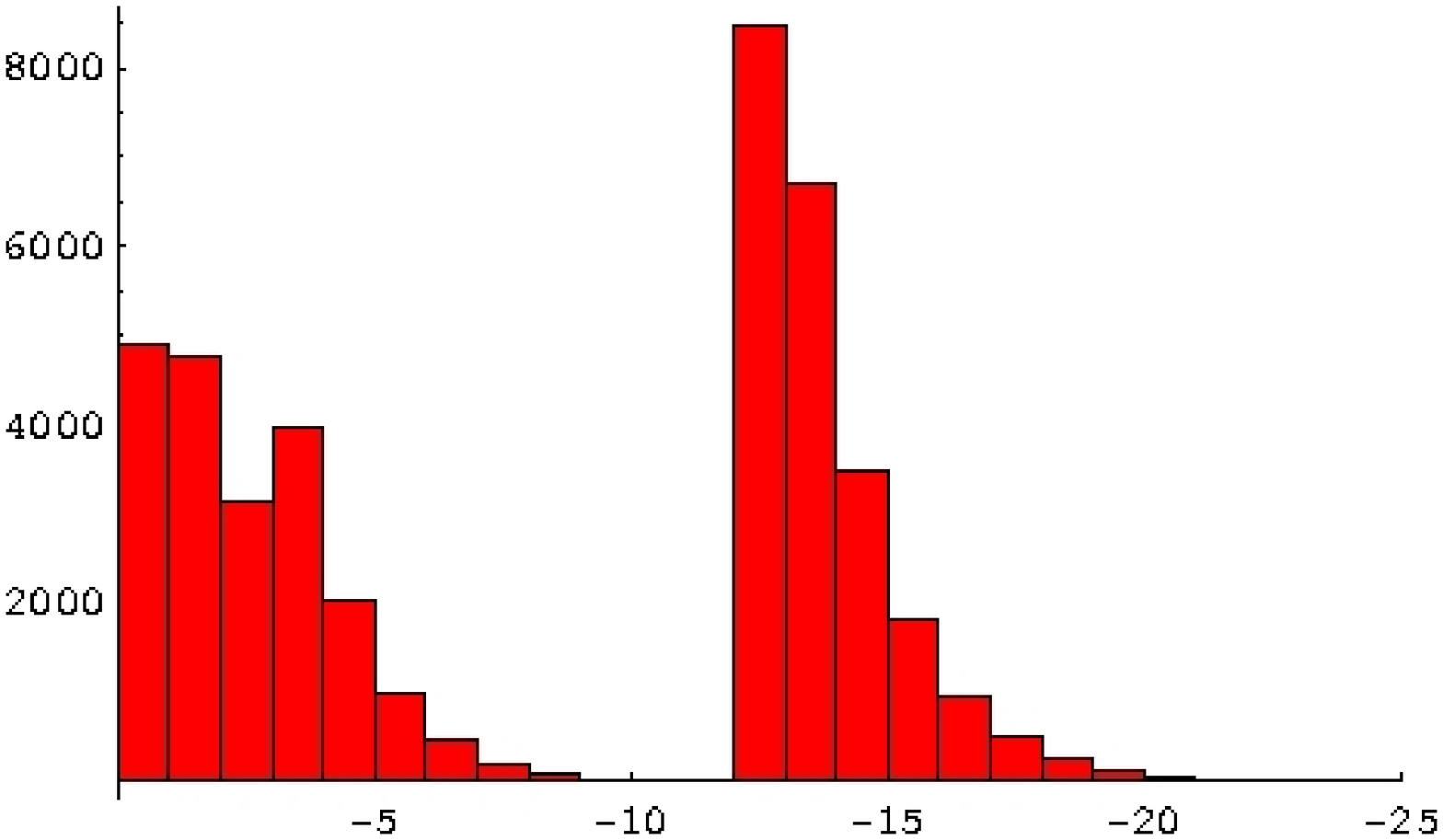}
    \hspace{0.0in}
    \epsfxsize=2.1in
    \epsffile{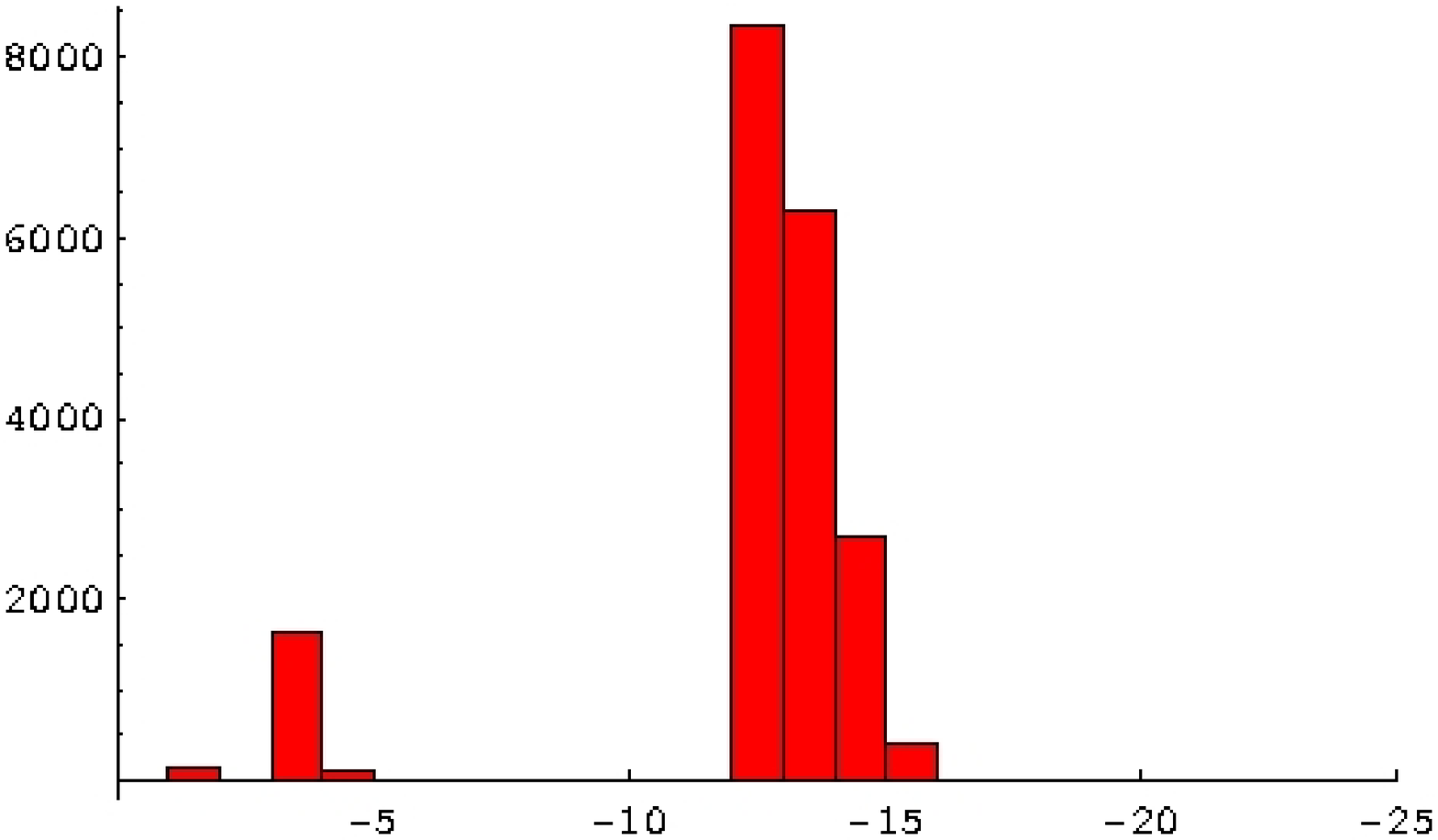}
    \hspace{0.0in}
    \epsfxsize=2.1in
    \epsffile{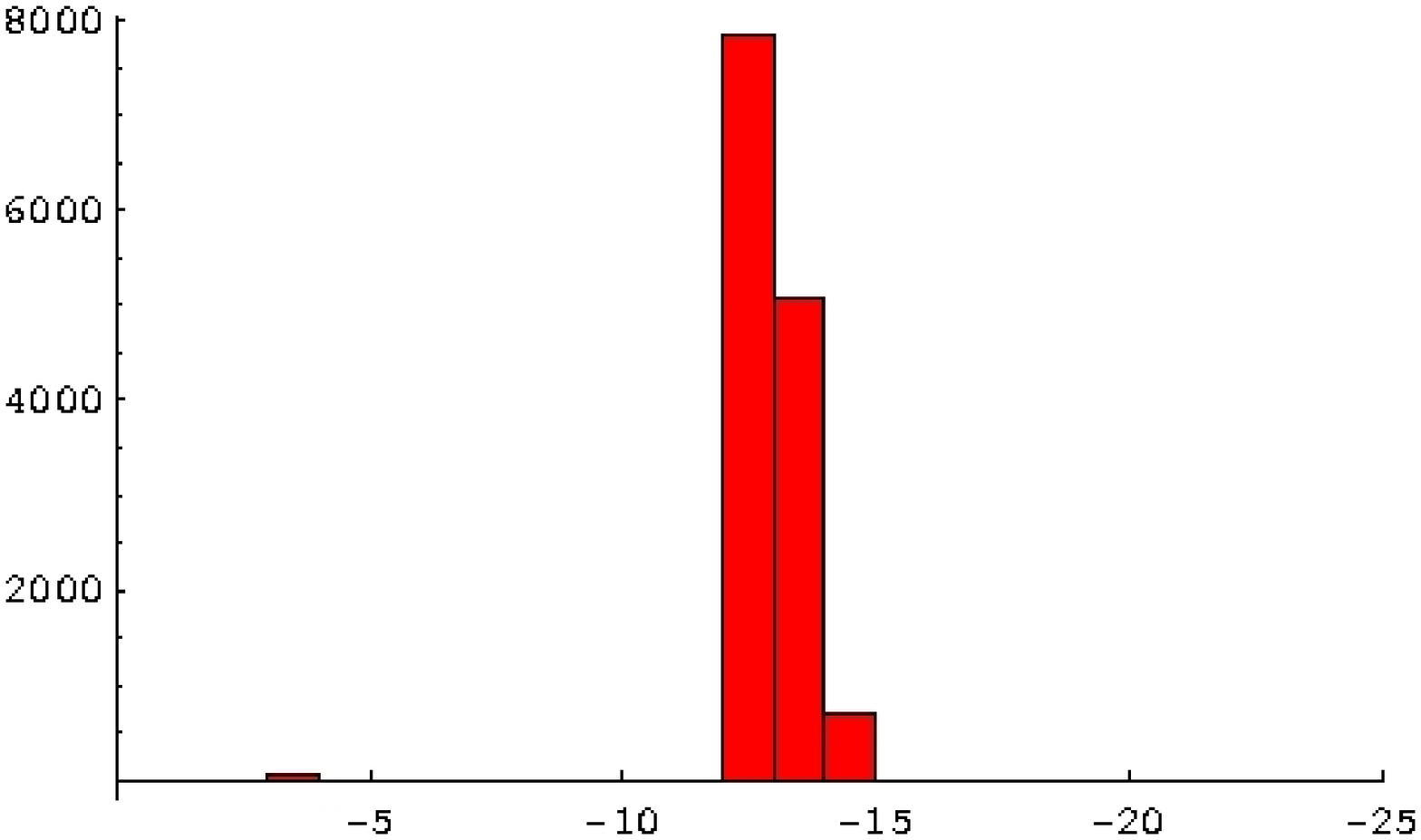}
    }
  }
\caption{The gravitino mass distribution with the x-axis denoting
the logarithm of the gravitino mass (to base 10). Left:
Distribution corresponding to scan one in (\ref{scan1}). Middle:
Distribution corresponding to scan two in (\ref{scan2}) for which
manifolds with the number of moduli $N<50$ were excluded from the
scan. Right: Distribution corresponding to scan three in
(\ref{scan3}) for which manifolds with the number of moduli
$N<100$ were excluded from the scan.}
    \label{Plot79}
\end{figure}
In the first three scans scan we cover a broad range of values by
choosing $P_{max}=100$. Taking into account the SUGRA constraint
($s_i>1$), we have the following ranges of integers for the first
scan: \begin{equation}\label{scan1} 3\leq P\leq
100\,;\,\,\,\,\,\,\,\, 3\leq(Q-P)\leq 100-P\,;\,\,\,\,\,\,\,\,
2\leq N<\frac{14\,(P+(Q-P))}{\pi\,(3(Q-P)-8)}\,.
\end{equation} In the second scan we have excluded the small N region
and considered only the manifolds with $N\geq50$. Thus we have the
following ranges of constants for the second scan:
\begin{equation}\label{scan2} 3\leq P\leq 100\,;\,\,\,\,\,\,\,\,
3\leq(Q-P)\leq 100-P\,;\,\,\,\,\,\,\,\, 50\leq
N<\frac{14\,(P+(Q-P))}{\pi\,(3(Q-P)-8)}\,.
\end{equation} In the third scan we have
only considered manifolds with $N\geq100$. Thus we have the
following ranges of integers for the second scan:
\begin{equation}\label{scan3} 3\leq P\leq 100\,;\,\,\,\,\,\,\,\,
3\leq(Q-P)\leq 100-P\,;\,\,\,\,\,\,\,\, 100\leq
N<\frac{14\,(P+(Q-P))}{\pi\,(3(Q-P)-8)}\,.
\end{equation} The first two distributions in Figure \ref{Plot79}
clearly have several prominent peaks. Amazingly, in all three
plots one of the peaks landed right in the $m_{3/2}\sim {\mathcal
O}(1-100)\,{\rm TeV}$ range! The high scale peaks on the left plot
appear to be around $m_{3/2}\sim 10^{14}\,{\rm GeV}$ and the GUT
scales. However, for the middle plot the GUT scale peak almost
disappears. Recall that the middle plot corresponds to scan two in
(\ref{scan2}) where we excluded all the manifolds for which the
number of moduli $N$ is less than $50$. Therefore, the high scale
peaks are largely dominated by contributions from the $G_2$
manifolds with a small number of moduli $N<50$. As seen from the
right plot, when $G_2$ manifolds with $N<100$ are excluded from
the scan, the peak at the $m_{3/2}\sim 10^{14}\,{\rm GeV}$ scale
has all but disappeared, whereas the peak at $m_{3/2}\sim
{\mathcal O}(1-100)\,{\rm TeV}$ remains virtually unchanged.

\begin{figure}[hbtp]
  \centerline{\hbox{ \hspace{0.0in}
    \epsfxsize=2.1in
    \epsffile{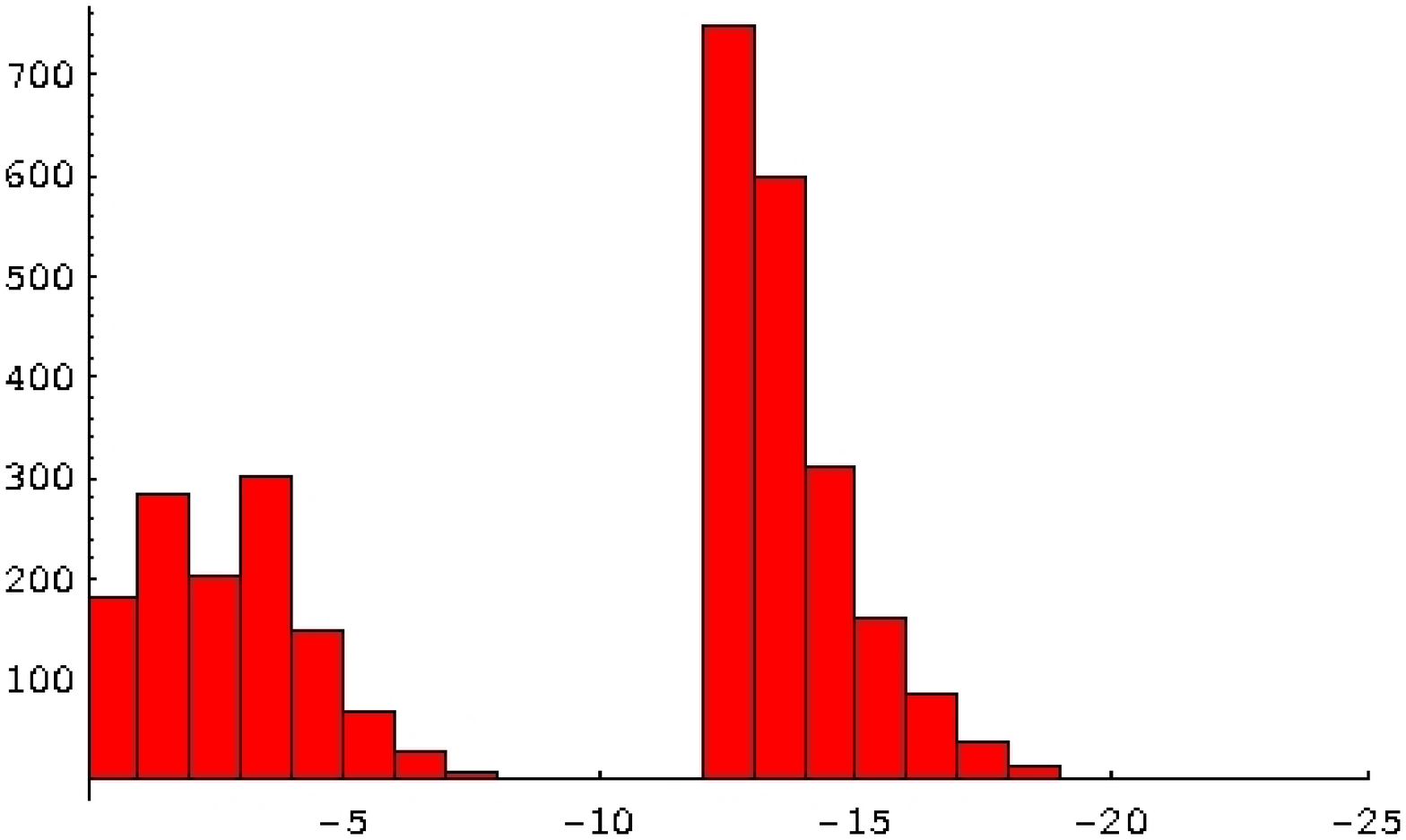}
    \hspace{0.0in}
    \epsfxsize=2.1in
    \epsffile{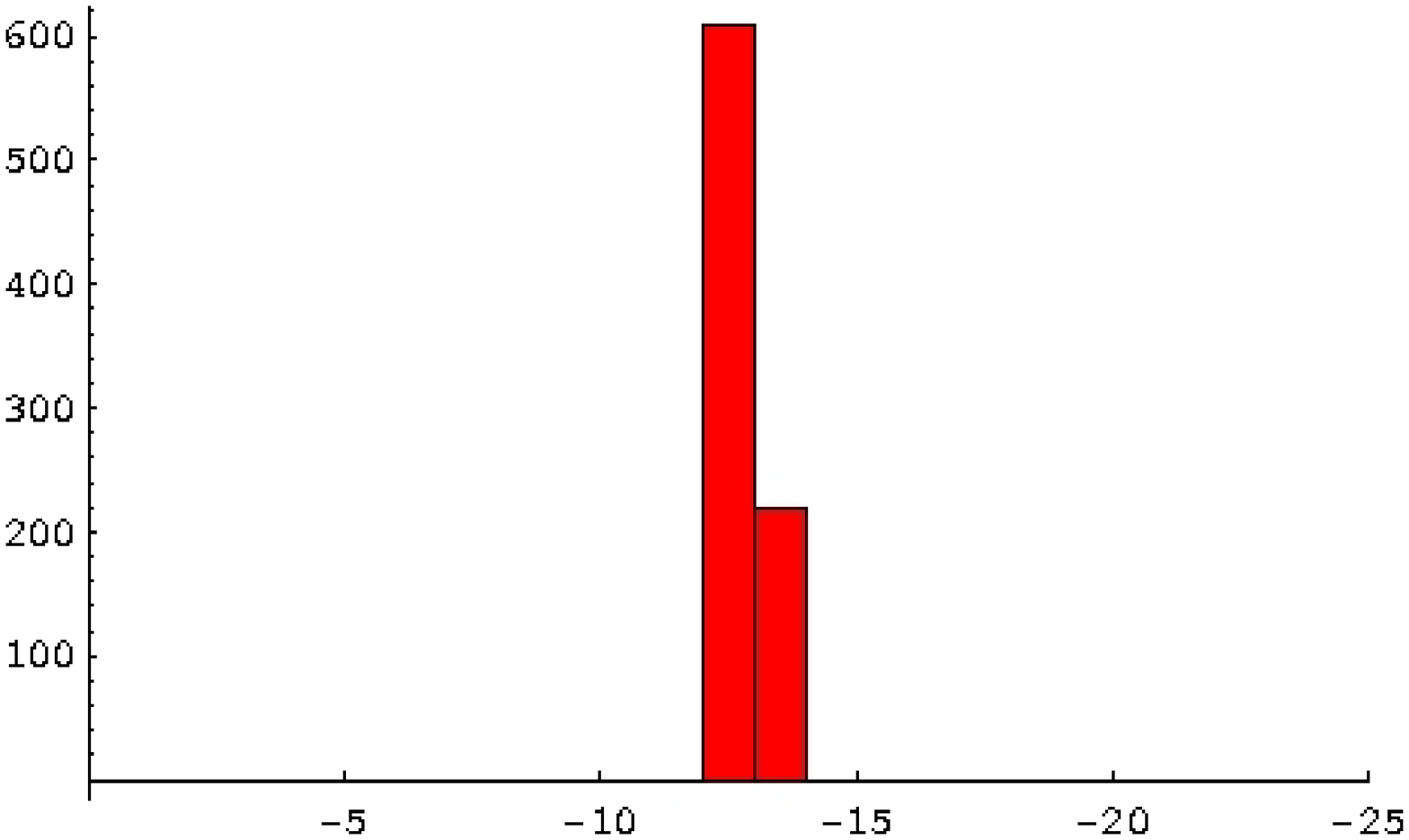}
    \hspace{0.0in}
    \epsfxsize=2.1in
    \epsffile{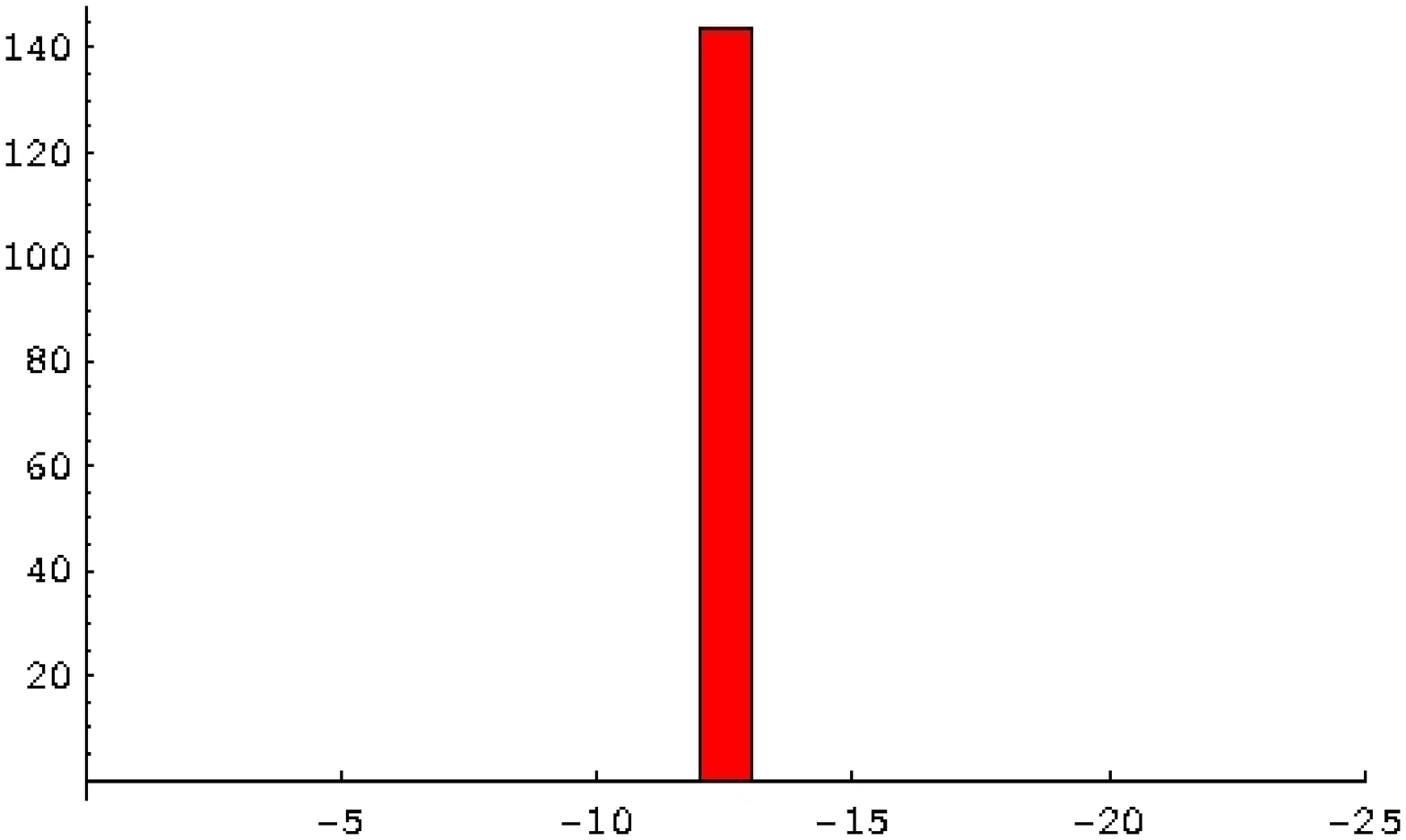}
    }
  }
\caption{The gravitino mass distribution with the x-axis denoting
the logarithm of the gravitino mass (to base 10). Left:
Distribution corresponding to scan four in (\ref{scan4}). Middle:
Distribution corresponding to scan five in (\ref{scan5}) for which
manifolds with the number of moduli $N<50$ were excluded from the
scan. Right: Distribution corresponding to scan six in
(\ref{scan6}) for which manifolds with the number of moduli
$N<100$ were excluded from the scan.} \label{Plot80}
\end{figure}

\begin{figure}[hbtp]
  \centerline{\hbox{ \hspace{0.0in}
    \epsfxsize=2.1in
    \epsffile{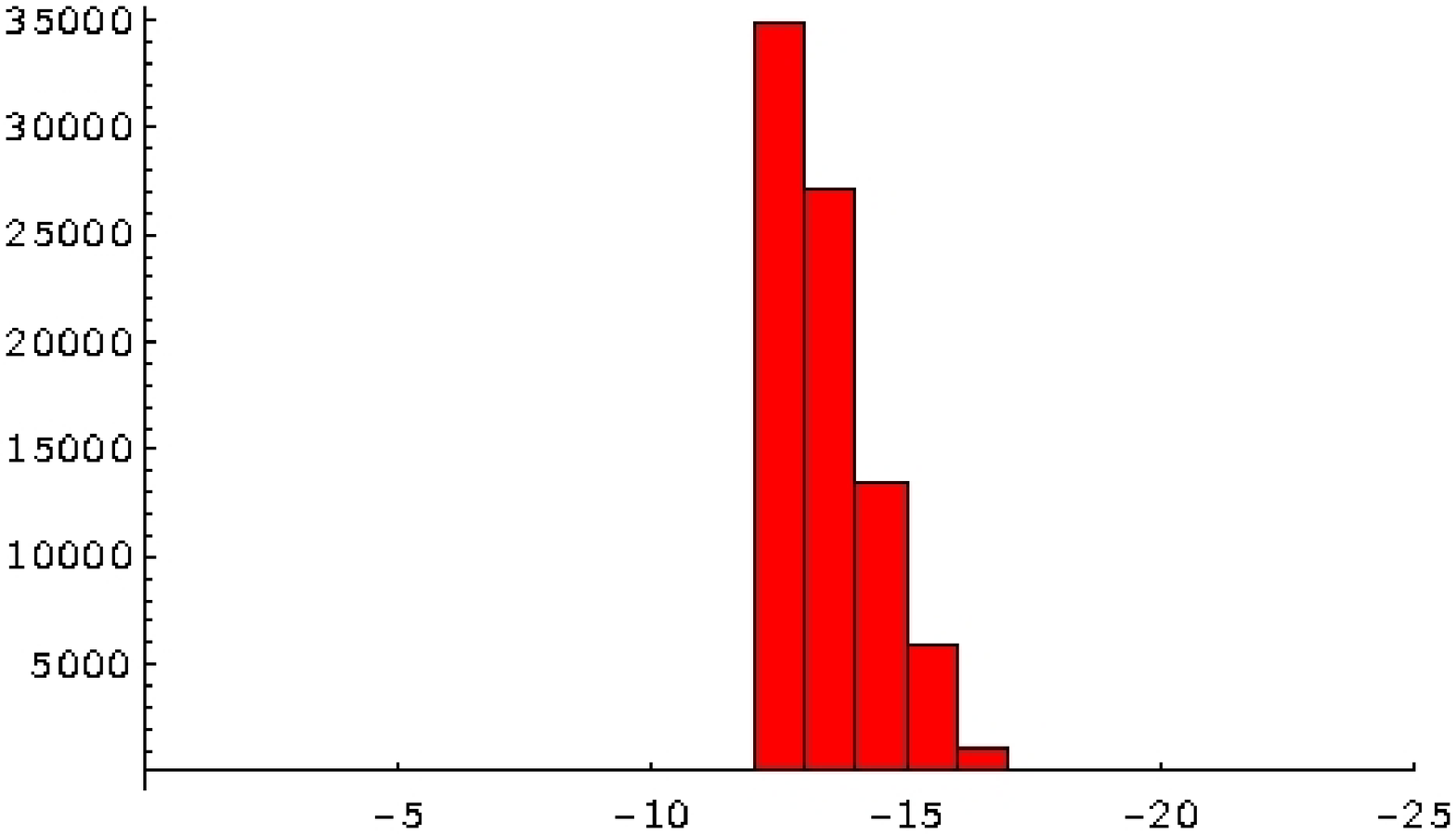}
    \hspace{0.0in}
    \epsfxsize=2.1in
    \epsffile{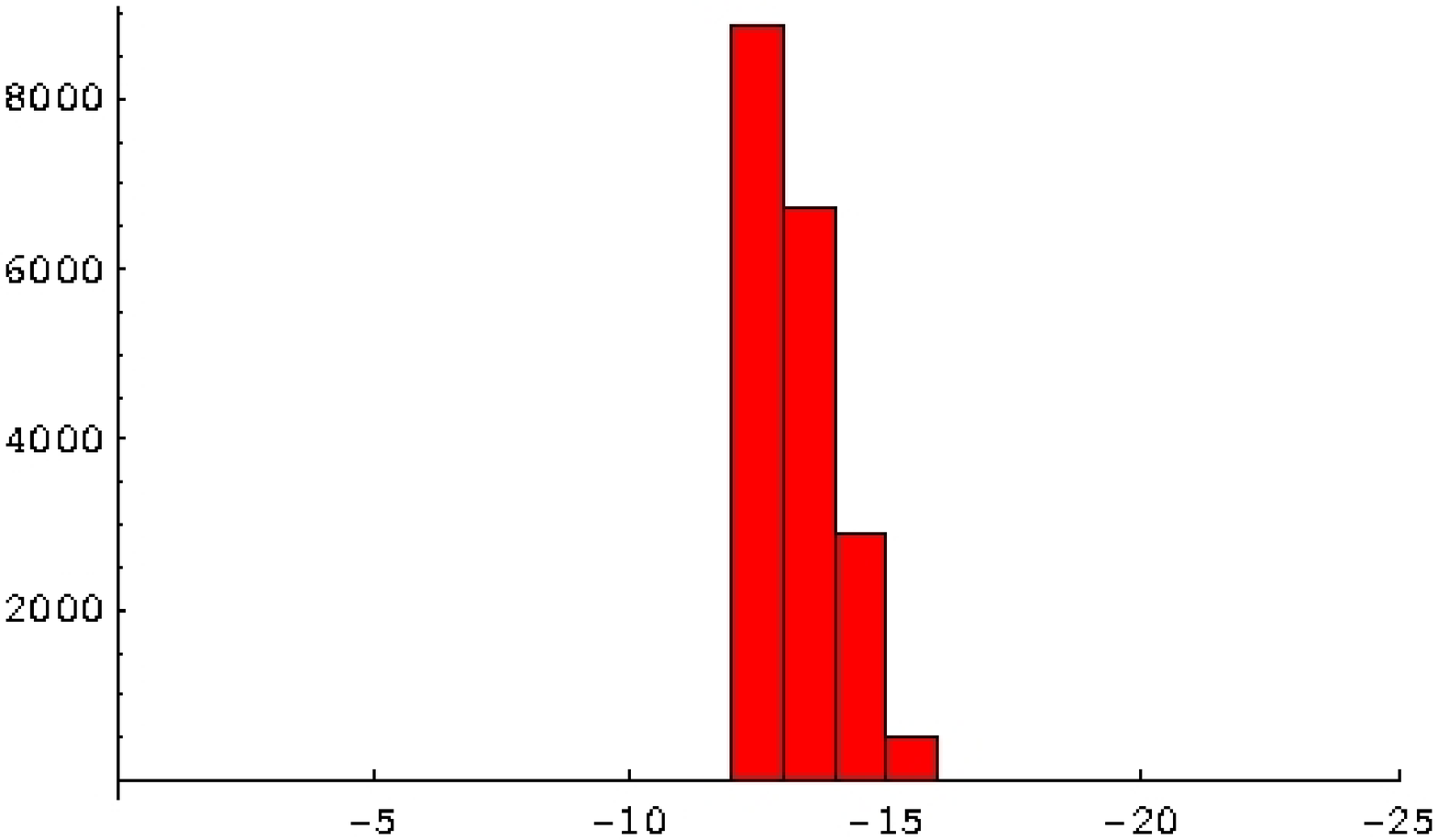}
    \hspace{0.0in}
    \epsfxsize=2.1in
    \epsffile{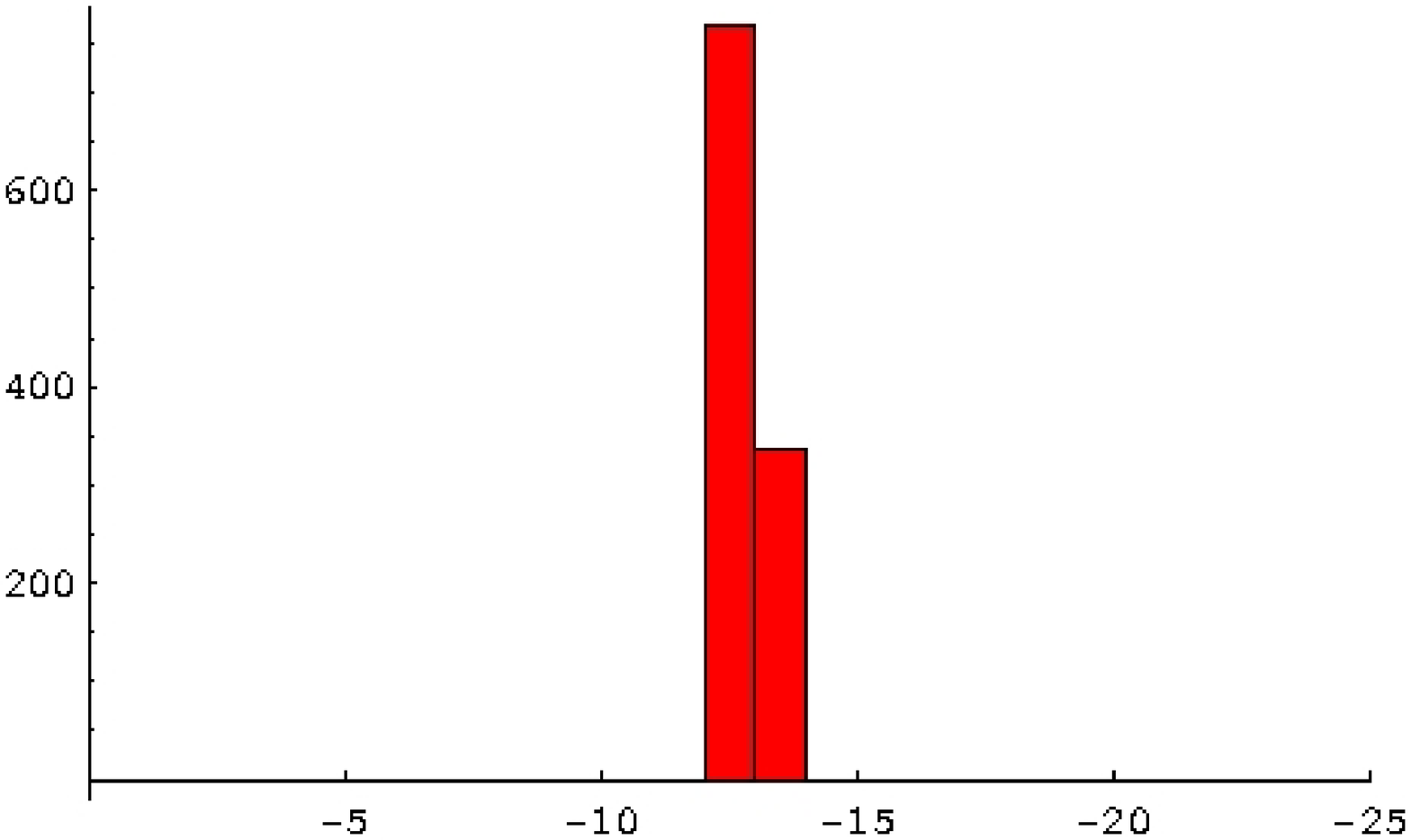}
    }
  }
\caption{The gravitino mass distribution with the x-axis denoting
the logarithm of the gravitino mass (to base 10). Scans for the
smallest possible choice $(Q-P)_{min}=3$. Left: Distribution
corresponding to the scan with $P_{max}=200$. Middle: Distribution
corresponding to the scan with $P_{max}=100$. Right: Distribution
corresponding to the scan with $P_{max}=30$.} \label{Plot81}
\end{figure}

In Figure \ref{Plot80} we included three more scans for which the
upper bound on $P$ was reduced to $P_{max}=30$. The fourth scan
has the following ranges:\begin{equation}\label{scan4} 3\leq P\leq
30\,;\,\,\,\,\,\,\,\, 3\leq(Q-P)\leq 30-P\,;\,\,\,\,\,\,\,\, 2\leq
N<\frac{14\,(P+(Q-P))}{\pi\,(3(Q-P)-8)}\,.
\end{equation} In the fifth scan we again excluded the small N region an
considered only the manifolds with $N\geq50$ and considered
$P_{max}=30$:
\begin{equation}\label{scan5}
3\leq P\leq 30\,;\,\,\,\,\,\,\,\, 3\leq(Q-P)\leq
30-P\,;\,\,\,\,\,\,\,\, 50\leq
N<\frac{14\,(P+(Q-P))}{\pi\,(3(Q-P)-8)}\,.
\end{equation} In the sixth scan we considered only the manifolds with $N\geq
100$ and $P_{max}=30$: \begin{equation}\label{scan6} 3\leq P\leq
30\,;\,\,\,\,\,\,\,\, 3\leq(Q-P)\leq 30-P\,;\,\,\,\,\,\,\,\,
100\leq N<\frac{14\,(P+(Q-P))}{\pi\,(3(Q-P)-8)}\,.
\end{equation}
Again, in Figure \ref{Plot80} we notice that the
${\mathcal O}(1-100)\,{\rm TeV}$ peak narrows around
$m_{3/2}\sim{\mathcal O}(100)\,{\rm TeV}$, as we exclude manifolds
with small number of moduli. As the same time, the peaks at the
high scale completely disappear for $G_2$ manifolds with $N>50$.
Finally, in Figure \ref{Plot81} we chose the smallest possible
value $Q-P=3$
and scanned integers $P$ and $N$ in the following ranges:
\begin{eqnarray}\label{scan7}
&&3\leq P\leq 200\,;\,\,\,\,\,50\leq N<\frac{14\,(P+3)}{\pi}\,,\\
&&3\leq P\leq 100\,;\,\,\,\,\,50\leq N<\frac{14\,(P+3)}{\pi}\,,\nonumber\\
&&3\leq P\leq 30\,;\,\,\,\,\,50\leq
N<\frac{14\,(P+3)}{\pi}\,.\nonumber
\end{eqnarray}
In all three plots in Figure \ref{Plot81} we see the same peak at
$m_{3/2}\sim {\mathcal O}(1-100)\,{\rm TeV}$, which narrows around
$m_{3/2}\sim{\mathcal O}(100)\,{\rm TeV}$ as $P_{max}$ is
decreased.

Therefore, from the above distributions we conclude that the peak
corresponding to $m_{3/2}\sim(1-100)\,{\rm TeV}$ is entirely due
to the smallest possible value $(Q-P)_{min}=3$. This can be
explained if we examine the gravitino mass formula in
(\ref{gravitino49}). In particular the constant factor
$e^{-28}\sim 10^{-12}$ is most crucial in lowering the gravitino
mass to the TeV scale. It is easy to trace the origin of this
factor to the constraint (\ref{er765}), imposed by the requirement
to have a zero cosmological constant (to leading order). When
(\ref{er765}) is used along with the requirement $Q-P=3$ we simply
get
\begin{equation}\label{ior} P\ln\left(\frac{A_1Q}{A_2P}\right)=84\,.
\end{equation}
When this is substituted into the gravitino mass
(\ref{gravitino3}), the corresponding suppression factor turns
into the constant
\begin{equation}\label{crucial}
\left(\frac {A_1Q}{A_2P}\right)^{-\frac{P}{Q-P}}=e^{-28}\,.
\end{equation}
Physically, this suppression factor corresponds to the hidden
sector gaugino condensation scale (cubed). Recall that for an
$SU(Q)$ hidden sector gauge group, the scale of gaugino
condensation is given by \be\label{gaugcond}
\Lambda_g=m_p\,e^{-\frac{8\pi^2}{3Q\,g^2}}=m_p\,e^{-\frac{2\pi}{3Q}{\rm
Im}f}\,. \ee The moduli vevs in (\ref{app247}) completely
determine the gauge kinetic function. Taking $Q-P=3$ we obtain
\be\label{gaugkinf} {\rm Im}f=\sum_{i=1}^N N_i
s_i=\frac{14\,Q}{\pi}. \ee Substituting (\ref{gaugkinf}) into
(\ref{gaugcond}) we obtain the following scale of gaugino
condensation \be \Lambda_g=m_p\,e^{-28/3}\approx 2.15\times
10^{14}\,{\rm GeV}\,. \ee It is important to note that the
expression in the R.H.S. of (\ref{er765}) is quite large ($=84$,
when $Q-P=3$) in the leading order, and the quantity ($A_1Q/A_2P$)
which is fixed by imposing the vacuum energy constraint is inside
a logarithm. Therefore, even when one incorporates all the higher
order corrections and tunes the ratio $A_1Q/A_2P$ inside the
logarithm to set the cosmological constant equal to the observed
value, the constant on the R.H.S. (=84), crucial in obtaining the
${\mathcal O}(100)\,{\rm TeV}$ scale peak, is hardly affected.

\begin{figure}[hbtp]
  \centerline{\hbox{ \hspace{0.0in}
    \epsfxsize=4.0in
    \epsfbox{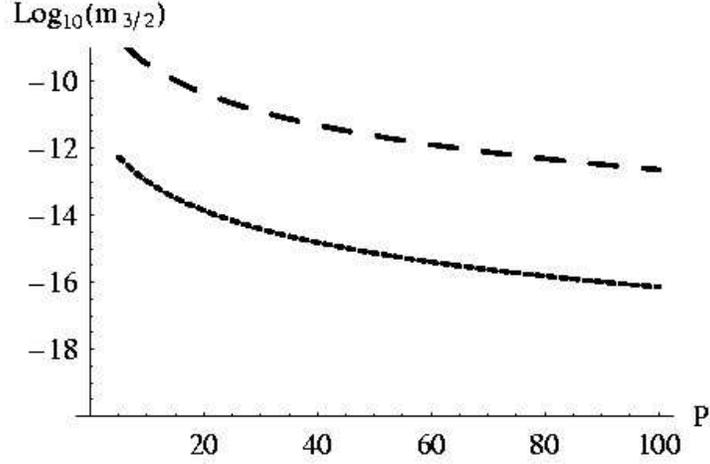}
    }
  }
\caption{Plot of $\log_{10}(m_{3/2})$ as a function of $P$ for
$Q-P=3$. Short-dashed curve corresponds to $N=50$. Long-dashed
curve corresponds to $N=500$.} \label{Plot82}
\end{figure}

The dominance of the ${\mathcal O}(1-100)\,{\rm TeV}$ range also
becomes clear from Figure \ref{Plot82}, where $\log_{10}(m_{3/2})$
as a function of $P$ for $Q-P=3$ is plotted for a manifold with
$N=50$ moduli - short-dashed curve, and a manifold with $N=500$
moduli - long-dashed curve.
Indeed, even when we do not impose the SUGRA constraint, from the
above plot we can see that the ${\mathcal O}(1-100)\,{\rm TeV}$
range is covered by a large swath of the vacuum space and it is
not so surprising that the corresponding distribution peaks at
that scale. This essentially follows from the formula for the
gravitino mass in (\ref{gravitino49}).

An important point which should be emphasized is that for $Q-P=3$,
the gravitino mass dependence on $P$ and $N$ appears {\em only}
through the volume $V_X$, as can be seen from (\ref{gr780}). Thus,
the distribution in Figure \ref{Plot81} directly correlates with
the corresponding distribution of the stabilized volume of the
seven-dimensional manifold $V_X$ as a function of $P$ and $N$.
Therefore, it is the dominance of the vacua with a relatively
small volume, which results in the peak at ${\mathcal
O}(100)\,{\rm TeV}$.

Also note that in the above analysis we simply set the constant
coefficient due to the threshold corrections $C_2$ to unity. It
would be interesting to get a handle on this quantity and include
its variation into the gravitino mass distribution study.

One could argue that even though $Q-P=(Q-P)_{min}=3$ gives a peak
for the gravitino mass distribution at around ${\mathcal
O}(100)\,{\rm TeV}$ scale, it seems plausible from a theoretical
point of view to have many examples of gauge singularities in
$G_2$ manifolds such that $Q-P>3$. However, by imposing the
supergravity constraint that all moduli $s_i$ are larger than
unity (which is the regime in which the entire analysis is valid),
one sees that having $Q-P>3$ drastically reduces the upper bound
on $N$ compared to that for $Q-P=3$ (see eqn(\ref{app247})).
Therefore, the peaks in the gravitino mass distribution obtained
for $-2\leq \log_{10}(m_{3/2})\leq-5$ in Figures \ref{Plot79} and
\ref{Plot80} come from vacua with a \emph{small} number of moduli
as well as $Q-P>3$, compatible with the analysis in the
supergravity regime. Since it is presumably true that the number
of $G_2$ manifolds with the required gauge singularities which
have a large number of moduli is much larger than those with a
small number of moduli, it seems reasonable to expect that the
peak of the gravitino mass distribution obtained at around
${\mathcal O}(100)\,{\rm TeV}$ scale is quite robust and is
representative of the most generic class of $G_2$ manifolds with
the appropriate gauge singularities.

One could also contrast the results obtained above with those
obtained for the Type IIB flux vacua. In Type IIB flux vacua, one
has to {\em independently} tune both the gravitino mass to a TeV
scale (if one requires low scale supersymmetry to solve the
hierarchy problem) as well as the cosmological constant to its
observed value. This is quite different to what we are finding
here.

Finally, we should emphasize that imposing the supergravity
approximation was crucial in obtaining low scale SUSY breaking.
Plausibly, the vacua which exist in the $M$ theoretic, small
volume regime will have a much higher SUSY breaking scale.
However, such vacua presumably also have the incorrect electroweak
scale (either zero or $M_{11}$).
%%%%%%%%%%%%%%%%%%%%%%%%%%%%%%%%%%%%%%%%%%%%%%%%
%%%%%%%%%%%%%%%%%%%%%%%%%%%%%%%%%
%%%%%%%%%%%%%%%%%%%%%%%%%%%%%%%%%

\subsubsection{Including more than one flavor of quarks in the hidden sector}

In the previous analysis we assumed a single flavor for the quarks
in the hidden sector, i.e $N_f=1$. In order to check that the way
we obtain a dS metastable minimum is robust and not dependent on a
particular choice of chiral hidden sector matter spectrum, we
would like to extend our analysis to include more than one flavor
so that the meson fields are given by: \be \phi^{\,\sigma}_{\,\bar
\sigma}\equiv \left(Q^{\sigma}\tilde Q_{\bar
\sigma}\right)^{1/2}\,, \ee where $\sigma,\bar
\sigma=\overline{1,N_f}$. It is convenient to keep the effective
K\"{a}hler potential for the hidden sector mesons in canonical
form. For this purpose we assume that the flavor space matrix
$\phi^{\,\sigma}_{\,\bar \sigma}$ is diagonal, i.e. \ba
&&<Q^{\sigma}\tilde Q_{\bar \sigma}>=0,\,\,\,\,\,\,\,{\rm for}\,\, \sigma\not =\bar \sigma\\
&&<Q^{\sigma}\tilde Q_{\bar \sigma}>\not=0,\,\,\,\,\,\,\,{\rm
for}\,\, \sigma =\bar \sigma\,.\nonumber \ea In this case, the
determinant appearing in (\ref{mattersup}) is simplified \be {\rm
det}(\phi^{\,\sigma}_{\,\bar
\sigma})=\prod_{\sigma=1}^{N_f}{\phi_\sigma}\,,\,\,\,\,\,\,{\rm
where} \,\,\,\, {\phi_\sigma}\equiv\phi^{\,\sigma}_{\,\sigma}\,.
\ee Thus, the nonperturbative superpotential and the K\"{a}hler
potential are then given by \ba \label{mattersuptwosectors45}
&&W=A_1\prod_{\sigma=1}^{N_f}{\phi_\sigma}^a\,e^{ib_1\,f}+A_2e^{ib_2\,f}\,\\
&&K =
-3\ln(4\pi^{1/3}\,V_X)+\sum_{\sigma=1}^{N_f}\phi_{\sigma}\bar\phi_{\sigma}\,,
\nonumber \ea where we again denoted $b_1\equiv 2\pi/P$,
$b_2\equiv 2\pi/Q$, $P\equiv N_c-N_f$ and $a\equiv -2/P$. After
minimizing with respect to the axions, the scalar potential is
given by \ba \label{potential_noaxions56}
V&=&\frac{e^{\sum_{\sigma=1}^{N_f}{(\phi_0)_{\sigma}^2}}}{48\pi
V_X^3}\,
[(b_1A_1\prod_{\sigma=1}^{N_f}(\phi_0)_{\sigma}^ae^{-b_1\vec\nu\cdot\,\vec
a}- b_2A_2e^{-b_2\vec\nu\cdot\,\vec
a})^2\sum_{i=1}^{N}a_i({\nu_i})^2+\\
&&
3(A_1\prod_{\sigma=1}^{N_f}(\phi_0)_{\sigma}^ae^{-b_1\vec\nu\cdot\,\vec
a}-A_2e^{-b_2\vec\nu\cdot\,\vec a})^2+3(\vec\nu\cdot\,\vec
a)(A_1\prod_{\sigma=1}^{N_f}(\phi_0)_{\sigma}^ae^{-b_1\vec\nu\cdot\,\vec
a}-A_2e^{-b_2\vec\nu\cdot\,\vec a})\nonumber \\
&
&(b_1A_1\prod_{\sigma=1}^{N_f}(\phi_0)_{\sigma}^ae^{-b_1\vec\nu\cdot\,\vec
a}-b_2A_2e^{-b_2\vec\nu\cdot\,\vec a}) +\sum_{\gamma=1}^{N_f}
\frac 3 4{(\phi_0)_{\gamma}^2}\times \nonumber
\\ && ((\frac{a}{(\phi_0)_{\gamma}^2}+1)A_1\prod_{\sigma=1}^{N_f}(\phi_0)_{\sigma}^a
e^{-b_1\vec\nu\cdot\,\vec a}-A_2e^{-b_2\vec\nu\cdot\,\vec
a})^2]\,.\nonumber \ea Instead of presenting a full analysis of
this more general case we would simply like to check that we have
a metastable dS vacuum, and that the main feature of the dS
vacuum, namely the emergence of the TeV scale when the tree-level
cosmological constant is set to zero, survives when $N_f>1$. For
this purpose we need to compute the scalar potential at the
minimum with respect to the moduli $s_i$ as a function of the
meson fields $\phi_{\sigma}$.

The generalization of the equations minimizing the scalar
potential is fairly straightforward. In particular, in the limit
when the size of the associative cycle ${\rm
Im}f=\vec\nu\cdot\,\vec a$ is large, for $A=1$ the generalization
of the second equation in (\ref{e47}), which determines $\tilde
L_{1,+}$ takes on the following form \be \label{e479} \frac
23\left(\tilde L_{1,+}\right)^2+\tilde
L_{1,+}+\sum_{\sigma=1}^{N_f}\frac {3{a\beta b_1}\hat y} {14 \hat
x\hat z} \left(\frac{a\beta} {(\phi_0)_{\sigma}^2\,\hat
x}+1\right)=0\, \nonumber \ee where we again defined $\beta\equiv
\frac{A_1}{A_2}\prod_{\sigma=1}^{N_f}(\phi_0)_{\sigma}^ae^{-(b_1-b_2)\vec\nu\cdot\,\vec
a}$, $\hat x\equiv \beta -1$, $\hat y\equiv b_1\beta -b_2$ and
$\hat z\equiv b_1^2\beta -b_2^2$. Thus, in the large three-cycle
limit we again have $\beta\approx b_2/b_1=P/Q$ so that $\hat
y\rightarrow 0$ and the leading order solution for $L_{1,+}$ is
again given by \be\label{lplus} L_{1,+}\approx -\frac 3 2\,. \ee
In this case the moduli are stabilized at the same values given by
(\ref{app24}). Since both the superpotential and the K\"{a}hler
potential are completely symmetric with respect to the meson
fields, it seems reasonable to expect that there is a vacuum where
all $\phi_{\sigma}$ are stabilized at the same value, i.e.
$(\phi_0)_{\sigma}=\tilde\phi_0$ for all
$\sigma=\overline{1,N_f}$. Using the solution for the moduli vevs
(\ref{app24}) and the above assumption we obtain the following
expression for the potential at the extremum with respect to the
moduli $s_i$ as a function of $\tilde\phi_0$ \be\label{po4590}
V_0=N_f\frac {(A_2\hat x)^2}{64\pi V_X^3}\left[\tilde\phi_0^4+
\left(\frac {2\,a\beta}{\hat x}-\frac
3{N_f}\right)\tilde\phi_0^2+\left(\frac {a\beta}{\hat
x}\right)^2\right]\frac
{e^{N_f\tilde\phi_0^2}}{\tilde\phi_0^2}\left(\frac
{A_1Q}{A_2P}\right)^{-\frac{2P}{Q-P}}\,. \ee

By setting the discriminant of the biquadratic polynomial in the
square brackets to zero we again obtain the leading order
condition on the tree-level cosmological constant to vanish:
\begin{equation}\label{e800}
\frac 3{N_f}-\frac
8{Q-P}-\frac{28}{P\ln\left(\frac{A_1Q}{A_2P}\right)}=0\,.
\end{equation}
Since the solutions for the moduli in the dS case correspond to
branch b) where $Q>P$ and ${A_1Q}>{A_2P}$, zero vacuum energy
condition (\ref{e800}) can be satisfied only when \be \frac
3{N_f}>\frac 8{Q-P}\,\,\,\,\Rightarrow\,\,\,\, (Q-P)>\frac 8
3N_f\,. \ee Therefore, a vanishing tree-level cosmological
constant in the leading order results in the following set of
conditions: \be\label{fr765}
{P\ln\left(\frac{A_1Q}{A_2P}\right)}=\frac{28(Q-P)N_f}{3(Q-P)-8N_f}\,,\,\,\,\,{\rm
and}\,\,\,\,\,(Q-P)>\frac 8 3N_f\,. \ee Recall that the key to
obtaining the TeV scale gravitino mass was the exponential
suppression factor $e^{-28}$ when $Q-P=(Q-P)_{min}=3$, related to
the scale of gaugino condensation. In the present case, up a
factor of order one, we have \be m_{3/2}\sim \frac
{m_p}{V_X^{3/2}}\left(\frac{A_1Q}{A_2P}\right)^{-\frac P{Q-P}}=
\frac {m_p}{V_X^{3/2}}e^{-\frac{28 N_f}{3(Q-P)-8N_f}} \ee Consider
a few examples where $N_f>1$. From (\ref{fr765}) we have the
following set \ba\label{exfl}
&&N_f=2,\,\,\,\,\,(Q-P)_{min}=6,\,\,\,\,\,{P\ln\left(\frac{A_1Q}{A_2P}\right)}=
168,\,\,\,\,\,\,\,\,m_{3/2}\sim\frac {m_p}{V_X^{3/2}}e^{-\frac{28
N_f}{3(Q-P)-8N_f}}=
\frac {m_p}{V_X^{3/2}}e^{-{28}} \\
\nonumber\\
&&N_f=3,\,\,\,\,\,(Q-P)_{min}=9,\,\,\,\,\,{P\ln\left(\frac{A_1Q}{A_2P}\right)}=252,
\,\,\,\,\,\,\,\,m_{3/2}\sim\frac {m_p}{V_X^{3/2}}e^{-\frac{28
N_f}{3(Q-P)-8N_f}}=
\frac {m_p}{V_X^{3/2}}e^{-{28}}\nonumber\\
\nonumber\\
&&N_f=4,\,\,\,\,\,(Q-P)_{min}=11,\,\,\,\,\,{P\ln\left(\frac{A_1Q}{A_2P}\right)}=1232,
\,\,\,\,\,\,\,\,m_{3/2}\sim\frac {m_p}{V_X^{3/2}}e^{-\frac{28
N_f}{3(Q-P)-8N_f}}=
\frac {m_p}{V_X^{3/2}}e^{-{112}}\nonumber\\
\nonumber\\
&&N_f=4,\,\,\,\,\,\,\,\,\,\,\,\,Q-P=12,\,\,\,\,\,\,\,\,\,\,\,\,
{P\ln\left(\frac{A_1Q}{A_2P}\right)}=336,\,\,\,\,\,\,\,\,m_{3/2}
\sim\frac {m_p}{V_X^{3/2}}e^{-\frac{28 N_f}{3(Q-P)-8N_f}}=\frac
{m_p}{V_X^{3/2}}e^{-{28}}\nonumber \ea

Remarkably, in all but one cases listed above we obtain {\em the
same} suppression factor $e^{-28}\approx 7\times 10^{-13}$ which
was the reason for the peak at $m_{3/2}\sim{\cal O}(1-100) {\rm
TeV}$! Note that the only example which did not fall into this
range was the third case for which the condition on the
cosmological constant to vanish was
${P\ln\left(\frac{A_1Q}{A_2P}\right)}=1232$, which is too
unrealistic anyway, as it requires either extremely large dual
coxeter numbers for the gauge groups $N_c,Q\sim{\cal O}(1000)$ or
an exponentially large ratio inside the logarithm. On a similar
note, as can be seen from the third entry in each line in
(\ref{exfl}), increasing the number of flavors $N_f$ even further
would again require either $P,Q\,>300$ or an extremely large ratio
$\left(\frac{A_1Q}{A_2P}\right)$, which appears inside the
logarithm. Therefore, limiting our analysis to the cases with
$N_f<5$ seems quite reasonable.

Recall that for $N_f=1$ the TeV scale appeared for the minimum
value $(Q-P)_{min}=3$ whereas the vacua corresponding to the
higher values of $Q-P$ generally failed to satisfy the SUGRA
constraint for more generic $G_2$ manifolds with a large number of
moduli. For this reason, considering larger values of $Q-P$ for
the examples listed above is probably unnecessary.

Thus, given that the assumptions we made in the beginning of this
subsection are reasonable, it appears that the connection of the
TeV scale SUSY breaking to the requirement that the tree-level
vacuum energy is very small is a fairly robust feature of these
vacua, independent of the number of flavors.

%%%%%%%%%%%%%%%%%%%%%%%%%%%%%%%%%%%%%%%%%%%%%%%%
%%%%%%%%%%%%%%%%%%%%%%%%%%%%%%%%%
%%%%%%%%%%%%%%%%%%%%%%%%%%%%%%%%%

\subsubsection{Including matter in both hidden sectors}
In the previous analysis we tried to be minimalistic and included
chiral matter in only one of the hidden sectors. Due to this
asymmetry, we obtained two types of solutions - a supersymmetric
AdS extremum when $P>Q$ corresponding to branch a) and a dS
minimum when $Q>P$ (when condition (\ref{e80} holds),
corresponding to branch b). Using this result it is then fairly
straightforward to figure out what happens when both hidden
sectors produce F-terms due to chiral matter. For the sake of
simplicity, we will again consider the case when $N_f=1$ in both
hidden sectors. In this case, the K\"{a}hler potential is given by
\be K = -3\ln(4\pi^{1/3}\,V_X)+\phi\bar\phi+\psi\bar\psi\,. \ee
After minimizing with respect to the axions, the non-perturbative
superpotential (up to a phase) is given by \ba
\label{mattersuptwosectors093}
W=-A_1{\phi}^{a_1}\,e^{-\frac{2\pi}P\,{\rm
Im}f}+A_2{\psi}^{a_2}\,e^{-\frac{2\pi}Q\,{\rm Im}f}\,, \ea where
$a_1\equiv-2/P$ and $a_2\equiv-2/Q$. We will now check to see if
it is still possible to obtain SUSY extrema when both hidden
sectors have chiral matter. Setting the moduli $F$-terms to zero
we obtain \be\label{nueq02} \nu_k=\nu=-\frac
{3\,PQ}{4\pi}\frac{\tilde\beta-1}{Q\tilde\beta-P}\,, \ee where
$\tilde\beta\equiv \frac{A_1{\phi}^{a_1}}{A_2{\psi}^{a_2}}
\,e^{-(\frac{2\pi}{P}-\frac{2\pi}Q)\,{\rm Im}f}$. At the same
time, setting the matter $F$-terms to zero results in the
following conditions: \be\label{mescon01}
\left(\frac{a_1}{\phi_0^2}+1\right)\tilde\beta-1=0\,. \ee
\be\label{mescon02} -\frac{a_2}{\psi_0^2}+\tilde\beta-1=0\,. \ee
Expressing $\tilde\beta$ from (\ref{nueq02}) and substituting it
into (\ref{mescon01}-\ref{mescon02}) and using the definitions for
$a_1$ and $a_2$, we obtain the following expressions for the meson
field vevs: \ba\label{phisol01}
&&\phi_0^2=\frac{2+3Q/(2\pi\nu)}{P-Q}\,,\,\\
&&\psi_0^2=\frac{2+3P/(2\pi\nu)}{Q-P}\,. \ea Since $\nu$ as well
as both $\phi_0^2$ and $\psi_0^2$ are positive definite, we have
the following two possibilities: \ba
&&a)\,\,\,\,\,P>Q:\,\,\,\,\,\,\Rightarrow\,\,\,\,\,\,F_{\phi}=0\,\,\,{\rm and}\,\,\,\,F_{\psi}\not=0,\,\\
&&b)\,\,\,\,\,P<Q:\,\,\,\,\,\,\Rightarrow\,\,\,\,\,\,F_{\phi}\not=0\,\,\,{\rm
and}\,\,\,\,F_{\psi}=0.\,\nonumber \ea Thus, when both hidden
sectors have chiral matter, supersymmetric extrema are absent.
Instead, when condition (\ref{e80}) holds (for branch a) we simply
swap $P$ and $Q$, $A_1$ and $A_2$ in (\ref{e80})), for each branch
we obtain a dS vacuum where only one of the matter $F$-terms is
non-zero. Keep in mind that although in the above analysis we used
condition (\ref{nueq02}) obtained by setting the moduli $F$-terms
to zero, even in the dS case when the moduli $F$-terms are
non-zero, one of the two mesons will be stabilized at a value such
that the corresponding matter $F$-term is zero. The zero $F$-term
has no effect on the analysis of the dS solution and apart from
replacing $\tilde\alpha$ with $\tilde\beta$ defined above, the
same solution obtained previously for the dS vacuum applies. In
this case, the only difference will be in the meson field vevs:
\ba a)\,\phi_0^2\approx\frac 2{P-Q}+\frac
7{P\,\ln\left(\frac{A_2P}{A_1Q}\right)},\,
\psi_0^2&\approx&1-\frac 2{P-Q}+\sqrt{1-\frac 2{P-Q}}- \nonumber\\
&&\frac
7{Q\,\ln\left(\frac{A_2\,P}{A_1 Q}\right)}\left(\frac 32+\sqrt{1-\frac 2{P-Q}}\right)\,\\
b)\,\psi_0^2\approx\frac 2{Q-P}+\frac
7{Q\,\ln\left(\frac{A_1Q}{A_2P}\right)},
\,\phi_0^2&\approx&1-\frac 2{Q-P}+\sqrt{1-\frac 2{Q-P}}- \nonumber
\\ && \frac 7{P\,\ln\left(\frac{A_1\,Q}{A_2 P}\right)}\left(\frac 3
2+\sqrt{1-\frac 2{Q-P}}\right)\nonumber \ea Therefore, the dS
solution obtained for the minimal case when only one of the hidden
sectors has chiral matter does not change even when we include
chiral matter in both hidden sectors.

%%%%%%%%%%%%%%%%%%%%%%%%%%%%%%%%%

%%%%%%%%%%%%%%%%%%%%%%%%%%%%%%%%%%%%%%%%%%%%%%%%

\subsection{Phenomenology}\label{pheno}

In this section, we will begin the analysis of more detailed
particle physics features of the vacua, with emphasis on the soft
supersymmetry breaking parameters, since we are particularly
interested in predicting collider physics observables that will be
measured at the LHC.

The low energy physics observables are determined by the
k\"{a}hler potential, superpotential and the gauge kinetic
function of the effective ${\cal N}=1,d=4$ supergravity. The gauge
kinetic function ($f$) has already been discussed. The K\"{a}hler
potential and the superpotential can be written in general as
follows: \ba \label{eq:KW} K &=& \hat{K}(s^i) +
\tilde{K}_{\bar{\alpha}\beta}(s^i) \,\bar{\Phi}^{\bar{\alpha}}
\Phi^{\beta} + Z_{\alpha\beta}(s^i,{\phi}^h)\,{\Phi}^{\alpha}{\Phi}^{\beta} + ... \\
 W &=& \hat{W}(z^i) +
{\mu}'{\Phi}^{\alpha}{\Phi}^{\beta} + Y'_{\alpha\beta\gamma}\,
{\Phi}^{\alpha}{\Phi}^{\beta}{\Phi}^{\gamma} + ...\nonumber \ea

\noindent where ${\Phi}^{\alpha}$ are the visible sector chiral
mater fields, $\tilde{K}_{\bar{\alpha}\beta}$ is their K\"{a}hler
metric and $Y'_{\alpha\beta\gamma}$ are their \emph{unnormalized}
Yukawa couplings. ${\phi}^h$ denote the hidden sector matter
fields. The first terms in $K$ and $W$ depend only on the bulk
moduli and have been already studied earlier. In general there can
be a mass term (${\mu}'$) in the superpotential, but as explained
in \cite{decon}, natural discrete symmetries can exist which
forbid it, in order to solve the doublet-triplet splitting
problem. The quantity $Z_{\alpha\beta}$ in the K\"{a}hler
potential will be important for generating an effective $\mu$ term
as we will see later.

Since the vacua have low scale supersymmetry, the effective
lagrangian must be equivalent to the MSSM plus couplings involving
possibly additional fields beyond the MSSM. For simplicity in this
section we will assume an observable sector which is precisely the
MSSM, although it would also be interesting to consider natural
$M$ theoretic extensions. The MSSM lagrangian is characterized by
the Yukawa and gauge couplings of the standard model and the soft
supersymmetry breaking couplings. These are the scalar squared
masses $m_i^2$, the trilinear couplings $A_{ijk}$, the $\mu$ and
$B\mu$ mass parameters and the gaugino masses. In $M$ theory all
of these couplings become functions of the various constants
$(A_i, N, P, Q, N_k)$ which are determined by the {\it particular}
$G_2$-manifold $X$. In addition, because we are now discussing the
observable sector, we have to explain the origin of observable
sector gauge, Yukawa and other couplings in $M$ theory. As we have
already explained, all gauge couplings are integer linear
combinations of the $N$ moduli, the $N$ integers determining the
homology class of the three dimensional subspace of $X$ which
supports that particular gauge group. Furthermore, the entire
superpotential is generated by membrane instantons, as we have
already discussed. Therefore mass terms and Yukawa couplings in
the superpotential are also determined by integer linear
combinations of the moduli fields. Hence, in addition to the
constants $(A_i, N, P, Q, N_k)$ which determine the moduli
potential, additional integers enter in determining the observable
sector superpotential. Generically, though, we do not expect these
integers to be large in the basis that the moduli K\"{a}hler
metric is given by $(6)$.

We determine the values of the soft SUSY breaking couplings at
$M_{unif}$ in the standard way : The moduli fields, hidden sector
matter fields as well as their auxiliary fields are replaced by
their {\it vevs} in the ${\cal N}=1,d=4$ SUGRA lagrangian. One
then takes the flat limit $M_p \rightarrow \infty$ with $m_{3/2}$
fixed. This gives a global SUSY lagrangian with soft SUSY breaking
terms \cite{Nilles:1983ge}. Unfortunately, in $M$ theory the
matter K\"{a}hler potential is difficult to compute. This leads to
theoretical uncertainties in the calculation of the scalar masses
and $A$, $B$ and $\mu$ parameters. Fortunately though we are able
to calculate the gaugino masses. Our main phenomenological result
is that the tree level gaugino masses are suppressed relative to
the gravitino mass. After explaining this, we will go on to
discuss the other soft terms in a certain, calculable limit.

\subsubsection{Suppression of Gaugino masses}\label{treegaugino}
Grand Unification is particularly natural in $G_2$ vacua of $M$
theory \cite{Friedmann:2002ty}. This implies that the gaugino
masses at tree level (at the unification scale) are
\emph{universal}, i.e. the gauginos of the three SM gauge groups
have the same mass. In order to compute the SM sector gaugino mass
scale at tree-level we need the Standard Model gauge kinetic
function, $f_{sm}$. In general this will be an integer linear
combination of the moduli, with integers $N^{sm}_i$, which is
linearly independent of the hidden sector gauge kinetic function
in general. The expression for the tree-level MSSM gaugino masses
in general ${\cal N}=1,d=4$ SUGRA is given by: \be\label{ga1}
M_{1/2}=m_p\frac{e^{\hat
K/2}K^{n\bar{m}}F_{\bar{m}}\partial_n\,f_{sm}}{2{i\,\rm
Im}f_{sm}}\,, \ee Note that the gauge kinetic function is
independent of the hidden sector matter fields. Therefore, the
large hidden sector matter $F$ term responsible for the dS minimum
does not contribute to the gaugino masses at tree level. We will
now proceed to evaluate this expression explicitly both for the
AdS and dS vacua. We will find that generically, the gaugino
masses are suppressed relative to the gravitino mass.

\subsubsection{Gaugino masses in AdS Vacua}

Choosing the hidden sector to be pure SYM with gauge groups
$SU(P)$ and $SU(Q)$, the normalized gaugino mass in these
compactifications can be expressed as \be M_{1/2}=-\frac
{m_p\,e^{-i\gamma_W}}{8\sqrt{\pi}V_X^{3/2}}\left[\frac {4\pi}
3\left(\frac{A_1}Pe^{-{\frac{2\pi}{P}{\rm
Im}f}}-\frac{A_2}Qe^{-{\frac{2\pi}{Q}{\rm
Im}f}}\right)\frac{\sum_{i=1}^{N}{N^{sm}_is_i\nu_i}}{\sum_{i=1}^{N}{N^{sm}_is_i}}+A_1e^{-{\frac{2\pi}{P}{\rm
Im}f}}-A_2e^{-{\frac{2\pi}{Q}{\rm Im}f}}\right] \ee where
$\gamma_W$ is the phase of the superpotential $W$. In the leading
order, the last two terms in the brackets can be combined as
\be\label{sup} A_1e^{-{\frac{2\pi}{P}{\rm
Im}f}}-A_2e^{-{\frac{2\pi}{Q}{\rm
Im}f}}=A_2\left[\frac{P-Q}Q\right]
\left[\frac{A_2P}{A_1Q}\right]^{-\frac P{P-Q}}\,. \ee On the other
hand, the two terms in the round brackets coming from the partial
derivative of the superpotential cancel in the leading order.
Therefore we need to take into account the first subleading order
contribution (\ref{eq46}). In this order, we obtain
\footnote{Recall that $\rm{Im}(f)_A^{(c)}\equiv{\cal T}^{\,(c)}_A$
(see (\ref{deft})).}: \be\label{part}
\left(\frac{A_1}Pe^{-{\frac{2\pi}{P}{\rm
Im}f}}-\frac{A_2}Qe^{-{\frac{2\pi}{Q}{\rm Im}f}}\right) =\frac
1{2\pi}A_2\left[\frac{P-Q}Q\right]
\left[\frac{A_2P}{A_1Q}\right]^{-\frac
P{P-Q}}\frac{B_A^{(c)}}{{\cal T}^{\,(c)}_A}\,. \ee From
(\ref{sup}) and (\ref{part}) we notice that the absolute value of
gaugino mass can now be conveniently expressed in terms of the
gravitino mass (for a given value of $A$ and $c$, as discussed in
previous sections) as \be\label{ga2} |M_{1/2}|^{(c)}_A=\frac2
3\frac{\sum_{i=1}^{N}a_i\,L_{A,\,k}^{(c)} \left(L_{A,\,k}^{(c)}+3/
2\right)({N^{sm}_i}/{N_i})}{\sum_{i=1}^{N}a_i\,L_{A,\,k}^{(c)}
({N^{sm}_i}/{N_i})}\times (m_{3/2})^{(c)}_A\,, \ee where we also
used (\ref{sol34}) and (\ref{eq43}). Finally, using (\ref{lk}) and
the first equation in (\ref{e20}), after some algebra we arrive at
the following expression for the gaugino mass:
\be\label{gauginomass} |M_{1/2}|^{(c)}_A=\left(\frac 4
3{T}^{(c)}_A+1\right)\frac{q-A}{q+\frac{{T}^{(c)}_A}{H^{(c)}_A}}\times
(m_{3/2})^{(c)}_A\,, \ee where we have introduced a new quantity
\be\label{q}
q=\frac{\sum_{i=1}^{N}m_ia_i({N^{sm}_i}/{N_i})}{\sum_{i=1}^{N}a_i({N^{sm}_i}/{N_i})}\,,
\ee such that the range of possible values for $q$ is \be -1\leq
q\leq 1\,. \ee

Note that the general formula (\ref{gauginomass}) which relates
the gravitino and gaugino masses is completely independent of the
number of moduli.

\begin{figure}[hbtp]
  \centerline{\hbox{ \hspace{0.0in}
    \epsfxsize=3.3in
    \epsffile{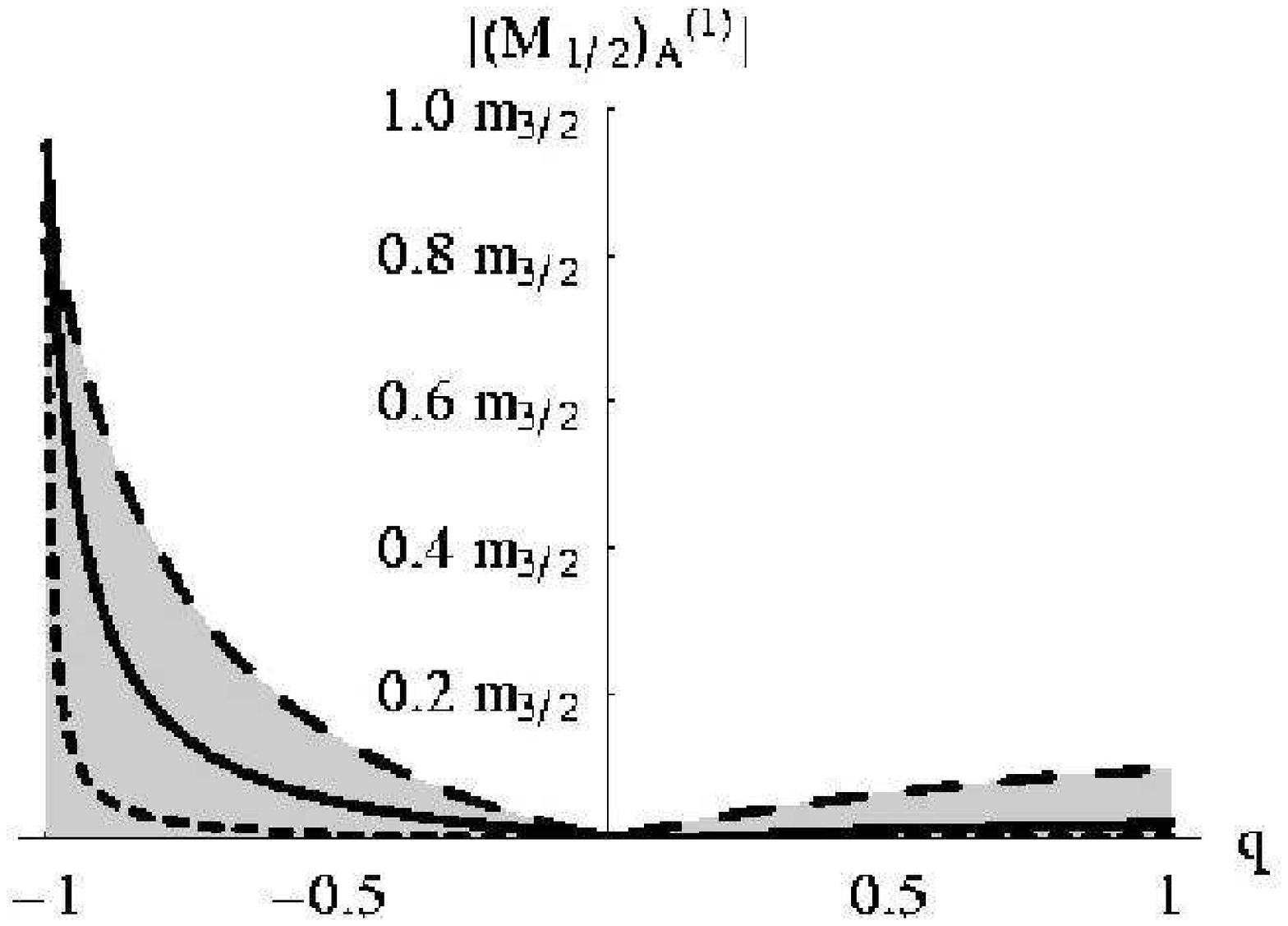}
    \hspace{0.0in}
    \epsfxsize=3.3in
    \epsffile{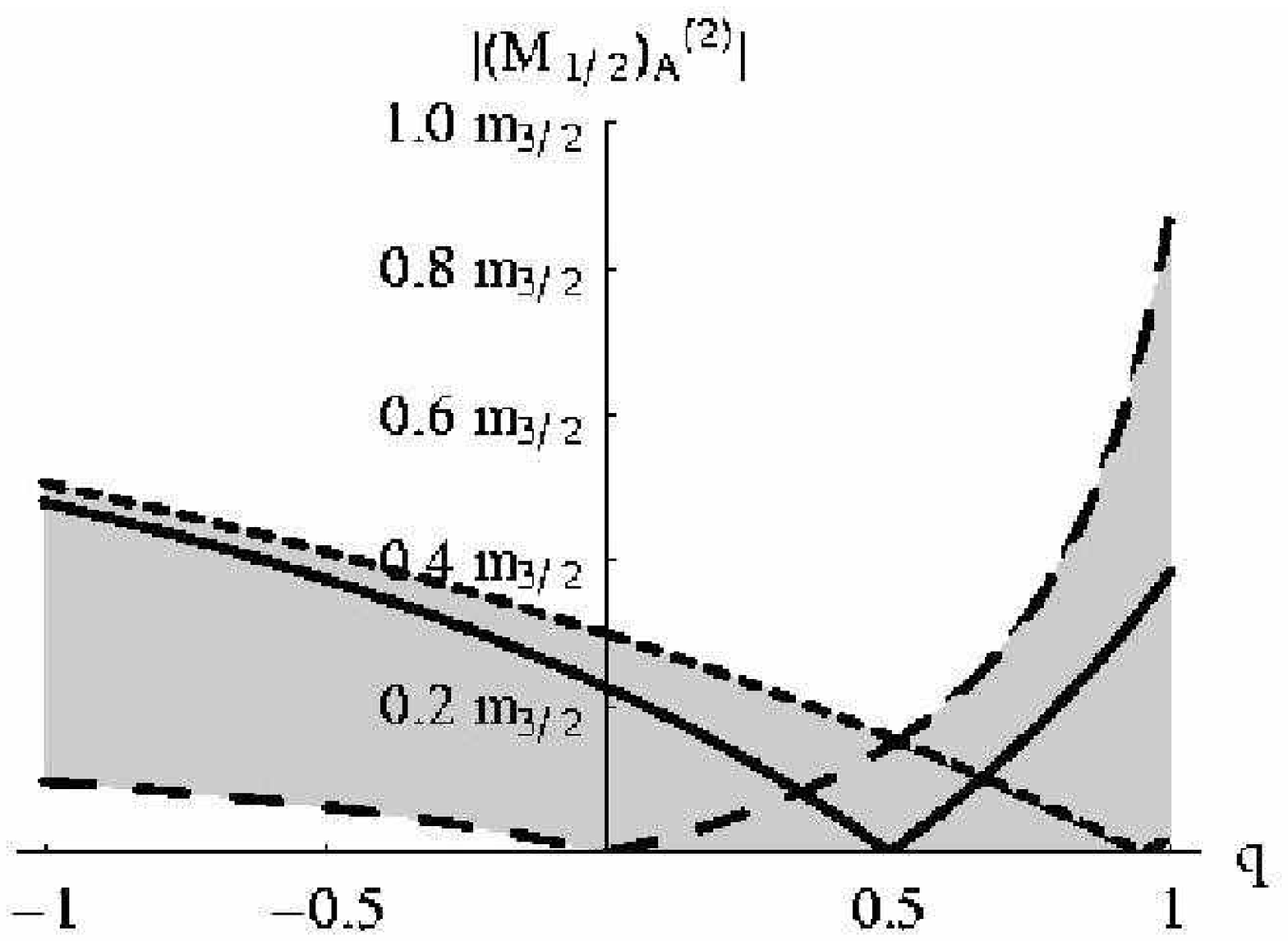}
    }
  }
\caption{Absolute values of $(M_{1/2})^{(1)}_A$-left and
$(M_{1/2})^{(2)}_A$ -right in units of gravitino mass as functions
of $q$. As parameter $A$ varies over $0\leq A < 1/7$ - on the left
and $0\leq A \leq 1$ - on the right the whole light grey region is
covered.
\newline
Left plot: $A=0$- long dashed line, $A=1/9$ - solid line, $A=5/36$
- short dashed line.
\newline
Right plot: $A=0$-long dashed line, $A=0.5$ - solid line, $A=0.95$
- short dashed line.} \label{twomoduliplot20}
\end{figure}

When all $m_k$ have the same sign the gaugino mass in
(\ref{gauginomass}) automatically vanishes. This is expected since
the solution when $A=\pm 1$ is the SUSY extremum. In Figure
{\ref{twomoduliplot20}} we have plotted absolute values of
$(M_{1/2})^{(1)}_A$ and $(M_{1/2})^{(2)}_A$ as functions of $q$.

For a significant fraction of the space in both plots we have
$(M_{1/2})^{(1,2)}_A\leq 0.2\,(m_{3/2})^{(1,2)}_A$, so the gaugino
masses are typically suppressed compared to the gravitino mass for
these AdS vacua. Note also that the suppression factor in
(\ref{gauginomass}) is independent of the gravitino mass. This
result is different from the small hierarchy between $M_{1/2}$ and
$m_{3/2}$ in the Type IIB flux vacua \cite{Conlon:2006us}, where
the gaugino mass is generically suppressed by ${{\rm
ln}\left(m_{3/2}\right)}$.

For the special case $A=0$, system (\ref{e20}) yields two
solutions with positive moduli. Therefore, there will be two
different values for the gaugino mass corresponding to these
solutions. After some algebra we obtain
\begin{eqnarray}\label{specialgaugino}
|M_{1/2}|^{(1,2)}_0&&=\left(\frac{5-{\sqrt{17}}}{4}\right)\left|\frac
q{q\pm\frac{-9+\sqrt{17}}{\sqrt{-26+10\sqrt{17}}}}\right|
\times (m_{3/2})^{(1,2)}_0 \nonumber\\
&&\sim 0.22\,\left|\frac q{q\mp 1.25}\right|\times
(m_{3/2})^{(1,2)}_0\,.
\end{eqnarray}
Again, this relation is valid for any AdS vacuum with broken SUSY
with $A=0$ with an arbitrary number of moduli.

To check the accuracy of the approximate gaugino mass formula we
again try the special case with two moduli $a_1=a_2=7/6$ with the
same choice of the constants as in (\ref{choice1}) and the integer
combination for the Standard Model gauge kinetic function
$\{N^{sm}_1=2,N^{sm}_2=1\}$. In this case equation
(\ref{specialgaugino}) for the absolute value of $M_{1/2}$ yields:
\be (M_{1/2})^{(1)}_0=164.4 \,{\rm
GeV},\,\,\,\,\,(M_{1/2})^{(2)}_0=95 \,{\rm GeV}\,, \ee whereas the
exact values computed numerically for the same choice of constants
are: \be (M_{1/2})^{(1)}_0=165.4\, {\rm
GeV},\,\,\,\,\,(M_{1/2})^{(2)}_0=97 \,{\rm GeV}\,. \ee This
demonstrates a high degree of accuracy of our approximation,
similar to that for the gravitino mass.

%%%%%%%%%%%%%%%%%%%%%%%%%%%%%%%%%

\subsubsection{Gaugino masses in dS Vacua}

From the formula for the gaugino mass in (\ref{ga1}), the gaugino
mass for the dS vacua in general can be expressed as \ba
\label{e59} M_{1/2} &=&
\frac{e^{-i{\gamma}_W}m_p\,e^{\phi_0^2/2}}{8\sqrt{\pi}V_X^{3/2}}\left[\frac
{2} 3\,\tilde
y\,\frac{\sum_{i=1}^{N}{N^{sm}_is_i\nu_i}}{\sum_{i=1}^{N}{N^{sm}_is_i}}+\tilde
x\right]A_2{e^{-b_2\vec\nu\cdot\vec a}} \nonumber \\
\implies M_{1/2} &=& -e^{-i{\gamma}_W}\left(\frac 2 3L_{1,+}
+1\right)\,m_{3/2}\,, \ea where in the second equality we used
(\ref{sol37}) and the fact that for these vacua $\nu_i=\nu$ for
all $i=\overline{1,N}$, independent of $i$. Also, by including the
minus sign we took into account that $m_{3/2}=e^{K/2}\left|\tilde
x\right|A_2{e^{-b_2\vec\nu\cdot\vec a}}$  but $\tilde x < 0$,
since $Q-P\geq3$. From (\ref{e51}) we can find $\tilde L_{1,+}$
including the first subleading contribution \be\label{e60} \tilde
L_{1,+}\approx -\frac 3 2+\frac {3{a\tilde\alpha b_1}\tilde y} {14
\tilde x\tilde z}\left(\frac{a\tilde\alpha}{\phi_0^2\,\tilde
x}+1\right)\,. \ee For $\tilde x$, $\tilde y$ and $\tilde z$ in
(\ref{e60}) we use the definitions (\ref{xyzw1}) and substitute
the approximate result (\ref{e57}) for $\tilde\alpha$. Then after
substituting(\ref{e60}) into (\ref{e59}) and assuming that
$Q-P\sim {\mathcal O}(1)$, in the limit when $P$ is large the
approximate tree level MSSM gaugino mass is given by
\be\label{gaugino33} M_{1/2} \approx -\frac {e^{-i{\gamma}_W}}
{P\,\ln\left(\frac {A_1Q}{A_2P}\right)}\left(1+\frac
2{\phi_0^2\,(Q-P)}+\frac 7{\phi_0^2\,P\,\ln\left(\frac
{A_1Q}{A_2P}\right)}\right)\times {m_{3/2}}\,, \ee where we use
(\ref{e58}) to substitute for $\phi_0^2$. It is important to note
some features of the above equation. First, equation
(\ref{gaugino33}) is completely independent of the choice of
integers $N_i^{sm}$ for the Standard Model gauge kinetic function
as well as the integers $N_i$ for the hidden sector. Second, it is
independent of the number of moduli $N$ and moreover, it is also
independent of the particular details of the internal manifold
described by the rational numbers $a_i$ appearing in the
K\"{a}hler potential (\ref{vol}). These properties imply that
relation (\ref{gaugino33}) is  universal for all $G_2$ holonomy
compactifications consistent with our aproximations, independent
of many internal details of the manifold. Furthermore, since the
denominator $P\ln(A_1Q/A_2P)$ is typically of
$\mathcal{O}$(10-100) and since the expression in the round
brackets in (\ref{gaugino33}) is slowly varying and for the range
under consideration is of $\mathcal{O}$(1), we see that {\it
gaugino masses are always suppressed relative to the gravitino for
these dS vacua}.

After one imposes the constraint equation (\ref{er765})
((\ref{ior}) when $Q-P=3$) to make the cosmological constant very
small, one can get rid of one of the constants in
(\ref{gaugino33}), and further simplify the expression for the
universal tree level gaugino mass parameters for the dS vacuum
with a very small cosmological
constant:\begin{equation}\label{gaugino36} M_{1/2}\approx-\frac
{e^{-i{\gamma}_W}} {84}\left(1+\frac 2{3\phi_0^2}+\frac
7{84\phi_0^2}\right)\times
{m_{3/2}}=-e^{-i{\gamma}_W}\frac{139+396\sqrt{3}}{34356}\times
{m_{3/2}}\approx -e^{-i{\gamma}_W}0.024\times {m_{3/2}}\,.
\end{equation} As in the more general case
(eqn.(\ref{gaugino33})), the tree level gaugino mass is suppressed
compared to the gravitino mass, and the suppression factor can
also be predicted.

One would like to understand the physical origin of the
suppression of the gaugino masses at tree level, especially for
the dS vacua which are phenomenologically relevant. As mentioned
earlier, since the matter $F$ term does not contribute to the
gaugino masses, the gaugino masses can only get contributions from
the moduli $F$ terms, which, as explained in the last paragraph in
section \ref{unique}, vanish in the leading order of our
approximation. The first subleading contribution is suppressed by
the inverse power of the volume of the associative three-cycle,
causing the gaugino masses to be suppressed relative to the
gravitino. Since the inverse volume of this three-cycle is
essentially $\alpha_{hidden}$ - the hidden sector gauge coupling
in the UV - the suppression is due to the fact that the hidden
sector is asymptotically free. In large volume type IIB
compactifications, the moduli $F$ terms also vanish in the leading
order, leading to suppressed gaugino masses as well
\cite{Conlon:2006us}. However, in contrast to our case, there the
subleading contribution is suppressed by the inverse power of the
volume of the compactification manifold. Note that a large
associative cycle on a $G_2$ manifold does \emph{not} translate
into a large volume compactification manifold. Thus, unlike the
large volume type IIB compactifications, these $M$ theory vacua
are consistent with standard gauge coupling unification.

\subsubsection{Other parameters and Flavor issues}\label{others}

The trilinears, scalars, anomaly mediated contributions to gaugino
masses and the $B\mu$ parameter depend more on the microscopic
details of the theory -- the Yukawa couplings, the ${\mu}'$
parameter and the K\"{a}hler metric for visible sector matter
fields. The flavor structure of the Yukawa matrices as well as
that of the K\"{a}hler metric for matter fields is crucial for
estimating flavor changing effects. We will comment on these at
appropriate places.

The (un-normalized) Yukawa couplings in these vacua arise from
membrane instantons which connect singularities where chiral
superfields are supported (if some singularities coincide, there
could also be order one  contributions). They are given by: \be
\label{yukawas1} Y'_{\alpha\beta\gamma} =
C_{\alpha\beta\gamma}\,e^{i2\pi\sum_i l^{\alpha\beta\gamma}_i
z^i}\ee \noindent where $C_{\alpha\beta\gamma}$ is an ${\mathcal
O}(1)$ constant and $l^{\alpha\beta\gamma}_i$ are integers.

The moduli dependence of the matter K\"{a}hler metric is
notoriously difficult to compute in generic string and $M$ theory
vacua, and the vacua under study here are no exception. The best
we can do here is to consider the Type IIA limit of these vacua.
The matter K\"{a}hler metric has been computed in type IIA
intersecting $D$6-brane vacua on toroidal orientifolds
\cite{Bertolini:2005qh} building on earlier work
\cite{Lust:2004cx}. Since chiral fermions living at intersections
of D6-branes lift to chiral fermions supported at conical
singularities in $M$ theory \cite{Atiyah:2001qf, Cvetic:2001kk},
we will simply uplift the IIA calculation to $M$ theory. Thus the
results of this section are strictly only valid in the Type IIA
limit.

Lifting the Type IIA result to $M$ theory, one gets (see Appendix
for details): \ba \label{kmetric} \tilde{K}_{\bar{\alpha}\beta}
&=& {\delta}_{\bar{\alpha}\beta}\,\prod_{i=1}^n
\left(\frac{\Gamma(1-\theta^{\alpha}_i)}
{\Gamma(\theta^{\alpha}_i)} \right)^{1/2}\nonumber \\
\tan(\pi \theta^{\alpha}_i)&=&
c^{\alpha}_i\,(s_i)^l;\;\;c^{\alpha}_i = {\mathrm
constant};\;l={\mathrm rational\;number\;of\;{\mathcal O}(1)}.\ea

In the type IIA toroidal orientifolds, the underlying symmetries
always allow us to have a diagonal K\"{a}hler metric
\cite{Bertolini:2005qh}. We have assumed for simplicity the
K\"{a}hler metric to be diagonal in the analysis below. Now, we
will write down the general expressions for the physical Yukawa
couplings and the soft parameters - the trilinears and the scalars
and then estimate these in $M$ theory compactifications. The $\mu$
and $B\mu$ parameters will be discussed in section \ref{muBmu}.

The K\"{a}hler potential for the chiral matter fields is
non-canonical for any compactification in general. In determining
physical implications however, it is much simpler to work in a
basis with a canonical K\"{a}hler potential.  So, to canonically
normalize the matter field K\"{a}hler potential, we introduce the
normalization matrix $\mathcal{Q}$ : \ba \label{norm} \Phi
\rightarrow {\cal Q} \cdot \Phi, \;\;\; s.t. \;\;\; {\cal
Q}^{\dag}\tilde{K}{\cal Q} = 1. \ea The $\mathcal{Q}$s are
themselves only defined up to a unitary transformation, i.e.
$\mathcal{Q}'=\mathcal{Q}\cdot\mathcal{U}$ is also an allowed
normalization matrix if $\mathcal{U}$ is unitary. If the
K\"{a}hler metric is already diagonal
($\tilde{K}_{\bar{\alpha}\beta}=\tilde{K}_{\alpha}\delta_{\bar{\alpha}\beta}$),
the normalization matrix can be simplified :
$\mathcal{Q}_{\bar{\alpha}\beta}=(\tilde{K}_{\alpha})^{-1/2}\delta_{\bar{\alpha}\beta}$.
The normalized (physical) Yukawa couplings are
\cite{Nilles:1983ge}: \ba \label{ymu} Y_{\alpha\beta\gamma} &=&
e^{\hat{K}/2}\,\frac{\hat{W}^{\star}}{|\hat{W}|}Y'_{{\alpha}'{\beta}'{\gamma}'}\,
\mathcal{Q}_{{\alpha}'\alpha}\mathcal{Q}_{{\beta}'\beta}\mathcal{Q}_{{\gamma}'\gamma}\ea

It was shown in \cite{King:2004tx} that in the class of theories
with a hierarchical structure of the un-normalized Yukawa
couplings (in the superpotential), the K\"{a}hler corrections to
both masses and mixing angles of the SM particles are subdominant.
Since in these compactifications, it is very natural to obtain a
hierarchical structure of the un-normalized Yukawa couplings due
to their exponential dependence on the various moduli and also
because of some possible family symmetries, therefore one expects
the effects of the K\"{a}hler corrections which are less under
control, to be subdominant. The expressions for the
\emph{un-normalized} trilinears and scalar masses are given by
\cite{Nilles:1983ge}: \label{m'A'} \ba m'^2_{\bar{\alpha}\beta}
&=& (m_{3/2}^2 + V_0)\,\tilde{K}_{\bar{\alpha}\beta} - e^{\hat
K}F^{\bar{m}}(\partial_{\bar{m}}\partial_n\,\tilde{K}_{\bar{\alpha}\beta}-\partial_{\bar{m}}\,
\tilde{K}_{\bar{\alpha}\gamma}\,\tilde{K}^{\gamma\bar{\delta}}\partial_n\,\tilde{K}_{\bar{\delta}\beta})F^n
\\
A'_{\alpha\beta\gamma} &=& \frac{\hat{W}^{\star}}{|\hat{W}|}
e^{\hat{K}/2}\,F^m\,[\hat{K}_m\,Y'_{\alpha\beta\gamma}+\partial_m
Y'_{\alpha\beta\gamma}-(\tilde{K}^{\delta\bar{\rho}}\,\partial_n\tilde{K}_{\bar{\rho}\alpha}\,Y'_{\alpha\beta\gamma}+
\alpha \leftrightarrow \gamma + \alpha \leftrightarrow \beta)]
\nonumber \ea The \emph{normalized} scalar masses and trilinears
are thus given by: \ba m^2_{\bar{\alpha}\beta} &=&
(\mathcal{Q}^{\dag}\cdot
m'^2 \cdot \mathcal{Q})_{\bar{\alpha}\beta} \\
\tilde{A}_{\alpha\beta\gamma} &=& A'_{{\alpha}'{\beta}'{\gamma}'}
\mathcal{Q}_{{\alpha}'\alpha}\mathcal{Q}_{{\beta}'\beta}\mathcal{Q}_{{\gamma}'\gamma}\nonumber
\ea

Let us discuss the implications for the soft terms, beginning with
the Anomaly mediated corrections to gaugino masses.

\vspace{0.5cm} \hspace{3.5cm} {\it Anomaly mediated contributions
to Gaugino Masses} \vspace{0.5cm}

We saw in section \ref{treegaugino} that the gauginos are
generically suppressed relative to the gravitino. Since anomaly
mediated gaugino masses are also suppressed relative to the
gravitino (by a loop factor), they are non-negligible compared to
the tree level contributions and have to be taken into account.
Also, since anomaly mediated contributions for the three gauge
groups are \emph{non-universal}, these introduce non-universality
in the gaugino masses at the unification scale.

The general expression for the anomaly mediated contributions is
given by \cite{Gaillard:1999yb}: \be \label{anom123}(M)^{am}_a =
-\frac{g_a^2}{16\pi^2}[-(3C_a-\sum_{\alpha}C_a^{\alpha})e^{\hat
K/2}W^{*}+(C_a-\sum_{\alpha}C_a^{\alpha})e^{\hat K/2}
F^mK_m+2\sum_{\alpha}(C_a^{\alpha}e^{\hat
K/2}F^m\partial_m\ln(\tilde{K}_{\alpha}))]\ee where $C_a$ and
$C_a^{\alpha}$ are the casimir invariants of the $a^{th}$ gauge
group and $\alpha$ runs over the number of fields charged under
the $a^{th}$ gauge group. For a given spectrum such as that of the
MSSM, $C_a$ and $C_a^{\alpha}$ are known.

We first compute the F-term contributions \ba
&&e^{\hat K/2}F^iK_i=\frac{14}3e^{-i\gamma_W}\left(L_{1,+}+\frac 3 2\right)\times m_{3/2}\,\\
&&e^{\hat
K/2}F^{\phi}K_{\phi}=e^{-i\gamma_W}\left(\frac{a\tilde\alpha}{\tilde
x}+\phi_0^2\right)\times m_{3/2}\,. \nonumber \ea Then, equation
(\ref{anom123}) gives \ba \label{anomalyds} (M)^{am}_a
&=&-e^{-i\gamma_W}\frac {\alpha_{GUT}}{4\pi}
[-(3C_a-\sum_{\alpha}C_a^{\alpha})+\frac {14}3(C_a-\sum_{\alpha}C_a^{\alpha})\left(\tilde L_{1,+}+\frac 3 2\right) \\
&+&(C_a-\sum_{\alpha}C_a^{\alpha})\left(\frac{a\tilde\alpha}{\tilde
x}+\phi_0^2\right)-\frac 4 3\left(\tilde L_{1,+}+\frac 3
2\right)\sum_{\alpha}C_a^{\alpha}\,\sum_{i}\,\frac{1}{2\pi}\left(l\,\psi^{\alpha}_i\,\sin(2\pi{\theta}^{\alpha}_i)
\right)]\times m_{3/2}\,. \nonumber \ea where \be
\left(\alpha_{GUT}\right)^{-1}=\sum_{i=1}^{N}s_iN_i^{sm}\,,\ee and
we have defined the quantity: \ba
{\psi}^{\alpha}_i({\theta}^{\alpha}_i) \equiv \frac{d
\ln(\tilde{K}_{\alpha})}{d\,\theta^{\alpha}_i}\,,\ea where
${\theta}^{\alpha}_i$ implicitly depends on the moduli. However,
it is much simpler to keep the dependence as a function of
${\theta}^{\alpha}_i$, as is explained in the Appendix. Depending
on the values of the Casimir invariants $C_a$ and $C_a^i$ for the
three gauge groups, the anomaly mediated contribution can either
add to or cancel the tree level contributions. Here we also took
into account that $m_{3/2}=e^{K/2}\left|\tilde
x\right|A_2{e^{-b_2\vec\nu\cdot\vec a}}$  but $\tilde x < 0$,
since $Q-P\geq3$. Using the expression for $L_{1,+}$ in
(\ref{e61}) along with the definitions of $\tilde x$, $\tilde y$
and $\tilde z$ in (\ref{xyzw1}) in terms of $\tilde\alpha$ in
(\ref{e57}) and assuming that $Q-P\sim {\mathcal O}(1)$, in the
limit when $P$ is large we obtain \be\label{ltplus} \tilde
L_{1,+}=-\frac 3 2+\frac
3{2\,P\,\ln\left(\frac{A_1Q}{A_2P}\right)} \left(1+\frac
2{(Q-P)\,\phi_0^2}+\frac 7{\phi_0^2\,P\,\ln\left(\frac
{A_1Q}{A_2P}\right)}\right)\,. \ee Using (\ref{eqab}) and
substituting for $\nu$ from (\ref{app24}) into (\ref{eqab})
together with $a=-2/P$ we can express
\begin{equation}\label{e890}
\frac{a\tilde\alpha}{\tilde x}+\phi_0^2=\phi_0^2\left(1+\frac
2{(Q-P)\,\phi_0^2}+\frac{7}{\phi_0^2\,P\ln\left(\frac{A_1Q}{A_2P}\right)}\right)\,.
\end{equation}

Substituting (\ref{ltplus}) and (\ref{e890}) into
(\ref{anomalyds}) we obtain \ba \label{anomalyds1} (M)^{am}_a
&=&-e^{-i\gamma_W} \frac {\alpha_{GUT}}{4\pi}
[-(3C_a-\sum_{\alpha}C_a^{\alpha})+\left(1+\frac
2{(Q-P)\,\phi_0^2}+\frac 7{\phi_0^2\,P\,\ln\left(\frac
{A_1Q}{A_2P}\right)}\right)\\
&\times&\left((C_a-\sum_{\alpha}C_a^{\alpha})\left(\phi_0^2 +\frac
7{P\,\ln\left(\frac{A_1Q}{A_2P}\right)}\right)
-\frac{2\sum_{\alpha}C_a^{\alpha}\,\sum_{i}\,\frac{1}{2\pi}\left(l\,\psi^{\alpha}_i\,\sin(2\pi{\theta}^{\alpha}_i)
\right)}{P\,\ln\left(\frac{A_1Q}{A_2P}\right)}\right)]\times
m_{3/2}\,.\nonumber \ea

Note that these $M$ theory vacua do not have a no-scale structure.
Therefore, the anomaly mediated gaugino masses are only suppressed
by loop effects, in contrast to the type IIB compactifications,
which exhibit a no-scale structure in the leading order
\cite{Conlon:2006us}, leading to an additional suppression of the
anomaly mediated gaugino masses.

As before, when one imposes the constraint (\ref{ior}), the
anomaly mediated gaugino mass contribution can be simplified
further and is given by:\begin{eqnarray} \label{anomalyds13}
(M)^{am}_a &=&-\frac {\alpha_{GUT}{(e^{-i\gamma_W})}}{4\pi}
[-(3C_a-\sum_{\alpha}C_a^{\alpha})+\frac{29055+11374\sqrt{3}}{29448}{(C_a-\sum_{\alpha}C_a^{\alpha})}\nonumber \\
&-&\frac{139+396\sqrt{3}}{17178}{\sum_{\alpha}C_a^{\alpha}\,\sum_{i}\,\frac{1}{2\pi}\left(l\,\psi^{\alpha}_i\,\sin(2\pi{\theta}^{\alpha}_i)
\right)}]\times m_{3/2}\nonumber\\
(M)^{am}_a &\approx&-\frac {\alpha_{GUT}{(e^{-i\gamma_W})}}{4\pi}
[-(3C_a-\sum_{\alpha}C_a^{\alpha})+1.6556\,{(C_a-\sum_{\alpha}C_a^{\alpha})} \\
&-&0.048{\sum_{\alpha}C_a^{\alpha}\,\sum_{i}\,\frac{1}{2\pi}\left(l\,\psi^{\alpha}_i\,\sin(2\pi{\theta}^{\alpha}_i)\right)}]\times
m_{3/2}\,. \nonumber
\end{eqnarray}
From the left plot in Figure \ref{Am2} of the Appendix we note
that
$\left|\frac{1}{2\pi}\left(l\,\psi^{\alpha}_i\,\sin(2\pi{\theta}^{\alpha}_i)
\right)\right|<0.5$. In a generic case, we expect that parameters
$\theta^{\alpha}_i$ are all different and, as a result, the terms
appearing inside the corresponding sum over $i$ partially cancel
each other. Thus, in a typical case we expect that \be
\label{genericcond}
\left|{\sum_{\alpha}C_a^{\alpha}\,\sum_{i}\,\frac{1}{2\pi}\left(l\,\psi^{\alpha}_i\,\sin(2\pi{\theta}^{\alpha}_i)
\right)}\right|<1\,. \ee Neglecting the corresponding contribution
in (\ref{anomalyds13}), taking $\alpha_{GUT}=1/25$, and
substituting the Casimirs for an MSSM spectrum, we obtain the
following values in the leading order, up to an overall phase
$e^{-i\gamma_W}$: \be\label{ano456}
(M)^{am}_{U(1)}\approx\,\,0.01377\times m_{3/2},\,\,\,\,\,\,\,
(M)^{am}_{SU(2)}\approx\,\,0.02317\times m_{3/2},\,\,\,\,\,\,\,
(M)^{am}_{SU(3)}\approx\,\,0.02536\times m_{3/2}\,. \ee Finally,
combining the tree-level (\ref{gaugino36}) plus anomaly mediated
(\ref{ano456}) contributions, we obtain the following {\em
non-universal} gaugino masses at the unification scale: \be
M_1\approx\,\,-10.24\times 10^{-3}\,m_{3/2},\,\,\,\,\,\,\,
M_2\approx\,\,-0.84\times 10^{-3}\, m_{3/2},\,\,\,\,\,\,\,
M_3\approx\,\,+1.35\times 10^{-3}\, m_{3/2}\,. \ee We immediately
notice remarkable cancellations for $M_2$ and $M_3$ between the
tree-level and the anomaly mediated contributions. Recall that
since the distribution of $m_{3/2}$ peaked at $m_{3/2}\sim{\cal
O}(100)\,{\rm TeV}$, the possible range of gaugino masses is in
the desirable range $m_{1/2}\sim {\cal O}(0.1-1)\,{\rm TeV}$. One
of the consequences of these cancellations is a comparatively
lighter gluino. Furthermore, since $M_2$ is a lot smaller than
$M_1$, the neutralino LSP is wino-like.

One should be extremely cautious however, since the predictive
expressions above are only true if (\ref{genericcond}) is
satisfied. The extra contribution neglected in the above estimates
is given by: \be \Delta_a=0.15\times
10^{-3}\,{\sum_{\alpha}C_a^{\alpha}\,\sum_{i}\,\frac{1}{2\pi}\left(l\,\psi^{\alpha}_i\,\sin(2\pi{\theta}^{\alpha}_i)\right)}\times
m_{3/2}\,. \ee Because of the large cancellations between the
tree-level and anomaly mediated contributions, it may happen that
these corrections become important in a relatively small region of
the overall parameter space, leading to a deviation from the above
result thereby altering the pattern of gaugino masses. Further
corrections may also come from varying $\alpha_{GUT}$ as well as
taking into account the subleading corrections to the condition
for the cosmological constant to be very small. We will study
these issues in detail in \cite{ToAppear}.

\vspace{0.5cm} \hspace{7.0cm} {\it Trilinears} \vspace{0.5cm}

\noindent The normalized trilinear can be written as : \ba
\label{trilinears2}\tilde{A}_{\alpha\beta\gamma} &=&
\frac{\hat{W}^{\star}}{|\hat{W}|} e^{\hat{K}/2}\,
(\tilde{K}_{\alpha}\tilde{K}_{\beta}\tilde{K}_{\gamma})^{-1/2}
\,\left(\sum_{i} e^{\hat
K/2}F^m\,[\hat{K}_m\,Y'_{\rho\delta\kappa}+\partial_m
Y'_{\rho\delta\kappa}-\partial_m\ln(\tilde{K}_{\alpha}\tilde{K}_{\beta}\tilde{K}_{\gamma})]\right) \nonumber\\
&=& Y_{\alpha\beta\gamma}\,\left(\sum_{i} e^{\hat
K/2}F^m\,[\hat{K}_m+\partial_m\,\ln
(Y'_{\alpha\beta\gamma})-\partial_m\ln(\tilde{K}_{\alpha}\tilde{K}_{\beta}\tilde{K}_{\gamma})]\right)
\ea As stated earlier, the subscripts $\{\alpha,\beta,\gamma\}$
stand for the visible chiral matter fields. For example, $\alpha$
can be the left-handed up quark doublet, $\beta$ can be the
right-handed up quark singlet and $\gamma$ can be the up-type
higgs doublet. Our present understanding of the microscopic
details of these constructions does not allow us to compute the
three individual trilinear parameters -- corresponding to the
up-type Yukawa, the down-type Yukawa and the lepton Yukawa
matrices, explicitly. One can only estimate the rough overall
scale of the trilinears.

We see from (\ref{trilinears2}) that the normalized trilinears are
proportional to the Yukawas since the K\"{a}hler metric is
diagonal. If instead the off-diagonal entries in the k\"{a}hler
metric are small but non-zero, it would lead to a slight deviation
from the proportionality of the trilinears to the Yukawa
couplings. In most phenomenological analyzes, the trilinears
$\tilde{A}$ are taken to be proportional to the Yukawas and the
\emph{reduced} trilinear couplings
$A_{\alpha\beta\gamma}\equiv\tilde{A}_{\alpha\beta\gamma}/Y_{\alpha\beta\gamma}$
are used. We expect this to be true in these compactifications
from above. If the Yukawa couplings are those of the Standard
Model, then from (\ref{trilinears2}) the normalized reduced
trilinear coupling $A_{\alpha\beta\gamma}$ for de Sitter vacua in
general is given by \ba
\label{tri46}A_{\alpha\beta\gamma}&=&e^{-i\gamma_W}\left({1+\frac
2{(Q-P)\,\phi_0^2}+\frac 7{\phi_0^2\,P\,\ln\left(\frac
{A_1Q}{A_2P}\right)}}\right)(\phi_0^2+ \frac 1
{\,P\,\ln\left(\frac{A_1Q}{A_2P}\right)}[7
+2\ln\left|\frac{C_{\alpha\beta\gamma}}{Y'_{\alpha\beta\gamma}}\right|\nonumber\\
&+&\sum_i\,\frac{1}{2\pi}(l\,\psi^{\alpha}_i\,\sin(2\pi{\theta}^{\alpha}_i)+{\alpha}\rightarrow
\beta + \alpha \rightarrow \gamma)])\times m_{3/2}\,.\ea If we
then use (\ref{ymu}) together with (\ref{kmetric}) and
$\mathcal{Q}_{\bar{\alpha}\beta}=(\tilde{K}_{\alpha})^{-1/2}\delta_{\bar{\alpha}\beta}$,
we obtain the following expression for the trilinears \ba
\label{tri49}A_{\alpha\beta\gamma}&=&m_{3/2}\,e^{-i\gamma_W}\left({1+\frac
2{(Q-P)\,\phi_0^2}+\frac 7{\phi_0^2\,P\,\ln\left(\frac
{A_1Q}{A_2P}\right)}}\right)(\phi_0^2+ \frac 1
{\,P\,\ln\left(\frac{A_1Q}{A_2P}\right)}\times \nonumber \\ && [7
+2\ln\left|\frac{C_{\alpha\beta\gamma}}{Y_{\alpha\beta\gamma}}\right|-3\ln(4\pi^{1/3}V_X)+\phi_0^2
-\sum_i\,[\left\{\frac 1
2\ln\left(\frac{\Gamma(1-\theta^{\alpha}_i)}{\Gamma(\theta^{\alpha}_i)}\right)
-\frac{1}{2\pi}l\,\psi^{\alpha}_i\,\sin(2\pi{\theta}^{\alpha}_i)\right\}\nonumber
\\ && +{\alpha}\rightarrow \beta + \alpha \rightarrow \gamma\,]])\,.\ea Imposing the constraint equation
(\ref{ior}) on the expression above, the reduced trilinears for a
dS vacuum with a tiny cosmological constant are simplified to: \ba
\label{trides} A_{\alpha\beta\gamma}&=&e^{-i\gamma_W}
(\frac{69+22\sqrt{3}}{72}+\frac{139+396\sqrt{3}}{34356}[7
+2\ln\left|\frac{C_{\alpha\beta\gamma}}{Y_{\alpha\beta\gamma}}\right|-7\,\ln\left(\frac{14(P+3)}{N}\right)\nonumber
\\ && +\frac
1{72}\left(15+22\sqrt{3}\right)-6\,\ln\left(\frac
2\pi\right)-\sum_i\,[\left\{\frac 1
2\ln\left(\frac{\Gamma(1-\theta^{\alpha}_i)}{\Gamma(\theta^{\alpha}_i)}\right)-\frac{1}{2\pi}l\,\psi^{\alpha}_i\,\sin(2\pi{\theta}^{\alpha}_i)\right\}
\nonumber \\ & & +{\alpha}\rightarrow \beta + \alpha \rightarrow \gamma\,]])\times m_{3/2}\nonumber\\
A_{\alpha\beta\gamma}&\approx& e^{-i\gamma_W}(1.4876+0.024\,[10.45
+2\ln\left|\frac{C_{\alpha\beta\gamma}}{Y_{\alpha\beta\gamma}}\right|-7\,\ln\left(\frac{14(P+3)}
{N}\right)\\&&-\sum_i\left(\left\{\frac 1
2\ln\left(\frac{\Gamma(1-\theta^{\alpha}_i)}
{\Gamma(\theta^{\alpha}_i)}\right)-\frac{1}{2\pi}l\,\psi^{\alpha}_i\,\sin(2\pi{\theta}^{\alpha}_i)\right\}
+{\alpha}\rightarrow\beta + \alpha \rightarrow
\gamma\right)])\times m_{3/2}\,.\nonumber \ea

We see that compared to the gauginos, the trilinears depend on
more constants. The quantity $\left\{\frac 1
2\ln\left(\frac{\Gamma(1-\theta^{\alpha}_i)}{\Gamma(\theta^{\alpha}_i)}\right)-
\frac{1}{2\pi}(l\,\psi^{\alpha}_i\,\sin(2\pi{\theta}^{\alpha}_i)\right\}$
is of $\mathcal{O}$(1). Therefore, in a generic situation, we
expect the terms inside the sum in $\sum_i \,\left\{\frac 1
2\ln\left(\frac{\Gamma(1-\theta^{\alpha}_i)}{\Gamma(\theta^{\alpha}_i)}\right)-
\frac{1}{2\pi}(l\,\psi^{\alpha}_i\,\sin(2\pi{\theta}^{\alpha}_i)\right\}$
to partially cancel each other and give an overall contribution
much smaller than the first three terms inside the square
brackets. Then, for known values of the physical Yukawa couplings
and reasonable values of $P$ and $N$, the trilinears generically
turn out to slightly larger than $m_{3/2}$.

\vspace{0.5cm} \hspace{7.0cm} {\it Scalar Masses} \vspace{0.5cm}

\noindent For an (almost) diagonal K\"{a}hler metric, the
normalized scalar masses reduce to : \be\label{sc567}
(m^2_{\bar{\alpha}\beta}) = [m_{3/2}^2+V_0-e^{\hat
K}F^{\bar{m}}F^n{\partial}_{\bar{m}}\partial_{n}\ln(\tilde{K}_{\alpha})]
\,{\delta}_{\bar{\alpha}\beta} \ee where we have used
(\ref{norm}). Using (\ref{ltplus}) in (\ref{sc567}), we obtain the
following expression for the scalar mass squared \ba
\label{scalarsds} (m_{\alpha}^2)&=& V_0 + (m^2_{3/2})\,[1-\frac
9{4\,P^2\left(\ln\left(\frac{A_1Q}{A_2P}\right)\right)^2}
\left(1+\frac 2{(Q-P)\,\phi_0^2}+\frac
7{\phi_0^2\,P\,\ln\left(\frac {A_1Q}{A_2P}\right)}\right)^2
 \nonumber \\ & &
\times\frac{1}{4\pi}\sum_i\,\{l^2\,{\psi}^{\alpha}_{\bar{i}i}\,\sin^2(2\pi{\theta}^{\alpha}_i)+l^2\,{\psi}^{\alpha}_i\,\sin(4\pi{\theta}^{\alpha}_i)
-2l\,{\psi}^{\alpha}_i\,\sin(2\pi{\theta}^{\alpha}_i)\}]\,. \ea
where we have defined another quantity:\ba
{\psi}^{\alpha}_{\bar{i}i}({\theta}^{\alpha}_i)\equiv
\frac{d{\psi}^{\alpha}_i}{d{\theta}^{\alpha}_i}\ea As in the case
of the trilinears, only the overall scale of the scalars can be
estimated, not the individual masses of different flavors of
squarks and sleptons. Once the cosmological constant is made small
by imposing the constraint (\ref{ior}), the scalars are given by
\begin{eqnarray}
\label{scalarsds2} (m_{\alpha}^2)&=& (m^2_{3/2})\,[1-
\frac{\left(139+396\sqrt{3}\right)^2}{524593216}\frac{1}{4\pi}\sum_i\,\{l^2\,{\psi}^{\alpha}_{\bar{i}i}\,
\sin^2(2\pi{\theta}^{\alpha}_i)+l^2\,{\psi}^{\alpha}_i\,
\sin(4\pi{\theta}^{\alpha}_i) \nonumber \\
& &-2l\,{\psi}^{\alpha}_i\,\sin(2\pi{\theta}^{\alpha}_i)\}]\,\nonumber\\
&\approx&(m^2_{3/2})\,[1-\frac{0.0013}{4\pi}\sum_i\,\{l^2\,
{\psi}^{\alpha}_{\bar{i}i}\,\sin^2(2\pi{\theta}^{\alpha}_i)+l^2\,{\psi}^{\alpha}_i\,\sin(4\pi{\theta}^{\alpha}_i)
-2l\,{\psi}^{\alpha}_i\,\sin(2\pi{\theta}^{\alpha}_i)\}]\,\nonumber\\
&\approx&m^2_{3/2}\,.
\end{eqnarray}
Thus, to a high degree of accuracy, in the IIA limit, the scalar
masses for de Sitter vacua are flavor universal as well as flavor
diagonal and independent of the details of the matter K\"{a}hler
metric described by parameters $\theta_i$. Moreover, to a very
good approximation, they are equal to the gravitino mass. A
natural expectation away from the IIA limit is that the squark and
slepton masses are always of order $m_{3/2}$. Since $m_{3/2}$ is
of several TeV, the scalars are quite heavy, naturally suppressing
flavor changing neutral currents (FCNCs).

%%%%%%%%%%%%%%%%%%%%%%%%%%%%%%%%%%%%%%%%%%%%%%%%%%%%%%%%%%%%%%%%%%%%%%%%%%%%%%%%%%%%%%%%%%%%%
%%%%%%%%%%%%%%%%%%%%%%%%%%%%%%%%%%%%%%%%%%%%%%%%%%%%%%%%%%%%%%%%%%%%%%%%%%%%%%%%%%%%%%%%%%%%%

\subsubsection{Effects of tuning the cosmological constant on the phenomenology}\label{tuning}

In this subsection we would like to give a rough estimate of how
the amount of tuning of the cosmological constant might affect the
values of the soft parameters. In fact, the constraint (\ref{ior})
which sets the cosmological constant to zero in the leading order
still results in a very large value of the cosmological constant
$V_0\sim 0.01\times m_{3/2}^2m_p^2$, once the subleading terms are
taken into account. Exact numerical computations for a manifold
with two moduli reveal that the subleading order corrections can
change the right hand side in (\ref{ior}) by as much as $\sim
5\%$.

The quantity most sensitive to such corrections is the gravitino
mass, since it is  proportional to
$\left(\frac{A_1Q}{A_2P}\right)^{-\frac P{Q-P}}$ and can therefore
change by a factor of order one. Of course, this hardly affects
the distributions of scales of $m_{3/2}$ and the emergence of the
TeV scale peak remains very robust.

Since the tree-level gaugino mass is suppressed in the leading
order as $m_{1/2}\sim
\frac{m_{3/2}}{P\ln\left(\frac{A_1Q}{A_2P}\right)}$, it is
somewhat sensitive to the above corrections and can change by a
few percent. However, such variation in $m_{1/2}$ is very mild,
considering that as these corrections are taken into account the
cosmological constant changes by many orders of magnitude.

Finally, the anomaly mediated gaugino mass contribution
(\ref{anomalyds1}), the trilinear couplings (\ref{tri49}) and
especially the scalar mass squares (\ref{scalarsds}) stay
virtually unaffected by the few percent corrections to
(\ref{ior}). This happens because the dominant contributions to
these soft parameters are completely {\em independent} of
${P\ln\left(\frac{A_1Q}{A_2P}\right)}$ and therefore stay the
same, whereas the contributions which do get affected by the
corrections are largely subdominant since they are proportional to
either $\frac 1{P\ln\left(\frac{A_1Q}{A_2P}\right)}$ or $\frac
1{P^2\ln^2\left(\frac{A_1Q}{A_2P}\right)}$.

These considerations indicate that one should not be concerned
that tuning of the cosmological constant will affect particle
physics phenomenology.
%%%%%%%%%%%%%%%%%%%%%%%%%%%%%%%%%%%%%%%%%%%%%%%%%%%%%%%%%%%%%%%%%%%%%%%%%%%%%%%%%%%%%%%%%%
%%%%%%%%%%%%%%%%%%%%%%%%%%%%%%%%%%%%%%%%%%%%%%%%%%%%%%%%%%%%%%%%%%%%%%%%%%%%%%%%%%%%%%%%%%%

\subsubsection{Radiative Electroweak Symmetry Breaking (REWSB)}\label{rewsb}

It is very important to check whether the soft supersymmetry
breaking parameters in these vacua naturally give rise to
radiative electroweak symmetry breaking (REWSB) at low scales. In
order to check that, one has to first RG evolve the scalar higgs
mass parameters $m_{H_u}^2$ and $m_{H_d}^2$ from the high scale to
low scales. Then one has to check whether for a given $\tan\beta$,
there exists a value of $\mu$ which satisfies the EWSB conditions.
At the one-loop level, we find that EWSB occurs quite generically
in the parameter space. This can be understood as follows. The
gaugino mass contributions to the RGE equation for $m_{H_u}^2$
push the value of $m_{H_u}^2$ up while the top Yukawa coupling,
third generation squark masses and the top trilinear pull it down.
The suppression of the gaugino mass relative to the gravitino mass
causes it to have a negligible effect on the RGE evolution of
$m_{H_u}^2$. On the other hand, the masses of squarks and
$A$-terms are both of ${\cal O}(m_{3/2})$, which guarantees that
$m_{H_u}^2$ is negative at the low scale. Typically, $m_{H_u}^2$
is proportional to $-m_{3/2}^2$, up to a factor less than one
depending on $\tan\beta$. Thus, the EWSB condition can be easily
satisfied with a $\mu$ parameter also of the order $m_{3/2}$. Note
that large $A$-terms (of ${\cal O}(m_{3/2})$) are crucial for
obtaining EWSB. Having large squark masses and small $A$-terms
cannot guarantee EWSB, as is known from the focus point region in
mSUGRA. By low scale effective theory criteria, the EWSB maybe
fine-tuned, but those criteria may change with an underlying
theory. Also, one has to ensure that the third generation squarks
have positive squared masses, which we have checked. We will
report a detailed analysis of these issues in \cite{ToAppear}.

\subsubsection{The $\mu$ and $B\mu$ problem}\label{muBmu}

We will not have much to say about the $\mu$ terms here, leaving a
detailed phenomenological study for our future work
\cite{ToAppear}. We will however, take this opportunity to
highlight the main theoretical issues.

The normalized $\mu$ and $B\mu$ parameters are : \ba \label{mubmu}
\mu &=&
(\frac{\hat{W}^{\star}}{|\hat{W}|}e^{\hat{K}/2}{\mu}'+m_{3/2}Z-e^{\hat{K}/2}F^{\bar{m}}\partial_{\bar{m}}Z)\,
(\tilde{K}_{H_u}\tilde{K}_{H_d})^{-1/2} \\
B\mu &=& (\tilde{K}_{H_u}\tilde{K}_{H_d})^{-1/2}\{
\frac{\hat{W}^{\star}}{|\hat{W}|} e^{\hat{K}/2}
{\mu}'(e^{\hat{K}/2}F^m\,[\hat{K}_m\,+\partial_m\,\ln{\mu}']-m_{3/2})+(2m_{3/2}^2+V_0)\,Z\}
\nonumber\ea

We see from above that the value of the physical $\mu$ and $B\mu$
parameters depend crucially on many of the microscopic details eg.
if the theory gives rise to a non-zero superpotential ${\mu}'$
parameter and/or if a non-zero bilinear coefficient $Z$ is present
in the K\"{a}hler potential for the Higgs fields.  From section
\ref{rewsb}, we see that one requires a $\mu$ term of ${\cal
O}(m_{3/2})$ to get consistent radiative EWSB. This is possible
for eg. when one has a vanishing $\mu'$ parameter and an ${\cal
O}(1)$ higgs bilinear coefficient $Z$, among other possibilities.

\subsubsection{Dark Matter}

For dS vacua with a small cosmological constant, $M_2 << M_1$ at
low scale. In addition, since $\mu$ should be of ${\cal
O}(m_{3/2})$ for consistent EWSB as seen in section \ref{rewsb},
both $M_2$ and $M_1$ are much less than $\mu$. Hence, the LSP is
wino-like. As was discussed in section \ref{tuning}, the tuning of
the cosmological constant has little effect on the gaugino masses,
thereby preserving the gaugino mass hierarchy. It is well known
that winos coannihilate quite efficiently as the universe cools
down. Since the wino masses in these vacua are ${\cal O}(100)$
GeV, the corresponding relic density after they freeze out is very
small. However there could be non-thermal contributions to the
dark matter as well, e.g. the decay of moduli fields into the LSP
after the LSP freezes out. In addition, one should remember that
the above result for a wino LSP is obtained after imposing the
requirement of a small cosmological constant. It would be
interesting to analyze the more general case where the results may
change. We leave a full analysis of these possibilities for the
future \cite{ToAppear}.

\subsubsection{Correlations}

As we have seen, the parameters of the MSSM depend on the
``microscopic constants'' determined by a given $G_2$ manifold and
can be explicitly calculated in principle. Therefore, the
parameters obtained are correlated with each other in general. For
instance we saw that the gaugino and gravitino masses are related.
By scanning over the allowed values of the ``microscopic
constants'', by scanning the space of $G_2$ manifolds, one obtains
a particular subspace of the parameter space of the MSSM at the
unification scale. For a given spectrum and gauge group, the RG
evolution of these parameters to low scales can also be determined
unambiguously, leading to correlations in soft parameters at the
low scale Finally, these correlations in the soft parameters will
lead to correlations in the space of actual observables (for eg,
the LHC signature space) as well. In other words, the predictions
of these vacua will only occupy a \emph{finite} region of the
observable signature space at say the LHC. Since two different
theoretical constructions will have different correlations in
general, this will in turn lead to different patterns of
signatures at the LHC, allowing us to distinguish among different
classes of string/$M$ theory vacua (at least in principle). These
issues, in particular the systematics of the distinguishing
procedure have been explained in detail in \cite{Kane:2006yi}.

\subsubsection{Signatures at the LHC}

The subject of predicting signatures at the LHC for a given class
of string vacua requires considerable analysis. Here, we will make
some preliminary comments, with a detailed analysis to appear in
\cite{ToAppear}.

The scale of soft parameters is determined by the gravitino mass.
We saw in section \ref{CClowsusy} that requiring the cosmological
constant to be very small by imposition of a constraint equation
(eqn. (\ref{er765})) fixes the overall scale of superpartner
masses to be of ${\mathcal O}(1-100)$TeV. Once the overall scale
is fixed, the pattern of soft parameters at $M_{unif}$ is crucial
in determining the signatures at the LHC. As explained earlier,
these $M$ theory vacua give rise to a specific pattern of soft
parameters at $M_{unif}$. We find non-universal gaugino masses
which are suppressed relative to the gravitino mass. We
furthermore expect that the scalar masses and trilinears are of
the same order as the gravitino. The $\mu$ and $B\mu$ parameters
are not yet understood. For phenomenological analysis however, we
may fix them by imposing consistent EWSB.

One can get a sense of the broad pattern of signatures at the LHC
from the pattern of soft parameters. Since gaugino masses are
suppressed and the fact that the anomaly contribution to the
gluino mass parameter approximately {\it cancels} the tree-level
contribution, one would generically get comparatively light
gluinos in these constructions, much lighter than the scalars,
which would give rise to a large number of events for many
signatures, in particular many events with same-sign dileptons and
trileptons in excess of the SM and many events with large missing
energy, even for a modest luminosity of $10\;fb^{-1}$. Since the
gauginos are lighter than the squarks and sleptons, gluino pair
production is likely to be the dominant production mechanism. The
LSP will be a neutralino for the same reason.

It is also possible to distinguish the class of vacua obtained
above from those obtained in Type IIB compactifications, by the
pattern of signatures at the LHC. For the large volume type IIB
vacua, the scalars are lighter than the gluino
\cite{Conlon:2006wz} while for the KKLT type IIB vacua, the
scalars are comparable to the gluino \cite{Choi:2005ge}. This
implies that squark pair production and squark-gluino production
are respectively the dominant production mechanisms at the LHC.
Since the LHC is a $pp$ collider, up-type squarks are
preferentially produced when they are kinematically allowed,
leading to a charge asymmetry which is preserved in cascade decays
all the way to the final state with leptons. On the other hand,
for the class of $M$ theory vacua described here, since gluino
pair production is the dominant mechanism and the decays of the
gluino are charge symmetric (it is a Majorana particle), the $M$
theory vacua predict a much smaller charge asymmetry in the number
of events with one or two leptons and $\geq$ 2 jets compared to
the Type IIB vacua.

\subsubsection{The Moduli and Gravitino Problems}

The cosmological moduli and gravitino problems can exist if moduli
and gravitino masses are too light in gravity mediated SUSY
breaking theories. Naively, after the end of inflation, the moduli
fields coherently oscillate dominating the energy density of the
universe. Since the interactions of the moduli are suppressed by
the Planck scale $(m_p)$, their decay rates are extremely small
leading to the onset of a radiation dominated universe at very low
temperature ($T_R \sim {\mathcal O}(10^{-3})$ MeV for moduli of
${\mathcal O}$(100 GeV-5TeV)), compared to what is required for
successful BBN.

To check if the moduli and gravitino problem can be resolved in
these $M$ theory compactifications, one has to first compute the
masses of the moduli. After doing this, we will use the results to
discuss the moduli and gravitino problems.

The geometric moduli $s_i$ appear in the lagrangian with a kinetic
term given by \be\label{kinetic}
\sum_{i=1}^{N}{\frac{3\,a_i}{4\,s_i^2}\partial_{\mu}s_i\partial^{\mu}s_i}\,,
\ee which is non canonical. The canonically normalized moduli
$\chi_i$ are \be \chi_i\equiv\sqrt{\frac{3\,a_i}2}{\ln}\,s_i\,.
\ee Their mass matrix is \be
\left(m_{\chi}^2\right)_{i\,j}=\frac{2\,\nu^2(a_i\,a_j)^{\frac
1{\,2}}}{3\,N_iN_j}\frac{\partial^2\,V}{\partial s_i\partial
s_j}\,, \ee where we took into account the fact that
$\frac{\partial\,V}{\partial s_i}=0$ at the extremum and that
$\nu_i=\nu$ for all $i=\overline{1,N}$. A fairly straightforward
but rather tedious computation yields the following structure of
the mass matrix \be\label{mmat}
\left(m_{\chi}^2\right)_{i\,j}=\left((a_i\,a_j)^{\frac
1{\,2}}\,K_1+\delta_{ij}\,K_2\right)\times m_{3/2}^2\,, \ee where,
in the large $Q$ approximation keeping $Q-P\sim {\mathcal O}(1)$,
$K_1$ and $K_2$ are given by \ba
&&K_1\approx\frac{112}{27}\left(\frac{\tilde z}{\tilde
x}\right)^2\nu^4\approx\frac{48
\left(Q\,\ln\left(\frac{A_1Q}{A_2P}\right)\right)^4}{343\left(Q-P\right)^4}\,\\
\,\nonumber\\
&&K_2=-\frac{40}9\left(\tilde
L_{1,+}\right)^2-\frac{56}3L_{1,+}-8-2\phi_0^2\left(\frac{a\tilde\alpha}
{\tilde x\phi_0^2}+1\right)^2\approx 10-\frac 8{Q-P}-\frac
8{(Q-P)^2\phi_0^2}-2\phi_0^2\,.\nonumber \ea Using condition
$Q-P=3$, we obtain from (\ref{e58}) in the leading order \be
\phi_0^2\approx(1+\sqrt 3)/3\,, \ee which results in \ba
&&K_1\approx\frac{16}{9261}\left(Q\,\ln\left(\frac{A_1Q}{A_2P}\right)\right)^4\,\\
\,\nonumber\\
&&K_2\approx8-2\sqrt{3}\,.\nonumber \ea Diagonalizing the $N\times
N$ matrix $\left(m_{\chi}^2\right)_{i\,j}$ in (\ref{mmat}) is hard
in general. However, for the ``all equal'' choice when
$a_i=7/(3N)$ for all $i=\overline{1,N}$ this task actually becomes
quite simple.

We use the fact that for an $N\times N$ matrix $A$ whose elements
are \be\label{mata} A_{ij}=a\,\,\,{\rm for}\,\,\,i\not
=j\,,\,\,\,{\rm and}\,\,\,\, A_{ii}=a+b\,, \ee the eigenvalues are
given by \ba
&&\lambda_i=b\,\,\,\,{\rm for}\,\,\,i=1,...,N-1\,,\\
&&\lambda_N=N\,a+b. \ea For the ``all equal'' choice the mass
matrix (\ref{mmat}) is precisely of the form (\ref{mata}) and
therefore we can explicitly determine the entire moduli mass
spectrum. The $N-1$ equal masses are smaller than the remaining
mass. All eigenvalues are positive, confirming we have a de Sitter
minimum. The $N-1$ equal moduli masses are given
by\be\label{modul1} (m_{\chi})_k\approx\sqrt{8-2\sqrt{3}}\,\times
m_{3/2}\approx2\, m_{3/2}\,,\,\,\,\,\,k=1,...,N-1\,, \ee The
remaining heavier modulus has \be\label{modul2}
M_{\chi}\approx\frac{4}{63}\left(Q\,\ln\left(\frac{A_1Q}{A_2P}\right)\right)^2\times
m_{3/2}\,. \ee Note that these masses are independent of the
number of moduli $N$.

After imposing the constraint (\ref{ior}) to make the cosmological
constant small, the mass of the heavy modulus is given by \be
M_{\chi} \approx 448\,m_{3/2} \ee

Since the gravitino mass distribution peaks at ${\mathcal O}(100)$
TeV, which is also in the phenomenologically relevant range, and
the light moduli are roughly twice as heavy compared to the
gravitino, the moduli masses are heavy enough to be consistent
with BBN constraints.

\subsection{Summary of Results}

We have shown that in fluxless $M$ theory vacua the entire
effective potential is generated by non-perturbative effects and
depends upon all the moduli. In this work we have studied such
this potential in detail when the non-perturbative effects are
dominated by strong gauge dynamics in the hidden sector and when
such vacua are amenable to the supergravity approximation. In the
simplest case, we studied $G_2$-manifolds giving rise to two
hidden sectors. The resulting scalar potential has $AdS$ vacua -
most of them with broken supersymmetry and one supersymmetric one.
Then we studied the cases in which there was also charged matter
in the hidden sector under the plausible assumption that the
matter K\"{a}hler potential has weak moduli dependence. In these
cases the potential receives positive contributions from
non-vanishing $F$-term vevs for the hidden sector matter leading
to a unique de Sitter minimum. In all cases we have explicitly
shown that all moduli are stabilized by the potential generated by
strong dynamics.

In the de Sitter  minimum we computed $m_{3/2}$ and found that a
significant fraction of solutions have $m_{3/2}$ in the TeV
region, even though the Planck scale is the only dimensionful
parameter in the theory. The suppression of $m_{3/2}$ is due to
the traditional dimensional transmutation as in the heterotic
theory; what is different here is that the moduli are all
stabilized, which has been difficult in the heterotic case. \ No
small parameters or fine-tunings occur, and the eleven dimensional
$M$ theory scale is slightly above the gauge unification scale but
below the Planck scale. The absence of fluxes is significant for
having $m_{3/2} \sim$ TeV with no small parameters or tuning, and
simultaneously $M_{11}$ not far below the Planck scale.

The problem of why the cosmological constant is not large is of
course not solved by this approach. We do however understand to a
certain extent what properties of $G_2$-manifolds are required in
order to solve it and we suggest that one can set the value of the
potential at the minimum to zero at tree level and proceed to do
phenomenology with the superpartners whose masses are described by
the softly broken Lagrangian. One particularly nice feature is
that we are able to explicitly demonstrate that the soft-breaking
terms are not sensitive to the value of the potential at the
minimum.

When we set the value of the potential at the minimum to zero at
tree level a surprising result occurs. Doing so gives a
non-trivial condition on the solutions. When this condition is
imposed on $m_{3/2}$, for generic $G_{2}$ manifolds it turns out
that the resulting values of $m_{3/2}$ are all in the TeV region.
Thus we do not have to independently set $V_{0}$ to zero
\emph{and} set  $m_{3/2}$ to the TeV region as has been required
in previous approaches.

A more detailed study of the phenomenology of these vacua,
particularly for LHC and for dark matter, is underway and will be
reported in the future. In the present work we presented the
relevant soft-breaking Lagrangian parameters and mentioned a few
broad and generic features of the phenomenology, for both our
generic solutions and for the case where $V_{0}$ is set to zero at
tree level. We presented a standard supergravity calculation of
the soft breaking Lagrangian parameters, and found that the scalar
masses $m_{\alpha },$ and also the trilinears, are approximately
equal to $m_{3/2}$, to the extent that our assumptions about the
matter Kahler potential are valid. Remarkably, the tree-level
gaugino masses are suppressed by a factor of ${\cal O}(10-100)$.
This suppression is present for all $G_2$-manifolds giving the de
Sitter minimum. \ For calculating the tree level gaugino masses
the matter K\"{a}hler potential does not enter, so the obtained
values at tree level are reliable. Because the gaugino suppression
is large, the anomaly mediated mass contributions are comparable
to the tree level ones, and significant cancellations can occur.
Gluinos are generically quite light, and should be produced
copiously at LHC and perhaps even at the Tevatron -- this is an
unavoidable prediction of our approach. We have also checked that
radiative electroweak symmetry breaking occurs over a large part
of the space of $G_2$ manifolds, and that the lightest neutralino
is a good dark matter candidate. It will be exciting to pursue a
number of additional phenomenological issues in our approach,
including inflation, baryogenesis, flavor and CP-violation
physics, Yukawa couplings and neutrino masses.

The approach we describe here apparently offers a framework that
can simultaneously address many important questions, from formal
ones to cosmological ones to phenomenological ones (apart from the
cosmological constant problem,which might be solved in a different
way). Clearly, however much work remains to be done. In
particular, a much deeper understanding of $G_2$-manifolds is
required to understand better some of the assumptions we made
about the K\"{a}hler potential of these vacua.

\chapter{Distinguishing String Constructions from Experimental
Observables}\label{distinguishing}

In the previous sections, we studied two different regions of the
full $M$ theory moduli space from a top-down perspective and
looked at some of their consequences for low energy physics.
However, with the arrival of the LHC, as well as forthcoming data
from other fields such as cosmology, flavor physics, etc. it is
imperative to analyze underlying theoretical models to the extent
that one could make predictions for real physics observables,
signatures at the LHC in particular. Therefore, in the future, we
plan to study the two constructions to the extent that predictions
for physics observables could be made. Moreover, once we have
data, the problem facing us would be the ``Inverse Problem",
\emph{viz.}, ``how to go from data to underlying theory?" One has
to first answer some ``zeroth order" questions in order to address
the Inverse Problem meaningfully, as was explained in the
Introduction. We would like to propose and explore an approach
which allows us to answer them successfully for many specific
classes of string constructions. The basic idea was first proposed
in \cite{Binetruy:2003cy}. Our study here shows that the idea is
very promising and it is possible to realize it in a concrete way.

By studying the pattern of signatures (signatures that are real
experimental observables) for many classes of realistic
microscopic constructions, one may be able to rule out some
classes of underlying theory constructions giving rise to the
observed physics beyond the standard model, and be pointed towards
others. Our results suggest that a lot of this can be done with
limited data and systematically improved with more data and better
techniques.

For concreteness, we focus on traditional low-scale supersymmetry
as new physics beyond the standard model and the underlying
theoretical framework of string theory with different
constructions giving rise to low-scale supersymmetry. While there
exist other possibilities for new physics beyond the standard
model such as technicolor \cite{Weinberg:1975gm}, large extra
dimensions \cite{Arkani-Hamed:1998rs}, warped extra dimensions
\cite{Randall:1998uk}, higgsless models \cite{Nomura:2003du},
composite higgs models \cite{Kaplan:1983sm}, little higgs models
\cite{Arkani-Hamed:2002qx}, split supersymmetry
\cite{Arkani-Hamed:2004fb}, etc., low-scale supersymmetry remains
the most appealing - both theoretically and phenomenologically. In
addition, even though some of these other possibilities may be
embedded in the framework of string theory, low-scale
supersymmetry is perhaps the most natural and certainly the most
popular possibility arising from string constructions. Having said
the above, we would like to emphasize that the proposed technique
is completely general and can be used for any new physics arising
from any theoretical framework whatsoever.

\section{Examples}\label{examples}
We want to focus on compactifications to four dimensions
preserving $\mathcal{N}$=1 supersymmetry, which stabilize all
(most) moduli and have a mechanism of generating the electroweak
scale, allowing us to connect these constructions to real
experimental observables. In addition, we want to focus on
compactifications which are valid within the supergravity
approximation, so that effective four dimensional supergravity
techniques are valid. The motivations for all the above were
already explained in section \ref{hierarchy}. Two very good
examples which show the above features as well as illustrate our
approach, are -- KKLT compactifications \cite{kklt03} and Large
Volume compactifications \cite{Balasubramanian:2005zx}. These
compactifications preserve ${\cal N}=1$ supersymmetry, so they can
be described by a superpotential, k\"{a}hler potential and gauge
kinetic function. Even though both examples fall under the broad
class of Type IIB flux compactifications, they arise in two very
different regions of the flux superpotential, resulting in rather
distinct phenomenological consequences. So, for the purposes of
this work, they will be treated differently. In the following, we
briefly list their main features relevant for phenomenology.

\vspace{0.5cm}

\hspace{4cm} \textit{Type IIB KKLT compactifications} (IIB-K)

\vspace{0.5cm}

This class of constructions is a part of the IIB landscape with
all moduli stabilized\cite{kklt03}. Closed string fluxes are used
to stabilize the dilaton and complex structure moduli at a high
scale and non-perturbative corrections to the superpotential are
used to stabilize the lighter K\"{a}hler moduli. One obtains a
supersymmetric anti-deSitter vacuum and D terms
\cite{Burgess:2003ic} or anti D-branes are used to break
supersymmetry and to lift the vacuum to a deSitter one.
Supersymmetry breaking is then mediated to the visible sector by
gravity. The flux superpotential ($W_0$) has to be tuned very
small to get a gravitino mass of $\mathcal{O}$(1-10 TeV). By
parameterizing the lift from a supersymmetric anti-deSitter vacuum
to a non-supersymmetric deSitter vacuum, one can calculate the
soft terms \cite{Choi:2004sx}. The soft terms depend on the
following microscopic input parameters -- \{$W_0,\alpha,n_i$\} or
equivalently \{$m_{3/2},\alpha,n_i$\}, where $\alpha$ is the ratio
$\frac{F^T/(T+\bar{T})}{m_{3/2}}$ and $n_i$ are the modular
weights of the matter fields \cite{Choi:2004sx}. In addition,
$\tan(\beta)$ and sign($\mu$) are fixed by electroweak symmetry
breaking. A feature of this class of constructions is that the
tree level soft terms are comparable to the anomaly mediated
contributions, which are always present and have been calculated
in \cite{GaNeWu99}.

\vspace{0.5cm}

\hspace{3cm} \textit{Type IIB Large Volume Compactifications}
(IIB-L)

\vspace{0.5cm}

This class of constructions also form part of the IIB landscape
with all moduli stabilized. In this case, the internal manifold
admits a large volume limit with the overall volume modulus very
large and all the remaining moduli small
\cite{Balasubramanian:2005zx}. Fluxes again stabilize the complex
structure and dilaton moduli at a high scale, but the flux
superpotential $W_0$ in this case can be $\mathcal{O}$(1). Higher
order corrections in Large Volume compactifications have been
computed in \cite{Berg:2007wt}. One also incorporates perturbative
contributions to the K\"{a}hler potential in addition to
non-perturbative contributions to the superpotential to stabilize
the K\"{a}hler moduli. A consequence of $W_0=\mathcal{O}$(1) is
that the conclusions are qualitatively different compared to the
KKLT case. Now one gets a non-supersymmetric anti-de Sitter vacuum
in contrast to the KKLT case, which can be lifted to a de Sitter
one by similar mechanisms as in the previous case. Since the
volume is very large, the string scale turns out to be quite low.
Assuming a natural value of $W_0$ to be $\mathcal{O}$(1)
\footnote{we actually varied it roughly from 0.1 to 10.}, to get a
gravitino mass of $\mathcal{O}$(1-10 TeV), one needs the string
scale of $\sim 10^{11}$ GeV. Since the string scale is much
smaller than the unification scale, one cannot have standard gauge
unification in these compactifications with $W_0$ as $O(1)$.
Supersymmetry breaking is again mediated to the visible sector by
gravity and soft terms can be calculated \cite{Conlon:2006us}.
Anomaly mediated contributions turn out to be important for some
soft parameters and have to be accounted for. The soft terms
depend on the following microscopic input parameters - \{${\cal
V},n_i$\} or equivalently \{$m_{3/2},n_i$\}, where $\mathcal{V}$
denotes the volume of the internal manifold and $n_i$ denote the
modular weight of the matter fields. $\tan(\beta)$ and sign($\mu$)
are fixed by electroweak symmetry breaking.

\vspace{0.3cm}

These two classes of compactifications are good for the following
two reasons :

\begin{itemize}
\item These compactifications stabilize \emph{all} the moduli,
making them massive at acceptable scales. This is good for two
reasons -- a) Light scalars (moduli) are in conflict with
astrophysical observations, and b) Since particle physics masses
and couplings explicitly depend on the moduli, one cannot compute
these couplings unless the moduli are stabilized.

\item They have a mechanism for generating a small gravitino mass
($O(1-10)$ TeV). This is essential to deal with the {\it hierarchy
problem}. The mechanisms available for generating a small
gravitino mass may not be completely satisfactory though. For
example, the KKLT vacua require an enormous amount of tuning,
while the Large Volume vacua (with $W_0 = O(1)$) do not have
standard gauge unification at $2\times10^{16}$ GeV. The class of M
theory vacua discussed in the previous chapter though, stabilize
all the moduli, naturally explain the hierarchy and are also
consistent with standard gauge unification \cite{Acharya:2006ia}.
\end{itemize}

There do not exist MSSM-like matter embeddings in the KKLT and
Large Volume classes of vacua at present. However, since many
examples of MSSM-like matter embeddings have been constructed in
simpler type II orientifold constructions, one hopes that it will
be possible to also construct explicit MSSM-like matter embeddings
in these vacua as well in the future. Therefore, we take the
following approach in our analysis -- we {\it assume} the
existence of an MSSM matter embedding on stacks of D7
branes\footnote{for concreteness.} in these vacua and analyze the
consequences for low energy observables. Having said that, it is
important to understand that the assumption of an MSSM-like matter
embedding has been made only for conceptual and computational
simplicity -- a) Any model of low energy supersymmetry must at
least have the MSSM matter spectrum for consistency, so assuming
the MSSM seems to be a reasonable starting point. b) In addition,
most of the software tools and packages available are optimized
for the MSSM. In principle, the approach advocated is completely
general and can be applied to any theoretical construction. The
main point we want to emphasize is that \emph{it is possible to
complete the first steps towards addressing the deeper inverse
problem}. Choosing a different class of vacua or the above vacua
with a different matter embedding will change the results, but not
the properties that it is possible to go from classes of
semi-realistic string vacua to experimental observables and that
classes of string vacua can be distinguished on the basis of their
experimental observables.

In order to illustrate better the fact that our approach works for
any given theoretical construction, we also include some other
classes of constructions in our analysis. These constructions are
{\it inspired} from microscopic string constructions and include
some of their model building and some of their moduli
stabilization features, although not in a completely convincing
and comprehensive manner. Also, in these constructions the
supersymmetry breaking mechanism is not specified explicitly, it
is only parameterized. These constructions serve as nice toy
constructions making it easy to connect these constructions to low
energy phenomenology quickly and efficiently. Therefore, even
though from a strictly technical point of view they only have
educational significance, they are still very helpful in bringing
home the point we want to emphasize.

All the string constructions studied in this work have a thing in
common -- the soft supersymmetry breaking terms at the string
scale are determined in terms of a few parameters. This is in
stark contrast to completely phenomenological models such as
mSUGRA or minimal gauge mediation, where the soft supersymmetry
breaking terms are chosen \emph{by hand} instead of being
determined from a few underlying parameters.

Although for all studied string constructions the soft terms are
determined in terms of a few underlying parameters, the KKLT and
Large Volume constructions differ from the others in the sense
that for these constructions, the parameters which determine the
soft terms are intimately connected to the underlying microscopic
theoretical structure compared to the other constructions. From a
practical point of view though, once the soft terms are
determined, then one can treat all the constructions at par as far
as the analysis of low energy observables is concerned. This also
applies if one is only interested in understanding the origin of
the specific pattern of signatures of a given construction from
its spectrum and soft parameters, as is done in sections
\ref{spectrum} and \ref{fromsoft}. However, in order to understand
the origin of the soft parameters from the structure of the
underlying theoretical construction, it makes more sense to
analyze the KKLT and Large Volume constructions as they are
microscopically better defined\footnote{in the sense that they
provide an explicit mechanism of supersymmetry breaking and moduli
stabilization.} and because they provide a better representation
of phenomenological characteristics of classes of string vacua.
This will be done in section \ref{fromtheory}.

The string-motivated constructions considered in the analysis are
the following:
\begin{itemize}\footnotesize{
\item HM-A -- Heterotic M theory constructions with one modulus.
\item HM-B -- Heterotic M theory constructions with five-branes.
\item HM-C -- Heterotic M theory constructions with more than one
moduli. \item PH-A -- Weakly coupled heterotic string
constructions with non-perturbative corrections to the K\"{a}hler
potential. \item PH-B --  Weakly coupled heterotic string
constructions with a tree level K\"{a}hler potential and multiple
gaugino condensates. \item II-A -- Type IIA constructions on
toroidal orientifolds with Intersecting D branes.}
\end{itemize}

\section{The ``String" Benchmark Pattern Table - Results} \label{results}

Before we explain, we briefly summarize the results so that the
reader can see the goals. The results for the pattern table are
summarized in Table \ref{resultstable}. The rows and columns
constitute eight ``string'' constructions\footnote{One should be
aware of the qualifications made in the previous section.}
analyzed in our study.

\begin{table}[h!]
{\begin{center}
\begin{tabular}{|l|c|p{1.5cm}|p{1.5cm}|p{1.5cm}|p{1.5cm}|p{1.5cm}|p{1.5cm}|p{1.5cm}|p{1.5cm}|}
\hline
  && HM-A & HM-B & HM-C & PH-A & PH-B & II-A & IIB-K & IIB-L
\\ \hline\hline HM-A & & -- & PY & PY & Yes & Yes & Yes & Yes & Yes
\\ HM-B && & -- & PY & Yes & Yes & Yes & Yes & Yes
\\ HM-C && &  & -- & PY & Yes & Yes & PY & PY
\\ PH-A && &  &  & -- & Yes & Yes & Yes & PY
\\ PH-B && &  &  &  & -- & Yes & Yes & Yes
\\ II-A && &  &  &  &  & -- & Yes & Yes
\\
\hline\hline IIB-K && &  &  &  &  &  &  -- & Yes
\\ IIB-L && &  &  &  &  &  &  &  --
\\
\hline
\end{tabular}
\end{center}}
\caption{{\bf The String Pattern Table Results} \newline A ``Yes"
for a given pair of constructions indicates that the two
constructions are distinguishable in a robust way, while a ``No"
indicates that the two models are not distinguishable with
available data (5 $fb^{-1}$ in this case). A "Probably Yes (No)"
means that the two models are (aren't) distinguishable in large
regions of their parameter spaces.}\label{resultstable}
\end{table}

For each construction, we go from the ten or eleven dimensional
string/M theory to its four dimensional effective theory and then
to its LHC signatures. For the string constructions we study, the
task of deducing the effective four-dimensional lagrangian has
already been accomplished. Therefore, we use results for the
description of the effective four-dimensional theories from
literature. However, barring the KKLT constructions \footnote{The
collider phenoemnology of Large Volume constructions had not been
studied at the time of writing, it was studied soon afterwards.},
none of the other constructions have been studied to the extent
that predictions for LHC observables can be made. In this work, we
have studied the phenomenological consequences of each of these
constructions in detail, and computed their LHC signatures. An LHC
signature by definition is one that is really observable at a
hadron collider, e.g. number of events for $n$ leptons, $m$ jets
and $\notEt$, and various ratios of numbers of events, but not
(for example) masses of superpartners or $\tan \beta$. Signatures
are typically of two kinds - counting signatures, as mentioned in
the examples above and distribution signatures, e.g. the effective
mass distribution, invariant mass distribution of various objects,
etc.

The results shown in Table \ref{resultstable} are deduced by
calculating signatures for the various constructions and looking
for signatures that are particularly useful in distinguishing
different constructions, shown in Table \ref{patterntable}. The
details of the procedure involved and the kind of signatures used
will be explained unambiguously in later sections. The results are
shown here so that the interested reader can see the goals. The
rows depict the string constructions used in our study while the
columns consist of useful signatures, which will be defined
precisely later. The \emph{pattern table} (Table
\ref{patterntable}) has been constructed for 5 $fb^{-1}$ of data
at the LHC, which is roughly two years' worth of initial LHC
running. Since everyone is eager to make progress, we focus on
getting early results. More data will allow doing even better.
From Table \ref{resultstable}, it can be seen that most of the
pairs can be distinguished from each other, encouraging optimism
about the power and usefulness of this analysis.

\begin{table}[h!]
{\begin{center}
\begin{tabular}{|l|c|p{1.7cm}|p{1.7cm}|p{1.7cm}|p{1.7cm}|p{1.7cm}|p{1.7cm}|p{1.7cm}|} \cline{1-9}
Signature && A & B & C & D & E & F & G
\\ \hline \hline Condition && $> 1200$ & $> 25$ & $> 1.6 $ & $> 0.54 $ & $> 0.05 $ & $ > 160 $GeV & $> 0.58 $
\\\hline \hline HM-A && OC& OC& OC& OC & OC & Both & OC
\\ \cline{1-9} HM-B && Both & Both & Both & Both & Both& Both &
Both
\\ \cline{1-9} HM-C && Both & Both & OC & Both & Both & Both & Both
\\ \cline{1-9} PH-A && ONC & N.O. & OC & Both & ONC & ONC & Both
\\ \cline{1-9} PH-B && N.O. & N.O. & N.O. & N.O. & N.O. & Both & N.O.
\\ \cline{1-9} II-A && ONC & N.O. & ONC & OC & ONC & ONC & ONC
\\ \cline{1-9} IIB-K && ONC & ONC & OC & ONC & Both & OC & ONC
\\ \cline{1-9} IIB-L && ONC & N.O. & OC & Both  & ONC & ONC & Both
\\ \cline{1-9}
\end{tabular}
\end{center}}
\caption{{\bf The String Pattern Table}\newline An ``$OC$'' for
the $i^{th}$ row and $j^{th}$ column means that the signature is
observable for many models of the $i^{th}$ construction. The value
of the $j^{th}$ signature for the $i^{th}$ construction is
(almost) always consistent with the condition in the second row
and $j^{th}$ column of the Table. An ``$ONC$'' also means that the
signature is observable for many models of $i^{th}$ construction.
However, the value of the signature (almost) always does
\emph{not} consistent with the condition as specified in the
second row and $j^{th}$ column of the Table. A ``$Both$'' means
that some models of the $i^{th}$ construction have values of the
$j^{th}$ signature which are consistent the condition in the
second row and the $j^{th}$ column while other models of the
$i^{th}$ construction have values of the $j^{th}$ signature which
are not consistent with the condition. An ``N.O.'' for the
$i^{th}$ row and $j^{th}$ column implies that the $j^{th}$
signature is \emph{not} observable for the $i^{th}$ construction,
i.e. the values of the observable for all (most) models of the
construction are always below the observable limit as defined by
(\ref{observability}), for the given luminosity (5 $fb^{-1}$). So,
the construction is not observable in the $j^{th}$ signature
channel with the given amount of ``data''.}\label{patterntable}
\end{table}

The logically simplest way to distinguish constructions on the
basis of their signature pattern would be to construct a
multi-dimensional plot which shows that all constructions occupy
different regions in the multi-dimensional space. Since this is
not practically feasible, we construct two dimensional projection
plots for various pairs of signatures. For simplicity in this
initial analysis, each observable signature has been divided into
two classes, based on the value the observable takes. The
observable value dividing the two classes is chosen so as to yield
good results. For a given two dimensional plot for two signatures,
we will have clusters of points representing various
constructions. Each point will represent a set of parameters for a
given construction, which we call a ``model". The cluster of
points representing a given construction may form a connected or
disconnected region. To distinguish any two given constructions,
we essentially look for conditions in this two dimensional plane
which are satisfied by all (most) models of one construction
(represented by one cluster of points) but not satisfied by all
(most) models of the other construction (represented by the other
cluster of points). In this way, it will be possible to
distinguish the two given constructions.

\begin{figure}[h!]
\center \epsfig{file=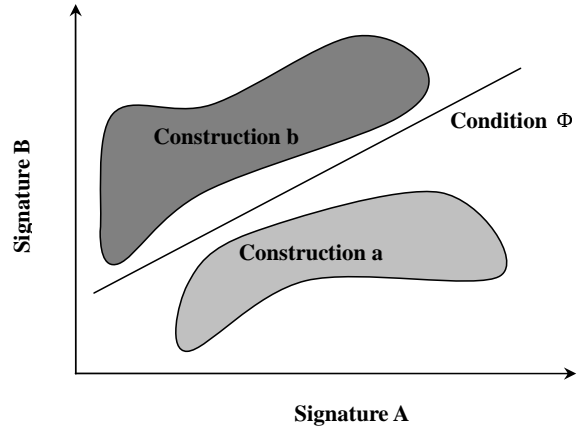,height=6cm,angle=0}
\caption{\footnotesize Cartoon to illustrate the method used to
distinguish constructions. } \label{cartoon}
\end{figure}

The cartoon in Figure \ref{cartoon} illustrates the above point in
a clear way. In a given two dimensional plot with axes given by
signatures $A$ and $B$, we will in general have two clusters of
points for two given constructions $a$ and $b$, as shown by the
light and dark regions respectively. If we define a condition
$\Phi$ on the signatures $A$ and $B$ such that it is given by the
line (or curve in general) shown in the cartoon, then it is
possible to distinguish constructions $a$ and $b$ by the above set
of signatures. To be clear, the above method of distinguishing
theoretical constructions has some possible technical limitations,
which will be addressed in section \ref{limitations}. Since the
purpose here is to explain the overall approach in a simple
manner, we have used the above method. One can make the approach
more sophisticated to tackle more complicated situations, as is
mentioned in section \ref{limitations}.

The counting signatures in Table \ref{patterntable} denote numbers
of events in excess of the Standard Model (SM). A description of
some of the SM backgrounds included is in the next section.

\vspace{0.2cm}

\hspace{3.5cm}\emph{Simple Criterion for Distinguishing
Constructions}

\vspace{0.2cm}

Based on the above pattern table, our criterion for distinguishing
any given pair of constructions (a ``Yes'' in the pattern table)
is that their respective entries are very different in at least
one column, such as an $OC$ for one construction while an $ONC$ or
$N.O.$ for another construction. A $Both$ for one construction
while an $OC$ or $ONC$ or $N.O.$ for another does not distinguish
the constructions cleanly. If there is only a small region of
overlap between the two constructions for all signatures, then the
two constructions can be distinguished in the regions in which
they don't overlap. This would give a ``Probably Yes (PY)'' in the
pattern table, otherwise it would give a ``Probably No (PN)".
Similarly, an $ONC$ for one construction and an $N.O.$ for another
also does not distinguish the two constructions cleanly, and would
give a ``PY'' or ``PN'' depending on their overlap. Carrying out
this procedure for all constructions and signatures gives the
result in Table \ref{resultstable}. It should be kept in mind
though that the result shown in Table \ref{resultstable} is only
for a simple set of signatures. Using more sophisticated
signatures and analysis techniques could give better resulst.
Also, there are typically other useful signatures present than
what is listed in the Table. We have only shown the most useful
ones. In section \ref{distinguishibility}, we give a description
of the useful signatures and explain why these particular
signatures are useful in distinguishing the various constructions
in terms of the spectrum, the soft terms and in turn from the
underlying theoretical structure.

\section{Procedural Details}\label{procedure}

In this section  we enumerate the procedure to answer question (A)
in the Introduction, namely, how to go from a string construction
to the space of LHC signatures.

The first step concerns the spectrum of a given construction. Many
of the string-motivated constructions considered give a
semi-realistic spectrum which contains the MSSM, and perhaps also
some exotics. However for simplicity, in this initial analysis we
only consider the MSSM fields because mechanisms may exist which
project the exotic fields out or make them heavy. As already
explained, for the KKLT and Large Volume vacua, we just assume the
existence of an MSSM matter embedding. The weakly and strongly
coupled heterotic string constructions are naturally compatible
with gauge coupling unification at $M_{unif} \sim 2 \times
10^{16}$ GeV, but type II-A and type II-B constructions are not in
general. One can however try to impose that as an additional
constraint even for type II constructions since gauge unification
provides a very important clue to beyond-the-Standard Model
physics. Therefore, all constructions except the Large Volume
compactifications (IIB-L) used in our analysis either naturally
predict or consistently assume the existence of gauge coupling
unification at $M_{unif}$. The IIB-L construction does not have
the possibility of gauge coupling unification at $M_{unif}$
compatible with having a supersymmetry breaking scale of
$\mathcal{O}$(TeV) \cite{Balasubramanian:2005zx}. The string scale
for the IIB-L constructions is taken to be of ${\cal
O}(10^{10}-10^{11})$ GeV in order to have a supersymmetry breaking
scale of $\mathcal{O}$(TeV), making it incompatible with standard
gauge unification.

In order to connect to low energy four-dimensional physics, one
has to write the effective four dimensional action at the String
scale /11 dim Planck scale, which we will denote by $M_s$. $M_s$
will be equated to $M_{unif}$ for all constructions except for the
Large Volume compactifications (IIB-L). The four-dimensional
effective action for each of these constructions is determined by
a set of microscopic ``input" parameters of the underlying theory.
Soft supersymmetry breaking parameters for the MSSM fields are
calculated at $M_s$ as functions of these underlying input
parameters. The input parameters are taken to vary within
appropriate ranges, as determined by theoretical and
phenomenological considerations. In a more realistic construction,
some of these parameters may actually be fixed by the theory. Our
approach therefore is broad in the sense that we include a wide
range of possibilities without restricting too much to a
particular one.

Each of the ``models'' for a particular construction is thus
defined by a list of input parameters which in turn translate to a
\emph{parameter space} of soft parameters at the unification
scale. In other words, the boundary conditions for the soft
parameters are determined by the underlying microscopic
constructions. As emphasized earlier, this is very different from
an \emph{ad hoc} choice of boundary conditions for the soft
parameters, as in many models like mSUGRA or minimal gauge
mediation.

Once the soft parameters are \emph{determined} from the underlying
microscopic parameters, then the procedure to connect these soft
parameters to low-scale physics is standard, namely, the soft
parameters are evolved through the Renormalization Group (RG)
evolution programs (SuSPECT \cite{Djouadi:2002ze} and SOFTSUSY
\cite{Allanach:2001kg}) to the electroweak scale, and the spectrum
of particles produced is calculated. For concreteness and
simplicity, we assume that no intermediate scale physics exists
between the electroweak scale and the unification scale, though we
wish to study constructions with intermediate scale physics and
other subtleties in the future.

Of all the models thus generated, only some will be consistent
with low energy experimental and observational constraints. Some
of the most important low energy constraints are :

\begin{itemize}\footnotesize{
\item Electroweak symmetry breaking (EWSB). \item Experimental
bounds for superpartner masses. \item Experimental bound for the
Higgs mass. \item Constraints from flavor and CP physics.\item
Upper bound on the relic density from WMAP.}
\end{itemize}

In this analysis, we use $\tan \beta$ as an input parameter and
the RGE software package determines $\mu$ and $B\mu$ by requiring
consistent EWSB. This is because none of the constructions studied
are sufficiently well developed so as to predict these quantities
{\it a priori}. In the future, we hope to consider constructions
which predict $\mu$, $B\mu$ and the Yukawa couplings allowing us
to deduce $\tan \beta$ at the electroweak scale and explain EWSB.
The models we consider are consistent with constraints on particle
spectra, $b \rightarrow s\gamma$ \cite{Abe:2001hk}, $(g_{\mu}-2)$
\cite{Bennett:2002jb} and the upper bound on relic density. One
should not impose the lower bound on relic density since
non-thermal mechanisms can turn a small relic density into a
larger one. Also, the usually talked about lower bound on the LSP
mass ($\sim 50$ GeV) is only present for models with gaugino mass
unification. There is no general lower bound on the LSP mass,
especially if one also relaxes the constraint of a thermal relic
density. In our analysis, we have not imposed any lower bound on
the LSP mass. In our study at present, we have generated $\sim 50$
models for most constructions ($\sim 100$ models for the PH-A
construction)\footnote{However, not all 50 (or 100) models will be
above the observable limit in general.}. This might seem too small
at first. However from our analysis it seems that the results we
obtain are robust and do not change when more points are added.
For a purely statistical analysis this could be a weak point, but
because in each case we know the connection between the theory and
signatures and understand why the points populate the region they
do, we expect stable results. We simulated many more models for
two particular constructions and found that the qualitative
results do not change, confirming our expectations. This will be
explained in section \ref{limitations}.

Once one obtains the spectrum of superpartners at the low scale,
one calculates matrix elements for relevant physics processes at
parton level which are then evolved to ``long-distance" physics,
accounting for the conversion of quarks and gluons into jets of
hadrons, decays of tau leptons,etc. We carry out this procedure
using PYTHIA 6.324 \cite{Sjostrand:2003wg}. The resulting hadrons,
leptons and photons have to be then run through a detector
simulation program which simulates a real detector. This was done
by piping the PYTHIA output to a modified CDF fast detector
simulation program PGS \cite{pgs}. The modified version was
developed by John Conway, Stephen Mrenna and others and
approximates an ATLAS or CMS-like detector. The output of the PGS
program is in a format which is also used in the LHC Olympics
\cite{LHCO}. It consists of a list of objects in each event
labelled by their identity and their four-vector. Lepton objects
are also labelled by their charges and b-jets are tagged. With
this help of these, one can construct a wide variety of
signatures.

The precise definitions of jets and isolated leptons, criteria for
hadronically decaying taus, efficiencies for heavy flavor tagging
as well as trigger-level cuts imposed on objects are the same as
used for the ``blackbox" data files in the LHC Olympics
\cite{LHCO}. In addition, we impose event selection cuts as the
following:
\begin{itemize}
\item If the event has photons, electrons, muons or hadronic taus,
we only select the particles which satisfy the following -- Photon
$P_T\;> 10$ GeV; Electron, Muon $P_T\;
> 10$ GeV; (hadronic) Tau $P_T\; > 100$ GeV. \item For any event with jets, we
only select jets with $P_T\; > 100$ GeV. \item Only those events
are selected with $\notE_T\;$ in the event $> 100$ GeV.
\end{itemize}
These selection cuts are quite simple and standard. We have used a
simple and relatively ``broad-brush" set of cuts since for a
preliminary analysis, we want to analyze many constructions
simultaneously. As mentioned before, we have simulated 5 $fb^{-1}$
of data at the LHC for each model. A small luminosity was chosen
for two reasons -- first, in the interest of computing time and
second, in order to argue that our proposed technique is powerful
enough to distinguish between different constructions even with a
limited amount of data. Of course, to go further than
distinguishing classes of constructions broadly from relatively
simple signatures, such as getting more insights about particular
models within a given construction, one would in general require
more data and could sharpen the approach by imposing more
exclusive cuts and signatures.

In order to be realistic, one has to take the effects of the
standard model background into account. In our analysis, we have
simulated the $t\bar{t}$ background and the diboson ($WW,ZZ$)+
jets background. We have not included the uniboson ($W,Z$) + jets
background in the interest of time. However, from
\cite{Baer:1995va} we know that for a $\notE_T$ threshold of 100
GeV, as has been used in our analysis, the $t\bar{t}$ background
is either the largest background or comparable to the largest one
($W$ + jets typically). Therefore, we expect our analysis and
results to be robust against addition of the $W$ + jets
background. The criteria we employ for an observable signature is
:
\begin{equation}
\frac{N_{signal}}{\sqrt{N_{bkgd}}} > 4; \;\;
\frac{N_{signal}}{N_{bkgd}} > 0.1; \;\; N_{signal} > 5.
\label{observability}
\end{equation}

These conditions are quite standard. A signature is observable
only when the most stringent of the three constraints is
satisfied.

Although the steps used in our analysis are quite standard, there
are at least two respects in which our analysis differs from that
of previous ones in the literature. First, our work shows for the
first time that with a few reasonable assumptions, one can study
string theory constructions to the extent that reliable
predictions for experimental observables can be made, and more
importantly, different string constructions give rise to
overlapping but distinguishable footprints in signature space.
Moreover, it is possible to understand \emph{why} particular
combinations of signatures are helpful in distinguishing different
constructions, from the underlying theoretical structure of the
constructions. Therefore, even though we have done a simplified
(but reasonably realistic) analysis in terms of trigger and
selection level cuts, detection efficiencies of particles,
detector simulation and calculation of backgrounds, we expect that
doing a more sophisticated analysis will only change some of the
details but not the qualitative results. In particular, it will
not affect the properties that predictions for experimental
observables can be made for many classes of realistic string
constructions, and that patterns of signatures are sensitive to
the structure of the underlying string constructions, making it
possible to distinguish among various classes of string
constructions.

\section{Distinguishibility of Constructions} \label{distinguishibility}

\subsection{General Remarks}\label{remarks}

For convenience, we present the signature pattern table again. As
was also mentioned in section \ref{results}, each signature has
been broadly divided into two main classes for simplicity. The
value of the observable dividing the two classes is chosen so as
to yield the best results. A description of the most useful
signatures is given below:

\begin{table}[h!]
{\begin{center}
\begin{tabular}{|l|c|p{1.7cm}|p{1.7cm}|p{1.7cm}|p{1.7cm}|p{1.7cm}|p{1.7cm}|p{1.7cm}|} \cline{1-9}
Signature && A & B & C & D & E & F & G
\\ \hline \hline Condition && $> 1200$ & $> 25$ & $> 1.6 $ & $> 0.54 $ & $> 0.05 $ & $ > 160 $GeV & $> 0.58 $
\\\hline \hline HM-A && OC& OC& OC& OC & OC & Both & OC
\\ \cline{1-9} HM-B && Both & Both & Both & Both & Both& Both &
Both
\\ \cline{1-9} HM-C && Both & Both & OC & Both & Both & Both & Both
\\ \cline{1-9} PH-A && ONC & N.O. & OC & Both & ONC & ONC & Both
\\ \cline{1-9} PH-B && N.O. & N.O. & N.O. & N.O. & N.O. & Both & N.O.
\\ \cline{1-9} II-A && ONC & N.O. & ONC & OC & ONC & ONC & ONC
\\ \cline{1-9} IIB-K && ONC & ONC & OC & ONC & Both & OC & ONC
\\ \cline{1-9} IIB-L && ONC & N.O. & OC & Both  & ONC & ONC & Both
\\ \cline{1-9}
\end{tabular}
\end{center}}
\caption{{\bf The String Pattern Table}\newline An ``$OC$'' for
the $i^{th}$ row and $j^{th}$ column means that the signature is
observable for many models of the $i^{th}$ construction. The value
of the $j^{th}$ signature for the $i^{th}$ construction is
(almost) always consistent with the condition in the second row
and $j^{th}$ column of the Table. An ``$ONC$'' also means that the
signature is observable for many models of $i^{th}$ construction.
However, the value of the signature (almost) always does
\emph{not} consistent with the condition as specified in the
second row and $j^{th}$ column of the Table. A ``$Both$'' means
that some models of the $i^{th}$ construction have values of the
$j^{th}$ signature which are consistent the condition in the
second row and the $j^{th}$ column while other models of the
$i^{th}$ construction have values of the $j^{th}$ signature which
are not consistent with the condition. An ``N.O.'' for the
$i^{th}$ row and $j^{th}$ column implies that the $j^{th}$
signature is \emph{not} observable for the $i^{th}$ construction,
i.e. the values of the observable for all (most) models of the
construction are always below the observable limit as defined by
(\ref{observability}), for the given luminosity (5 $fb^{-1}$). So,
the construction is not observable in the $j^{th}$ signature
channel with the given amount of ``data''.}\label{patterntable2}
\end{table}

\begin{itemize}
\footnotesize{ \item A -- Number of events with trileptons and
$\geq 2$ jets. The value of the observable dividing the signature
into two classes is 1200. \item B -- Number of events with clean
(not accompanied by jets) dileptons. The value of the observable
dividing the signature into two classes is 25. \item C -- $(Y/X)$
\footnote{The ratio $(Y/X)$ is computed only when both signatures
$X$ and $Y$ are above the observable limit.}; Y= Number of events
with 2 leptons, 0 b jets and $\geq 2$ jets, X= Number of events
with 0 leptons, 1 or 2 b jets and $\geq 6$ jets\footnote{This
signature is not very realistic in the first two years. Please
read the discussion in this subsection as to why this signature is
still used.}. The value of the observable dividing the signature
into two classes is 1.6. \item D -- $(Y/X)$; Y= Number of events
with 2 leptons, 1 or 2 b jets and $\geq 2$ jets, X= Number of
events with 2 leptons, 0 b jets and $\geq 2$ jets. The value of
the observable dividing the signature into two classes is 0.54.
\item E -- The charge asymmetry in events with one electron or
muon and $\geq$ 2 jets
($A_{c}^{(1)}=\frac{N(l^+)-N(l^-)}{N(l^+)+N(l^-)}$). The value of
the observable dividing the signature into two classes is 0.065.
\item F -- The peak of the missing energy distribution. The value
of the observable dividing the signature into two classes is 160
GeV. \item G -- $Y/X$; Y = Number of events with same sign
different flavor (SSDF) dileptons and $\geq 2$ jets, X = Number of
events with 1 tau and $\geq$ 2 jets. The value of the observable
dividing the signature into two classes is 0.5.}\end{itemize}

Although there exist other signatures which can distinguish among
some of the above constructions, the above set of signatures turn
out to be the most economic and useful in distinguishing all
constructions considered\footnote{It is important to understand
that these signatures were useful in distinguishing constructions
which have at least some models giving rise to observable
signatures with the given luminosity (5 $fb^{-1}$). With more
luminosity, many more models of these constructions would give
rise to observable signatures, so one would in general have a {\it
different} set of useful distinguishing signatures.}. We
understand that some of these signatures are not very realistic.
For example, the signature which counts the number of events with
0 leptons, 1 or 2 b jets and $\geq$ 6 jets is not very realistic
initially because of difficulties associated with calibrating the
fake missing $\notE_T$ from jet mismeasurement in events with six
or more jets. However, we have used this signature in our analysis
at this stage because it helps in explaining our results and the
approach in an economic way. Also, the fact that there are other
useful signatures\footnote{although they may be less economical in
the sense that one would need more signatures to distinguish the
same set of constructions.} which distinguish these constructions
gives us additional confidence about the robustness of our
approach. These signatures were hand-picked by experience and by
trial-and-error. Once the set of useful signatures was collected,
the next task was to understand \emph{why} the above set of
signatures were useful in distinguishing the constructions based
on their spectrum, soft parameters and their underlying
theoretical setup. This is the subject of sections \ref{spectrum},
\ref{fromsoft} and \ref{fromtheory} respectively. We hope that
carrying out the same exercise for other constructions can help
build intuition about the kind of signatures which any given
theory can produce. This can eventually help in building a
dictionary between structure of underlying theoretical
constructions and their collider signatures.

The above list of signatures consists of counting signatures and
distribution signatures at the LHC. The counting signatures denote
number of events in excess of the Standard Model. Naively, one
would think that the number of signatures is very large if one
includes lepton charge and flavor information, and $b$ jet
tagging. However, it turns out that not all signatures are
independent. In fact, they can be highly correlated with each
other, drastically decreasing the effective dimensionality of
signature space. Thus, in order to effectively distinguish
signatures, one needs to use signatures sufficiently orthogonal to
each other. This has been emphasized recently in
\cite{Arkani-Hamed:2005px}. We will see that having an underlying
theoretical construction allows us to actually find those useful
signatures. Even though we have only listed a few useful
signatures in Table \ref{patterntable2}, there are typically more
than one (sometimes many) signatures which distinguish any two
particular constructions. This is made possible by a knowledge of
the structure of the underlying theoretical constructions.

\subsection{Why is it possible to distinguish different Constructions?}\label{whypossible}

In view of the above comments, one would like to understand why it
is possible to distinguish different constructions in general and
why the signatures described in the previous section are useful in
distinguishing the various constructions in particular.

To understand the origin of distinguishibility of constructions,
one should first understand why each construction gives rise to a
specific pattern of soft supersymmetry breaking parameters and in
turn to a specific pattern of signatures. This is mainly due to
\emph{correlations in parameter space as well as in signature
space}. Let's explain this in detail. A construction is
characterized by its spectrum and couplings in general. These
depend on the underlying structure of the theoretical
construction, such as the form of the four-dimensional effective
action, the mechanism to generate the hierarchy, the details of
moduli stabilization and supersymmetry breaking, mediation of
supersymmetry breaking etc. At the end of the day, the theoretical
construction is defined by a small set\footnote{if they are indeed
``good" theoretical constructions.} of microscopic input
parameters in terms of which \emph{all} the soft supersymmetry
breaking parameters are computed. Since \emph{all} soft parameters
are calculated from the \emph{same} set of underlying input
parameters, this gives rise to correlations in the space of soft
parameters for any given construction. These correlations carry
through all the way to low energy experimental observables, as
will be explicitly seen later. The fact that there exist
correlations between different sets of parameters which have their
origin in the underlying theoretical structure allows us to gain
insights about the underlying theory, and is much more powerful
than completely phenomenological parameterizations such as mSUGRA,
minimal gauge mediation, etc.

Since any two \emph{different} theoretical constructions will
differ in their underlying structure in some way \emph{by
definition}, the correlations obtained in their parameter and
signature spaces will also be different in general. All these will
in general have \emph{different} effects on issues which influence
low energy phenomenology in an important manner, such as the scale
of supersymmetry breaking, unification of gauge couplings (or
not), flavor physics, origins of CP violation, etc. These factors
combined with experimental constraints allow different string
constructions to be distinguished from each other in general.

We now wish to understand why the particular signatures described
in section \ref{remarks} are useful in distinguishing the studied
constructions. In order to successfully do so, one has to
understand the relevant features of the various constructions and
their implications to hadron collider phenomenology, and devise
signatures which are sensitive to those features.

In the following subsections, we explain how to distinguish the
above constructions. In principle, one could directly try to
connect patterns of signatures to underlying string constructions.
However, in practice it is helpful to divide the whole process of
connecting patterns of signatures to theoretical constructions in
a few parts -- first, the results of the pattern table for each
construction are explained based on the spectrum of particles at
the low scale; second, important features of the spectrum of
superpartners at the low scale for the different constructions
(which give rise to their characteristic signature patterns) are
explained in terms of the soft supersymmetry breaking parameters
at the high scale; and third, the structure of the soft parameters
is explained in terms of the underlying theoretical structure of
the constructions.

For readers not interested in the details in the next subsection,
the main points to take away are that any given theoretical
construction only gives rise to a specific pattern of observable
signatures, and that one can understand and trust the regions of
signature space that are populated by a given theoretical
construction and that such regions are quite different for
different constructions, illustrating in detail that LHC
signatures can distinguish different theoretical constructions.

\subsection{Explanation of Signatures from the Spectrum}\label{spectrum}

In this subsection, we take the spectrum pattern for different
constructions as given and explain patterns of signatures based on
them. Then it is possible to treat all the constructions equally
as far as the explanation of the pattern of signatures from the
spectrum is concerned. The characteristic features of the spectrum
for the constructions considered are as follows:

\begin{itemize}{\footnotesize
\item HM-A  –- Universal soft terms. Bino LSP (``coannihilation
region"\footnote{Explained in
\ref{fromtheory}.\label{footnote6}}). Moderate gluinos (550-650
GeV), slightly lighter scalars. \item HM-B -- Universal soft
terms. Has two (disconnected) regions. Region I similar to HM-A.
Region II either ``focus point region"\footnote{Explained in
\ref{fromtheory}} or ``funnel region"\footnote{Explained in
\ref{fromtheory}.} Scalars much heavier ($> 800$ GeV) than
gauginos in region II. \item HM-C -- Non-universal soft terms.
Occupies a big region in signature space encompassing the two
regions mentioned for the HM-B construction. Heavy scalars. Can
have bino, wino or higgsino LSP. Spectrum and signature pattern
quite complicated. \item PH-A  --- Non-universal soft terms. Bino,
higgsino or mixed bino-higgsino LSP. Bino LSP has light gluino ($<
600$ GeV). Higgsino or mixed bino-higgsino LSP have gluinos
ranging from moderately heavy to heavy ($600-1200$ GeV). Heavy
scalar masses ($\geq 2$ TeV). \item PH-B -- Non-universal soft
terms. Wino and bino LSP. Light gluinos (200-550 GeV) always have
wino LSPs while heavier gluinos can have bino or wino LSPs.
Comparatively heavy LSP (can be upto 1 TeV), heavy scalar masses
($\geq 1$ TeV) except stau which is relatively light ($\geq 500$
GeV). \item II-A -- Non-universal soft terms. Can have bino, wino,
higgsino or mixed bino-higgsino LSP. Light or moderately heavy
gluino ($300-600$ GeV), scalar masses heavier than gluinos but not
very heavy ($< 1$ TeV). Stops can be as light as 500 GeV. Spectrum
and signature pattern quite complicated. \item IIB-K --
Non-universal soft terms. Heavy spectrum ($\geq$ 1 TeV) in
general, but possible to have light spectrum ($\leq$ 1 TeV). Bino
LSP\footnote{We have only analyzed $\alpha > 0$.}. Gluinos are
greater than about $450$ GeV while the lightest squarks
($\tilde{t}_1$) are greater than about $200$ GeV. For some models,
$\tilde{\tau}_1$ can be light as well. \item IIB-L --
Non-universal soft terms. Mixed bino-higgsino LSP. The gluinos
have a lower bound of about $350$ GeV, while the lightest squark
($\tilde{t}_1$) has a lower bound of about 700 GeV. }
\end{itemize} \label{spectra}

\begin{figure}[h!]
\begin{center}
 \epsfig{file=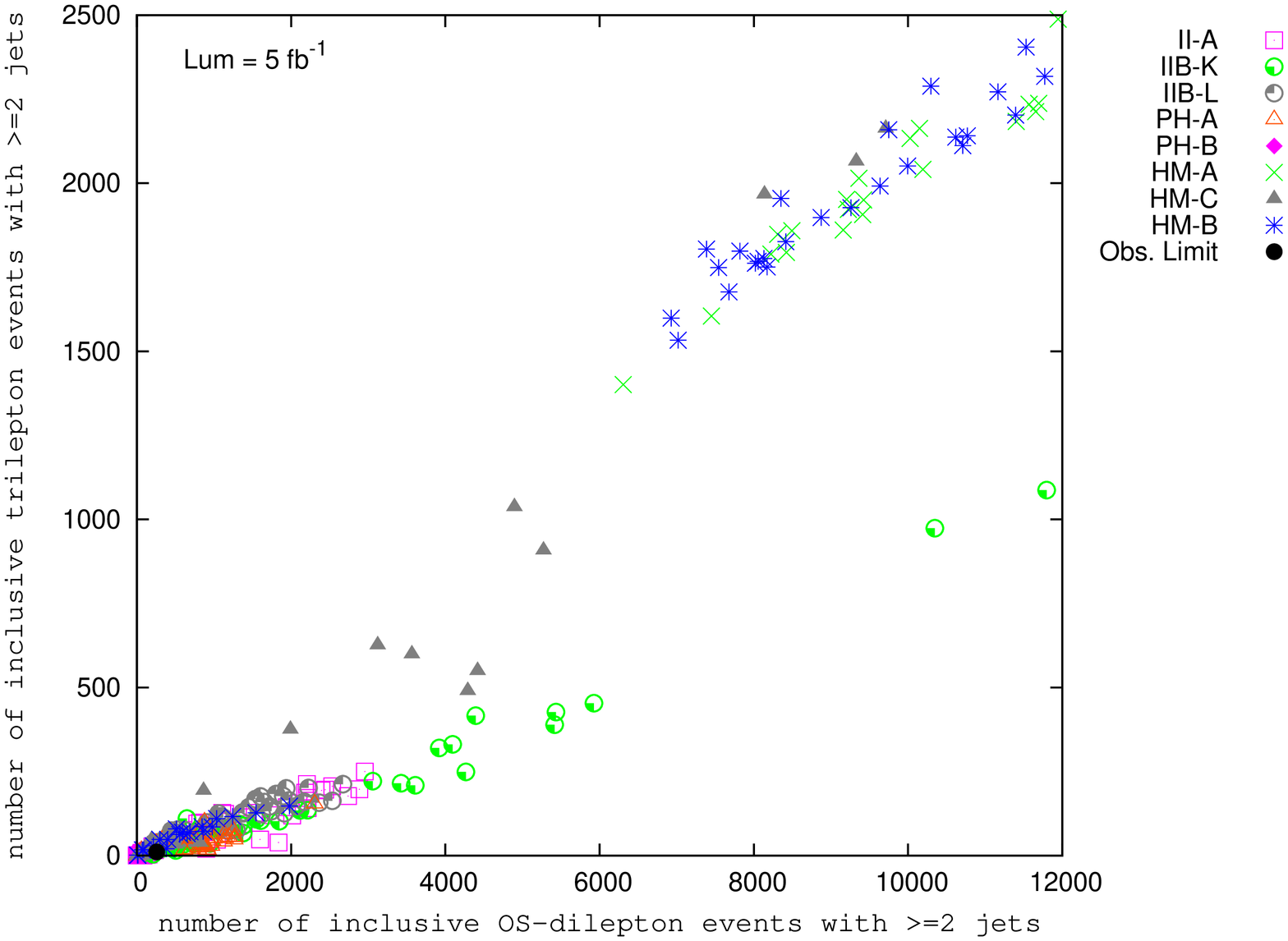,height=16cm, angle=-90}
\end{center}
\caption{Plot of number of events with opposite-sign dileptons and
$\geq$ 2 jets and number of events with three leptons and $\geq$ 2
jets. The black dot represents the lower limit of observability of
the two signatures, according to conditions in equation
(\ref{observability}). Note that the HM-A and overlaping HM-B and
HM-C construction can be distinguished easily from the PH-A, PH-B,
II-A, IIB-K and IIB-L constructions, as they occupy very different
regions. The plots are best seen in color.} \label{OS}
\end{figure}

As mentioned in the list of characteristic features of the
spectrum above, the HM-B and HM-C constructions roughly occupy two
regions in signature space, as can be seen from Figure \ref{OS}.
One of these regions overlaps with the HM-A construction. This is
because the HM-B and HM-C constructions contain the HM-A
construction as a subset. Since the three constructions have the
same theoretical structure in a region of their high scale
parameter space, the models of the three constructions in that
particular region cannot be distinguished from each other from
their signature pattern as their signatures will always overlap.
However, since the HM-B and HM-C constructions have a bigger
parameter space, they also occupy a bigger region of signature
space compared to the HM-A construction. Thus, it is possible to
distinguish the HM-A, HM-B and HM-C constructions in regions in
which they don't overlap, i.e. in regions in which their
underlying theoretical structure is different. This is the origin
of ``Probably Yes (PY)'' in Table \ref{resultstable}.

\begin{figure}[h!]
\begin{center}
\epsfig{file=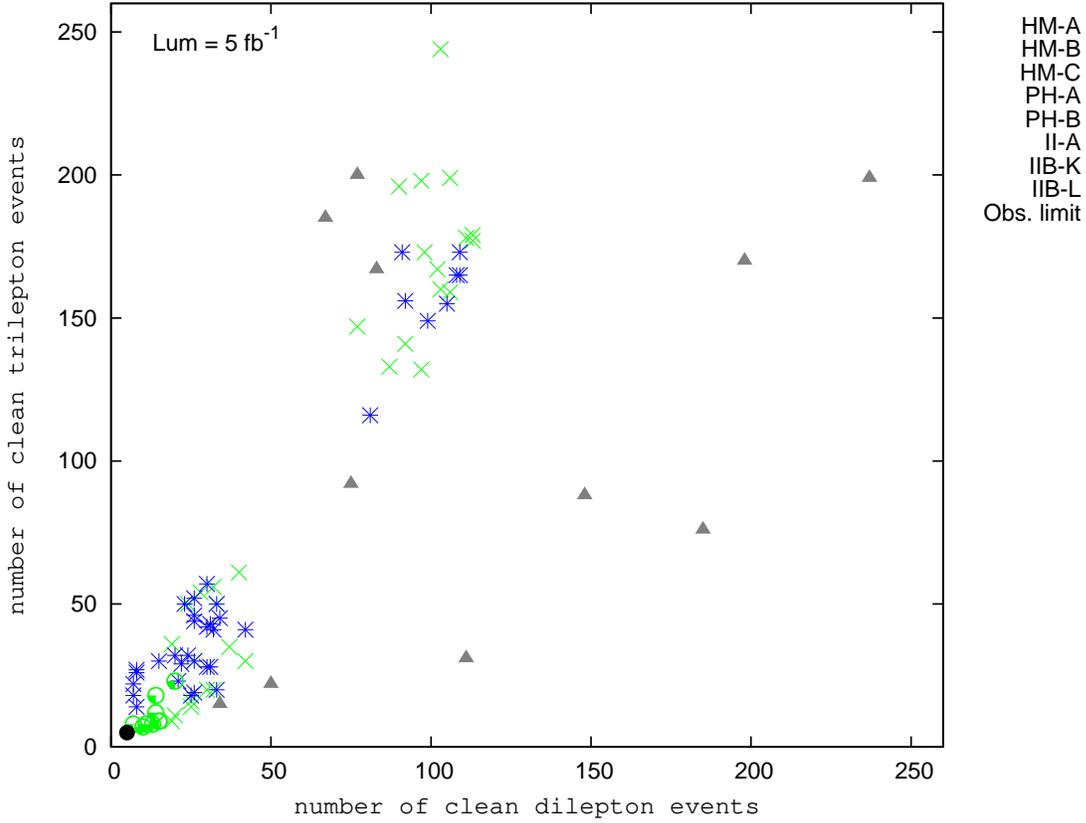,height=16cm, angle=-90}
\end{center}
\caption{Plot of number of events with clean dileptons and number
of events with clean trileptons. ``clean" means not accompanied by
jets. The black dot represents the lower limit of observability of
the two signatures, according to conditions in equation
(\ref{observability}). The models below the observable limit have
not been shown. Note that the HM-A and overlapping HM-B and HM-C
constructions can be distinguished from the PH-A, PH-B, II-A and
IIB-L constructions, since the latter are not observable with the
given luminosity. The plots are best seen in color.} \label{dilep}
\end{figure}

Figures \ref{OS} and \ref{dilep} shows two (very)\footnote{since
we do not take into account the charge and flavor information for
leptons, and the flavor information for jets (whether the jets are
of heavy flavor or not). } inclusive signature plots\footnote{The
figures are best seen in color.}. One can try to explain the
differences in these signatures among the HM-A construction and
the overlapping HM-B and HM-C constructions on the one hand and
the PH-A, PH-B, II-A, IIB-K and IIB-L constructions on the other,
from their spectra.

The HM-A construction and the overlapping HM-B, HM-C constructions
have a comparatively light spectrum at the low scale. These give
rise to a subset of the well known mSUGRA boundary conditions, so
after imposing constraints from low energy physics as in section
\ref{procedure}, one finds that the allowed spectrum consists of
light and moderately heavy gluinos, slightly lighter squarks and
light sleptons. Thus, $\tilde{g}\tilde{q}$ production and
$\tilde{q}\tilde{q}$ pair production are dominant with direct
$\tilde{N_2}\,\tilde{C_1}$ production also quite important.

Both squarks and gluinos ultimately decay to $\tilde{N_2}$ and
$\tilde{C_1}$ and since most sleptons (including selectrons,
smuons) are accessible, both $\tilde{N_2}$ and $\tilde{C_1}$ decay
to leptons and the LSP via sleptons. Since the mass difference
between $\tilde{N_2}$, $\tilde{C_1}$ and LSP ($\tilde{N_1}$) is
big (because of universal boundary conditions, $\Delta
M_{\tilde{N_2}-\tilde{N_1}}, \Delta M_{\tilde{C_1}-\tilde{N_1}}
\sim M_{\tilde{N_1}}$), most of the leptons produced pass the
cuts. The $\tilde{N_2}$ and $\tilde{C_1}$ by themselves are also
comparatively light ($< 200$ GeV). On the other hand, the PH-A,
PH-B, II-A and IIB-L constructions are required to have heavy
scalars and gluinos varying in mass from light to heavy. So, the
$\tilde{N_2}$ and $\tilde{C_1}$ produced from gluinos decay to the
LSP mostly through a virtual $Z$ and $W$ respectively, which makes
their branching ratio to leptons much smaller. Because of
\emph{non-universal} soft terms, $\Delta
M_{\tilde{N_2}-\tilde{N_1}}$ and $\Delta
M_{\tilde{C_1}-\tilde{N_1}}$ can be bigger or smaller than in the
universal case. In the PH-A, PH-B, II-A and IIB-L constructions,
they are required to be comparatively smaller, leading to leptons
which are comparatively softer on average, many of which do not
pass the cuts.

For clean dilepton events, direct production of $\tilde{N_2}$ and
$\tilde{C_1}$ is required. The HM-A construction and overlapping
HM-B and HM-C constructions have comparatively lighter
$\tilde{N_2}$ and $\tilde{C_1}$, so
 $\tilde{N_2}$ and $\tilde{C_1}$ are directly produced. On the other hand, most models of the
PH-B and II-A constructions have heavier $\tilde{N_2}$ and
$\tilde{C_1}$ compared to the HM-A construction, making it harder
to produce them directly. The PH-A and IIB-L constructions have
some models with light $\tilde{N_2}$ and $\tilde{C_1}$, but the
other factors (decay via virtual W and Z, and smaller mass
separation between $\tilde{N_2}$, $\tilde{C_1}$ and LSP) turn out
to be more important, leading to no observable clean dilepton
events. Therefore, the result is that \emph{none} of the models of
the PH-A, PH-B, II-A and IIB-L constructions have observable clean
dilepton events. Thus, it is possible to distinguish the HM-A
construction and overlapping HM-B and HM-C constructions from the
PH-A, PH-B, II-A and IIB-L constructions by signatures A and B in
Table \ref{patterntable2} (shown in Figures \ref{OS} and
\ref{dilep}).

The case with the IIB-K construction is slightly different. These
constructions have many models with a heavy spectrum which implies
that those models do not have observable events with the given
luminosity of 5 $fb^{-1}$.  However, these constructions can also
have light gluinos and squarks with staus also being light in some
cases. So $\tilde{g}\tilde{q}$ production is typically dominant
for these models. The gluinos and squarks decay to $\tilde{N_2}$
and $\tilde{C_1}$ as for other constructions. Since for many IIB-K
models, the lightest stau is heavier than $\tilde{N_2}$ and
$\tilde{C_1}$ (even though it is relatively lighter than in the
PH-A, PH-B, II-A and IIB-L constructions), the $\tilde{N_2}$ and
$\tilde{C_1}$ decay to the LSP through a virtual $Z$ and $W$
respectively, making the branching fraction to leptons much
smaller than for the HM-A and overlapping HM-B and HM-C
constructions. In addition, the mass differences $\Delta
M_{\tilde{N_2}-\tilde{N_1}}$ and $\Delta
M_{\tilde{C_1}-\tilde{N_1}}$ are required to be smaller for the
IIB-K construction in general compared to that for the HM-A and
overlapping HM-B and HM-C constructions, making it harder for the
leptons to pass the cuts. Some IIB-K models have comparable mass
differences $\Delta M_{\tilde{N_2}-\tilde{N_1}}$ and $\Delta
M_{\tilde{C_1}-\tilde{N_1}}$ as the HM-A construction, but they
have much heavier gluinos compared to those for the HM-A
construction, making their overall cross-section much smaller.
Therefore, the IIB-K construction has fewer events for leptons in
general (in particular for trileptons) compared to that for the
HM-A and overlapping HM-B and HM-C constructions, as seen from
Figure \ref{OS}.

\begin{figure}[h!]
\begin{center}
 \epsfig{file=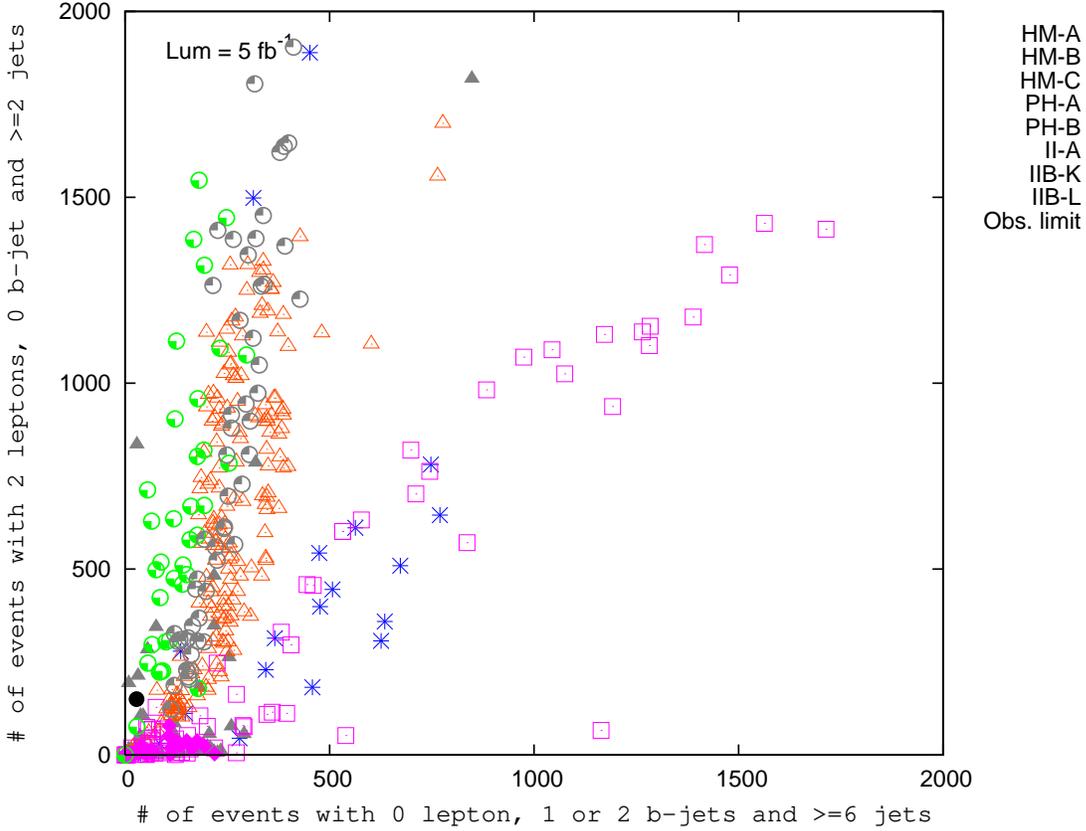,height=16cm, angle=-90}
\end{center}
\caption{Plot of number of events with 0 leptons, 1 or 2 b jets
and $\geq$ 6 jets and number of events with 2 leptons, 0 b jets
and $\geq$ 2 jets. The black dot represents the lower limit of
observability of the two signatures, according to conditions in
equation (\ref{observability}). The models below the observable
limit have also been shown to emphasize that the II-A construction
has very different number of events for these signatures compared
to other constructions even {\it without} imposing the
observability constraint. Note that the PH-B and II-A
constructions can be distinguished from the PH-A, IIB-K and IIB-L
constructions because they have very different slopes. The plots
are best seen in color.} \label{0l12b6j}
\end{figure}

Region II of the HM-B construction and the non-overlapping region
of the HM-C construction (with HM-A) cannot be cleanly
distinguished from the PH-A, PH-B, II-A, IIB-K and IIB-L
constructions from the above signatures. Region II of the HM-B
construction (the ``focus point'' or ``funnel region'' of mSUGRA)
however, can be distinguished from these constructions with the
help of other signatures\footnote{For example, the signatures
shown in Figure \ref{0l12b6j} can distinguish Region II of the
HM-B construction (the HM-B models distinct from the
PH-A,PH-B,IIB-K and IIB-L region all belong to Region II) with the
PH-A, IIB-K and IIB-L constructions. As another example, the ratio
of number of events with 0 b jets and $\geq$ 2 jets and number of
events with $\geq$ 3 b jets and $\geq$ 2 jets can distinguish
Region II of HM-B with PH-B, II-A and IIB-K constructions. These
can be explained on the basis of their spectra, but has not been
done here for simplicity. Also, the HM-B row in Table
\ref{patterntable2} has not been divided into two parts (to
account for the two regions) to avoid clutter.}. The
non-overlapping region of the HM-C construction is a very big
region in signature space because of its big parameter space,
making it relatively harder to distinguish it from some of the
other constructions. Since we have not found signatures cleanly
distinguishing the \emph{whole} msoft-II region from some of the
other constructions, we have put a ``Probably Yes'' in the
corresponding rows and columns in Table \ref{resultstable}.

\begin{figure}[h!]
\begin{center}
 \epsfig{file=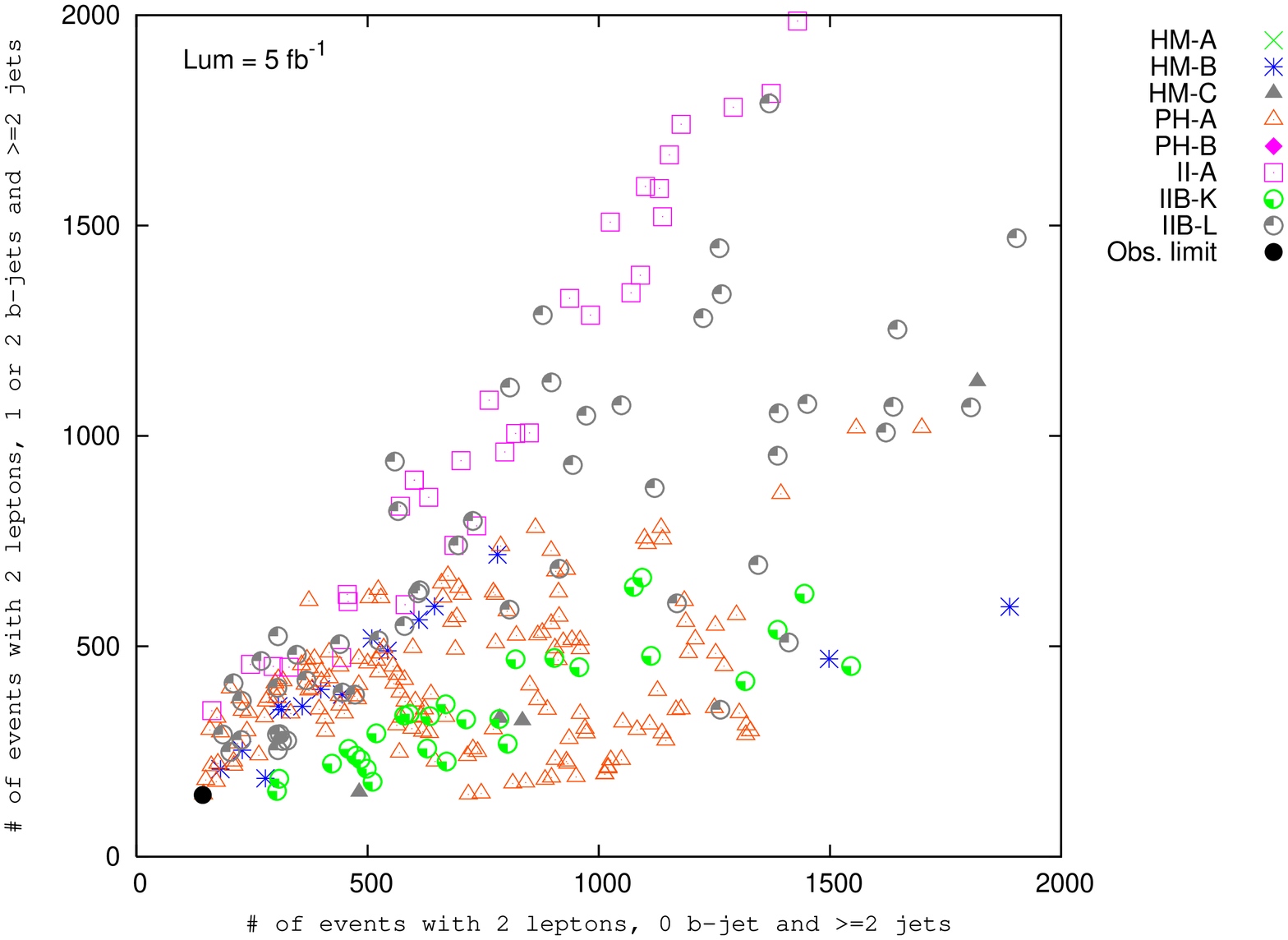,height=16cm, angle=-90}
\end{center}
\caption{Plot of number of events with 2 leptons, 0 b jets and
$\geq$ 2 jets and number of events with 2 leptons, 1 or 2 b jets
and $\geq$ 2 jets. The black dot represents the lower limit of
observability of the two signatures, according to conditions in
equation (\ref{observability}). The models below the observable
limit have not been shown. Note that the PH-B and II-A
constructions can be distinguished from each other since the
former is not observable while the latter is observable. One can
also partially distinguish the PH-A and IIB-L constructions. The
plots are best seen in color.}\label{lepton}
\end{figure}

Now we explain the distinguishibility of the PH-A, PH-B, II-A,
IIB-K and IIB-L constructions. Figure \ref{0l12b6j} shows that the
PH-B and II-A constructions can be distinguished from the PH-A,
IIB-K and IIB-L constructions (signature $C$ in Table
\ref{patterntable2})\footnote{This signature may not be very
realistic. However, as explained in section \ref{remarks}, it has
been used here because it is very economical and illustrates the
approach in a simple manner.}. Figure \ref{lepton} shows that the
PH-B and II-A constructions can be distinguished from each other
and that the PH-A and IIB-L constructions can be {\it partially}
distinguished from each other (signature $D$ in Table
\ref{patterntable2}). To understand why it is possible to do so,
we look at these constructions in detail.

Let's start with the IIB-K construction. As explained earlier, the
IIB-K construction typically gives a heavy spectrum, although it
is possible to have a light spectrum with light gluinos and
squarks (stops) with the stau also being light in some cases. So,
$\tilde{g}\tilde{q}$ production is typically dominant. First and
second generation squarks are copiously produced. The first and
second generation squarks decay to non b-jets and the gluino also
decays more to non b jets than to b jets because the IIB-K
construction always has a bino LSP. Therefore, as seen from Figure
\ref{lepton}, the IIB-K construction has more events with 2
leptons, 0 b jets and $\geq$ 2 jets compared to those with 2
leptons, 1 or 2 b jets and $\geq$ 2 jets. Also, since the mass
difference between $\tilde{N_2},\tilde{C_1}$ and $\tilde{N_1}$ is
only big enough for leptons to pass the cuts but not for jets to
pass the cuts, the number of events with 0 leptons, 1 or 2 b jets
and $\geq$ 6 jets is smaller than those with 2 leptons, 0 b jets
and $\geq$ 2 jets, as seen from Figure \ref{0l12b6j}.

The PH-B construction has scalars which are quite heavy ($> 1$
TeV). So, gluino pair production is dominant.  The branching ratio
of gluinos to $\tilde{C_1}$ + jets is typically the largest for
$\tan \beta \geq 20$, if it is kinematically allowed
\cite{Bartl:1994bu}, followed by $\tilde{N_i}$ + jets, $i=1,2,3$,
which are smaller. This is generally true for this construction.

When the gluino is light ($\leq 550$ GeV), the PH-B construction
has wino LSP. Since $\mu$ is large ($> 1.3$ TeV), $m_{\tilde{C_1}}
\sim m_{\tilde{N_1}}$. Since $\tilde{C_1}$ and $\tilde{N_1}$ are
wino and $\tilde{N_2}$ is bino, the gluino decays mostly to non b
jets. Also, the leptons and jets coming from the decay of
$\tilde{C_1}$ to $\tilde{N_1}$ are very soft and do not pass the
cuts. When the gluino directly decays to $\tilde{N_1}$, there are
no leptons and only two jets. When the gluino is heavier ($> 550$
GeV), the PH-B construction can have wino as well as bino LSPs.
Because of a heavier gluino, the overall cross-section goes down.
For the wino LSP case, the same argument as above applies in
addition to the small cross-section implying even fewer lepton
events. For the bino LSP case, the gluino again decays mostly to
non b jets as $\tilde{C_1}$ and $\tilde{N_2}$ are wino and
$\tilde{N_1}$ is bino. Also, $m_{\tilde{C_1}} \sim
m_{\tilde{N_2}}$ but at the same time $\Delta
M_{\tilde{N_2}-\tilde{N_1}}$ and $\Delta
M_{\tilde{C_1}-\tilde{N_1}}$ are quite small ($\leq 20 $ GeV),
leading to soft leptons and jets many of which don't pass the
cuts. So the result is that PH-B models have very few events with
leptons and/or b jets. Therefore, as seen from Figures
\ref{0l12b6j} and \ref{lepton}, the PH-B construction does not
give rise to observable events for signatures $D$ and $E$ in Table
\ref{patterntable2}.

The PH-A construction is required to have bino, higgsino or mixed
bino-higgsino LSP with very heavy scalar masses ($\geq 2$ TeV).
Light gluinos ($\leq 600$ GeV) have bino LSP, while higgsino or
mixed bino-higgsino LSPs have heavier gluinos ($600-1200$ GeV).
$\tilde{N_2}$ and $\tilde{C_1}$ are also quite light ($< 250$
GeV).

For the bino LSP case, gluinos typically decay to $\tilde{N_1}$,
$\tilde{N_2}$ and $\tilde{C_1}$ and non b jets most of the time
compared to b jets, as $\tilde{N_1}$ is bino and $\tilde{N_2}$ and
$\tilde{C_1}$ are wino. The decays of $\tilde{N_2}$ and
$\tilde{C_1}$ can give rise to leptons passing the cuts. In
addition, direct production of $\tilde{N_2}$ and $\tilde{C_1}$ is
also important as they are light. They can also give rise to 2
leptons, 0 b jets and $\geq 2$ jets. Therefore, in this case, one
has more events with (two) leptons and non b jets compared to
those with (two) leptons and b jets. This can be seen clearly from
Figure \ref{lepton}. The decays of $\tilde{N_2}$ and $\tilde{C_1}$
also give jets (as they decay through virtual Z and W). Since the
gluino is light, the cross-section is quite big implying that
there are also a fair number of events with 0 leptons, 1 or 2 b
jets and $\geq 6$ jets. However, the number of events with 2
leptons, 0 b jets and $\geq 2$ jets is more than with  0 leptons,
1 or 2 b jets and $\geq 6$ jets due to the dominant branching
ratio of gluinos to non b jets. This can be seen from Figure
\ref{0l12b6j}.

For the higgsino LSP case, the gluino is comparatively heavier
($600-1200$ GeV), leading to a significant decrease in
cross-section. Now $\tilde{N_1}$, $\tilde{N_2}$ and $\tilde{C_1}$
are mostly higgsino, leading to a lot of production of b jets as
the relevant coupling is proportional to the mass of the
associated quark. The fact that $\tilde{N_1}$, $\tilde{N_2}$ and
$\tilde{C_1}$ are mostly higgsino also makes their masses quite
close to each other, implying that leptons and jets produced from
the decays of $\tilde{N_2}$ and $\tilde{C_1}$ to $\tilde{N_1}$ are
very soft and do not pass the cuts. Thus, these models have very
few events with leptons. Since the jets coming from the decays of
$\tilde{N_2}$ and $\tilde{C_1}$ to $\tilde{N_1}$ are also very
soft, the PH-A higgsino LSP models also have very few events with
0 leptons, 1 or 2 b jets and $\geq 6$ jets. Therefore, these do
not give rise to observable events for the signatures in Figures
\ref{0l12b6j} and \ref{lepton}.

For the mixed bino-higgsino LSP case, the gluino is again quite
heavy ($600-1200$ GeV), making the cross section much smaller
compared to the bino LSP case. Since $\tilde{N_1}$, $\tilde{N_2}$
and $\tilde{C_1}$ have a significant higgsino fraction, the
gluinos again decay more to b jets compared to non b jets. The
mass separation between $\{\tilde{N_2},\tilde{N_3},\tilde{C_1}\}$
and $\tilde{N_1}$ is such that the decays of $\tilde{N_2}$ and
$\tilde{C_1}$ to $\tilde{N_1}$ produce leptons which typically
pass the cuts and jets which only sometimes pass the cuts.
Therefore, these PH-A models have few events with 2 leptons, 0 b
jets and $\geq 2$ jets. They give rise to events with 2 leptons, 1
or 2 b jets and $\geq 2$ jets but since the overall cross-section
is much smaller than for the bino LSP case, the number of events
for the above signature for these PH-A models is just a little
above the observable limit, as can be seen from Figure
\ref{lepton}. This is the origin of the ``Both'' entry for
signature $D$ in the row for the PH-A construction. Because of the
small overall cross-section as well as the fact that jets produced
from the decays of $\tilde{N_2}$ and $\tilde{C_1}$ only sometimes
pass the cuts, the number of events for 0 leptons, 1 or 2 b jets
and $\geq 6$ jets is also small for these PH-A models, as seen
from Figure \ref{0l12b6j}.

The IIB-L construction always has a mixed bino-higgsino LSP, for
both light and heavy gluino models. The light gluino IIB-L models
have a large overall cross-section. The gluinos decay both to non
b jets and b jets owing to the mixed bino-higgsino nature of the
LSP. Also, the mass separation between $\tilde{N_2}$,
$\tilde{C_1}$ and $\tilde{N_1}$ is not large which means that the
leptons produced from the decays of  $\tilde{N_2}$ and
$\tilde{C_1}$ pass the cuts but the jets produced seldom pass the
cuts. So, the IIB-L construction has many events with 2 leptons, 0
b jets and $\geq$ 2 jets as well as with 2 leptons 1, or 2 b jets
and $\geq$ 2 jets but not as many with 0 leptons, 1 or 2 b jets
and $\geq$ 6 jets, as seen from Figures \ref{0l12b6j} and
\ref{lepton}.

From Figure \ref{lepton}, one sees that the IIB-L construction can
be distinguished \emph{partially} from the PH-A construction,
leading to a ``Probably Yes (PY)'' in the pattern table. One can
understand it as follows - as mentioned above, the IIB-L
construction always has a mixed bino-higgsino LSP while the PH-A
construction has a mixed bino higgsino LSP only when the gluino is
heavy (i.e. for a heavy spectrum). For light gluino models, as
mentioned before, the PH-A construction has a bino LSP. Therefore,
{\it for light gluino models}, the ratio of number of events with
2 leptons, 1 or 2 b jets and $\geq$ 2 jets and number of events
with 2 leptons 0 b jets and $\geq$ 2 jets is much more for the
IIB-L construction compared to the PH-A construction. These are
the models which differentiate the IIB-L and PH-A constructions in
Figure \ref{lepton}. It turns out that mixed bino-higgsino LSP
models with heavy gluinos in both constructions have very similar
spectra\footnote{This has been explicitly checked.}, leading to
very similar signatures in all studied channels. Therefore, the
IIB-L construction and the PH-A construction are not
distinguishable in this special region of spectrum and signature
space with the present set of signatures. Using more sophisticated
signatures may help distinguish these signatures more cleanly. As
already mentioned before, the PH-A construction also has models
with a pure higgsino LSP. Those models have very heavy gluinos
however, leading to no observable events in Figures \ref{0l12b6j}
and \ref{lepton}.

Moving on to the II-A construction, we note that it can have a
bino, wino, higgsino or mixed bino-higgsino LSP with light to
moderately heavy gluino ($300-600 $ GeV) and moderately heavier
scalars (stops can be specially light). The spectrum and signature
pattern are quite complicated. Let's analyze all possible cases.

In this construction, the branching ratio of gluinos to
$\tilde{C_1}$ + jets is the largest as mentioned before, since
$\tan \beta \geq 20$. For the wino LSP case, since
$M_{\tilde{C_1}} \sim M_{\tilde{N_1}}$ the leptons and jets from
the decays of $\tilde{C_1}$ to $\tilde{N_1}$ are very soft and do
not pass the cuts. The decay of the gluino to $\tilde{C_1}$ is
accompanied by non b jets since $\tilde{N_1}$ and $\tilde{C_1}$
are wino and $\tilde{N_2}$ is bino. So, the II-A models with wino
LSP do not give rise to observable events with leptons and/or b
jets. This implies that the signatures in Figures \ref{0l12b6j}
and \ref{lepton} are not observable for these II-A models.

For the bino LSP case, $\tilde{N_2}$ and $\tilde{C_1}$ are quite
heavy ($> 350$ GeV), sometimes being even heavier than the gluino,
in which case only the decay of gluino to $\tilde{N_1}$ is allowed
leading to no leptons. Even when the decays of gluino to
$\tilde{C_1}$ and $\tilde{N_2}$ are allowed, they are mostly
accompanied by comparatively soft non b jets (due to kinematic
reasons). Since these II-A models are required to have
$\tilde{N_2}$ and $\tilde{C_1}$ much heavier than the PH-A bino
LSP models, the direct production of $\tilde{N_2}$ and
$\tilde{C_1}$ which could be a source of events with 2 leptons, 0
b jets and $\geq 2$ jets, is also relatively suppressed. Therefore
these models do not give rise to observable events with 2 leptons
0 b jets and $\geq 2$ jets as well as with 2 leptons, 1 or 2 b
jets and $\geq 2$ jets. However, there are some bino LSP II-A
models which also have light squarks (stops mostly) and light
gluinos in addition to having heavy $\tilde{N_2}$ and
$\tilde{C_1}$ as above. For these bino LSP models,
$\tilde{q}\tilde{q}$ pair production is quite important. These
squarks mostly decay to a gluino and quarks, followed by the decay
of the gluino to mostly the LSP and jets (both b and non b jets).
Thus, these bino LSP II-A models have many events with 0 leptons,
1 or 2 b jets and $\geq 6$ jets but no observable events with 2
leptons, 1 or 2 b jets and $\geq 2$ jets.

For the higgsino LSP case, since $\tan \beta \geq 20$, the gluino
mostly decays to $\tilde{C_1}$ + b jets as the associated coupling
is proportional to the mass of the relevant quark. Also, in this
case $\tilde{N_1}$, $\tilde{N_2}$ and $\tilde{C_1}$ are all
higgsino like and very close to each other. So, leptons from the
decays of $\tilde{N_2}$ and $\tilde{C_1}$ to $\tilde{N_1}$ are
very soft and do not pass the cuts. Therefore II-A models with
higgsino LSP do not give rise to observable events with leptons
and/or non b jets. For some of these higgsino LSP II-A models,
there are still a fair number of events with 0 leptons 1 or 2 b
jets and $\geq 6$ jets. This is because even though
$\tilde{q}\tilde{q}$ pair production is less important compared to
those in bino LSP II-A models, the branching ratio of gluinos to b
jets is much bigger (due to a higgsino LSP).

For the mixed bino-higgsino LSP case, the decay of gluino to
$\tilde{C_1}$ + b jets is dominant since $\tilde{C_1}$ is mostly
higgsino. The next important decays are to $\tilde{N_1}$,
$\tilde{N_2}$ and $\tilde{N_3}$ + b jets followed by a small
fraction to non b jets. The mass separation between
$\{\tilde{N_2},\tilde{N_3},\tilde{C_1}\}$ and $\tilde{N_1}$ is
such that leptons produced can easily pass the cuts, while the
jets produced sometimes pass the cuts. Some of these mixed
bino-higgsino LSP II-A models also have comparatively light
squarks, implying that $\tilde{q}\tilde{q}$ and
$\tilde{q}\tilde{g}$ production are also important. The squark
decays to a quark and a gluino, followed by the usual decays of
the gluino. For these models, $\tilde{N_2}$ and $\tilde{C_1}$ are
also light, implying that in such cases direct production of
$\tilde{N_2}$ and $\tilde{C_1}$ is also possible. The decays of
$\tilde{N_2}$ and $\tilde{C_1}$ can give rise to events with 2
leptons, 0 b jets and $\geq 2$ jets.

Therefore, the conclusion is that for II-A models with mixed
bino-higgsino LSP and light squarks, there are observable events
with 2 leptons, 1 or 2 b jets and $\geq 2$ jets; with 0 leptons, 1
or 2 b jets and $\geq 6$ jets as well as with 2 leptons 0 b jets
and $\geq 2$ jets. The number of events with 2 leptons, 1 or 2 b
jets and $\geq 2$ jets is greater than those with 2 leptons, 0 b
jets and $\geq 2$ jets because of the dominant branching fraction
of gluinos to b jets. Therefore, signature $D$ (ratio of the above
two type of events - Figure \ref{lepton}) can distinguish the II-A
and PH-B constructions as the II-A construction has observable
events while the PH-B construction does not give rise to
observable events. The number of events for 0 leptons 1 or 2 b
jets and $\geq 6$ jets will be larger than those with 2 leptons, 0
b jets and $\geq 2$ jets, again due to the dominant branching
ratio of the gluino to b jets. So, signature $C$ (ratio of the
above two type of events - Figure \ref{0l12b6j}) can distinguish
the PH-A,IIB-K and IIB-L constructions from the II-A and PH-B
constructions. The above results can be seen from Figures
\ref{0l12b6j} and \ref{lepton}, where the qualitative difference
between the constructions is clear.
\begin{figure}[h!]
\begin{center}
 \epsfig{file=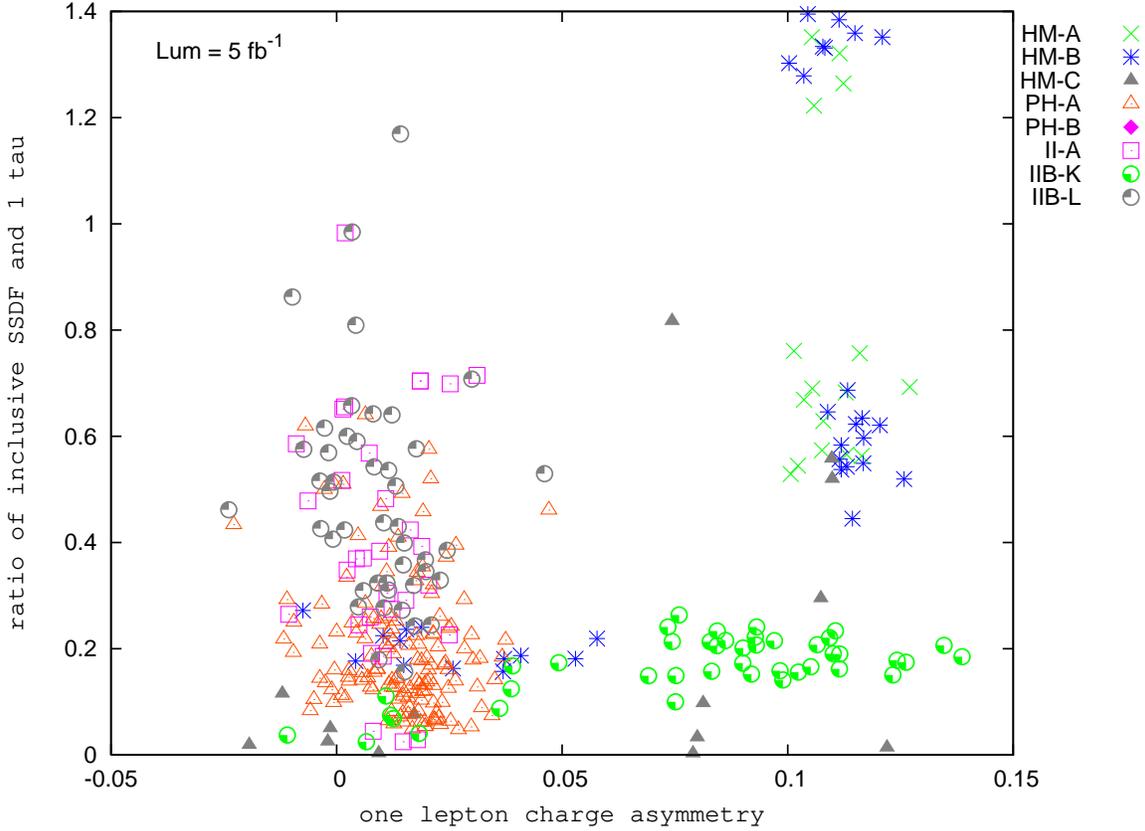,height=16cm, angle=-90}
\end{center}
\caption{Plot of the charge asymmetry in events with a single
electron or muon $\&\;\geq$ 2 jets $\&$ the ratio of number of
events with same sign different flavor (SSDF) dileptons and $\geq$
2 jets and number of events with 1 tau and $\geq$ 2 jets. The
models which are below the observable limit as defined by
(\ref{observability}) are not shown. Note that the IIB-K
construction can be distinguished from the PH-A and IIB-L
constructions, as the former occupies a mostly horizontal region
while the latter occupy a mostly vertical region. The overlapping
IIB-K and PH-A models can be distinguished from Figure
\ref{metpk}. The plots are best seen in color.}\label{ac1-SSDF}
\end{figure}
The II-A models shown above the observable limit have mixed
bino-higgsino LSP with lighter squarks than in other II-A cases.

\begin{figure}[h!]
\begin{center}
 \epsfig{file=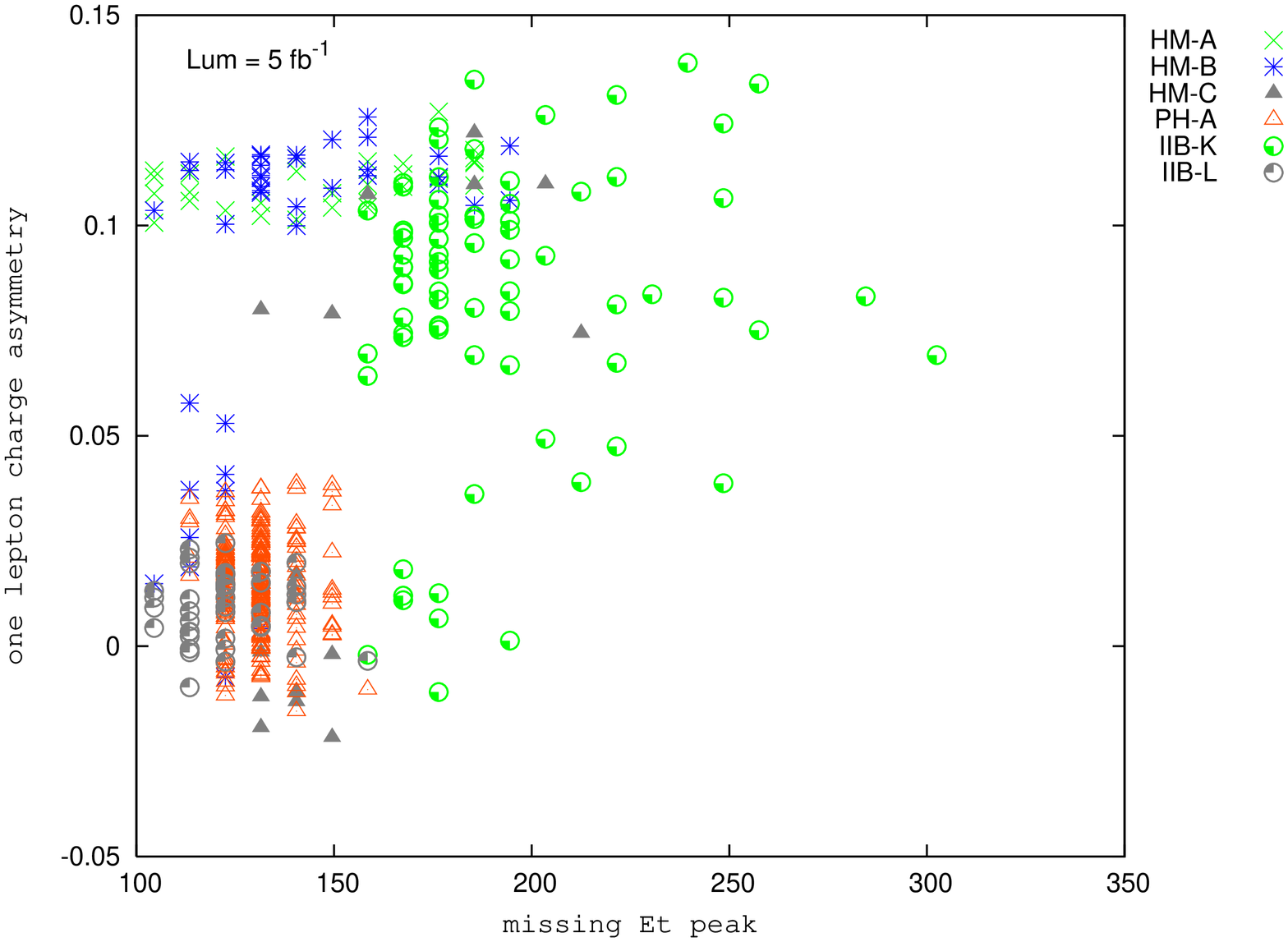,height=16cm, angle=-90}
\end{center}
\caption{Plot of the peak of the $\notE_T$ distribution and the
charge asymmetry in events with a single electron or muon
$\&\;\geq$ 2 jets. The models which are below the observable limit
as defined by (\ref{observability}) are not shown. Note that this
plot distinguishes the overlapping PH-A and IIB-K models in Figure
\ref{ac1-SSDF}. The plots are best seen in color.}\label{metpk}
\end{figure}

We are now left with explaining the distinguishibility of the PH-A
and IIB-L constructions on the one hand and the IIB-K construction
on the other. Figures \ref{ac1-SSDF} and \ref{metpk} show that the
IIB-K construction can be distinguished from the PH-A and IIB-L
constructions. The reason is as follows - As explained earlier,
the IIB-K construction can have a light spectrum with light
gluinos, light stop and sometimes a light stau. So,
$\tilde{g}\tilde{q}$ production is typically dominant. Since
up-type squarks are produced preferentially at the LHC (as it is a
$pp$ collider), they decay preferentially to a positive chargino
$\tilde{C}_1^+$, which in turn decays preferentially to a
positively charged lepton $l^+$ (in its leptonic decays).
Therefore, the asymmetry in number of events with a single
electron or muon and $\geq$ 2 jets
($A_c^{(1)}$)\footnote{$A_c^{(1)}
\equiv\frac{N(l^+)-N(l^-)}{N(l^+)+N(l^-)}$, where for example
$N(l^+)$ is the number of events with a single positively charged
electron or muon and $\geq$ 2 jets.} is much greater than in the
case of PH-A and IIB-L constructions where $\tilde{g}\tilde{g}$
pair production is dominant. There are a few IIB-K models which
have a small $A_c^{(1)}$ and which overlap with some PH-A models
(seen in Figure \ref{ac1-SSDF}) even though $\tilde{g}\tilde{q}$
production is dominant. This is due to some special features of
their spectrum, such as the lightest stop and/or the lightest stau
being very light. These features either suppress the production of
$\tilde{C}_1^+$ or suppress the decay of $\tilde{C}_1^+$ to
electrons or muons. However, as seen from Figure \ref{metpk},
these overlapping IIB-K models can be distinguished from the PH-A
and IIB-L constructions by the peak of the $\notE_T$ distribution.
This is related to the mass of the LSP. The IIB-K constructions
(which are observable with 5 $fb^{-1}$) have a comparatively
heavier LSP than the IIB-L constructions in general, making the
peak of the $\notE_T$ distribution larger than those for the IIB-L
constructions\footnote{This is because the IIB-K models with a
light spectrum have the lightest stop correlated with the mass of
the LSP ($m_{\tilde{t}_1} \geq m_{\tilde{N}_1}$)
\cite{Choi:2004sx}; $\tilde{t}_1$ cannot be too light, else it
would be directly seen at the Tevatron.}. The PH-A models which
overlap with the small $A_c^{(1)}$ IIB-K models have bino LSPs
which are lighter than that of the IIB-K models. So, the PH-A
models have a smaller $\notE_T$ peak than the overlapping IIB-K
models in Figure \ref{ac1-SSDF}, as can be seen from Figure
\ref{metpk}.

We have thus explained the distinguishibility of all constructions
based on the spectrum at the low scale. Typically, there are more
than one (sometimes many) signatures which can distinguish any two
given constructions. This redundancy gives us confidence that our
analysis is robust and that the conclusions will not be affected
with more sophisticated analysis. For simplicity, we have only
explained one signature distinguishing a pair of constructions but
all such signatures can be understood similarly.

\subsection{Explanation of Spectrum from the Soft
Parameters}\label{fromsoft}

We now turn to understanding the origin of the spectrum of
particles at the low scale (which are responsible for the
signature pattern) for the constructions in terms of pattern of
soft parameters. For illustrative purposes, we carry out this
exercise for two constructions -- the HM-A construction and the
PH-B construction. If we take the soft parameters at the string
(or unification) scale as given, then it does not matter that
these are really only toy constructions. The kind of analysis
carried out for these constructions below can also be carried out
for more well motivated constructions such as the KKLT and Large
Volume ones, as well. However, to go to the final step, i.e. from
the soft parameters to the structure of the underlying theoretical
construction, it makes sense to stick to the more well motivated
constructions -- the KKLT and Large Volume constructions, as we
will in the following subsection.

Starting with the HM-A construction, one would like to understand
its characteristic spectrum, viz. $m_{\tilde{g}} \sim
m_{\tilde{q}}
> m_{\tilde{l}}$. Why does the gluino mass lie in the range $550-650$ GeV?
We note that the HM-A construction is a heterotic M theory
construction compactified on a Calabi-Yau with only one K\"{a}hler
modulus. This implies that the soft terms obtained at the
unification scale are universal \cite{Choi:1997cm}. Thus, the soft
terms obtained at the unification scale are a special case of the
well studied mSUGRA boundary condition. Now, phenomenological
studies of the mSUGRA boundary condition have shown that in order
to get a small relic density (satisfying the WMAP upper bound
\footnote{Typically, a lower bound on the relic density is also
imposed. However, we have only used the upper bound in our
analysis, as explained in section \ref{procedure}. The area
covered by the three regions can change depending on whether a
lower bound is also imposed.}), there are three allowed regions in
the $m-M$ plane \cite{Ellis:2003cw} \footnote{One usually assumes
$A_0$ = 0 in these plots.}. Here $m$ stands for the universal
scalar mass parameter while $M$ stands for the universal gaugino
mass parameter. These three regions are the following:
\begin{itemize}
\item The stau coannihilation region -- In this region, the stau
is almost degenerate with the LSP which is a bino. One gets an
acceptable relic density because of coannihilation of the stau and
the LSP to a tau. This requires $m<M$ with $m$ roughly between 100
and 150 GeV and $M$ roughly between 150 and 300 GeV (assuming
$A_0$ = 0). \item The focus point region -- This region requires a
large scalar mass parameter ($m > M$) at the unification scale,
and gives rise to a higgsino LSP with acceptable relic density.
\item The funnel region -- In this region, the LSP is annihilated
by a $s$-channel pole, with $m_{LSP}
 \approx m_A/2$. This also requires $m>M$.
\end{itemize}

\noindent In the case of the HM-A construction, the soft mass
parameters always have the hierarchy $M>m$ \cite{Choi:1997cm},
which implies that only the stau coannihilation region is possible
for the HM-A construction. Also, the allowed ranges for the $m$
and $M$ parameters roughly explains the mass scale of the gluino
and squarks at the low scale from standard RG evolution.
Therefore, one has to now understand the origin of the allowed
values of the $m$ and $M$ parameters from the nature of the
expressions for soft terms and the ``theory'' input parameters.

The expressions for the soft terms depend on three input
parameters -- the goldstino angle $\theta$, the gravitino mass
$m_{3/2}$ and the parameter $\alpha(t+\bar{t})$ with $t$ as the
K\"{a}hler modulus, in addition to $\tan{\beta}$. For futher
details, the reader is referred to \cite{Choi:1997cm}. The
expressions for the soft parameters are given by : \ba
\label{soft-msoftI} M &=&
\frac{\sqrt{3}Cm_{3/2}}{(s+\bar{s})+\alpha(t+\bar{t})}\{(s+\bar{s})\sin(\theta)e^{-i\gamma_s}+
\frac{\alpha(t+\bar{t})}{\sqrt{3}}\cos(\theta)e^{-i\gamma_t}\} \\
m^2&=&
V_0+m_{3/2}^2-\frac{3m_{3/2}^2C^2}{3(s+\bar{s})+\alpha(t+\bar{t})}\,\{\alpha(t+\bar{t})\,
(2-\frac{\alpha(t+\bar{t})}{3(s+\bar{s})+\alpha(t+\bar{t})})\sin^2(\theta)
+ \nonumber\\
& &
(s+\bar{s})\,(2-\frac{3(s+\bar{s})}{3(s+\bar{s})+\alpha(t+\bar{t})})\cos^2(\theta)-
\nonumber \\
& &
\frac{2\sqrt{3}\alpha(t+\bar{t})(s+\bar{s})}{3(s+\bar{s})+\alpha(t+\bar{t})}
\sin(\theta)\cos(\theta)\cos(\gamma_s-\gamma_t)\}
\nonumber\\
A&=&
\sqrt{3}Cm_{3/2}\{-1+\frac{3\alpha(t+\bar{t})}{3(s+\bar{s})+\alpha(t+\bar{t})}\sin(\theta)e^{-i\gamma_s}
+\nonumber
\\ & & \sqrt{3}(-1+\frac{3(s+\bar{s})}{3(s+\bar{s})+\alpha(t+\bar{t})})\cos(\theta)e^{-i\gamma_t}\}\nonumber
\ea

\begin{figure}
  \begin{center}
    \begin{tabular}{cc}
      \resizebox{80mm}{!}{\includegraphics{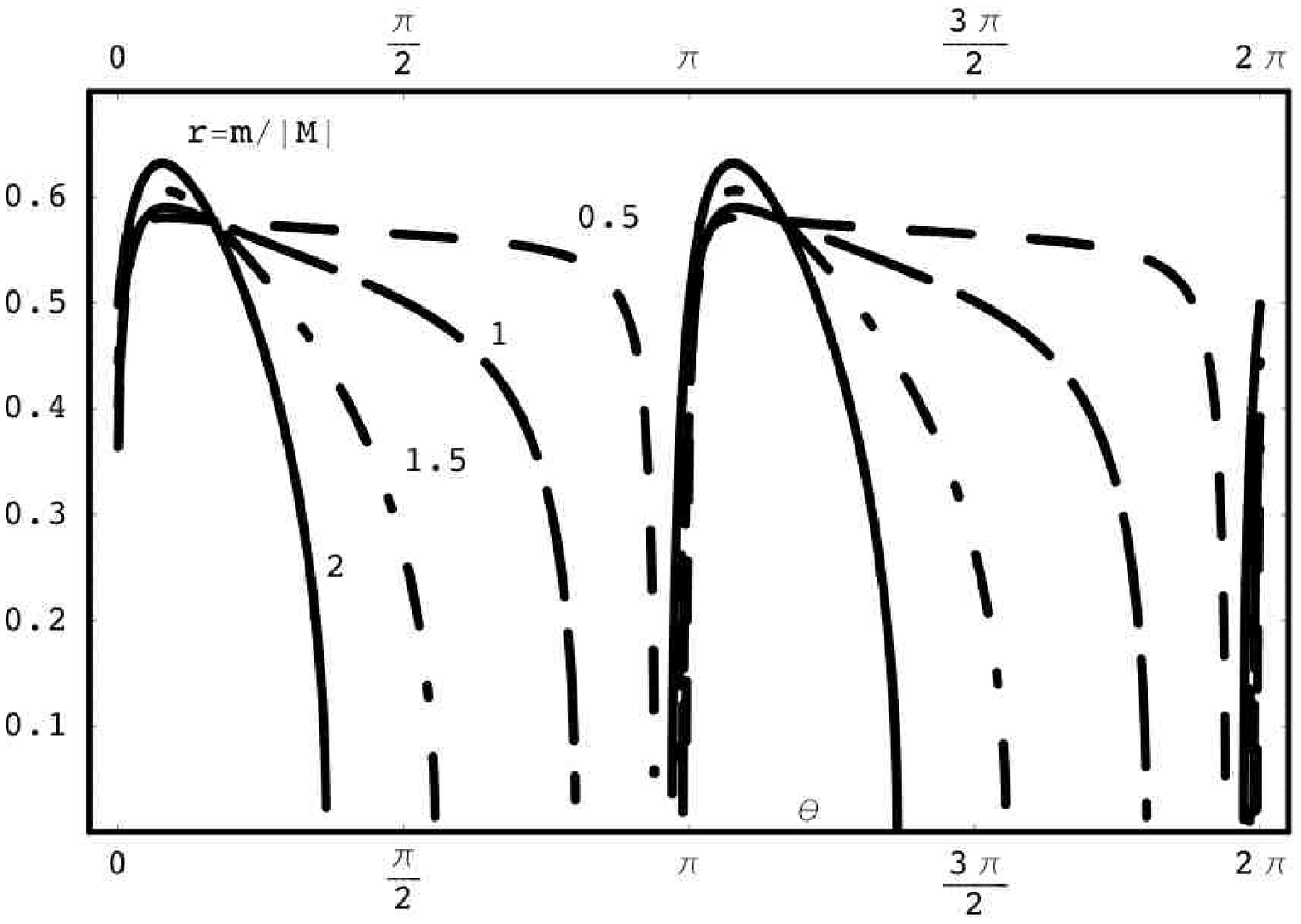}} &
      \resizebox{80mm}{!}{\includegraphics{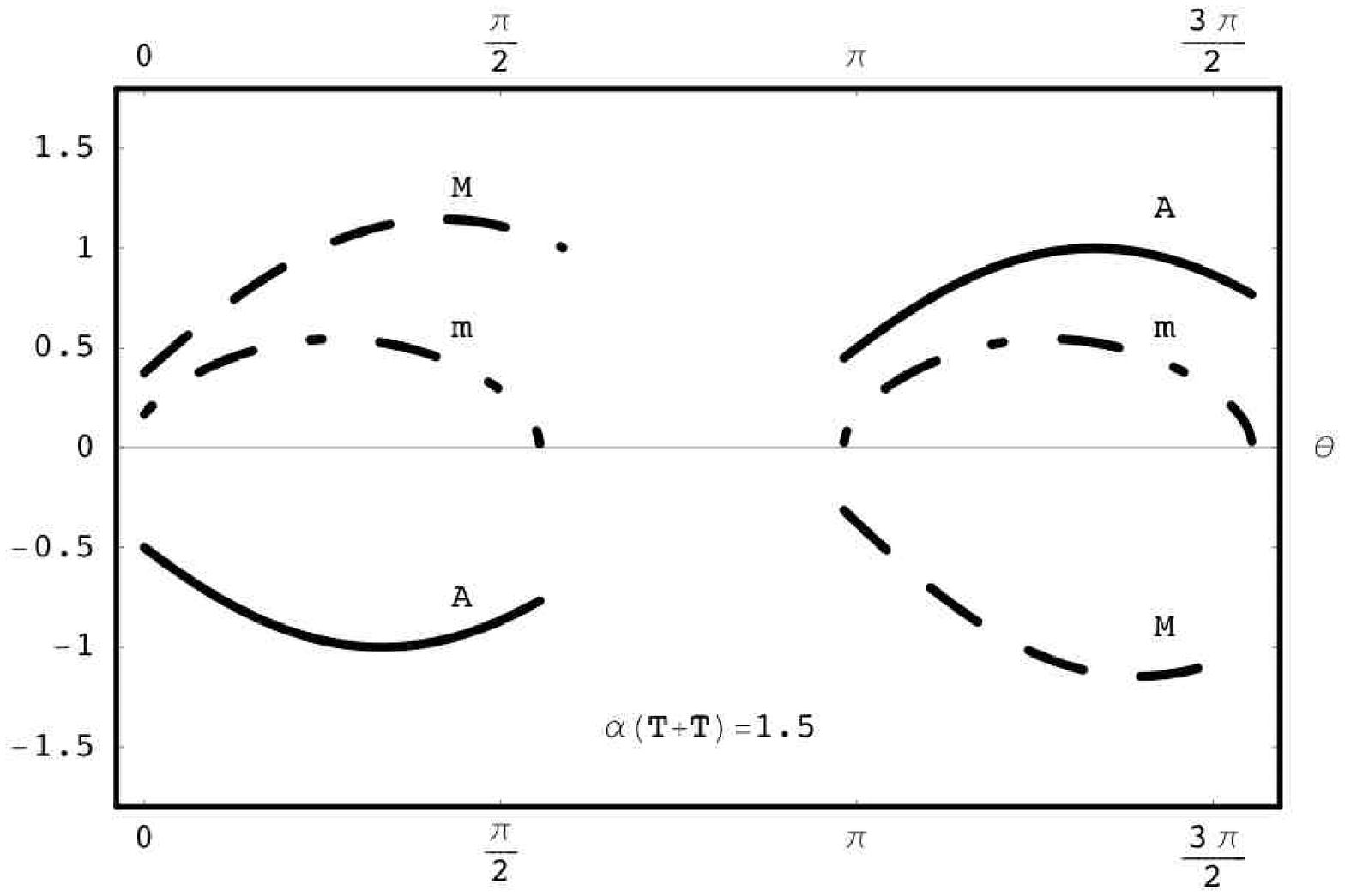}} \\
    \end{tabular}
    \caption{Left : The ratio $r\equiv m/|M|$ as a
function of $\theta$ for four different values of
$\alpha(T+\bar{T})$ represented by various curves. Right: The
universal soft parameters as a function of $\theta$ for
$\alpha(T+\bar{T}) = 1.5$. The solid curve stands for trilinears,
dotted dashed curve for scalars and dashed curve for gauginos. }
    \label{msoftI}
  \end{center}
\end{figure}

\noindent where we are using the following parameterization, which
define $F$ terms for the moduli \cite{Brignole:1997dp}: \ba
F^s&=&\sqrt{3}m_{3/2}C(s+\bar{s})\sin(\theta)e^{-i\gamma_s}\\
F^t&=&m_{3/2}C(t+\bar{t})\cos(\theta)e^{-i\gamma_t}\nonumber\\
C^2&=&1+\frac{V_0}{3m_{3/2}^2}\nonumber \ea The ratio of the
scalar to the gaugino mass parameter $r \equiv m/|M|$ is shown in
the first plot in Figure \ref{msoftI} as a function of the
goldstino angle $\theta$. For the stau coannihilation region, the
ratio $r$ has to be roughly $0.5-0.6$. We see that for this value
of $r$, one set of allowed values of $\theta$ lie near
$0,\pi,2\pi$ for all allowed values of $\alpha(T+\bar{T})$. This
means that the supersymmetry breaking is moduli dominated ($F_s
\approx 0$). In addition, there also exist other values of
$\theta$ which are closer to ($\frac{1}{2}\pi,\frac{3}{2}\pi$)
rather than to ($0,\pi,2\pi$). However, these values are ruled out
by constraints on low energy phenomenology, as the trilinear
parameter $A_0$ for these values is pretty large, as seen from the
second plot in Figure \ref{msoftI}. This is because a very large
value of the trilinear parameter makes the scalar mass squared run
negative at the low scale and also causes problems for EWSB. Once
the correct ratio $r$ of the gaugino mass parameter to the scalar
mass parameter is obtained, one can get their correct absolute
scales by tuning $m_{3/2}$ as all the soft parameters are
proportional to them. One thus gets a gluino in the $550-650$ GeV
range. The allowed values of $m_{3/2}$ lie in the TeV range.

Moving on to the PH-B construction, one would again like to
understand the origin of the characteristic features of its
spectrum, viz. heavy squarks ($\geq 1$ TeV), moderately heavy
sleptons except the stau which is considerably lighter, and
gluinos which can vary from being light ($250-450$ GeV) to heavy
($\geq 1000$ GeV). Light gluinos ($< 450 $GeV) in this
construction \emph{always} give rise to a wino LSP while the
heavier ones give rise to bino or wino LSPs, as we explain below.

The PH-B construction is a weakly coupled heterotic string
construction with a tree level K\"{a}hler potential and two
gaugino condensates. The soft terms for this construction depend
on the ``theory'' input parameters -- the gravitino mass
$m_{3/2}$, the Green-Schwarz coefficient $\delta_{GS}$ and the
$vev$ of the K\"{a}hler modulus $t$, in addition to $\tan{\beta}$.
In these kind of constructions, it was further noted that a
minimum with $F_s=0$  and $F_t \neq 0$ is preferred with $t$ being
stabilized at values slightly greater than 1. For details, refer
to \cite{Casas:1990qi}. The result is that all soft terms are zero
at tree level. The expressions for the soft parameters are
approximately given by \cite{Kane:2002qp}: \ba
\label{soft-racetrack} M_a &\approx&
\frac{g_a^2}{2}\,[(2\frac{\delta_{GS}}{16{\pi}^2}+b_a)\,G_2(t,\bar{t})+2b_am_{3/2}];
\;\;G_2(t,\bar{t})\equiv (2\zeta(t)+\frac{1}{t+\bar{t}}) \\
m_i^2 &\approx& \gamma_i\,m_{3/2}^2 \nonumber \\
A_{ijk} &\approx& m_{3/2}\,(\gamma_i+\gamma_j+\gamma_k) \nonumber
\ea \noindent where $\zeta(t)$ is the Riemann zeta function. The
dominant contribution to the soft scalar mass parameters is
proportional to the gravitino mass with the proportionality
constant being the anomalous dimension of the respective fields
($\gamma_i$). Since the anomalous dimension of the quarks is
bigger than that of the leptons, the squarks turn out to be
heavier than the sleptons. Also, the anomalous dimension of the
stau ($\tilde{\tau}$) is the smallest (smaller by a factor of
about 3 compared to that for $\tilde{Q}_3$), leading to the stau
as the lightest slepton. To get the absolute scale correct, one
has to realize that soft terms in this case arise from one loop
contributions. Thus, they are suppressed and therefore need a much
heavier $m_{3/2}$ $(\sim 20 $ TeV) in order to evade the chargino,
neutralino and higgs mass lower bounds. This is the reason for the
heavy squarks, moderately heavy sleptons and a light stau at the
low scale.
\begin{figure}
  \begin{center}
    \begin{tabular}{cc}
      \resizebox{70mm}{!}{\includegraphics{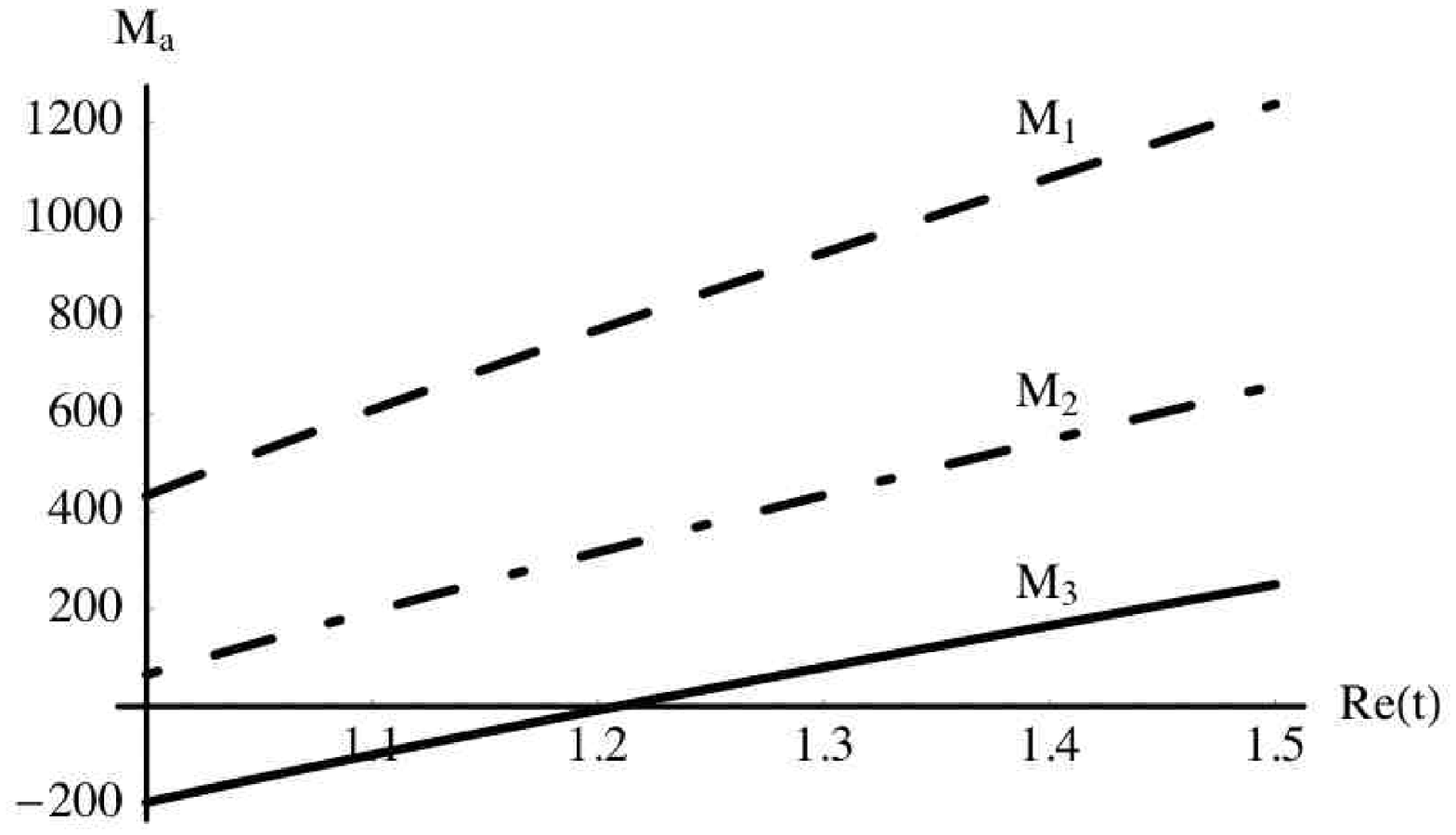}} &
      \resizebox{70mm}{!}{\includegraphics{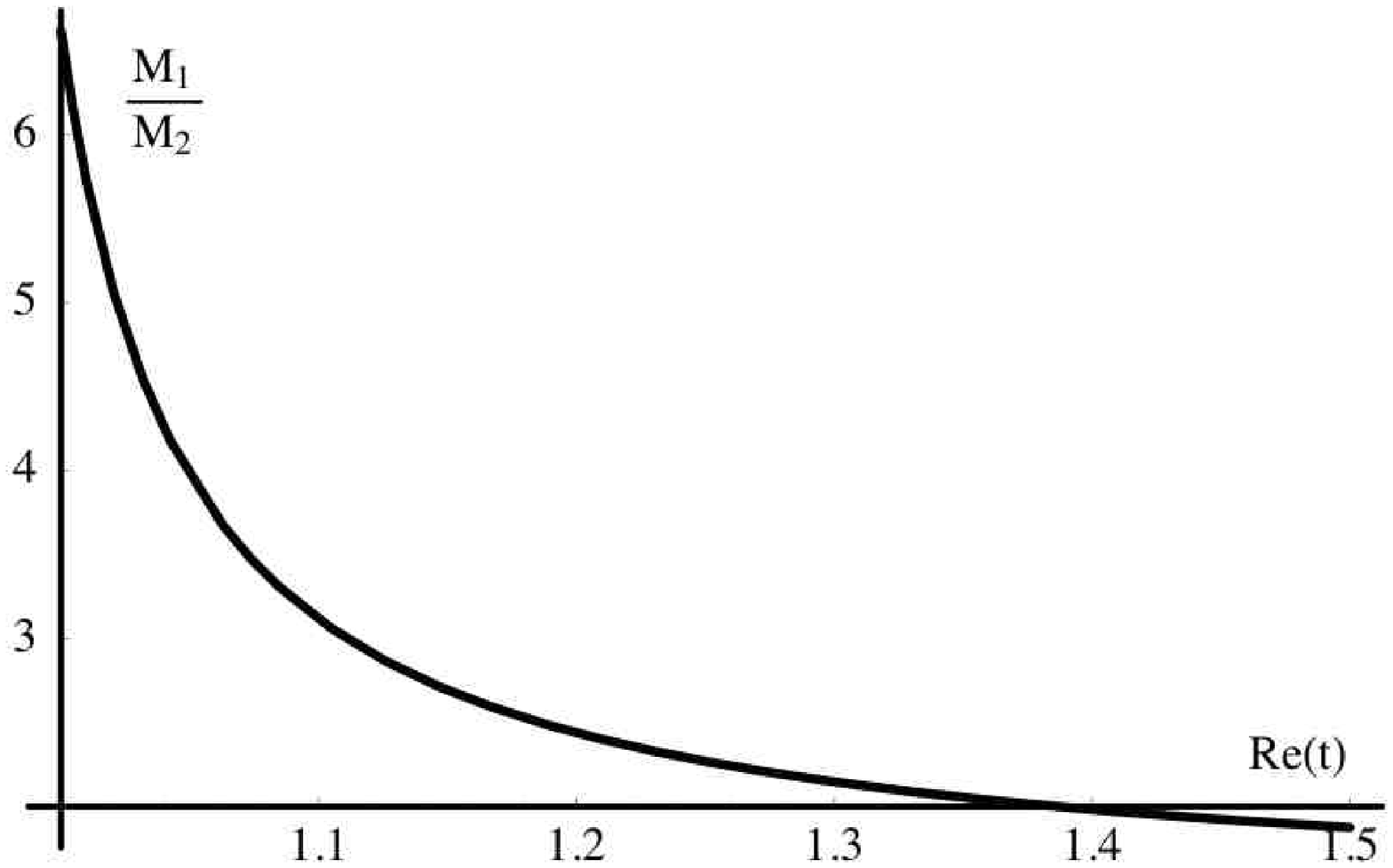}} \\
    \end{tabular}
    \caption{Left : Plots for gaugino mass parameters as a function of Re($t$):
The solid curve stands for $M_3$, dotted-dashed for $M_2$ and
dashed for $M_1$. Right : The ratio $M_1/M_2$ as a function of
$Re(t)$. Re($t$) varies from 1 to 1.5. The plots are shown for a
given value of $\delta_{GS}$ and $m_{3/2}$. $\delta_{GS}$ is -15
and $m_{3/2}$ is 20 TeV.}
    \label{racetrack}
  \end{center}
\end{figure}

For the gaugino sector, it is instructive to look at the plots of
the variation of gaugino mass parameters as a function of Re($t$)
and the ratio of bino and wino mass parameter ($\frac{M_1}{M_2}$)
as functions of Re($t$) for a given value of $\delta_{GS}$ and
$m_{3/2}$. Choosing different values of $\delta_{GS}$ does not
change the qualitative feature of the plots. Since all gaugino
mass parameters are proportional to $m_{3/2}$, changing $m_{3/2}$
changes the overall scale of all gaugino mass parameters. We first
explain why light gluinos give rise to a wino LSP while heavier
ones to a bino LSP for a fixed $m_{3/2}$. We then consider the
effects of changing $m_{3/2}$. The first plot in Figure
\ref{racetrack} shows that the gluino mass parameter is the
smallest of the three (taking the sign into account) at the
unification scale. This arises from the fact that the combination
($G_2(t,\bar{t})+2m_{3/2}$) in (\ref{soft-racetrack}) is negative
and the one loop beta function coefficient $b_3$ is the largest
for $M_3$ \cite{Kane:2002qp}. From the second plot, we see that
ratio $\frac{M_1}{M_2}$ is greater than 2 for Re($t$) smaller than
a certain value, which is around 1.4 for the value of
$\delta_{GS}$ chosen in the figure. Since we roughly have :

\begin{eqnarray}
M_{1_{low}} \approx 0.45\,M_{1_{unif}}; \;\; M_{2_{low}} \approx
0.9\,M_{2_{unif}},
\end{eqnarray}

\noindent it implies that if the ratio
$\frac{M_{1_{unif}}}{M_{2_{unif}}}$ is greater than 2, we have
$M_{1_{low}} > M_{2_{low}}$, leading to a wino LSP. Therefore for
Re($t$) smaller than a certain value (1.4 in the figure), one
obtains a wino LSP while for greater values of Re($t$) one obtains
a bino LSP. From the first plot, one now sees that for values of
Re($t$) smaller than the critical value, the gluino mass is quite
small at the unification scale. Thus for a given $m_{3/2}$, PH-B
models with small gluino masses have wino LSPs while those with
heavy gluinos have bino LSPs. If we now change the gravitino mass,
we change the overall scale of the gaugino mass parameters. Since
the scalars are also proportional to $m_{3/2}$, it is not possible
to make $m_{3/2}$ very small as the higgs mass bound will be
violated. But one can have a large gravitino mass giving rise to a
large gluino mass, with both wino and bino LSPs. Bino LSP models
however have heavier gluino masses than those with wino LSP as
$M_3$ is bigger for the bino LSP models, as explained above.  One
also finds that for Re($t$) around a particular value (1.2 in the
above figure), the gluino mass almost vanishes leading to a gluino
LSP at the low scale, which is not considered in our analysis.
Therefore that region in Re($t$) is not allowed. Another region
which is not allowed by low energy constraints is the region near
Re($t$)=1, where $M_2$ becomes very small, leading to
incompatibilities with the lightest chargino and LSP bounds.

\subsection{Explanation of Soft Parameters from the Underlying Theoretical Construction}\label{fromtheory}

One finally needs to explain the structure of soft terms (which
explains the spectrum pattern and hence the signature pattern)
from the structure of the underlying theoretical construction.
This would complete the sequence of steps to go from LHC
signatures to string theory. As explained before, we carry out
this exercise for the KKLT and Large Volume constructions, since
they are well defined from a microscopic point of view, and have a
reasonably well understood mechanism of supersymmetry breaking and
moduli stabilization.

Both of the constructions have complex structure moduli and
dilaton stabilized by turning on generic fluxes. The K\"{a}hler
moduli are stabilized by including non-perturbative corrections to
the superpotential. In Large Volume (IIB-L) vacua, a certain kind
of $\alpha'$ correction is also taken into account in the scalar
potential unlike that in the KKLT (IIB-K) vacua. In type IIB-K
constructions, the flux superpotential has to be fine-turned so as
to give a small ($\sim$ TeV) gravitino mass. For the IIB-L
construction however, no fine-tuning of the flux superpotential is
required. This gives rise to a relatively low (intermediate scale)
string scale if one wants a small gravitino mass.

The common feature of these two constructions is that the
K\"{a}hler moduli are stabilized mostly by non-peturbative
corrections. This leads to a particular feature in the gaugino
sector. It was shown in \cite{Conlon:2006us} that the gaugino
masses are suppressed relative to the gravitino mass $(\sim
m_{3/2}/\ln(m_{pl}/m_{3/2}))$ in all type IIB vacua with matter
residing on stacks of D7-branes and with all K\"{a}hler moduli
stabilized mostly by non-perturbative corrections to the
superpotential.

For the IIB-K constructions studied mostly in the literature,
there is only one overall K\"{a}hler modulus, the F-term of which
is suppressed. Since both the scalar masses and trilinear terms
are proportional to this F-term, they are both suppressed relative
to the gravitino in the IIB-K construction \cite{Choi:2004sx}.
Anomaly mediated contributions to soft terms have to be added as
they are comparable with those at tree level, leading to mixed
modulus-anomaly soft supersymmetry breaking terms. The above
feature survives for cases with more K\"{a}hler moduli, if all of
them are stabilized mostly by non-perturbative effects
\cite{Denef:2005mm}. On the other hand, the IIB-L constructions
require the presence of a large volume limit -- this means that
the Calabi-Yau manifold must have at least two K\"{a}hler moduli-
one of which is small ($T_s$) and the other is big
($T_b$)\cite{Balasubramanian:2005zx}. The presence of the
perturbative $\alpha'$ correction in the K\"{a}hler potential
gives a contribution to the scalar potential of the same order as
the non-perturbative corrections for the ``big'' K\"{a}hler
modulus, in contrast to that in the IIB-K construction. The F-term
of the small K\"{a}hler modulus ($F_s$) is suppressed by
$\ln(m_{pl}/m_{3/2})$, while that of the big K\"{a}hler modulus
($F_b$) is not suppressed. Since only D7-branes wrapping the small
4-cycle (represented by the small modulus) give a reasonable gauge
coupling, the visible sector gaugino masses are proportional to
$F_s$ and are suppressed relative to $m_{3/2}$. However both $F_s$
and $F_b$ enter into the expression for scalar masses and
trilinear terms. Since $F_b$ is not suppressed, therefore for the
IIB-L construction, only the gaugino sector has mixed
modulus-anomaly terms, with the scalars and trilinears generically
of the same order as the gravitino mass. This characteristic
feature is also true for Calabi-Yaus with more K\"{a}hler moduli
provided they admit a large volume limit, though the explicit soft
terms are hard to obtain. Another difference between the IIB-K and
IIB-L constructions is that the soft terms for the IIB-K
construction are first computed at the unification scale ($\sim
10^{16}$ GeV) while those for the IIB-L construction are computed
at the intermediate string scale ($\sim 10^{11}$ GeV).

The above analysis thus explains the origin of the pattern of soft
parameters in terms of the structure of the underlying theoretical
construction for the two constructions which leads to a
distinguishable signature pattern at the LHC. The important thing
to take home from this analysis is that different constructions
lead to different effective actions and therefore to different
expressions for the soft terms in terms of the underlying
microscopic input parameters. In addition, the relations
\emph{among} the different soft parameters
($M_a,m^2_i\,\&\,A_{ijk}$) also change for different
constructions. Therefore, a proper understanding of these
relations and their implications for relevant features of the
phenomenology is the key to relating high scale theory and data.
In this sense, we think that the approach advocated here is likely
to work even if one has much more realistic constructions from
different parts of the M-theory amoeba which stabilize all the
moduli, generate a stable hierarchy and also give a realistic
spectrum and couplings.

\section{Distinguishing Theories Qualitatively}\label{qualitative}

In the previous section, we analyzed the eight constructions in
great detail -- in particular, we computed the LHC signatures of
these constructions and understood the origin of these signatures
from features of the theoretical constructions. It is worthwhile
to ask whether one can abstract important lessons from this
exercise so that one could use them to analyze other classes of
constructions, and to draw qualitative reliable conclusions from
data.

For example, one could try to first extract relevant
phenomenological features of the effective beyond-the-Standard
Model theory from data and then focus on classes of M theory vacua
which give rise to those particular features. This alternative may
also be more helpful to people who are interested primarily in
understanding general features of beyond-the-Standard Model (BSM)
physics from LHC data rather than connecting it to an underlying
high scale theory like string theory. From our studies, we find
that a combination of features of any construction crucially
determine the broad pattern of LHC signatures. For concreteness,
we write our results in the framework of low-scale supersymmetry
as BSM physics, similar results will hold true for other
approaches as well. The important features we find\footnote{There
could be more such features.} are:

\begin{itemize}
\item The universality (or not) of gaugino masses at the
unification (or compactification) scale. \item If gauginos are
non-universal - the origin of the non-universality, i.e. whether
the non-universality is present at tree-level itself as opposed to
arising mostly due to one-loop anomaly mediated contributions.
\item If gauginos are non-universal - the hierarchy between $M_1$,
$M_2$,  $M_3$ and $\mu$. \item The relative hierarchy between the
scalars and gauginos at the string scale, i.e. whether the scalars
are of the same order as the gauginos as opposed to being heavier
or lighter than the gauginos. \item Nature and content of the LSP.
\item Hierarchy among scalars at the string scale, particularly
third family {\it vs} the first and second families.
\end{itemize}

Some comments are in order. These features are not always
independent of each other. For example, the hierarchy between
$M_1$, $M_2$ and $M_3$ determines the nature of the LSP (combined
with a knowledge of $\mu$). Also, if tree-level gaugino masses are
small so that non-universality arises only due to the anomaly
mediated contributions, then the gauginos are typically suppressed
relative to the scalars and the hierarchy between $M_1$, $M_2$ and
$M_3$ is fixed. Another important fact which should be kept in
mind is that a combination of all the features above gives rise to
the observed \emph{pattern} of LHC signatures, not just a
particular one. Therefore, once one obtains data, the task boils
down to figuring out the correct combination of ``relevant
features" which reproduces the data (at least roughly). Let's
explain this with two examples - the HM-A construction and the
IIB-L construction. Since we are only concerned with relevant
features of the effective BSM theory, all constructions considered
can be treated equally.

The HM-A and overlapping HM-B and HM-C constructions have
universal gaugino masses at the unification scale and have a bino
LSP. They also have scalars of the same order as the gauginos at
the unification scale, so RGE effects make the scalars (the third
generation in particular) lighter than the gauginos at the low
scale. This combination of ``relevant features" determines the
broad pattern of LHC signatures for these constructions. Since the
gauginos are universal at the unification scale, the ratio of
$M_1$, $M_2$ and $M_3$ at the low scale is $1 : 2 : 7$, which
controls the lower bound on the gluino mass and the LSP mass due
to experimental constraints on the chargino mass. Also, since
scalars are slightly lighter than gluinos at the low scale, both
$\tilde{g}\tilde{q}$ and $\tilde{q}\tilde{q}$ pair production are
comparable. The fact that the LSP is bino-like is also due to
universal gaugino masses as well as the fact that the scalars are
comparable to gauginos at the unification scale. A bino LSP can
then reduce its relic density by stau coannihilation, which can
only happen if the stau is light and almost degenerate with the
LSP. All these factors give rise to a very specific set of
signature pattern, as analyzed in section \ref{spectrum}.

In contrast, the IIB-L construction has a different set of
``relevant features" which determine its broad pattern of LHC
signatures and also allow it to be distinguishable from the other
constructions. The gaugino masses are non-universal at the string
scale\footnote{One does not have standard gauge unification at
$2\times10^{16}$ GeV in these models with $W_0=O(1)$
\cite{Balasubramanian:2005zx}.} and the scalars are heavier than
the gauginos at the low scale\footnote{This implies that the
scalars are also heavier than the gauginos at the string scale.}.
One also a mixed bino-higgsino LSP in this case. Since the scalars
are heavier than the gauginos, $\tilde{g}\tilde{g}$ pair
production is the dominant production mechanism. Also, the only
way to decrease the relic density of a bino-like LSP is to have a
significant higgsino fraction as the stau (or stop) coannihilation
channel is not open. Again, these features result in a very
specific set of signature pattern, as analyzed in section
\ref{spectrum}.

We therefore see that the features mentioned above are crucial in
determining the pattern of signatures at the LHC. Having said
that, it is important to remember that these features only
determine the broad pattern of LHC signatures and one needs more
inputs to explain the entire signature pattern in detail.

\section{Possible Limitations}\label{limitations}

\begin{figure}[h!]
  \begin{center}
      \epsfig{file=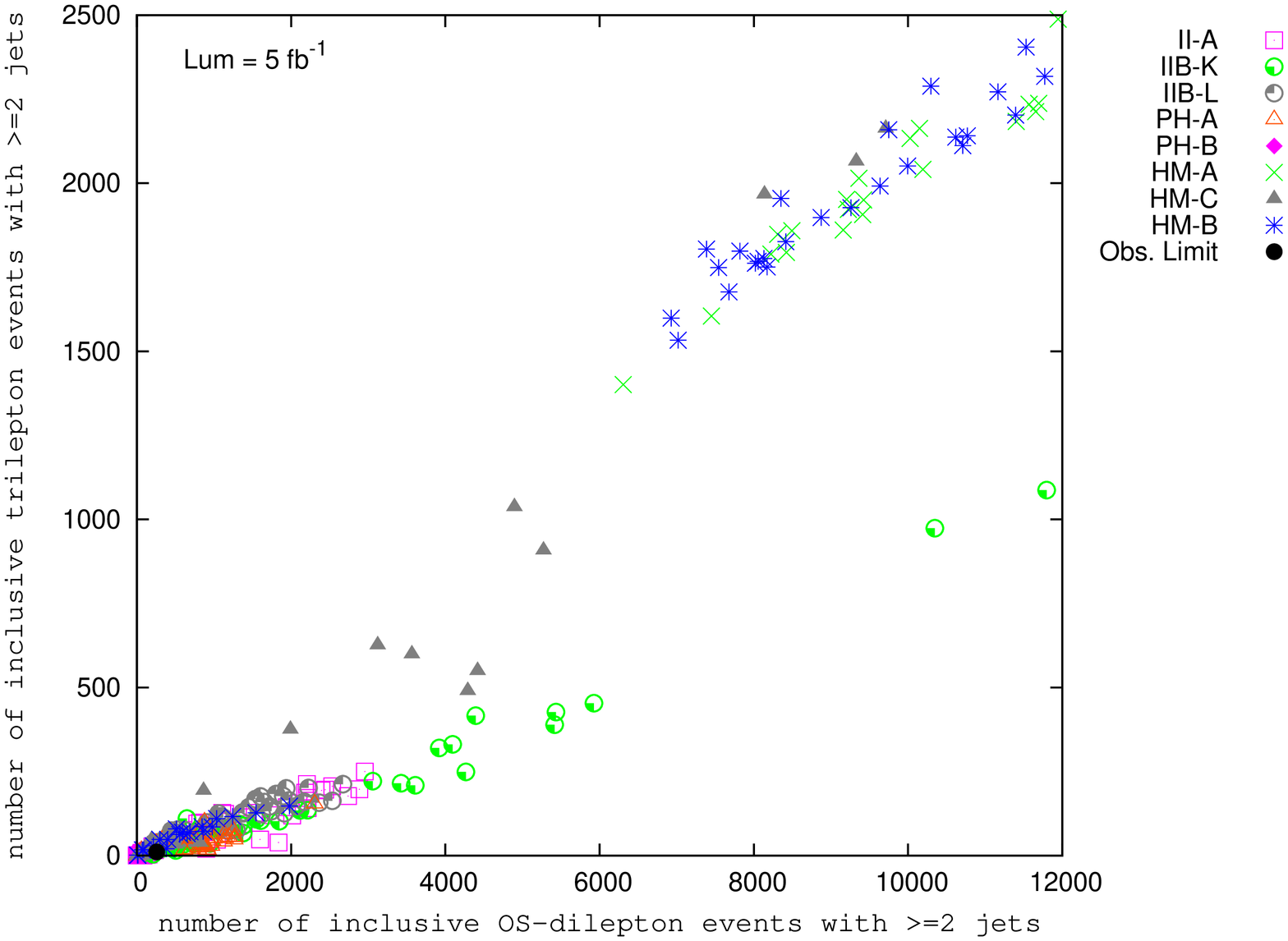,height=16cm, angle=-90}
    \caption{Plot of number of events with 1 lepton and $\geq$ 2 jets and number of
events with opposite sign dileptons and $\geq$ 2 jets, each
sampled with $\sim 50$ models, except PH-A ($\sim$100 models).}
    \label{OS-3l}
  \end{center}
\end{figure}
\begin{figure}[h!]
  \begin{center}
 \epsfig{file=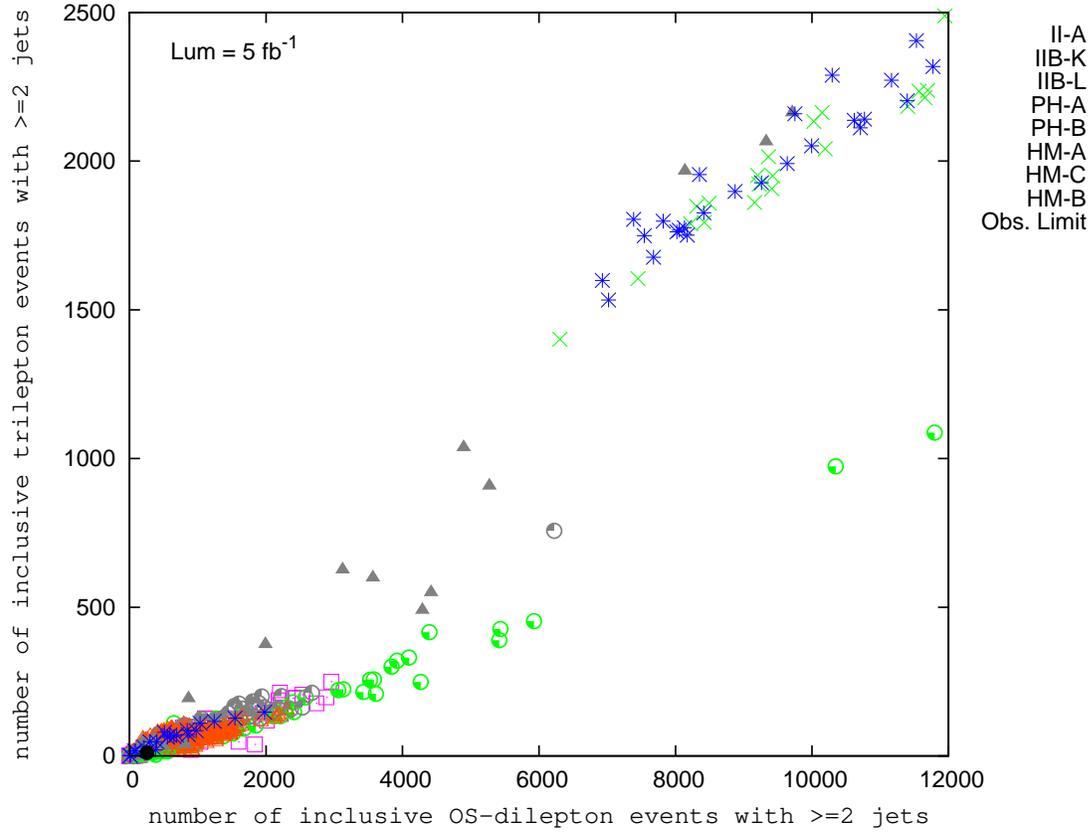,height=16cm, angle=-90}
 \caption{The same
    plot as in Figure \ref{OS-3l}, in which the IIB-K construction is
sampled with $\sim 100$ models and the PH-A construction with
$\sim$400 models.}\label{OS-3l-new}
  \end{center}
\end{figure}
\begin{figure}[h!]
  \begin{center}
      \epsfig{file=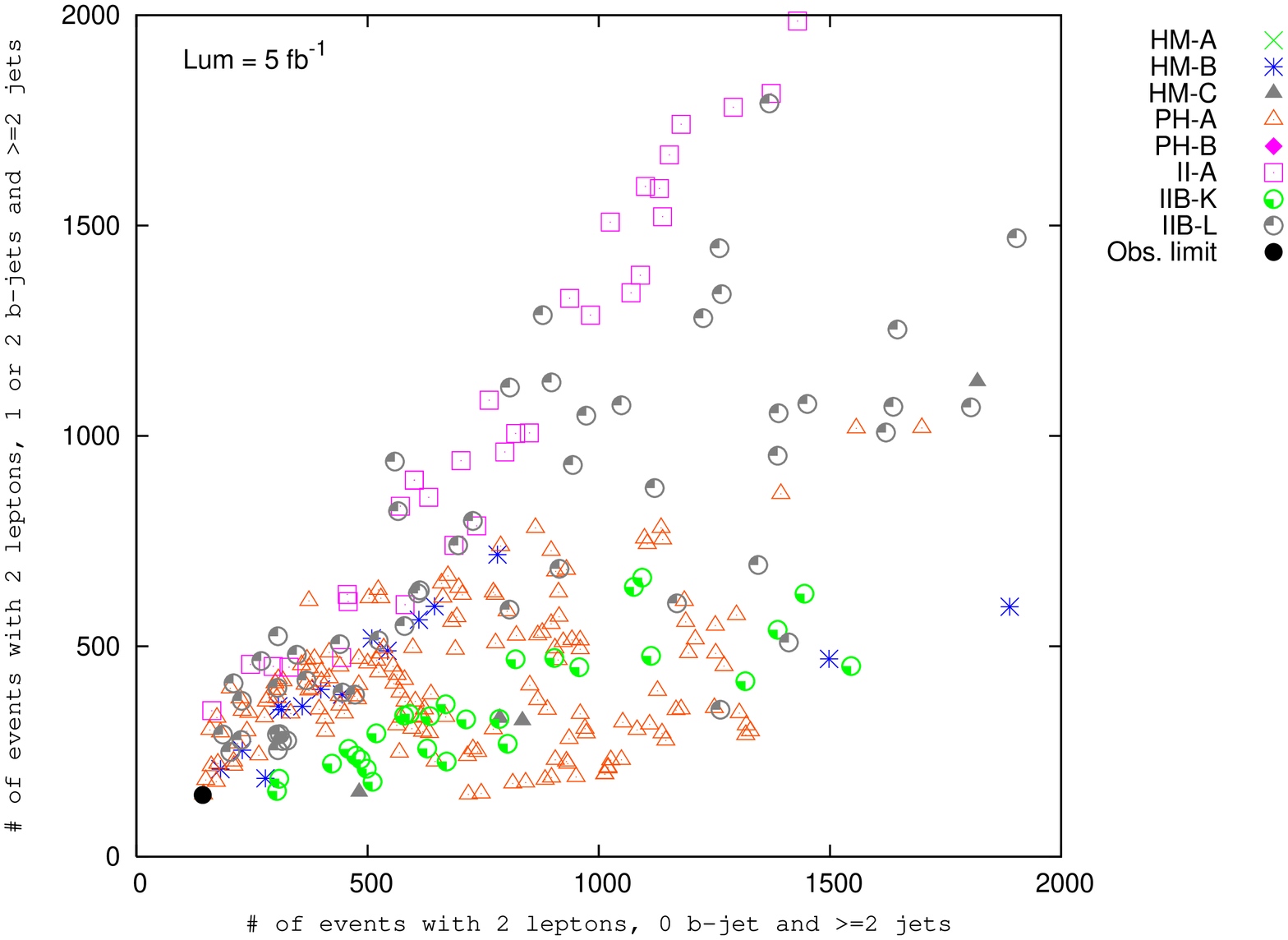,height=16cm, angle=-90}
    \caption{Plot with x axis showing number of events with 2 leptons, 0 b jets and $\geq$2 jets,
      and y axis showing the number of events with 2 leptons, 1 or 2 b jets and $\geq 2$ jets
     each sampled with $\sim 50$ models, except PH-A ($\sim$100 models).}
    \label{ratio1}
  \end{center}
\end{figure}
\begin{figure}[ht]
\begin{center}
 \epsfig{file=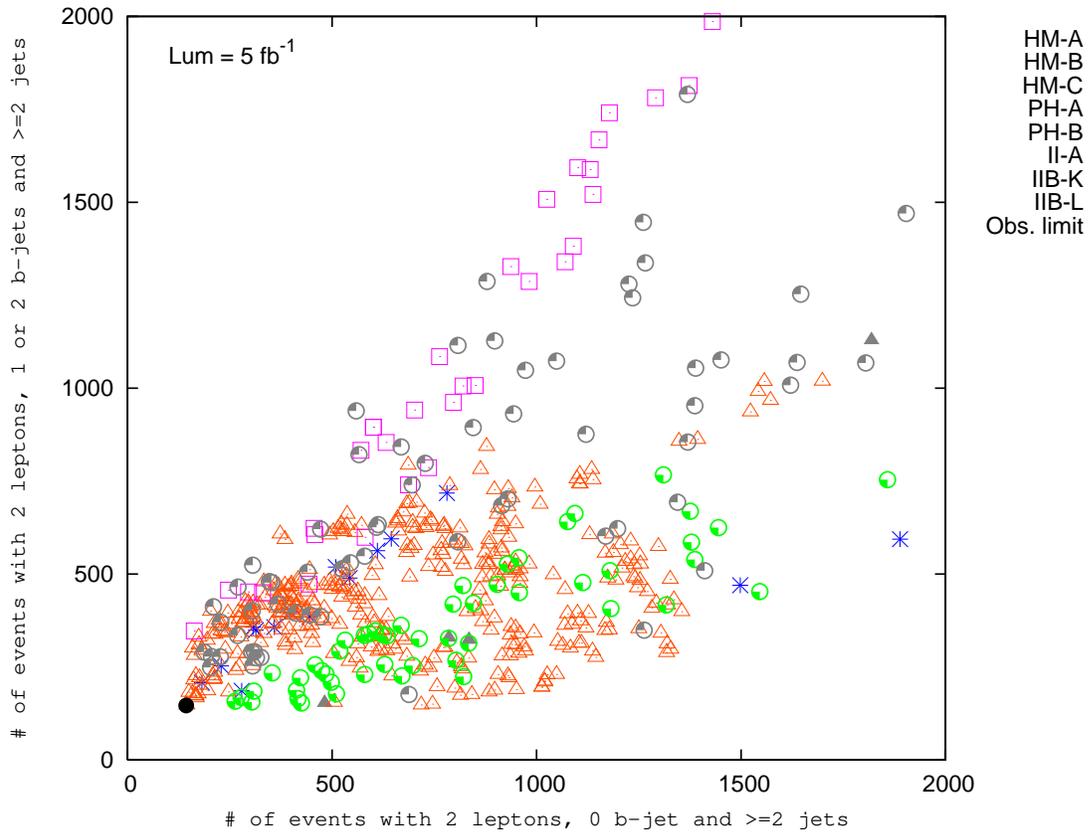,height=16cm, angle=-90}
 \caption{The same
    plot as in Figure \ref{ratio1}, in which the IIB-K construction is
sampled with $\sim 100$ models and the PH-A construction with
$\sim$400 models.}\label{ratio2}
  \end{center}
\end{figure}

One may raise questions about a few aspects of our analysis. The
first concerns the sampling of the parameter space of each
construction. One may worry that by only considering $\sim$50
models ($\sim$100 for the PH-A construction)\footnote{Not all
50(or 100) models simulated will be above the observable limit in
general.}, the parameter space of each construction is sampled
very sparsely and adding more models could qualitatively change
the overall signature pattern of the constructions. We think
however, that it is reasonable to expect that that is not true.
This is because, as explained in the previous sections, we have
outlined the origin of the pattern of signatures of the various
constructions on the basis of their spectrum and in turn on the
basis of their underlying theoretical setup. Since the dependence
of the soft terms on the microscopic input parameters as well as
relations between the soft terms are known and have been
understood, we expect our results for the pattern table to be
robust even when the parameter space is sampled more densely. This
will be strictly true only if one understands the theoretical
construction well enough so that one has a ``representative
sample" of the entire parameter space of that construction. We
expect this to be true for our constructions. In order to confirm
our expectation, we simulated $\sim$400 models for the PH-A
construction and $\sim$100 for the IIB-K construction and we found
that the results obtained with $\sim$100 (and $\sim$50) models did
not change when other models were added in our analysis. This can
be seen from Figures \ref{OS-3l} and \ref{OS-3l-new} as well as
Figures \ref{ratio1} and \ref{ratio2}. The other signatures also
do not change the final result. In the future, we plan to do a
much more comprehensive analysis with a dense sampling of the
parameter space for all the other constructions.

Another possible objection could be that the procedure of dividing
a signature into two classes arbitrarily and distinguishing them
on the basis of falling into one class or the other is too naive
and may lead to misleading results arising from intrinsic
statistical uncertainties and background uncertainties in the
value of each signature and impreciseness of the boundary. While
this is a valid concern in general, we think that this does not
affect the main results of our analysis at this level. This can
again be attributed to the fact that the pattern of signatures is
understood on the basis of their underlying theoretical structure.
We are also encouraged as there are typically more than one
(sometimes many) signatures distinguishing any two particular
constructions.

Another possible limitation which one could point out is that
distinguishing theoretical constructions on the basis of two
dimensional signature plots is not very powerful. Since we are
only looking at various two-dimensional projections of a
multi-dimensional signature space, it is possible that two
different theoretical constructions occupy different regions in
the multi-dimensional signature space even though they overlap in
all the two-dimensional projections. One would then not be able to
cleanly distinguish two constructions by this approach even though
they are intrinsically distinguishible. However, one can get
around this limitation by tagging individual models of each
theoretical construction. It would then be possible to figure out
if two different constructions are distinguishible even if they
overlap in all two dimensional signature plots. As already noted
in section \ref{results}, our purpose was to outline the approach
in a simple manner. It is clear that the approach has to be made
more sophisticated for more complicated situations.

For a mathematically precise way of distinguishing pairs of
constructions, one could use the following procedure. Imagine
dividing the parameter space of the two constructions into a
coarse grid with coordinates given by their parameter vectors
$\vec{x}$. For example, for the PH-A construction, $\vec{x} =
\{m_{3/2},a_{np},\tan(\beta)\}$ \cite{Kane:2002qp}. We can then
construct a ${\chi}^2$-like variable defined as follows:
\begin{eqnarray} \label{chisq}
(\Delta S)_{AB} =
\mathrm{min}_{\{\vec{x},\vec{y}\}}\sum_{i=1}^{n_{sig}}\left(\frac{s_{A}^i(\vec{x})-s_{B}^i(\vec{y})}{\sigma_{AB}^i}\right)^2
\end{eqnarray}
\noindent where $A$ and $B$ stand for the two constructions, $s_i$
stands for the $i$-th signature and $\sigma_{AB}^i$ stands for the
error bar assigned between the $A$ and $B$ constructions for the
$i$-th signature. $\sigma_{AB}^i$ can be determined from
statistical errors of the $i$-th signature for constructions $A$
and $B$ as well as the standard model background error for the
$i$-th signature.

Since we minimize with respect to the parameter vectors $\vec{x}$
and $\vec{y}$ of the two constructions, equation (\ref{chisq}) can
be geometrically visualized as the ``minimum distance squared" in
the full multi-dimensional signature space between the two
constructions $A$ and $B$ with an appropriate ``metric"
\footnote{The inverse square of the error $\sigma^i_{AB}$ acts
like the metric in signature space.}, and serves as a measure of
the difference between the two constructions $A$ and $B$. Carrying
out the minimization procedure in practice is quite a non-trivial
task, especially when the parameter vectors have many components,
as the time required to complete the minimization procedure in a
satisfactory way is too large. However techniques have been
introduced to get around this problem, for example see
\cite{Allanach:2004my}. In the future, we plan to do a
comprehensive analysis with more statistics and a precise method
to distinguish models, as explained above.

Finally, one might object to the approach of figuring out
experimental predictions for various classes of underlying
theoretical constructions and distinguishing different theoretical
constructions from each other before actual data, instead of the
more ``standard" approach of comparing each theoretical
construction with actual data. Many reasons can be given in this
regard. From a conceptual point of view, this approach fits will
with the philosophy of addressing the Inverse Problem in a general
framework. It is also crucial to addressing the question of
predictivity of an underlying theory like string theory in general
and the particular string vacuum we live in, in particular. From a
more practical point of view, studying the various subtle aspects
of connecting an underlying theoretical construction with collider
signatures is quite a non-trivial and subtle exercise and requires
considerable time and investment. It is therefore very helpful to
build knowledge and intuition in this regard and be prepared for
actual data. Carrying out this exercise could help discover
important properties of the low-energy implications of various
classes of theoretical constructions, in turn pointing to new
classes of collider observables as well as helping design new
analysis techniques.

\section{Summary and Future Directions}\label{conclude}

In this work, we have tried to address the goal of learning about
the underlying theory from LHC data - the deeper Inverse Problem.
We have proposed an approach by which it can be shown that the two
prerequisites to addressing the deeper Inverse Problem, namely, a)
To reliably go from a microscopic theory to the space of
experimental observables, and b) To distinguish among the various
classes of microscopic theoretical constructions on the basis of
their experimental signatures, can be satisfied for many
semi-realistic string constructions which can be described within
the supergravity approximation. In our opinion, the work is
seminal in the sense that it proposes a new way of thinking about
fundamental theory, model-building and collider phenomenology such
that there is a better synergy between each of these subfields.
Perhaps the most important result, which has never before been
presented, is that different classes of string constructions give
\emph{finite} footprints in signature space and that different
string constructions give practically overlapping but different
and distinguishable footprints.

The reason it is possible to distinguish theoretical constructions
is that patterns of experimental observables (for eg. signatures
at the LHC) are \emph{sensitive} to the structure of the
underlying theoretical constructions, because of correlations (see
section \ref{whypossible}). More precisely, this means that a
given theoretical construction occupies a {\it finite} region of
signature space which is in general different from another
theoretical construction. Moreover, the origin of this difference
can be understood from the underlying structure of the
construction. Therefore, even though we have carried out a
simplified analysis in terms of the imposition of cuts, detection
efficiencies of particles, detector simulation and calculation of
backgrounds, we still have confidence in the robustness of our
results. We have analyzed two classes of string vacua and six
other string-motivated constructions in detail. The point of this
exercise was to illustrate our approach, the same procedure should
be carried out for more classes of realistic vacua so that the
procedure becomes more-and-more useful. If the approach fails, it
will not be because of these and similar issues discussed above,
but rather because many regions of the entire M theory amoeba
cannot yet be analyzed by the approach we use. However, we think
that rather than giving up ahead of time, the best attitude is to
continue to expand both theoretical understanding as well as our
approach, and confront them with data.

There are two directions in which the approach advocated in this
work can be generalized and sharpened further. The first concerns
theoretical issues -- efforts should be made to go beyond toy
models focussing on few aspects of theory and phenomenology to
more holistic ones that address (if not solve) all of the issues
an underlying string theoretic construction might be expected to
explain. For instance, a better understanding of the theory can
fix some (or all) of the microscopic input parameters of a given
construction, increasing the predictivity of the construction. The
approach advocated by M. Douglas and W. Taylor, {\it viz.} to look
for correlations in the space of observables by analyzing
different classes of vacua is very similar to our approach in
principle.

The second concerns the analysis and interpretation of data and
its connection to the underlying theory. Creative thinking is
needed in identifying collider observables which more directly
probe the key features of the Beyond-the-Standard Model (BSM)
lagrangian and its connection to the underlying theory. We were
able to identify some useful observables by examining specific
constructions. In addition, one should find ways in which
observables from all fields -- collider physics, flavor physics,
cosmology, etc. could be used in conjunction to distinguish among,
and favor or exclude, many classes of string constructions in a
quick and robust manner. Our proposed technique is very useful in
this regard as it is very easy to add non-collider observables --
such as from flavor physics, cosmology, etc. to the collider
observables such that they are all treated in a uniform manner.

It is important to understand that the proposed approach should be
applied at various stages, with different tools and techniques
useful for each stage. The first stage would consist of
distinguishing many classes of constructions with limited amount
of data by using simple signatures and simple analysis techniques.
This has the advantage that one can rule out various classes of
constructions with relative ease. However, in order to go further,
it is important to develop more specialized analysis techniques
and use more exclusive signatures. This is best done in subsequent
stages, when one zooms in to a more limited set of constructions
and also obtains more data. Since one has better statistics, one
can use optimized and more exclusive signatures as well as use
more sophisticated analysis techniques to get more detailed
information about the constructions. Many of these sophisticated
analysis techniques already exist in the literature
\cite{Kneur:1998gy}, although they have been applied to very
special scenarios like minimal supergravity, minimal gauge
mediation, etc. One would now need to apply similar techniques
(suitably modified) to the set of constructions consistent with
limited data. Also, in the past year, a lot of progress has been
made towards uncovering the low energy spectrum and parameters
from (simulated) LHC signatures in the form of ``blackboxes''
constructed by some groups. This has been the program of the
\emph{LHC Olympics} Workshops in the past year \cite{LHCO-talks}.

Combining these sophisticated techniques with (some) knowledge of
the connection between theoretical constructions and data obtained
in the first stage, we hope that one can further distinguish the
remaining constructions, learn more about underlying theoretical
issues, like supersymmetry breaking and mediation, moduli
stabilization, inflation, etc.

\chapter{From Low Scale Data to High Scale Theory - Obstacles and
Resolutions}\label{lowtohigh}

In the previous chapter, we analyzed low energy predictions for
physical observables - signatures at the LHC, for many theoretical
models of supersymmetry breaking arising in string theory or
strongly motivated from string theory. In this chapter, we would
like to analyze a more bottom-up approach to the Inverse Problem,
{\it viz.} to go from data to theory in a more model-independent
way. Of course, addressing the problem in complete generality is
extremely challenging. So, we will make some assumptions about the
nature of new physics which are well motivated from a theoretical
point of view and also make the problem more tractable.

Following our arguments for the solution of the hierarchy problem
in chapter \ref{hierarchy} and similar to chapter
\ref{distinguishing}, we will assume that the new physics fits in
the framework of low energy supersymmetry. Since the apparent
unification of gauge couplings in the MSSM suggests an underlying
grand unified theoretical framework, we will consider new physics
consistent with gauge unification. Finally, we will assume that
``forthcoming" LHC data provides us with at least rough values of
the superpartner spectrum. Even with these assumptions, the
problem of unravelling the underlying high-scale theory is
complicated if the low scale spectrum is not measured completely
or precisely, or if there is new physics at heavy scales
(consistent with gauge unification) beyond the reach of collider
experiments. In this chapter we will study some of these obstacles
to running up, and we investigate how to get around them.  Our
main conclusion is that even though such obstacles can make it
very difficult to accurately determine the values of all the soft
parameters at the high scale, there exist a number of special
combinations of the soft parameters that can avoid these
difficulties.  We also present a systematic application of our
techniques in an explicit example.

Taking gauge unification as being more than accidental from above,
we obtain significant constraints on the types of new physics that
can arise between the electroweak and the grand unification~(GUT)
scales. Any new phenomenon that enters the effective theory in
this energy range ought to maintain the unification of couplings,
and should be consistent with a (possibly generalized) GUT
interpretation. The simplest scenario is a \emph{grand desert}, in
which there is essentially no new physics at all below the
unification scale $M_{GUT}$.  In this case, if supersymmetry is
discovered at the Tevatron or the LHC, it will be possible to
extrapolate the measured soft supersymmetry breaking parameters to
much higher scales using the renormalization group~(RG). Doing so
may help to reveal the details of supersymmetry breaking, and
possibly also the fundamental theory underlying it.

If supersymmetry is observed in a collider experiment, it will be
challenging to extract all the supersymmetry breaking parameters
from the collider signals.  While some work has been put into
solving this problem~\cite{Zerwas:2002as}, there is still a great
deal more that needs to be done. The parameters extracted in this
way will be subject to experimental uncertainties, especially if
the supersymmetric spectrum is relatively heavy. There will also
be theoretical uncertainties from higher loop corrections in
relating the physical masses to their running
values~\cite{Pierce:1996zz}. These uncertainties in the
supersymmetry breaking parameters, as well as those in the
supersymmetric parameters, will complicate the extrapolation of
the soft masses to high
energies~\cite{Blair:2000gy,Martin:2001zw}. Much of the previous
work along these lines has focused on running the parameters of
particular models from the high scale down. This is useful only if
the new physics found resembles one of the examples studied.  Our
goal is to study the running from low to
high~\cite{Carena:1996km}.

Evolving the soft parameters from collider energies up to much
higher scales can also be complicated by new physics at
intermediate energies below $M_{GUT}$. The apparent unification of
gauge couplings suggests that if this new physics is charged under
the MSSM gauge group, it should come in the form of complete GUT
multiplets or gauge singlets. Indeed, the observation of very
small neutrino masses already suggests the existence of new
physics in the form of very heavy gauge singlet
neutrinos~\cite{Mohapatra:2005wg}. With new physics that is much
heavier than the electroweak scale, it is often very difficult to
study it experimentally, or to even deduce its existence.  If we
extrapolate the MSSM parameters without including the effects of
heavy new physics, we will obtain misleading and incorrect values
for the high scale values of these parameters~\cite{Baer:2000gf}.

  In the present work we study some of these potential obstacles
to the RG evolution of the MSSM soft parameters.  In
Section~\ref{sterm} we investigate how uncertainties in the
low-scale parameter values can drastically modify the extrapolated
high-scale values. We focus on the so-called $S$ term
(\emph{a.k.a.} the hypercharge $D$ term) within the MSSM, which
depends on all the soft masses in the theory, and can have a
particularly large effect on the running if some of these soft
masses go unmeasured at the LHC. In Sections~\ref{gutmult} and
\ref{neut}, we study two possible examples of heavy new physics.
Section~\ref{gutmult} investigates the effects of adding complete
vector-like GUT multiplets on the running of the soft parameters.
Section~\ref{neut} describes how including heavy Majorana
neutrinos to generate small neutrino masses can alter the running
of the MSSM soft parameters.  In Section~\ref{all} we combine our
findings and illustrate how they may be put to use with an
explicit example. Finally, Section \ref{summary} is reserved for
our conclusions. A summary of some useful combinations of scalar
soft masses is given in an Appendix.

  Our main result is that the high scale values of many Lagrangian
parameters can be very sensitive to uncertainties in their
low-scale values, or to the presence of heavy new physics.
However, in the cases studied we also find that there are
particular combinations of the Lagrangian parameters that are
stable under the RG evolution, or that are unaffected by the new
physics. These special parameter combinations are therefore
especially useful for making a comparison with possible high-scale
theories.

  Throughout our analysis, we simplify the RG equations
by setting all flavor non-diagonal soft terms to zero and keeping
only the (diagonal) Yukawa couplings of the third generation.
Under this approximation, we work to two loop order for the
running of the MSSM parameters, and interface with
Suspect~2.3.4~\cite{Djouadi:2002ze} to compute one-loop threshold
corrections at the low scale. For concreteness, we take this scale
to be $500\,\gev$. The additional running due to new physics
introduced at scales much larger than the weak scale is only
performed at one-loop. We also implicitly assume that the mass
scale of the messengers that communicate supersymmetry breaking to
the visible sector lies at or above the GUT scale, $M_{GUT} \simeq
2.5\times 10^{16}\,\gev$. Even so, our methods and general
analysis will also be applicable to scenarios that have lighter
messenger particles, such as gauge
mediation~\cite{Dine:1993yw,Giudice:1998bp}.  We also neglect the
effects of hidden sector running, which can be significant if
there are interacting states in the hidden sector significantly
lighter than $M_{GUT}$~\cite{Luty:2001jh,Dine:2004dv,
Cohen:2006qc}.  While this manuscript was in preparation, methods
similar to those considered in the present work were proposed in
Ref.~\cite{Cohen:2006qc} to deal with these additional
uncertainties in the high scale values of the soft parameters.

  In this work, we focus on low-energy supersymmetric models,
and particular forms of intermediate scale new physics. Despite
this restriction, we expect that our general techniques will be
applicable to other solutions of the gauge hierarchy problem, or
to more exotic forms of new intermediate scale physics.

%%%%%%%%%%%%%%%%%%%%%%%%%%%%%%%%%%%%%%%%%%%%%%%%%%%%%%%

\section{Uncertainties Due to the \emph{S} Term\label{sterm}}

  The one-loop renormalization group~(RG) equations of the MSSM soft
scalar masses have the form~\cite{Martin:1993zk} \be
(16\pi^2)\frac{dm_i^2}{dt} = \tilde{X}_i
-\sum_{a=1,2,3}8\,g_a^2C^a_i|M_a|^2 + \frac{6}{5}g_1^2Y_i\,S,
\label{1:msoft} \ee where $t=\ln(Q/M_Z)$, $\tilde{X}_i$ is a
function of the soft squared masses and the trilinear couplings,
$M_a$ denotes the $a$-th gaugino mass, and the $S$ term is given
by \bea
S &=& Tr(Y\,m^2)\label{1:sterm}\\
&=& m_{H_u}^2 - m_{H_d}^2 + tr(m_Q^2-2\,m_U^2+m_E^2 +
m_D^2-m_L^2)\nnmb
%\\
%&=& m_{h_u}^2 - m_{H_d}^2 + (m_{Q_{1}}^2+m_{Q_2}^2+m_{Q_3}^2)
%-2\,(m_{U_{1}}^2+m_{U_2}^2+m_{U_3}^2) + \ldots\nnmb
\eea where the first trace runs over all hypercharge
representations, and the second runs only over flavors.

  The $S$ term is unusual in that it connects the running of any single
soft mass to the soft masses of every other field with non-zero
hypercharge.  Taking linear combinations of the RG equations for
the soft scalar masses, the one-loop running of $S$ in the MSSM is
given by \be (16\pi^2)\,\frac{dS}{dt} = -2\,b_1\,g_1^2\,S,
\label{1:dels} \ee where $b_1 = -33/5$ is the one-loop
beta-function coefficient. Using Eq.~(\ref{1:dels}) in
Eq.~(\ref{1:msoft}) and neglecting the Yukawa-dependent terms
$\tilde{X}_i$ (which are expected to be small for the first and
second generations) the effect of $S\neq 0$ is to shift the high
scale value the soft mass would have had were $S(t_0)=0$ by an
amount \beq \Delta m_i^2(t) = \frac{Y_i}{Tr(Y^2)}\left[
\frac{g_1^2(t)}{g_1^2(t_0)}- 1\right]\,S(t_0). \label{1:delm2}
\eeq The one-loop RG equation for $S$ is homogeneous.  Thus, if
$S$ vanishes at any one scale, it will vanish at all scales (at
one-loop). In both minimal supergravity~(mSUGRA) and simple
gauge-mediated models, $S$ does indeed vanish at the (high) input
scale, and for this reason the effects of this term are often
ignored.

   From the low-energy perspective, there is no reason
for $S(t_0)$ to vanish, and in many cases its effects can be
extremely important.  Since $g_1$ grows with increasing energy,
the mass shift due to the $S$ term grows as well. For $t_0 \simeq
t_{M_Z}$ and $t\to t_{GUT}$, the prefactor in Eq.~(\ref{1:delm2})
is about $(0.13)\,Y_i$.  The value of $S(t_0)$ depends on all the
scalar soft masses, and the experimental uncertainty in its value
will therefore be set by the least well-measured scalar mass.  In
particular, if one or more of the soft masses aren't measured at
all, $S(t_0)$ is unbounded other than by considerations of
naturalness.  Fortunately, this uncertainty only affects the soft
scalar masses.  The $S$ term does not enter directly into the
running of the other soft parameters until three-loop
order~\cite{Martin:1993zk,Jack:1999zs}, and therefore its effects
on these parameters is expected to be mild.

  There is also a theoretical uncertainty induced by the $S$ term.
Such a term is effectively equivalent to a Fayet-Iliopoulos~(FI)
$D$ term for hypercharge~\cite{Jack:1999zs}. To see how this comes
about, consider the hypercharge $D$-term potential including a FI
term, \be {\cal L} = \ldots + \frac{1}{2}D_1^2 + \xi\,D_1 +
\sqrt{\frac{3}{5}}g_1\,D_1\,\sum_i\bar{\phi}_iY_i\phi^i - \sum_i
\tilde{m}_i^2|\phi^i|^2, \ee where we have also included the soft
scalar masses. Eliminating the $D_1$ through its equation of
motion, we find \beq {\cal L} = \ldots -\frac{1}{2}\xi^2 -
\frac{g_1^2}{2} \left(\sum_i\bar{\phi}_iQ_i\phi^i\right)^2 -
\sum_i\bar{\phi}_i\left(\tilde{m}_i^2+\sqrt{\frac{3}{5}}\,g_1\,
Y_i\,\xi\right)\phi^i. \eeq Thus, except for the constant addition
to the vacuum energy, the effect of the FI term is to shift each
of the soft squared masses by an amount \beq \tilde{m}_i^2 \to
m_i^2 := \tilde{m}_i^2+\sqrt{\frac{3}{5}}\,g_1\,Y_i\,\xi.
\label{1:mshift} \eeq The low-energy observable quantities are the
$\{m_i^2\}$, not the $\{\tilde{m}_i^2\}$.  Since we can't extract
the shift in the vacuum energy, the low-energy effects of the FI
term are therefore invisible to us, other than the shift in the
soft masses. This shift is exactly the same as the shift due to
the $S$ term. Let us also mention that the $S$ term, as we have
defined it in Eq.~(\ref{1:sterm}), runs inhomogeneously at
two-loops and above, so the exact correspondence between a
hypercharge $D$ term and the $S$ term of Eq.~(\ref{1:sterm}) does
not hold beyond one-loop order.

  A simple way to avoid both the large RG uncertainties in the soft
masses and the theoretical ambiguity due to the $S$ term is
apparent from Eq.~(\ref{1:delm2}).  Instead of looking at
individual soft masses, it is safer to consider the mass
differences \be Y_j\,m_i^2-Y_i\,m_j^2, \label{1:diff} \ee for any
pair of fields.  These differences are not affected by the mass
shifts of Eq.~(\ref{1:delm2}).  They are also independent of the
value of the FI term.

  In the rest of this section, we show how the $S$ term
can complicate the running of the soft masses to high energies
with a particular example.  If one of the scalar soft masses is
unmeasured, it is essential to use the linear mass combinations
given in Eq.~(\ref{1:diff}) instead of the individual masses
themselves. We also discuss how the special RG properties of the
$S$ term provide a useful probe of high scale physics if all the
scalar soft masses are determined experimentally.

\subsection{Example: SPS-5 with an Unmeasured Higgs Soft Mass}

  To illustrate the potential high-scale uncertainties in
the RG-evolved soft parameters due to the $S$ term, we study the
sample mSUGRA point SPS-5~\cite{Allanach:2002nj} under the
assumption that one of the scalar soft masses goes unmeasured at
the LHC.  If this is the case, the $S$ parameter is undetermined,
and the precise values of the high-scale soft terms are no longer
precisely calculable.  Even if the value of the $S$ term is
bounded by considerations of naturalness, the uncertainties in the
high-scale values of the soft scalar masses can be significant.

  The SPS-5 point is defined by the mSUGRA input values
$m_0 = 150$~GeV, $m_{1/2} = 300$~GeV, $A_0 = -1000$, $\tan\beta =
5$, and $sgn(\mu) > 0$, at $M_{GUT}$.  The mass spectrum for this
point has relatively light sleptons around 200~GeV, and somewhat
heavier squarks with masses near 400-600~GeV. The LSP of the model
is a mostly Bino neutralino, with mass close to 120~GeV.  The
perturbation we consider for this point is a shift in the
down-Higgs soft mass, $m_{H_d}^2$.

\begin{figure}[tb]
\center \epsfig{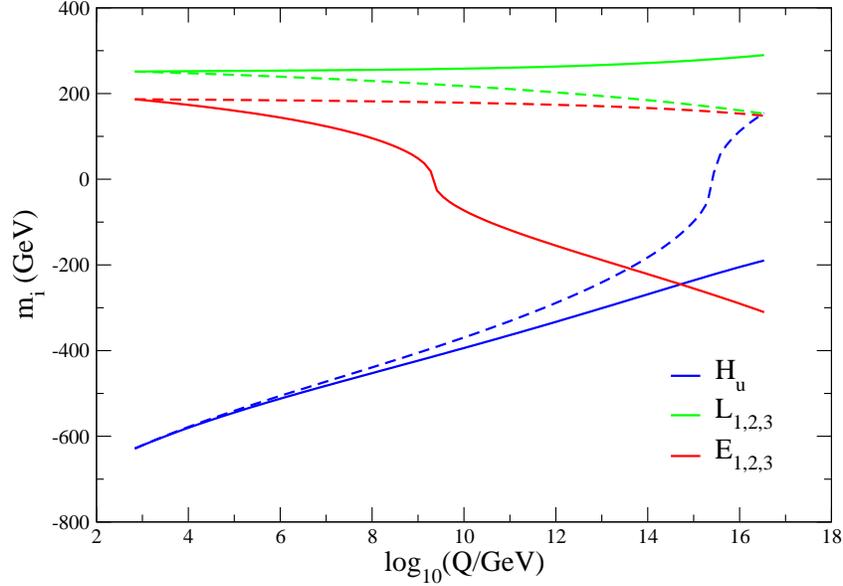}
\caption{ Deviations in the running of some of the SPS-5 soft
masses due to setting $m_{H_d}^2=(1000~\gev)^2$ at the low scale.
The solid lines show the running of $m_{H_u}^2$, $m_{E}^2$, and
$m_{L}^2$ with this perturbation, while the dashed lines show the
unperturbed running of these soft masses. The unperturbed
low-scale value of the down-Higgs soft mass is $m_{H_d}^2 \simeq
(235\,\gev)^2$. } \label{fig:sps5-a}
\end{figure}

  Of the soft supersymmetry breaking parameters in the MSSM,
the soft terms associated with the Higgs sector can be
particularly difficult to deduce from LHC measurements. At
tree-level, the independent Lagrangian parameters relevant to this
sector are~\cite{Martin:1997ns} \beq v,~~\tan\beta,~~\mu,~~M_A,
\label{hpar} \eeq where $v\simeq 174\,\gev$ is the electroweak
breaking scale, $\tan\beta = v_u/v_d$ is the ratio of the $H_u$
and $H_d$ VEVs, $\mu$ is the supersymmetric $\mu$-term, and $M_A$
is the pseudoscalar Higgs boson mass.  Other Higgs-sector
Lagrangian parameters, such as $m_{H_d}^2$ and $m_{H_u}^2$, can be
expressed in terms of these using the conditions for electroweak
symmetry breaking in the MSSM.

\begin{figure}[tb]
%{0.47\textwidth}
\begin{center}
\includegraphics[width = 0.65\textwidth]{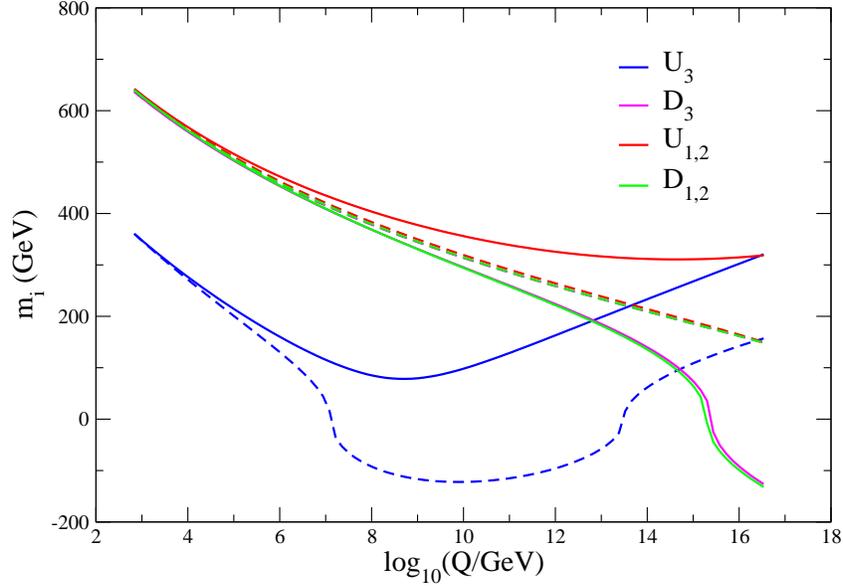}
\end{center}
\caption{ Deviations in the running of some of the SPS-5 soft
masses due to setting $m_{H_d}^2=(1000~\gev)^2$ at the low scale.
The solid lines show the running of $m_{U_{1,2}}^2$,
$m_{D_{1,2}}^2$, $m_{U_3}^2$, and $m_{D_3}^2$ with this
perturbation, while the dashed lines show the unperturbed running
of these soft masses.} \label{fig:sps5-b}
\end{figure}

  Among the Higgs sector parameters listed in Eq.~(\ref{hpar}),
only the value of $v$ is known. The value of $\mu$ can potentially
be studied independently of the Higgs scalar sector by measuring
neutralino and chargino masses and
couplings~\cite{Zerwas:2002as,Allanach:2002nj, neutralinomu},
although it is likely to be poorly determined if only hadron
colliders are available. A number of observables outside the Higgs
sector may also be sensitive to $\tan\beta$, especially if it is
large, $\tan\beta \geq 20$. For example, the dilepton invariant
mass distributions in the inclusive $2\ell+jets+\met$ channel can
vary significantly depending on the value of $\tan\beta$, but this
dependence is such that the value of $\tan\beta$ can at best only
be confined to within a fairly broad ranges~\cite{denegri}.
Determining $M_A$ at the LHC typically requires the discovery of
one of the heavier MSSM Higgs boson states. Finding these states
can also help to determine $\tan\beta$~\cite{Assamagan:2004ji}.
Unfortunately, the LHC reach for the heavier Higgs states is
limited, especially for larger values of $M_A$ and intermediate or
smaller values of $\tan\beta\leq 20$~\cite{heavyhiggs}.  If none
of the heavier Higgs bosons are found at the LHC, it does not
appear to be possible to determine both of the Higgs soft masses,
$m_{H_u}^2$ and $m_{H_d}^2$.

  For the low-scale parameters derived from SPS-5, the pseudoscalar Higgs
mass $M_A$ is about 700~GeV.  With such a large value of $M_A$,
and $\tan\beta = 5$, only the lightest SM-like Higgs boson is
within the reach of the LHC~\cite{heavyhiggs}. Motivated by this
observation, we examine the effect of changing the low-scale value
of $m_{H_d}^2$ on the running of the other soft parameters.  The
actual low-scale value of $m_{H_d}^2$ is about $(235~\gev)^2$.
The perturbation we consider is to set this value to
$(1000~\gev)^2$, while keeping $\tan\beta$ fixed. Such a
perturbation does not ruin electroweak symmetry breaking, and
tends to push the heavier higgs masses to even larger values. In
the present case, the heavier Higgs masses increase from about
$700~\gev$ to over $1200~\gev$.~\footnote{The values of $\mu$ and
$B\mu$ also change, although the variation in $\mu$ is very mild:
$\mu\simeq 640\,\gev \to 670\,\gev$.}

  The effects of this perturbation in $m_{H_d}^2$ on some of the soft
scalar masses are shown in Figs.~\ref{fig:sps5-a}
and~\ref{fig:sps5-b}. In these plots, we show $m_i =
m_i^2/\sqrt{|m_i^2|}$. The deviations in the soft masses are
substantial, and the $S$ parameter is the source of this
uncertainty. In addition to the $S$ term, varying $m_{H_d}^2$ can
also modify the running of the Higgs mass parameters and the
down-type squarks and sleptons through the Yukawa-dependent terms
$X_i$ in the RG equations, Eq.~(\ref{1:msoft}). In the present
case these Yukawa-dependent effects are very mild since for
$\tan\beta = 5$, the $b$ and $\tau$ Yukawas are still quite small.
This can be seen by noting the small difference between the
perturbed running of $m_{U_{1,2}}^2$ and $m_{U_3}^2$ in
Fig.~\ref{fig:sps5-b}. For larger values of $\tan\beta$, the
down-type Yukawa couplings can be enhanced and this non-$S$ effect
from $m_{H_d}^2$ can be significant. However, we also note that as
these Yukawa couplings grow larger, it becomes much easier for the
LHC to detect one or more of the heavier Higgs states.  To the
extent that the Yukawa-dependent shifts can be neglected, the
linear mass combinations of Eq.~(\ref{1:diff}) remove most of the
uncertainty due to an unknown $m_{H_d}^2$ in the running of the
soft masses that are measured. The effect of not knowing
$m_{H_d}^2$ has only a very small effect on the running of the
gaugino masses and the trilinear terms.

  In this example we have assumed that $m_{H_d}^2$ is the only
unmeasured soft scalar mass.  Several of the other soft scalar
masses may be difficult to reconstruct from LHC data as well. For
example, within many SUSY scenarios the third generation squarks
and some of the heavier sleptons have very small LHC production
rates. If there are other unmeasured soft scalar masses besides
$m_{H_d}^2$, the uncertainties due to the $S$ term in the
extrapolation of the measured scalar soft masses will be even
greater than what we have presented here. The mass combinations of
Eq.~(\ref{1:diff}) will be necessary to study the high scale
supersymmetry breaking spectrum in this case.

\subsection{Origins and Uses of the $S$ Term}

  While the $S$ term can complicate the extrapolation
of the soft scalar masses if one of them goes unmeasured, the
simple scale dependence of this term also makes it a useful probe
of the high scale theory if all the masses are determined. The
essential feature is the homogeneous RG evolution of the $S$ term,
given in Eq.~(\ref{1:dels}), which is related to the
non-renormalization of FI terms in the absence of supersymmetry
breaking.

  A non-vanishing $S$ term can arise from a genuine FI term
present in the high scale theory, or from non-universal scalar
soft masses at the high input scale.  The size of a fundamental
hypercharge FI term, $\xi$, is na\"ively on the order of the large
input scale.  Such a large value would either destabilize the
gauge hierarchy or lead to $U(1)_Y$ breaking at high energies.
However, the non-renormalization of FI terms implies that it is
technically natural for $\xi$ to take on much smaller values. In
this regard, a small value for $\xi$ is analogous to the $\mu$
problem in the MSSM.  Adding such a FI term to mSUGRA provides a
simple one-parameter extension of this model, and can have
interesting effects~\cite{deGouvea:1998yp,Falk:1999py}. Note,
however, that in a GUT where $U(1)_Y$ is embedded into a simple
group, a fundamental hypercharge FI term in the full theory is
forbidden by gauge invariance.

  It is perhaps more natural to have the $S$ term emerge from
non-universal scalar soft masses~\cite{dterm,dgut}. This is true
even in a GUT where $U(1)_Y$ is embedded into a simple group.
Within such GUTs, the contribution to $S$ from complete GUT
multiplets vanishes. However, non-zero contributions to $S$ can
arise from multiplets that are split in the process of GUT
breaking.  For example, in $SU(5)$ with $H_u$ and $H_d$ states
embedded in $\bf{5}$ and $\overline{\bf{5}}$ multiplets, a
non-zero low-energy value of $S$ can be generated when the heavy
triplet states decouple provided the soft masses of the respective
multiplets are unequal. Whether it is zero or not, the low-scale
value of the $S$ term provides a useful constraint on the details
of a GUT interpretation of the theory.

  So far we have only considered the $S$ term corresponding to
$U(1)_Y$.  If there are other gauged $U(1)$ symmetries, there will
be additional $S$ term-like factors for these too.  In fact, the
homogeneity of the $S$ term evolution also has a useful
implication for any non-anomalous \emph{global} $U(1)$ symmetry in
the theory.  The only candidate in the MSSM is $U(1)_{B-L}$, up to
linear combinations with $U(1)_Y$~\cite{u1x}.  Let us define
$S_{B-L}$ by the combination \bea
S_{B-L} &=& Tr(Q_{B-L}m^2)\label{1:sbl}\\
&=& tr(2m_Q^2 - m_U^2-m_D^2 - 2m_L^2+m_E^2),\nnmb \eea where the
second trace runs only over flavors.\footnote{ Up to flavor
mixing, we can also define an independent $S_{B-L}$ for each
generation.} We can think of $S_{B-L}$ as the effective $S$ term
for a gauged $U(1)_{B-L}$ in the limit of vanishing coupling. At
one loop order, the RG running of $S_{B-L}$ is given by \bea
(16\pi^2)\frac{dS_{B-L}}{dt} &=& \frac{3}{5}Tr(Q_{B-L}Y)g_1^2\,S\label{1:bls}\\
&=& n_g\frac{16}{5}\,g_1^2\,S,\nnmb \eea where $n_g=3$ is the
number of generations. If $S$ is measured and vanishes, $S_{B-L}$
provides a second useful combination of masses that is invariant
under RG evolution,\footnote{ This non-evolution of mass
combinations corresponding to non-anomalous global symmetries
persists at strong coupling. In models of conformal sequestering,
this can be problematic~\cite{Luty:2001jh}.} and yields an
additional constraint on possible GUT embeddings of the theory.

%%%%%%%%%%%%%%%%%%%%%%%%%%%%%%%%%%%%%%%%%%%%%%%%%%%%%%%

\section{New Physics: Complete GUT Multiplets\label{gutmult}}

  As a second line of investigation, we consider the effects of
some possible types of new intermediate scale physics on the
running of the MSSM soft terms.  If this new physics is associated
with supersymmetry breaking as in gauge
mediation~\cite{Dine:1993yw}, then it is of particular interest in
its own right. Indeed, in this case the low-energy spectrum of
soft terms may point towards the identity of the new physics after
RG evolution. On the other hand, there are many kinds of possible
new intermediate scale physics that are not directly related to
supersymmetry breaking. The existence of this type of new
phenomena can make it much more difficult to deduce the details of
supersymmetry breaking from the low-energy soft terms.

  A useful constraint on new physics is gauge coupling unification.
To preserve unification, the SM gauge charges of the new physics
should typically be such that all three MSSM gauge beta functions
are modified in the same way.\footnote{For an interesting
exception, see Ref.~\cite{Martin:1995wb}.}  This is automatic if
the new matter fills out complete multiplets of a simple GUT group
into which the MSSM can be embedded.  Such multiplets can emerge
as remnants of GUT symmetry breaking.

  As an example of this sort of new physics, we consider vector-like
pairs of complete $SU(5)$ multiplets.  For such multiplets, it is
possible to write a down a supersymmetric mass term of the form
\be \mathcal{W} \supset \tilde{\mu}\,\bar{X}\,X, \ee where $X$ and
$\bar{X}$ denote the chiral superfields of the exotic multiplets.
We also assume that the exotic multiplets have no significant
superpotential (Yukawa) couplings with the MSSM fields.  Under
these assumptions, the exotic $SU(5)$ multiplets can develop large
masses independently of the details of the MSSM. They will
interact with the MSSM fields only through their gauge
interactions.

  If the supersymmetric mass $\tilde{\mu}$ is much larger than the
electroweak scale, it will be very difficult to deduce the
presence of the additional GUT multiplets from low-energy data
alone. Moreover, an extrapolation of the measured soft masses
using the RG equations appropriate for the MSSM will lead to
incorrect values of the high-scale parameters.  In this section,
we characterize the sizes and patterns of the deviations in the
high scale soft spectrum induced by additional vector-like GUT
multiplets. We also show that even though the new matter
interferes with the running of the MSSM soft parameters, it is
often still possible to obtain useful information about the input
spectrum, such as the relative sizes of the gaugino masses and
whether there is inter-generational splitting between the soft
scalar masses.

\subsection{Shifted Gauge Running}

  The main effect of the exotic GUT multiplets is to shift the running
of the $SU(3)_c$, $SU(2)_L$, and $U(1)_Y$ gauge couplings.  Recall
that in the MSSM, the one-loop running of these couplings is
determined by \beq \frac{dg_a^{-2}}{dt} = \frac{b_a}{8\pi^2},
\label{rungauge} \eeq with $(b_1,~b_2,~b_3) = (-33/5,-1,3)$. The
presence of a massive GUT multiplet shifts each of the $b_a$ up by
an equal amount above the heavy threshold at $t = t_I =
\ln(\tilde{\mu}/M_Z)$, \beq \Delta b = - \nfive -
3\,N_{{\bf{10}\oplus\bar{\bf 10}}} + \ldots \eeq where $\nfive$ is
the number of additional $\bf{5}\!\oplus\!\overline{\bf{5}}$
representations and $N_{{\bf{10}\oplus\bar{\bf 10}}}$ is the
number of $\bf{10}\oplus\overline{\bf{10}}$'s. The modified
one-loop solution to the RG equations is therefore, \beq
g_a^{-2}(t) = \left\{
\begin{array}{lcc}
g_a^{-2}(t_0) +\frac{b_a}{8\pi^2}(t-t_0)&~~~~~&t<t_I,\\
g_a^{-2}(t_0) +\frac{b_a}{8\pi^2}(t-t_0)+\frac{\db}{8\pi^2}(t-t_I)
&~~~~~&t>t_I.
\end{array}\right.\label{grun}
\eeq It follows that the unification scale is not changed, but the
value of the unified gauge coupling is increased.  Note that the
number of new multiplets is bounded from above for a given value
of $\tilde{\mu}$ if gauge unification is to be
perturbative.\footnote{ Note that the parts of $\tilde{\mu}$
corresponding to the doublet and triplet components of the
$\bf{5}$ will run differently between $M_{GUT}$ and the
intermediate scale.  This will induce additional threshold
corrections that we do not include in our one-loop analysis.}

  The change in the gauge running shifts the running of all the soft
parameters, but the greatest effect is seen in the gaugino masses.
At one-loop, these evolve according to \beq \frac{dM_a}{dt} =
-\frac{b_a}{8\pi^2}g_a^2\,M_a. \label{rungino} \eeq It follows
that $M_a/g_a^2$ is RG invariant above and below the heavy
threshold.  If the threshold is also supersymmetric, $M_a$ will be
continuous across it at tree-level.  Since $g_a$ is also
continuous across the threshold at tree-level, the addition of
heavy vector-like matter does not modify the one-loop scale
invariance of the ratio $M_a/g_a^2$. This holds true whether or
not the new matter preserves gauge unification, but it is most
useful when unification holds. When it does, the measurement of
the low-energy gaugino masses immediately translates into a
knowledge of their ratio at $M_{GUT}$~\cite{Kane:2002ap}.
%{gaugino screening theorem...}

\begin{figure}[tbh]
\vspace{1cm}
\begin{center}
\includegraphics[width = 0.65\textwidth]{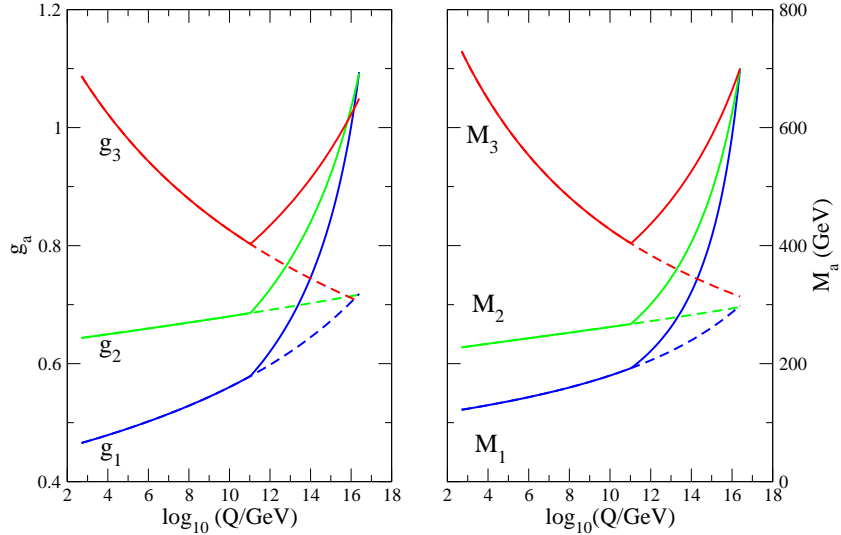}
\end{center}
\caption{ The shift in the running of the gauge couplings~(left)
and the gaugino masses~(right) due to 7 sets of ${\bf
5}\oplus\bar{\bf 5}$'s with mass $\tilde{\mu} = 10^{11}$~GeV.  The
universal gaugino mass is taken to be $700\,\gev$.}
\label{fig:gshift}
\end{figure}

  From Eq.~(\ref{grun}) and the one-loop scale invariance
of $M_a/g_a^2$, the shift in the gaugino masses due to the
additional matter is \beq M_a(t) =
\bar{M}_a(t)\,\left[1+\frac{\Delta b\,\bar{g}_a^2}{8\pi^2}
(t-t_I)\right]^{-1}, \eeq where $\bar{g}_a$ and $\bar{M}_a$ denote
the values these parameters would have for $\Delta b = 0$
(\emph{i.e.} the values obtained using the MSSM RG equations).
For $t=t_{GUT}$, the shift coefficient is identical for $a=1,2,3$
provided gauge unification occurs. The shift in the running of the
gauge couplings and the gaugino masses due to seven sets of ${\bf
5}\oplus\bar{\bf 5}$'s at $10^{11}$~GeV is illustrated in
Fig.~\ref{fig:gshift}. An unperturbed universal gaugino mass of
$M_{1/2} = 700\,\gev$ is assumed.  Both the values of the unified
gauge coupling and the universal gaugino mass at $M_{GUT}$ are
increased by the additional multiplets.

  The running of the soft masses also depends on the running of the
gauge couplings and gaugino masses, and is modified by the
appearance of new matter.  At one-loop order, in the limit that we
can neglect the Yukawa couplings, it is not hard to find the
shifts in the soft masses. We can divide these shifts into two
contributions, \beq m_i^2(t) = \bar{m}_i^2 + \Delta
m^2_{i_{\lambda}} + \Delta m^2_{i_{S}}, \eeq where $\bar{m}_i^2$
is the value of the soft mass obtained by running the measured
value up in the absence of the new matter, $\Delta
m^2_{i_{\lambda}}$ is the shift due to the modified gaugino
masses, and $\Delta m^2_{i_{S}}$ is due to the change in the
running of the $S$ term.

  The first shift, $\Delta m^2_{i_{\lambda}}$,
can be obtained straightforwardly from
Eqs.~(\ref{1:msoft},\ref{rungauge},\ref{rungino}), and is given by
\beq \Delta m_{i_{\lambda}}^2 =
\sum_a2\,C^a_i\left|\frac{M_a}{g_a^2}\right|^2 \Delta I_a
\label{delml} \eeq where $C^a_i$ the Casimir invariant of the
$a$-th gauge group and \beq \Delta I_a =
\frac{1}{(b_a+\db)}\,(g_a^4-g_{a_I}^4) -
\frac{1}{b_a}(\bar{g}_a^4-g_{a_I}^4), \eeq with $g_a$ the actual
gauge coupling at scale $t>t_I$ (including the extra matter),
$g_{a_I}$ the gauge coupling at the heavy threshold $t_I$, and
$\bar{g}_a$ the gauge coupling at scale $t>t_I$ in the absence of
new matter.
%Explicit expressions for the $\bar{g}_a(t)$ couplings follow
%directly from Eq.~(\ref{grun}) by setting $\Delta b = 0$.

  New chiral matter can also modify the running of the soft scalar masses
through the $S$ parameter.  The new matter shifts the running of
$S$ by changing the running of $g_1$, but it can also contribute
to the $S$ term directly at the heavy threshold. Combining these
effects, the value of $S$ above threshold is \bea S(t) &=&
\left(\frac{{g}_1}{g_{1_0}}\right)^2\,S(t_0)
+ \left(\frac{g_1}{g_{1_I}}\right)^2\,\Delta\,S\\
&=& \bar{S}(t) +
\left(\frac{{g}_1^2-\bar{g}_1^2}{g_{1_0}^2}\right)\,S(t_0) +
\left(\frac{g_1}{g_{1_I}}\right)^2\,\Delta\,S,\nnmb \eea where
$\Delta S$ is the shift in the value of $S$ at the threshold,
$g_{1_0}$ is the low-scale value of the gauge coupling, and
$\bar{S}(t)$ is the value the $S$ term would have in the absence
of the new matter. Inserting this result into Eq.~(\ref{1:msoft}),
the effect of the new matter on the running of the soft scalar
masses through the $S$ term is \bea \frac{5}{3Y_i}\Delta m^2_{i_S}
&=& \frac{1}{b_1}\left[ \left(\frac{\bar{g}_{1}}{g_{1_0}}\right)^2
- \left(\frac{g_{1_I}}{g_{1_0}}\right)^2\right]\,S_0
- \frac{1}{(b_1+\db)}\left[ \left(\frac{g_{1}}{g_{1_0}}\right)^2
- \left(\frac{g_{1_I}}{g_{1_0}}\right)^2\right]\,S_0\nnmb\\
&&\label{delms}\\
&&~~~- \frac{1}{(b_1+\db)}\left[
\left(\frac{g_{1}}{g_{1_I}}\right)^2 -1\right]\,\Delta S.\nnmb
\eea As before, the effects of the $S$ term on the soft masses are
universal up to the hypercharge prefactor.  Thus, they still
cancel out of the linear combinations given in Eq.~(\ref{1:diff}).

\subsection{Yukawa Effects and Useful Combinations}

  We have so far neglected the effects of the MSSM Yukawa
couplings on the modified running of the soft masses.  As a
result, the shifts in the running of the soft scalar masses
written above are family universal.  There are also non-universal
shifts in the soft scalar masses.  These arise from the
Yukawa-dependent terms in the soft scalar mass beta functions,
which themselves depend on the Higgs and third-generation soft
scalar masses. As a result, the low-energy spectrum obtained from
a theory with universal scalar masses at the high scale and
additional GUT multiplets can appear to have non-universal soft
masses at the high scale if the extra GUT multiplets are not
included in the RG evolution. These non-universal shifts are
usually a subleading effect, but as we illustrate below they can
be significant when the supersymmetric mass of the new GUT
multiplets is within a few orders of magnitude of the $\tev$
scale.

  Non-universal mass shifts obscure the relationship
between the different MSSM generations and the source of
supersymmetry breaking.  This relationship is closely linked to
the SUSY flavor problem~\cite{superckm}, and possibly also to the
origin of the Yukawa hierarchy.  For example, third generation
soft masses that are significantly different from the first and
second generation values is one of the predictions of the model of
Ref.~\cite{Nelson:2000sn}, in which strongly-coupled conformal
dynamics generates the Yukawa hierarchy and suppresses
flavor-mixing soft terms.  The relative sizes of the high scale
soft masses for different families is therefore of great
theoretical interest.

  Even when there are non-universal shifts from new physics,
it is sometimes still possible to obtain useful information about
the flavor structure of the soft scalar masses at the high scale.
To a very good approximation, the flavor non-universal
contributions to the RG evolution of the soft masses are
proportional to the third generation Yukawa couplings or the
trilinear couplings. There is also good motivation (and it is
technically allowed) to keep only the trilinear couplings for the
third generation. In this approximation, the Yukawa couplings and
the $A$ terms only appear in the one-loop RG equations for the
soft masses through three independent linear combinations.  Of the
seven soft masses whose running depends on these combinations, we
can therefore extract four mass combinations whose evolution is
independent of Yukawa effects at one-loop
order~\cite{iblop}.\footnote{ We also assume implicitly that the
soft masses are close to diagonal in the super CKM basis, as they
are quite constrained to be~\cite{superckm}.} They are: \bea
m^2_{A_3} &=& 2\,m_{L_3}^2-m_{E_3}^2\label{massflav1}\\
m^2_{B_3} &=& 2\,m_{Q_3}^2 - m_{U_3}^2 - m_{D_3}^2\nnmb\\
m^2_{X_3} &=& {2}\,m_{H_u}^2 - 3\,m_{U_3}^2\nnmb\\
m^2_{Y_3} &=& {3}\,m_{D_3}^2 + 2\,m_{L_3}^2 - 2\,m_{H_d}^2\nnmb
\eea The cancellation in the first two terms occurs because the
linear combinations of masses correspond to $L$ and $B$ global
symmetries.  They run only because these would-be symmetries are
anomalous under $SU(2)_L$ and $U(1)_Y$. The other mass
combinations can also be related to anomalous global symmetries of
the MSSM.

  These mass combinations have the same
one-loop RG running as certain combinations of masses involving
only the first and second generations.  For example, the
$m_{B_3}^2$ combination runs in exactly the same way at one-loop
as \beq m_{B_{i}}^2 = 2\,m^2_{Q_{i}} - m^2_{U_{i}}-m^2_{D_{i}},
\label{massflav2} \eeq for $i=1,2$. If these linear combinations
are unequal at the low scale, the corresponding soft masses will
be non-universal at the high scale. This holds in the MSSM, as
well as in the presence of any new physics that is flavor
universal and respects baryon number. On the other hand,
$m^2_{B_3} = m^2_{B_1}$ does not imply family-universal high scale
masses.  For example, within a $SO(10)$ GUT, a splitting between
the soft masses of the $\bf{16}$'s containing the first, second,
and third generations will not lead to a difference between the
mass combinations in Eq.~(\ref{massflav1}) at the low-scale. A
similar conclusion holds for the mass combinations $m^2_{A_i}$.

  In the case of $m^2_{X_3}$ and $m^2_{Y_3}$, it is less obvious
what to compare them to.  The trick here is to notice that in the
absence of Yukawa couplings, $m^2_{H_d}$ runs in the same way as
$m_{L_1}^2$ since they share the same gauge quantum numbers. If
the $S$ term vanishes as well, $m^2_{H_u}$ also has the same RG
evolution as $m^2_{L_1}$.  This motivates us to define \bea
m_{X_i}^2 &=& {2}\,m_{L_i}^2 - 3\,m_{U_i}^2,\label{massflav3}\\
m_{Y_i}^2 &=& {3}\,m_{D_i}^2,\nnmb \eea for $i=1,2$.  These mass
combinations can be compared with $m_{X_3}^2$ and $m_{Y_3}^2$ in
much the same way as for $m_{B_i}^2$ and $m_{A_i}^2$ (although
comparing the $m_{X_i}^2$'s is only useful for $S=0$). They also
correspond to anomalous global symmetries in the limit that the
first and second generation Yukawa couplings vanish.

  The mass combinations listed above can be useful if there
is heavy new physics that hides the relationships between the high
scale masses.  For instance, suppose the high scale soft spectrum
obtained using the MSSM RG equations applied to the  measured soft
scalar masses shows a large splitting between $m_{Q_3}^2$ and
$m_{Q_1}^2$. If the corresponding splitting between $m_{B_3}^2$
and $m_{B_1}^2$ (at any scale) is very much smaller, this feature
suggests that there is new physics that should have been included
in the RG running, or that there exists a special relationship
between $m_{Q_i}^2$, $m_{U_i}^2$, and $m_{D_i}^2$ at the high
scale.  A similar conclusion can be made for the other mass
combinations.

\subsection{Some Numerical Results}

  In our numerical analysis, we follow a similar procedure
to the one used in the previous section.  The MSSM running is
performed at two-loop order, and we interface with
Suspect~2.3.4~\cite{Djouadi:2002ze} to compute the low-scale
threshold corrections. New physics, in the form of vector-like GUT
multiplets at an intermediate scale is included only at the
one-loop level.  Unlike the previous section, we define our
high-energy spectrum using a simple mSUGRA model in the $\Delta
b\neq 0$ theory, and include the new physics effects in generating
the low-energy spectrum.  We then evolve this spectrum back up to
the unification scale using the MSSM RG evolution, with $\Delta b
=0$. Our goal is to emulate evolving the MSSM soft parameters in
the presence of unmeasured and unknown high-scale new physics.

\begin{figure*}[tbh]
\begin{center}
\vspace{1cm}
        \includegraphics[width = 0.65\textwidth]{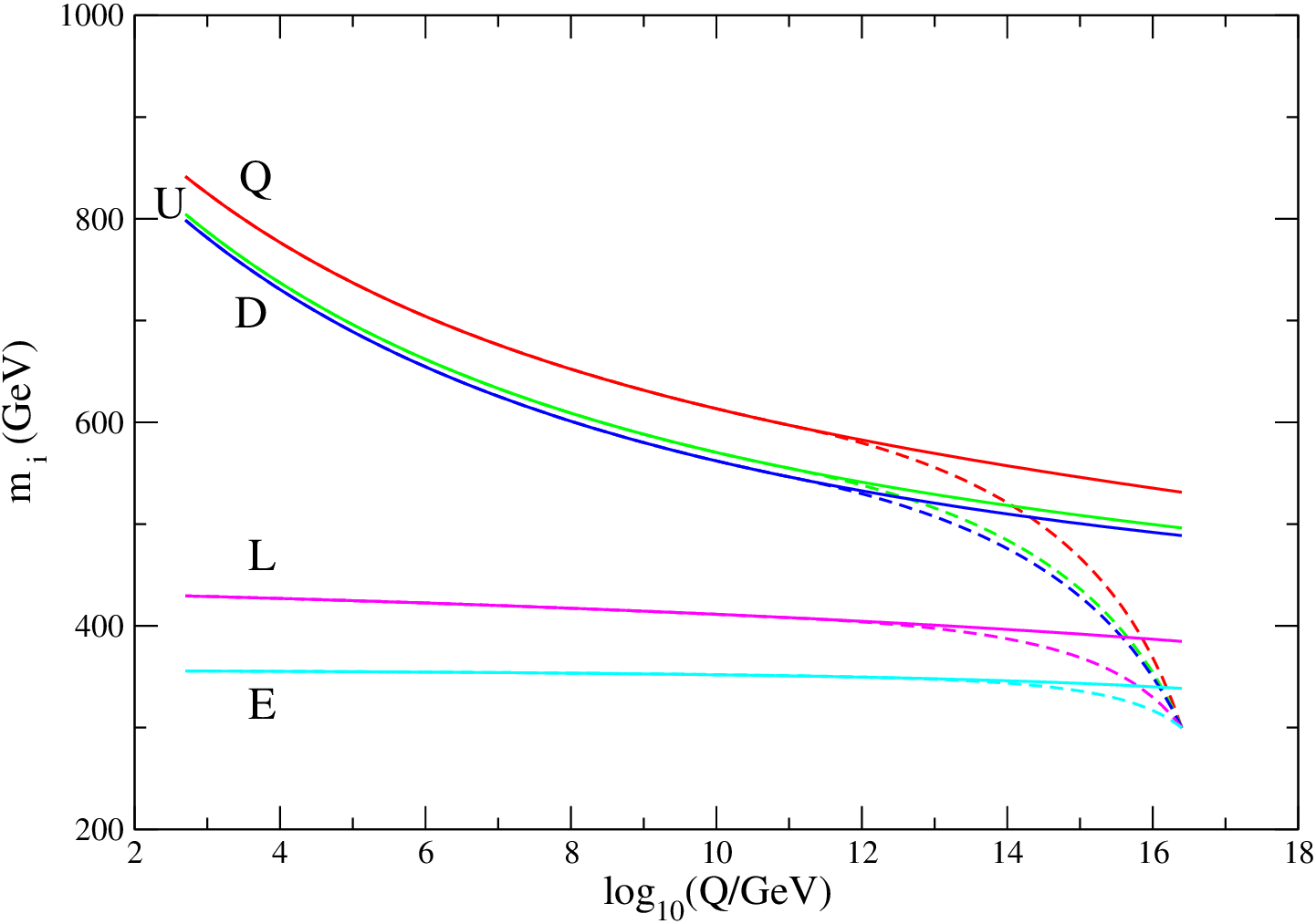}
\caption{ Running of the first generation soft scalar masses with
$\nfive = 7$ and $\tilde{\mu} = 10^{11}\,\gev$ for the mSUGRA
input parameters $m_0 = 300\,\gev$, $m_{1/2} = 700\,\gev$,
$\tan\beta = 10$, and $A_0 = 0$.  The dashed lines show the actual
running of these parameters, while the solid lines show the
running from low to high using the RG equations of the MSSM,
ignoring the additional heavy multiplets.} \label{fig:n5=7}
\end{center}
\end{figure*}

  The running of the soft masses with $\nfive = 7$ sets of $\five$
multiplets with an intermediate scale mass of $\tilde{\mu} =
10^{11}\,\gev$ is shown in Fig.~\ref{fig:n5=7} for the mSUGRA
parameters $m_0 = 300\,\gev$, $m_{1/2} = 700\,\gev$, $\tan\beta =
10$, and $A_0 =0$. These parameters are used to find the low
energy spectrum, which is then RG evolved back up to the high
scale with $\Delta b = 0$. From this figure, we see that a na\"ive
MSSM extrapolation of the soft parameters (\emph{i.e.} with
$\Delta b=0$) yields predictions for the high-scale soft scalar
masses that are significantly larger than the actual values.  In
the same way, the MSSM predicted values of the high-scale gaugino
masses are smaller than the correct values, as can be seen in
Fig.~\ref{fig:gshift}.  Note that for $\tilde{\mu} =
10^{11}\,\gev$, $\nfive = 7$ is about as large as possible while
still keeping the gauge couplings perturbatively small all the way
up to $M_{GUT}$.

\begin{figure*}[tbh]
\begin{center}
\vspace{1cm}
        \includegraphics[width = 0.65\textwidth]{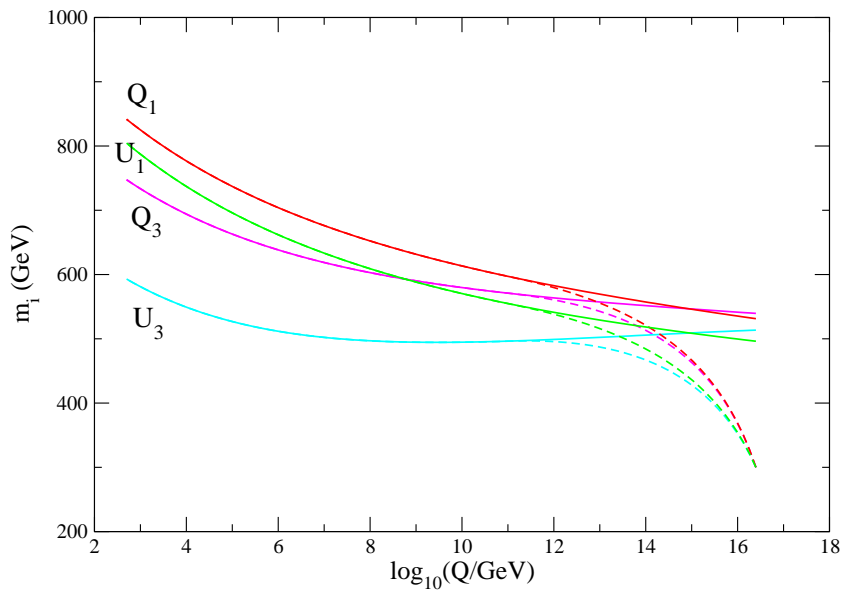}
\caption{ Running of the soft scalar masses of $Q_{1,3}$ and
$U_{1,3}$ with $\nfive = 7$ and $\tilde{\mu} = 10^{11}\,\gev$ for
the mSUGRA input parameters $m_0 = 300\,\gev$, $m_{1/2} =
700\,\gev$, $\tan\beta = 10$, and $A_0 = 0$.  The dashed lines
show the actual running of these parameters, while the solid lines
show the running from low to high using the RG equations of the
MSSM, ignoring the additional heavy multiplets. }
\label{fig:univ1}
%\vspace{1cm}
\end{center}
\end{figure*}

  Besides confusing the relationship between the high scale gaugino
and scalar soft masses, heavy GUT multiplets can also obscure the
comparison of the high scale scalar masses from different
generations. As discussed above, this arises from the backreaction
in the Yukawa-dependent terms in the RG equations for the third
generation soft scalar masses.  Numerically, we find that the
splitting is quite small compared to the absolute scale of the
masses for $\tilde{\mu} \geq 10^{11}\,\gev$.  This is illustrated
in Fig.~\ref{fig:univ1}. We also find that an approximate
preservation of universality persists for other values of $m_0$,
$A_0$, and $\tan\beta$ as well. The reason for this appears to be
that for $\tilde{\mu} \geq 10^{11}~\gev$, the Yukawa couplings are
smaller than the gauge couplings by the time the new physics
becomes relevant.

\begin{figure*}[tbh]
\begin{center}
\vspace{1cm}
        \includegraphics[width = 0.65\textwidth]{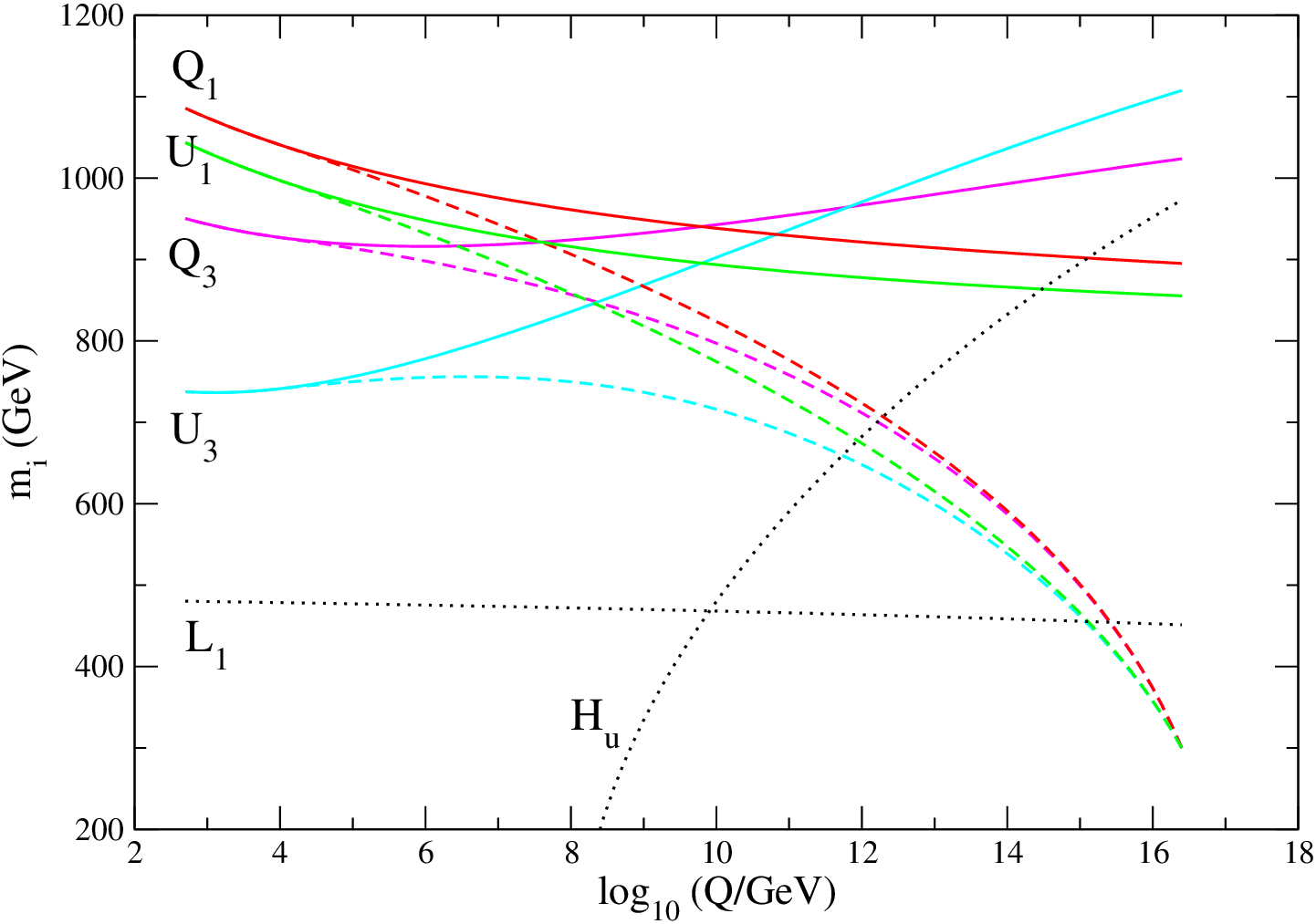}
\caption{ Running of the soft scalar masses of $Q_{1,3}$ and
$U_{1,3}$ with $\nfive = 3$ and $\tilde{\mu} = 10^{4}\,\gev$ for
the mSUGRA input parameters $m_0 = 300\,\gev$, $m_{1/2} =
700\,\gev$, $\tan\beta = 10$, and $A_0 = 0$.  The dashed lines
show the actual running of these parameters, while the solid lines
show the running from low to high using the RG equations of the
MSSM, ignoring the additional heavy multiplets. }
\label{fig:univ2}
%\vspace{1cm}
\end{center}
\end{figure*}

  A much greater splitting between the high scale values of
$m_{Q_1}^2$ and $m_{Q_3}^2$, and $m_{U_1}^2$ and $m_{U_3}^2$, is
obtained for lower values of $\tilde{\mu}$. This effect is shown
in Fig.~\ref{fig:univ2} for $\nfive = 3$ sets of $\five$
multiplets with an intermediate scale mass of $\tilde{\mu} =
10^{4}\,\gev$, and the mSUGRA parameters $m_0 = 300\,\gev$,
$m_{1/2} = 700\,\gev$, $\tan\beta = 10$, and $A_0 =0$. The value
of the gauge couplings at unification here is very similar to the
$\tilde{\mu} = 10^{11}\,\gev$ and $\nfive = 7$ case. As might be
expected, the Yukawa-dependent terms in the soft scalar mass RG
running become important at lower scales where the top Yukawa
approaches unity.

  It is also interesting to note that in both
Figs.~\ref{fig:univ1} and \ref{fig:univ2}, the soft masses appear
to take on family universal values, $m_{Q_1}^2 = m_{Q_3}^2$ and
$m_{U_1}^2 = m_{U_3}^2$, at the same scale, near $10^{15}\,\gev$
in Fig.~\ref{fig:univ1}, and close to $10^{10}\,\gev$ in
Fig.~\ref{fig:univ2}. It is not hard to show, using the mass
combinations in
Eqs.~(\ref{massflav1},\ref{massflav2},\ref{massflav3}), that this
feature holds exactly at one-loop order provided $S=0$, the high
scale masses are family-universal, and the only relevant Yukawa
coupling is that of the top quark. In this approximation, all the
family-dependent mass splittings are proportional to
$(m_{H_u}^2\!-\!m_{L_1}^2)$, and hence vanish when $m_{H_u}^2 =
m_{L_1}^2$.  This relationship can be seen to hold approximately
in Fig.~\ref{fig:univ2}, which also includes two-loop and bottom
Yukawa effects.

  \begin{figure*}[tbh]
\begin{center}
\vspace{1cm}
        \includegraphics[width = 0.65\textwidth]{300-700-10-mb.eps}
\caption{ Running of the soft scalar mass combinations $m_{B_3}$
and $m_{B_1}$ with $\nfive = 3$ and $\tilde{\mu} = 10^{4}\,\gev$
for the mSUGRA input parameters $m_0 = 300\,\gev$, $m_{1/2} =
700\,\gev$, $\tan\beta = 10$, and $A_0 = 0$.  The dashed lines
show the actual running of these parameters, while the solid lines
show the running from low to high using the RG equations of the
MSSM, ignoring the additional heavy multiplets. The small
deviations in these figures arise from higher loop effects. }
\label{fig:mb}
%\vspace{1cm}
\end{center}
\end{figure*}

\begin{figure*}[tbh]
\begin{center}
\vspace{1cm}
        \includegraphics[width = 0.65\textwidth]{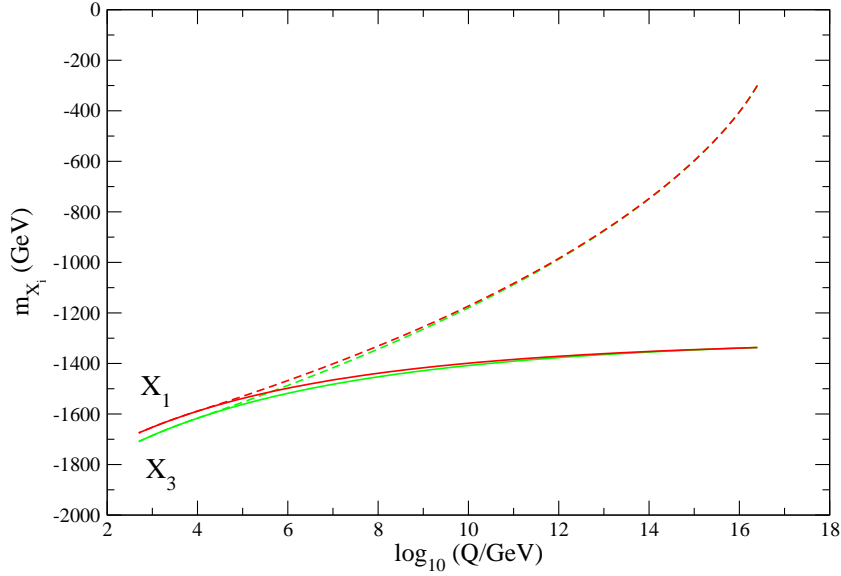}
\caption{ Running of the soft scalar mass combinations $m_{X_3}$
and $m_{X_1}$ with $\nfive = 3$ and $\tilde{\mu} = 10^{4}\,\gev$
for the mSUGRA input parameters $m_0 = 300\,\gev$, $m_{1/2} =
700\,\gev$, $\tan\beta = 10$, and $A_0 = 0$.  The dashed lines
show the actual running of these parameters, while the solid lines
show the running from low to high using the RG equations of the
MSSM, ignoring the additional heavy multiplets. The small
deviations in these figures arise from higher loop effects. }
\label{fig:my}
%\vspace{1cm}
\end{center}
\end{figure*}

  In Figs.~\ref{fig:mb} and \ref{fig:my} we show the running
of the mass combinations $m^{\phantom{2}}_{B_{1,3}}$ and
$m^{\phantom{2}}_{X_{1,3}}$ (where $m_i = m_i^2/\sqrt{|m_i^2|}$)
for $\nfive = 3$ and $\tilde{\mu} = 10^4\,\gev$ with the high
scale mSUGRA input values $m_0 = 300\,\gev$, $M_{1/2} =
700\,\gev$, $\tan\beta = 10$, and $A_0=0$.  These figures also
show the values of $m^{\phantom{2}}_{B_{1,3}}$ and
$m^{\phantom{2}}_{X_{1,3}}$ that would be obtained by running up
without including the effects of the heavy new physics. Comparing
these figures to Figs.~\ref{fig:univ1} and \ref{fig:univ2}, it is
apparent that the splittings between $m_{B_1}^2$ and $m_{B_3}^2$,
and $m_{X_1}^2$ and $m_{X_3}^2$, are very much less than the high
scale splittings between the $Q$ and $U$ soft masses.

  These relationship between the $B$ and $X$ soft mass combinations
from different families is a footprint left by the full theory
(including the heavy GUT multiplets) on the low-energy spectrum.
Since the scalar masses in the full theory are universal at the
high scale, the low scale splittings between the $B$ and $X$ soft
mass combinations are very small.  On the other hand, running the
low scale $Q$ and $U$ scalar soft masses up within the MSSM does
not suggest any form of family universality among these masses.
Therefore, as we proposed above, the low-energy values of these
particular combinations of soft scalar masses can provide evidence
for heavy new physics.\footnote{ We have checked that the small
splittings between $m^{\phantom{2}}_{B_1}$ and
$m^{\phantom{2}}_{B_3}$, as well as between
$m^{\phantom{2}}_{X_1}$ and $m^{\phantom{2}}_{X_3}$, arise from
higher loop effects.}

  The analysis in this section shows that even though
new physics in the form of additional heavy GUT multiplets can
significantly disrupt the predictions for the high scale soft
spectrum obtained by running in the MSSM, certain key properties
about the input spectrum can still be deduced using collider scale
measurements.  Most significantly, the low-scale values of the
gaugino masses and gauge couplings can be used to predict the
approximate ratios of the high-scale values, provided gauge
unification is preserved.

  The effect of
additional GUT multiplets on the scalar soft masses is more
severe. Extrapolating the soft masses without including the
contributions from the heavy GUT multiplets leads to a prediction
for the input soft masses that are generally too low.  The
splittings between the soft masses from different generations can
be shifted as well. Despite this, some of the flavor properties of
the input soft mass spectrum can be deduced by comparing the
evolution of the mass combinations in Eqs.~(\ref{massflav1}),
(\ref{massflav2}), and (\ref{massflav3}).  For example, $m_{B_3}^2
= m_{B_1}^2$ and $m_{L_1}^2 = m_{L_3}^2$ suggests some form of
flavor universality (or an embedding in $SO(10)$), even if the
scalar masses extrapolated within the MSSM do not converge at
$M_{GUT}$. We expect that these special mass combinations could
prove useful for studying other types of heavy new physics as
well.

%%%%%%%%%%%%%%%%%%%%%%%%%%%%%%%%%%%%%%%%%%%%%%%%%%%%%%%

\section{The (S)Neutrino Connection\label{neut}}

  We have seen in the previous sections that taking the unification of
the gauge couplings as a serious theoretical input still leaves
considerable room for experimental uncertainties and new physics
to modify the extrapolated values of the soft supersymmetry
breaking parameters at very high energies. For example, a
Fayet-Iliopoulos $D$-term for hypercharge or additional complete
GUT multiplets with intermediate scale masses will not disrupt
gauge coupling unification, but will in general change the running
of the parameters of the model.  In this section we wish to study
the effect of additional intermediate scale singlet matter with
significantly large Yukawa couplings to the MSSM matter fields. A
particularly well-motivated example of this, and the one we
consider, are heavy singlet neutrino multiplets.

  The observed neutrino phenomenology can be accommodated by extending
the matter content of the MSSM to include at least two right
handed~(RH) neutrino supermultiplets that are singlets under the
SM gauge group~\cite{Mohapatra:2005wg}. Throughout the present
work, we will assume there are three RH neutrino flavors. The
superpotential in the lepton sector is then given by \bea {\cal
W}_l={\bf y_e}\ LH_dE+{\bf y_\nu}\ LH_uN_R - {1\over2}{\bf M_R}
N_R N_R, \eea where $L,\,E$, and $N_R$ are respectively the
$SU(2)_L$ doublet, $SU(2)_L$ singlet, and neutrino chiral
supermultiplets, each coming in three families.  The quantities
${\bf y_{e}}$, ${\bf y_{\nu}}$, and $\bf{M_R}$ are $3\times 3$
matrices in lepton family space. The $H_d$ and $H_u$ fields
represent the usual Higgs multiplets. The gauge-invariant
interactions among leptons and Higgs superfields are controlled by
the family-space Yukawa matrices ${\bf y_e}$ and ${\bf y_\nu}$. As
is conventional, we shall implicitly work in a basis where ${\bf
y_e}$ is diagonal.  Since the $N_R$ are singlets, we can also add
to the superpotential a Majorana mass ${\bf M_R}$ for these
fields.

  Assuming the eigenvalues of ${\bf M_R}$ lie at a large
intermediate mass scale, $10^9-10^{14}$ GeV, we can integrate out
the RH neutrino superfields and obtain a term in the effective
superpotential that leads to small neutrino masses through the
see-saw mechanism, \beq {\cal W }_{m_\nu}=-{1\over2}{\bf y_\nu^T
M_R^{-1}y_\nu}\ LH_u LH_u. \eeq After electroweak symmetry
breaking, the neutrino mass matrix becomes \bea {\bf(m_\nu)}={\bf
y_\nu^T M_R^{-1}y_\nu}\, v_u^2, \eea where $v_u =
\left<H_u\right>$.  The mass matrix can be conveniently
diagonalized by the transformation \bea {\bf(m_\nu^{diag})=U^T
(m_\nu) U }, \eea with $\bf{U}$ a unitary matrix.  This matrix
$\bf{U}$ is the usual PMNS matrix that describes lepton mixing
relative to the flavor basis where the charged lepton Yukawa
matrix ${\bf y_e}$ is diagonal.

  One can also write the neutrino Yukawa matrix as~\cite{Casas:2001sr}
\beq {\bf y_\nu} = {1\over v_u} \bf \sqrt{M_R^{diag}}\ R\
\sqrt{m_\nu^{diag}}\ U^\dagger \label{ynucasas} \eeq where
$v_u=\langle H_u\rangle$ and $R$ is a complex orthogonal matrix
that parametrizes our ignorance of the neutrino Yukawas. As an
estimate, Eq.~(\ref{ynucasas}) shows that the size of the neutrino
Yukawa couplings will be on the order of \beq y_\nu \simeq
\frac{0.57}{\sin\beta} \left({ M_R\over 10^{14}
\gev}\right)^{1\over 2} \left({m_\nu\over 0.1 \eev}\right)^{1\over
2}. \eeq Thus, the neutrino Yukawa couplings can take large ${\cal
O}$(1) values comparable to the top Yukawa coupling for $M_R \sim
10^{14}\,\gev$. If this is the case, then above the see-saw mass
threshold the effects on the RG running of the MSSM soft
parameters due to the neutrino Yukawas can be substantial.

  The addition of RH neutrinos to the MSSM can lead to
lepton flavor violation~(LFV) through the RG running of the
off-diagonal slepton mass
terms~\cite{Borzumati:1986qx,Casas:1999tg}. In this work we will
only consider simple scenarios of neutrino phenomenology in which
the amount of lepton flavor violation induced by the heavy
neutrino sector is small. However, the observation of LFV signals
could potentially provide information about a heavy neutrino
sector~\cite{Borzumati:1986qx,Casas:1999tg,
Deppisch:2002vz,Petcov:2005yh}. Precision measurements of the
slepton mass matrices can also be used to constrain possible heavy
neutrino sectors~\cite{Baer:2000hx,Davidson:2001zk}. Heavy singlet
neutrinos may also be related to the source of the baryon
asymmetry through the mechanism of
leptogenesis~\cite{Buchmuller:2005eh}.

\subsection{Running Up}

  The strategy we use in this section is similar to the one followed
in the previous sections.  We assume a universal high scale mass
spectrum at $M_{GUT}$, and RG evolve the model parameters down to
the low scale $M_{low} = 500\,\gev$ including the additional
effects of the neutrino sector parameters.  The resulting low
scale spectrum is then run back up to $M_{GUT}$ using the RG
equations for the MSSM without including the neutrino sector
contributions. As before, we use this procedure to illustrate the
discrepancy between the extrapolated parameter values and their
true values if the new physics effects are not included in the
running.

  To simplify the analysis, we make a few assumptions about
the parameters in the neutrino sector.  We choose the orthogonal
matrix $R$ to be purely real, and we take the heavy neutrino mass
matrix ${\bf M_R}$ to be proportional to the unit matrix, ${\bf
M_R} = M_R\,\mathbf{I}$.  We also take the physical neutrino
masses to be degenerate.  This allows us to write \beq {\bf
y_{\nu}} = \frac{0.57}{\sin\beta} \left({ M_R\over
10^{14}\,\gev}\right)^{1\over 2} \left({m_\nu\over
0.1\,\eev}\right)^{1\over 2} {\bf R\,U}^{\dagger} := y_{\nu}\,{\bf
R\,U}^{\dagger}, \eeq so that the neutrino Yukawa couplings have
the form of a universal constant multiplying a unitary matrix.  To
fix the value of $y_{\nu}$, we will set $m_{\nu} =
0.1\,\eev$.\footnote{The diagonal neutrino mass matrix $\bf
m_\nu^{diag}$ and the lepton mixing matrix $\bf U$ are measured at
low scales, and one should really evaluate them at the
intermediate scale $M_R$ by running the Yukawa couplings
up~\cite{Petcov:2005yh}.  Since we are most interested in the
effect of the neutrino Yukawas after reaching the intermediate
scale, we will neglect this additional running below $M_R$.}. This
choice is close to being as large as possible while remaining
consistent with the cosmological bounds on the sum of the neutrino
masses, $\sum_{\nu} < 0.68\,\eev$~\cite{Spergel:2006hy}. Note that
larger values of the neutrino masses tend to maximize the size of
the resulting neutrino Yukawa couplings.

  We also need to impose boundary conditions on the RH sneutrino
masses and trilinear couplings.  We take these new soft parameters
to have universal and diagonal boundary conditions at the
unification scale, \beq m_{{N}_{ij}}^2 =
m_0^2\,\delta_{ij},~~~~~{\rm and }~~~~~ a_{{\nu}_{ij}} =
A_{0}\,y_{{\nu}_{ij}}, \label{nusoft} \eeq where $m_0$ and $A_0$
are the same universal soft scalar mass and trilinear coupling
that we will apply to the MSSM fields in the analysis to follow.
With these choices for the neutrino parameters, the effects of the
neutrino sector on the (one-loop) RG running of the MSSM soft
terms take an especially simple form. In particular, the amount of
leptonic flavor mixing induced is expected to be very small, and
the diagonal and universal form of the neutrino sector soft terms,
Eq.~(\ref{nusoft}), is approximately maintained at lower scales.

  To illustrate the effects of the neutrino sector, we set the
high scale spectrum to coincide with the SPS-5 benchmark point,
and we extend the corresponding soft terms to the neutrino sector.
The input values at $M_{GUT}$ for this point are $m_0 =
150\,\gev$, $m_{1/2} = 300\,\gev$, $A_0 = -1000\,\gev$, and
$\tan\beta=5$ with $\mu$ positive.  These input values tend to
magnify the effects of the neutrino sector because the large value
of $A_0$ feeds into the running of the soft masses. Large neutrino
Yukawa couplings alter the running of the top Yukawa coupling as
well.

\begin{figure*}[tbh]
\begin{center}
\vspace{1cm}
        \includegraphics[width = 0.65\textwidth]{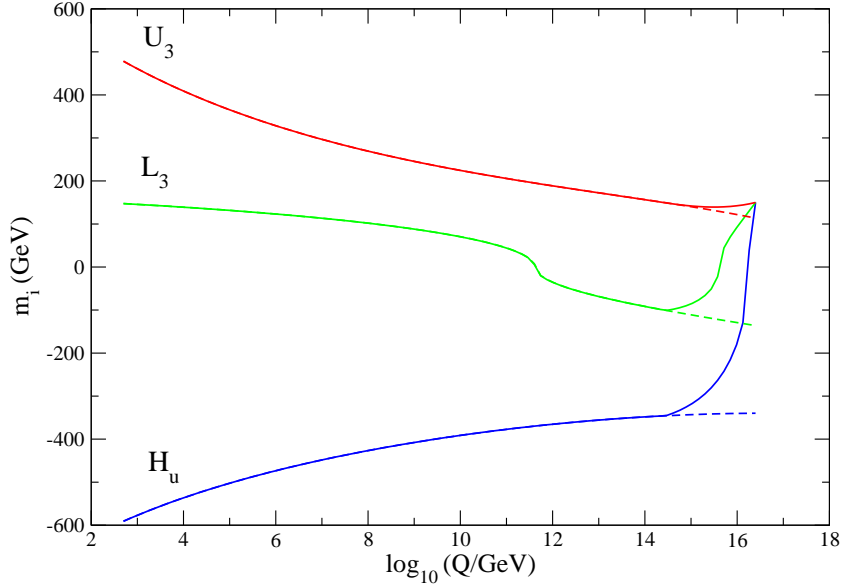}
\caption{ Running of the soft scalar masses of $H_u$, $L_3$, and
$U_3$ for SPS-5 input parameters at $M_{GUT}$ with three
additional heavy RH neutrinos of mass $M_R = 10^{14}\,\gev$. The
solid lines show the full running, including the neutrino sector
effects, while the dashed lines show the low-to-high running of
the soft masses in the MSSM, with the neutrino sector effects
omitted. } \label{fig:yn-sps5-1}
%\vspace{1cm}
\end{center}
\end{figure*}

  Of the MSSM soft parameters, the greatest effects of the heavy
neutrino sector are seen in the soft scalar masses and the
trilinear $A$ terms.  The gaugino masses are only slightly
modified. The evolution of the soft masses for the $H_u$, $L_3$,
and $U_3$ fields from low to high are shown in
Fig.~\ref{fig:yn-sps5-1}, both with and without including the
effects of the neutrino sector for $M_R = 3\times 10^{14}\,\gev$.
For this value of $M_R$ and with $\tan\beta = 5$, the Yukawa
coupling is close to being as perturbatively large as possible.
The extrapolated values of $m_{H_u}^2$ and $m_{L_3}^3$ deviate
significantly from the actual input values if the effects of the
neutrino sector are not taken into account in the RG evolution.
These fields are particularly affected because they couple
directly to the heavy neutrino states through the neutrino Yukawa
coupling. The shift in the running of $m_{U_3}^2$ arises
indirectly from the effect of the neutrino Yukawas on $m_{H_u}^2$
and the top Yukawa coupling $y_t$.

\begin{figure*}[tbh]
\begin{center}
\vspace{1cm}
        \includegraphics[width = 0.65\textwidth]{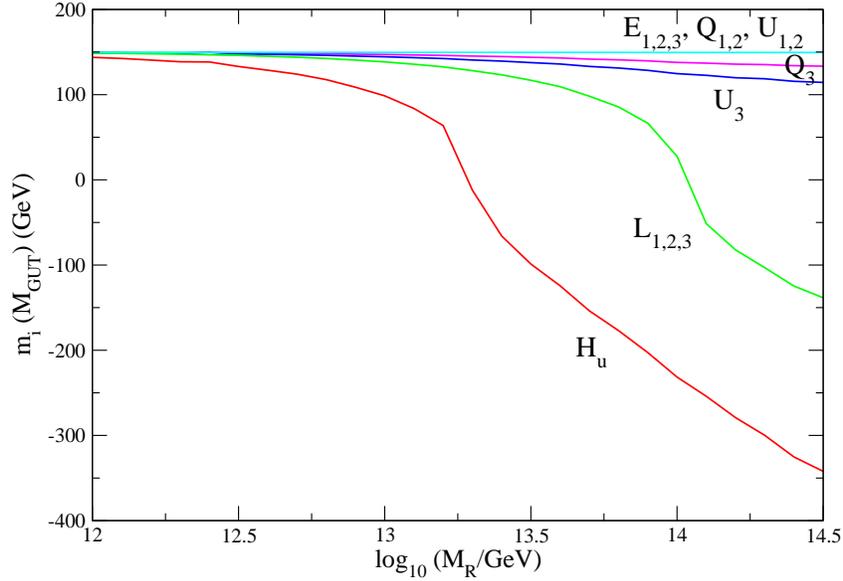}
\caption{ The high scale ($M_{GUT}$) values of the soft scalar
masses extrapolated using the MSSM RG equations, without including
neutrino sector effects.  The low-scale ($500\,\gev$) values of
the soft masses used in the extrapolation were obtained from SPS-5
input parameters at $M_{GUT}$ with three additional heavy RH
neutrinos of mass $M_R$.  The deviations from $m_i(M_{GUT}) =
150\,\gev$ represent the discrepancy between the MSSM extrapolated
values and the correct value in the full theory with heavy RH
neutrinos. These discrepancies are shown as a function of the
heavy neutrino scale $M_R$.} \label{fig:yn-sps5-2}
%\vspace{1cm}
\end{center}
\end{figure*}

  In Fig.~\ref{fig:yn-sps5-2} we show the size of the discrepancies
in the extrapolated high scale values of a few of the soft scalar
masses if the neutrino sector effects are not included in the
running. These discrepancies are plotted as a function of the
heavy neutrino mass scale $M_R$.  As above, the high scale input
spectrum consists of the SPS-5 values. Again, the soft masses
$m_{H_u}^2$ and $m_{L_i}^2$ are altered the most, although the
third generation squark soft masses also get shifted somewhat as a
backreaction to the changes in $m_{H_u}^2$ and the top Yukawa
coupling.  This plot also shows that the sizes of the
discrepancies remain quite small for $M_R$ less than
$10^{13}$~GeV. Values of $M_R$ considerably less than this are
favored if leptogenesis is to be the source of the baryon
asymmetry~\cite{Buchmuller:2005eh}. Neutrino masses well below
$0.1\,\eev$ would also lead to less pronounced deviations in the
extrapolated soft masses.

  The deviations induced by not including the new neutrino sector
physics in the running take a similar form to those obtained by
not taking account of heavy GUT multiplets.  For both cases, the
gaugino mass running is only modified in a very controlled way,
while the soft scalar masses and trilinear couplings deviate more
unpredictably.  In particular, the high scale flavor structure of
the soft masses can be obscured. With large neutrino Yukawa
couplings, the third generation squark masses receive additional
contributions to their running relative to the first and second
generations due to the potentially large effect of the neutrino
Yukawa couplings on $m_{H_u}^2$. The sizes of these additional
family-dependent shifts tend to be fairly small, as can be seen in
Fig.~\ref{fig:yn-sps5-2}.  Furthermore, these effects cancel out
in the mass combinations $m_{B_i}^2$ defined in
Eqs.~(\ref{massflav1}) and (\ref{massflav2}), and we find
$m_{B_1}^2 \simeq m_{B_3}^2$ at all scales, regardless of whether
of not the neutrino effects are included. On the other hand, the
running of $m_{A_3}^2$ relative to $m_{A_1}^2$, $m_{X_3}^2$
relative to $m_{X_1}^2$, and $m_{Y_3}^2$ relative to $m_{Y_1}^2$
need no longer coincide if there is a heavy neutrino sector.

  Finally, let us also mention that for more general neutrino sector
parameters than those we have considered, there can arise
significant lepton flavor mixing couplings in the MSSM slepton
soft terms from the RG running~\cite{Borzumati:1986qx}.
Measurements of this mixing in lepton flavor violating processes
can therefore provide an experimental probe of the heavy neutrino
multiplets~\cite{Borzumati:1986qx,Casas:1999tg,Deppisch:2002vz,Petcov:2005yh}.
A measured splitting among the three slepton masses $m^2_{L_i}$
would also constitute another indication of the existence of a
neutrino sector with sizeable Yukawa couplings and nontrivial
flavor structure. Both high and intermediate energy data may be
complementary and very useful in extrapolating the MSSM soft terms
to high energies.

%%%%%%%%%%%%%%%%%%%%%%%%%%%%%%%%%%%%%%%%%%%%%%%%%%%%%%%

\section{Putting it All Together: an Example\label{all}}

  In this section we summarize some of our previous results
with an explicit example.  We begin with a low energy spectrum of
MSSM soft supersymmetry breaking parameters that we assume to have
been measured at the LHC to an arbitrarily high precision. We then
attempt to deduce the essential features of the underlying high
scale structure by running the low energy parameters up and
applying some of the techniques discussed in the previous
sections.

  In our example we will make the following assumptions:
\begin{itemize}
\item The possible types of new physics beyond the MSSM are:
\begin{itemize}
\item Complete $\five$ GUT multiplets with a common (SUSY) mass
scale $\tilde{\mu}$. \item Three families of heavy singlet (RH)
neutrinos at the mass scale $M_R$. \item A fundamental hypercharge
$D$ term.
\end{itemize}
In the case of complete GUT multiplets, we will assume further
that there are no superpotential interactions with the MSSM states
as in Section~\ref{gutmult}. For heavy RH neutrinos, we will make
the same set of assumptions about the form of the mixing and mass
matrices as in Section~\ref{neut}.
\item The high scale spectrum has the form of a minimal SUGRA
model (up to the scalar mass shifts due to a hypercharge $D$-term)
at the high scale $M_{GUT} \simeq 2.5\times 10^{16}\,\gev$.
\item This mSUGRA spectrum also applies to the soft parameters
corresponding to any new physics sectors.  For example, a
trilinear $A$ term in the RH neutrino sector has the form ${\bf
a}_{\nu} = A_{0}\,{\bf y_{\nu}}$ at $M_{GUT}$, where $A_0$ is the
universal trilinear parameter.
\end{itemize}

  These assumptions are not entirely realistic, but they make the
analysis tractable.  Moreover, even though this exercise is highly
simplified compared to what will be necessary should the LHC
discover supersymmetry, we feel that it illustrates a number of
useful techniques that could be applied in more general
situations. With this set of assumptions, the underlying free
parameters of the theory are: \beq
\begin{array}{cclcccc}
m_0,&m_{1/2},&A_0,~& \xi,~&\nfive,&\tilde{\mu},~&M_R.
\end{array}
\eeq where $m_0$, $m_{1/2}$ and $A_0$ are common mSUGRA inputs at
$M_{GUT}$, $\xi$ is the fundamental hypercharge $D$ term, $\nfive$
is the number of additional $\five$ multiplets in the theory with
a supersymmetric mass $\tilde{\mu}$, and $M_R$ is the heavy
neutrino scale.

\subsection{Step 1: Running Up in the MSSM}

  As a first step, we run the low energy spectrum
up to the high scale $M_{GUT}$ using the RG equations for the
MSSM, without including any potential new physics effects. The low
energy MSSM soft spectrum we consider, defined at the low scale
$M_{low} = 500\,\gev$, is given in Table~\ref{lowscalespectrum}.
In addition to these soft terms, we also assume that $\tan\beta =
7$ has been determined, and that the first and second generation
soft scalar masses are equal. With this set of soft terms, we have
verified that the low energy superpartner mass spectrum is
phenomenologically acceptable using
SuSpect~2.3.4~\cite{Djouadi:2002ze}. The lightest Higgs boson mass
is $114\,\gev$ for a top quark mass of $m_t = 171.4\,\gev$.

  Even before extrapolating the soft parameters, it is possible
to see a number of interesting features in the spectrum. The most
obvious is that the low scale gaugino masses have ratios close to
$M_1:M_2:M_3 \simeq 1:2:6$. This suggests that the high scale
gaugino masses have a universal value $m_{1/2}$, and provides
further evidence for gauge unification. The low-energy value of
the $S$ term, as defined in Eq.~(\ref{1:sterm}), is also non-zero
and is in fact quite large, $S(M_{low}) \simeq (620\,\gev)^2$.
This indicates that there are significant contributions to the
effective hypercharge $D$ term in the high scale theory.  Since
the $S$ term is non-zero, it is also not surprising that $S_{B-L}
\simeq (446\,\gev)^2$, as defined in Eq.~(\ref{1:sbl}), is
non-zero as well.

\begin{figure}[hbtp]
  \centerline{\hbox{ \hspace{0.0in}
    \epsfxsize=4.0in
    \epsfbox{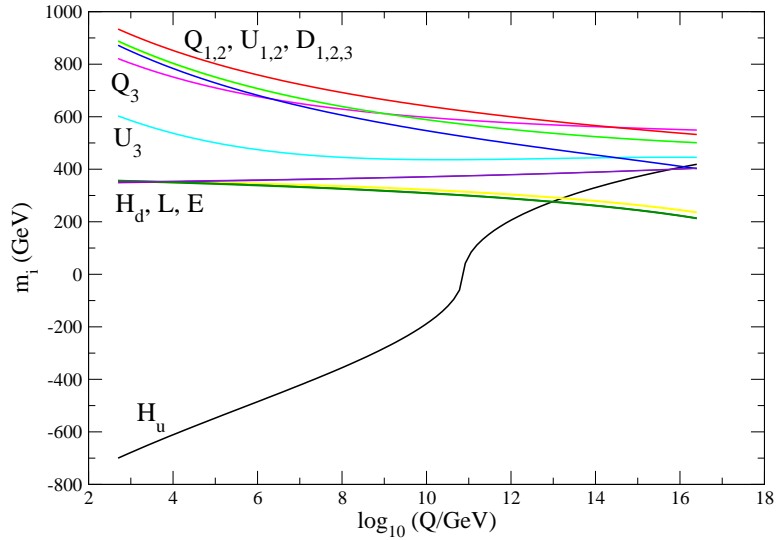}
    }
  }
\caption{ Scale dependence of the soft scalar masses for the input
soft parameters given in Table~\ref{lowscalespectrum}. No new
physics effects beyond the MSSM were included in the running. }
\label{fig:explot1}
%\vspace{1cm}
\end{figure}

The values of the soft parameters extrapolated to $M_{GUT}$ within
the MSSM are listed in Table~\ref{lowscalespectrum}. The MSSM
running of the soft scalar masses is also shown in
Fig.~\ref{fig:explot1}.  As anticipated, the gaugino masses unify
approximately to a value $M_1\simeq M_2\simeq M_3\simeq m_{1/2} =
350\,\gev$ at $M_{GUT}$.  The high scale pattern of the soft
scalar masses (and the trilinear $A$ terms) shows less structure,
and is clearly inconsistent with mSUGRA high scale input values.

Since $S(M_{low})$ is large and non-zero, we are motivated to look
for a hypercharge $D$ term contribution to the soft scalar masses.
Such a contribution would cancel in the mass combinations \beq
\Delta
m^2_{ij}=(Y_j\,m_i^2-Y_{i}\,m_j^2)/(Y_j-Y_i).\label{hypermasscomb}
\eeq If the high scale soft masses have the form $m_i^2 = m_0^2 +
\sqrt{\frac{3}{5}}g_1\,Y_i\,\xi$, as in mSUGRA with a hypercharge
$D$ term, these combinations will all be equal to $m_0^2$ at this
scale. The high scale soft masses here, extrapolated within the
MSSM, exhibit no such relationship.  Even so, these mass
combinations will prove useful in the analysis to follow.

\begin{table}[h!]
\centering
\begin{tabular}{|c|c||c|}\hline
Soft Parameter &Low Scale Value&High Scale Value\\
&(GeV)&(GeV)\\
\hline
$M_1$ & 146&356\\
$M_2$ & 274&355\\
$M_3$ & 859&370\\ \hline\hline
$A_t$ & -956&-766\\
$A_b$ & -1755&-818 \\
$A_{\tau}$ & -737&-524\\ \hline\hline
$m_{H_u}$ & -700&419\\
$m_{H_d}$ & 350&236\\  \hline
$m_{Q_3}$ & 821& 549\\
$m_{U_3}$ & 603& 445\\
$m_{D_3}$ & 884& 501\\
$m_{L_3}$ & 356& 213\\
$m_{E_3}$ & 349& 404\\ \hline
$m_{Q_1}$ & 934& 532\\
$m_{U_1}$ & 872& 402\\
$m_{D_1}$ & 888& 501\\
$m_{L_1}$ & 357& 213\\
$m_{E_1}$ & 352& 404\\ \hline
\end{tabular}
\caption{ The low-energy scale ($M_{low} = 500\,\gev$) soft
supersymmetry breaking spectrum used in our analysis. The soft
scalar masses listed in the table correspond to the signed square
roots of the actual masses squared. In this table we also the high
scale values of these soft parameters obtained by running them up
to $M_{GUT} \simeq 2.5\times 10^{16}\,\gev$ using the RG evolution
appropriate for the MSSM. } \label{lowscalespectrum}
\end{table}
It is also interesting to compare the pairs of mass combinations
$m_{A_i}^2,\,m_{B_i}^2,\,m_{X_i}^2$, and $m_{Y_i}^2$ for
$i=1,\,3$, as defined in Section~\ref{gutmult}. Of these, the most
useful pair is $m_{B_1}^2$ and $m_{B_3}^2$. At the low and high
scales (extrapolating in the MSSM) this pair obtains the values
\beq
\begin{array}{cccccc}
m_{B_1}^2(M_{low}) &\simeq& (441\,\gev)^2,~~~&
m_{B_3}^2(M_{low}) &\simeq& (452\,\gev)^2,\\
m_{B_1}^2(M_{GUT}) &\simeq& (392\,\gev)^2,~~~& m_{B_3}^2(M_{GUT})
&\simeq& (392\,\gev)^2.
\end{array}
\eeq
The near equality of $m_{B_1}^2$ and $m_{B_3}^2$ at the high scale
is particularly striking.  By comparison, there is a significant
inter-family splitting that occurs between the high scale soft
masses of $U_1$ and $U_3$, \beq m_{U_1}^2(M_{GUT}) =
(402\,\gev)^2,~~~~~m_{U_3}^2(M_{GUT}) = (445\,\gev)^2. \eeq This
apparent fine-tuning among the soft masses that make up the
$B$-type combinations is suggestive of an underlying structure in
the theory.  However, based on the values of the individual soft
masses extrapolated within the MSSM, such a structure is not
obvious.  Instead, we can interpret this as a hint for new
intermediate scale physics.

  Without our guiding assumptions, the high scale spectrum listed
in Table~\ref{lowscalespectrum} obtained by running up in the MSSM
does not exhibit any particularly remarkable features aside from
the universality of the gaugino masses. Even so, the curious
relationship between the $m_{B_1}^2$ and $m_{B_3}^2$ mass
combinations provides a strong hint that we are missing something.
It is not clear how strong this hint would have been had we also
included reasonable uncertainties in the low scale parameter
values.

\subsection{Step 2: Adding GUT Multiplets}

  As a first attempt to fit the low energy soft spectrum to the
class of models outlined above, let us consider adding additional
vector-like GUT multiplets the the theory at the scale
$\tilde{\mu}$. We try this first because, as we found in
Sections~\ref{gutmult} and \ref{neut}, the contributions from such
multiplets are potentially much larger than those due to heavy
singlet neutrinos.

  In adding the new GUT multiplets, we will make use of our starting
assumptions about the possible forms of new physics.  Given the
large value of $S(M_{low})$, there appears to be significant
hypercharge $D$ term.  Also from our assumptions, this $D$ term
will contribute to the soft scalar masses of the heavy GUT
multiplets, which will in term feed into the running of the MSSM
scalar masses through the $S$ term above the scale $\tilde{\mu}$.
For this reason, it is safer to work with the mass differences
defined in eq.~(\ref{hypermasscomb}) whose running (to one-loop)
does not depend on the S-term.

  Among the low scale soft masses listed in  Table~\ref{lowscalespectrum},
we expect the slepton soft mass $m^2_{E_1}$ to be among the
easiest to measure, and the least susceptible to new physics
effects. Thus, we will use it as a reference mass in all but two
of the differences we choose. The mass differences we consider are
\beq
\begin{array}{cccc}
\Delta m^2_{Q_1E_1},&\Delta m^2_{U_1E_1},&
\Delta m^2_{D_1E_1},&\Delta m^2_{L_1E_1}\\
\Delta m^2_{H_dE_1},&\Delta m^2_{H_uE_1},&
\Delta m^2_{Q_1D_1},&\Delta m^2_{Q_3D_1},\\
\Delta m^2_{Q_3E_1},&\Delta m^2_{U_3E_1},& \Delta
m^2_{D_3E_1},&\Delta m^2_{L_3E_1}.
\end{array}
\label{hypermasscombbasis} \eeq These depend on several
independent mass measurements.

\begin{figure*}[tbh]
\begin{center}
\vspace{1.2cm}
        \includegraphics[width = 0.85\textwidth]{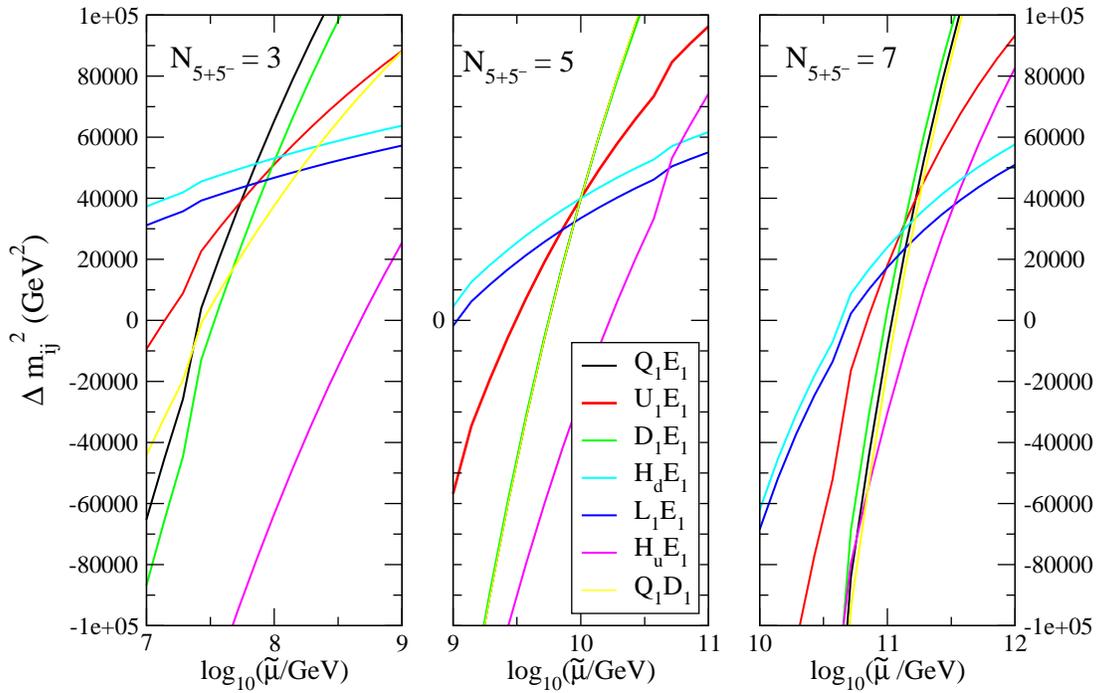}
\caption{ High scale values for the soft scalar mass differences
$\Delta{m}_{ij}^2$ defined in Eq.~(\ref{hypermasscombbasis}) as a
function of the mass scale $\tilde{\mu}$ of the $\nfive$ heavy GUT
multiplets for $\nfive = 3,\,5,\,7$.
%Among these mass differences, only those for $H_uE_1$ and $L_iE_1$
%are expected to be sensitive to a heavy singlet neutrino sector.
} \label{fig:n5scan}
%\vspace{1cm}
\end{center}
\end{figure*}

  In Fig.~\ref{fig:n5scan} we show the high scale values
of these mass differences obtained by running up the low scale
soft masses while including a given number of $\nfive$ additional
$\five$ GUT multiplets at the scale $\tilde{\mu}$.  For each of
the plots, nearly all the mass differences unify approximately, as
they would be expected to do if the underlying theory has a mSUGRA
spectrum. The best agreement with a mSUGRA model is obtained for
$\nfive = 5$ with $\tilde{\mu} = 10^{10}\,\gev$. (More precisely,
the agreement is obtained when a shift $\Delta b=-5$  is applied
to the gauge beta function coefficients $b_i$ at the scale
$\tilde{\mu} = 10^{10}\,\gev$).

  Taking $\nfive = 5$ and $\tilde{\mu} = 10^{10}\,\gev$,
the high scale values of the mass differences are \bea
&&\Delta{m}_{Q_1E_1}^2 ~~~=~~ (200\,\gev)^2 ~=~
\Delta{m}_{U_1E_1}^2 ~=~ \Delta{m}_{D_1E_1}^2
~=~\Delta{m}_{H_dE_1}^2\\
&&\begin{array}{cclccl}
\Delta{m}_{Q_3E_1}^2 &=& (197\,\gev)^2,&
\Delta{m}_{D_3E_1}^2 &=&(200\,\gev)^2\\
\Delta{m}_{H_uE_1}^2 &=& -(157\,\gev)^2,&
\Delta{m}_{L_{1,3}E_1}^2 &=& (183\,\gev)^2\\
%\Delta{m}_{U_3E_1}^2 = (197\,\gev)^2.\nnmb
\end{array}\nnmb
\eea Most of these high scale values coincide, suggesting a mSUGRA
value for the universal soft scalar mass of about $m_0 =
200\,\gev$.  On the other hand, the soft mass differences
involving the $H_u$ and $L$ fields show a significant deviation
from this near-universal value. Based on the results of
Section~\ref{neut}, these are precisely the scalar masses that are
most sensitive to a heavy singlet neutrino sector.

  Using this same choice of new physics parameters,
we can also estimate the values of the other mSUGRA parameters.
Heavy singlet neutrinos are not expected to significantly alter
the running of the gaugino soft mass parameters. If we run these
up to $M_{GUT}$ including $\nfive=5$ additional GUT multiplets at
$\tilde{\mu}=10^{10}\,\gev$, we find $m_{1/2} \simeq 700\,\gev$.
Doing the same for the trilinear couplings, we do not find a
unified high scale value for them. Instead, we obtain
$A_t(M_{GUT}) = -401\,\gev$, $A_{\tau} = -407\,\gev$, and $A_{b} =
-500\,\gev$.  This is not surprising since a heavy RH neutrino
sector would be expected to primarily modify $A_t$ and $A_{\tau}$,
while having very little effect on $A_b$. Thus, we also expect
$A_0 \simeq -500\,\gev$.

  It is possible to estimate the value of the hypercharge $D$ term as well.
Using the hypothesis $m_i^2 = m_0^2 +
\sqrt{\frac{3}{5}}g_1\,Y_i\xi$ at the high scale, we find \beq
\xi(M_{GUT}) =
\sqrt{\frac{5}{3}}\frac{1}{g_1}\,{\left(m_{E_1}^2-m_{H_d}^2\right)}/
{(Y_E-Y_{H_d})}\simeq (494\,\gev)^2.\label{xivalue} \eeq We obtain
similar values from the corresponding combinations of other mass
pairs with the exception of $L$ and $H_u$.  Based on our previous
findings, we suspect that the $L$ and $H_u$ soft masses are
modified by a heavy RH neutrino sector.

  Note that had we included experimental and theoretical uncertainties
it would have been considerably more difficult to distinguish
different values of $\nfive$ and $\tilde{\mu}$. Instead of finding
a single value for $\five$ and a precise value for $\tilde{\mu}$,
it is likely that we would have only been able to confine $\nfive$
and $\tilde{\mu}$ to within finite ranges.

\subsection{Step 3: Adding a Heavy Neutrino Sector}

  By adding $\nfive = 5$ complete $\five$ multiplets at the
scale $\tilde{\mu} = 10^{10}\,\gev$ and a hypercharge $D$ term, we
are nearly able to fit the low scale spectrum given in
Table~\ref{lowscalespectrum} to a mSUGRA model with $m_0 =
200\,\gev$, $m_{1/2} = 700\,\gev$, and $A_0 = -500\,\gev$.
However, there are several small deviations from this picture,
most notably in the soft masses for $H_u$ and $L$ as well as the
trilinear couplings $A_t$ and $A_{\tau}$.  We attempt to fix these
remaining discrepancies by including heavy RH neutrinos at the
scale $M_R$.

\begin{figure*}[tbh]
\begin{center}
\vspace{1.2cm}
        \includegraphics[width = 0.65\textwidth]{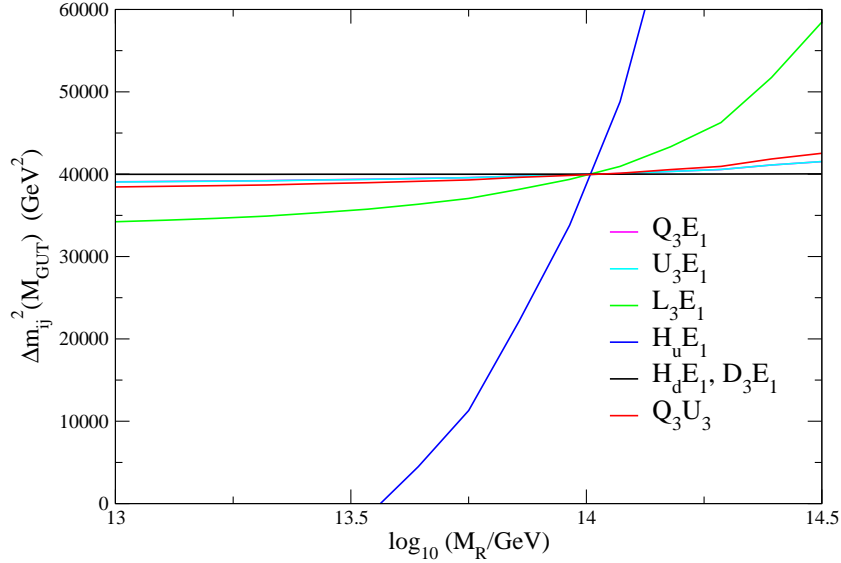}
\caption{ High scale values for third family and Higgs boson soft
scalar mass differences $\Delta{m}_{ij}^2$ (as defined in
Eq.~(\ref{hypermasscomb})) as a function of the mass scale $M_R$
of the heavy RH neutrinos. $\nfive = 5$ GUT multiplets were
included in the running above the scale $\tilde{\mu} =
10^{10}\,\gev$. } \label{fig:mrscan}
%\vspace{1cm}
\end{center}
\end{figure*}

  Given our initial assumptions about the form of a possible
RH neutrino sector, the only independent parameter in this sector
is the heavy mass scale $M_R$.  To investigate the effects of RH
neutrinos, we examine the high scale values of the mass
differences given in Eq.~(\ref{hypermasscomb}) for the third
generation scalars and $H_u$, using $m^2_{E_1}$ as a reference
mass. We add a RH neutrino sector with heavy mass $M_R$ and run
the low scale parameters listed in  Table~\ref{lowscalespectrum}
subject to the additional neutrino effects, as well as those from
$\nfive = 5$ heavy GUT multiplets with $\tilde{\mu} =
10^10\,\gev$. The result is shown in Fig.~\ref{fig:mrscan}, in
which we scan over $M_R$. This plot shows that if we include heavy
RH neutrinos at the mass scale near $M_R = 10^{14}\,\gev$, all the
mass differences will flow to a universal value of about $m_0 =
200\,\gev$ at the high scale.

\begin{figure*}[tbh]
\begin{center}
\vspace{1.2cm}
        \includegraphics[width = 0.65\textwidth]{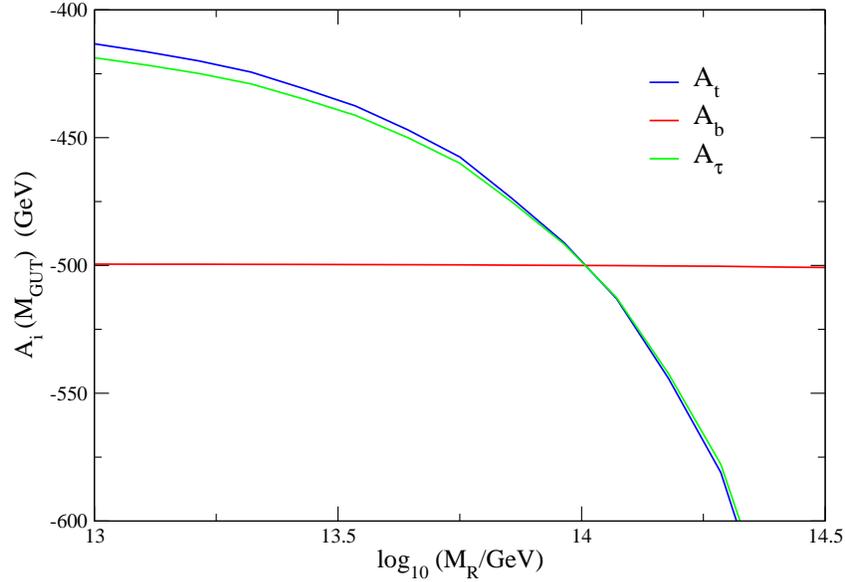}
\caption{ High scale values for the soft trilinear $A$ terms as a
function of the mass scale of the heavy RH neutrinos. $\nfive = 5$
GUT multiplets were included in the running above the scale
$\tilde{\mu} = 10^{10}\,\gev$. } \label{fig:mrascan}
%\vspace{1cm}
\end{center}
\end{figure*}

  We can further confirm this result by examining the
high scale trilinear couplings obtained by this procedure. These
also attain a universal value, $A_0 = -500\,\gev$, at the high
scale for $M_R =10^{14}\,\gev$ as shown in Figure
\ref{fig:mrascan}. These universal values are consistent with
those we hypothesized before the inclusion of heavy neutrino
sector effects. Similarly, we can also study the value of $\xi$
obtained using Eq.~(\ref{xivalue}), but using the high scale $L$
and $H_u$ soft masses computed by including heavy RH neutrinos in
their RG evolution. As before, we obtain a value $\xi \simeq
(494\,\gev)^2$.

\subsection{Results}

We have succeeded in deducing a high scale mSUGRA model augmented
by heavy new physics effects that reproduces the soft spectrum in
Table~\ref{lowscalespectrum}. The relevant parameter values are:
\beq
\begin{array}{cclcclccl}
m_0 &=& 200\,\gev,~~~&m_{1/2} &=& 700\,\gev,&A_0&=&-500\,\gev,\\
\nfive &=& 5,&\tilde{\mu} &=& 10^{10}\,\gev,\\
M_R &=& 10^{14}\,\gev,&\xi &=& (494\,\gev)^2.
\end{array}
\eeq Our example did not include potential uncertainties in the
input soft parameter values.  It is likely that such uncertainties
would make the analysis more challenging.

  Running the low scale soft parameters up to $M_{GUT}$ within the MSSM,
and without including any new physics, we obtained a reasonable
but mostly unremarkable high scale soft spectrum.  The most
obvious feature of this spectrum is the unification of the gaugino
masses. A more subtle aspect of the high scale spectrum is the
small splitting between $m_{B_1}^2$ and $m_{B_3}^2$ relative to
that between the $m_{Q_1}^2$ and $m_{Q_3}$, and $m_{U_1}^2$ and
$m_{U_3}^2$. This feature hinted at an underlying family-universal
flavor spectrum obscured by new physics effects. It is not clear
whether this hint would survive in a more complete treatment that
included uncertainties in the input parameter values.  By adding
new physics, in the form of heavy GUT multiplets and RH neutrinos,
a simple mSUGRA structure emerged.

\section{Summary}\label{summary}

  If supersymmetry is discovered at the LHC, the primary challenge
in theoretical particle physics will be to deduce the source of
supersymmetry breaking.  By doing so, we may perhaps learn about
the more fundamental theory underlying this source.  In most
models of supersymmetry breaking, the relevant dynamics take place
at energies much larger than those that will be directly probed by
the LHC. It is therefore likely that the soft supersymmetry
breaking parameters measured by experiment will have to be
extrapolated to higher scales using the renormalization group.
Given the apparent unification of gauge couplings in the MSSM only
slightly below $M_{\rm Pl}$, we may hope that there is little to
no new physics between the LHC scale and the supersymmetry
breaking scale so that such an extrapolation can be performed in a
straightforward way.

  Gauge unification still allows for some types of new
physics at intermediate scales such as complete GUT multiplets and
gauge singlets.  If this new physics is present, RG evolving the
MSSM soft parameters without including the new physics effects can
lead to an incorrect spectrum of soft parameters at the high
scale.  Even without new intermediate physics, if some of the MSSM
soft parameters are only poorly determined at the LHC, or not
measured at all, there can arise significant uncertainties in the
RG running of the soft masses that have been discovered.

  In the present work we have investigated both of these
potential obstacles to running up in the MSSM.  The soft scalar
masses are particularly sensitive to these effects, but we find
that the gaugino soft masses, and their ratios in particular, are
considerably more robust. If any one of the scalar soft masses
goes unmeasured at the LHC, the running of the remaining these
soft terms can be significantly modified by the effects of the
hypercharge $S$ term. These effects are especially severe for the
slepton soft masses, which otherwise do not tend to run very
strongly at all. We find that the uncertainties due to the $S$
term can be avoided if we consider the soft mass differences
$(Y_j\,m_i^2 - Y_i\,m_j^2)$, where $Y_i$ denotes the hypercharge
of the corresponding field.  If all the soft mass are measured,
the soft scalar mass combinations $S$ and $S_{B-L}$, defined in
Eqs.~(\ref{1:sterm}) and (\ref{1:sbl}), provide useful information
about potential GUT embeddings of the theory.

  We have also investigated the effects of two plausible types
of new physics beyond the MSSM that preserve consistency with
gauge unification; namely complete vector-like GUT multiplets and
heavy singlet neutrinos. In the case of complete GUT multiplets,
extrapolating the measured low-energy soft parameter values
without including the additional charged matter in the RG running
leads to high scale predictions for the gaugino masses that are
too small, and soft scalar masses that are too large. Even so, the
ratios of the gaugino masses at the high scale are not modified at
leading order, and can be predicted from the low-energy measured
values provided gauge unification occurs. The extrapolated values
of the scalar masses are shifted in more complicated ways, and
relationships such as family universality at the high-scale can be
obscured.  Despite this, certain hints about the underlying flavor
structure of the soft masses can still be deduced from the
properties of special linear combinations of the soft masses, such
as $(2\,m_{Q_3}^2-m_{U_3}^2-m_{D_3}^2)$ relative to
$(2\,m_{Q_1}^2-m_{U_1}^2-m_{D_1}^2)$.

  The running of the MSSM soft masses can also be modified if
there are heavy singlet neutrino chiral multiplets in the theory.
These can induce small masses for the standard model neutrinos
through the see-saw mechanism. If the singlet neutrino scale is
very heavy, greater than about $10^{13}\,\gev$, the corresponding
neutrino Yukawa couplings can be large enough to have a
significant effect on the running of the soft masses of $H_u$ and
$L$.  We have studied the size of these effects, as well as the
shifts in the other soft masses.  The extrapolated values of the
gaugino masses and the squark soft masses are only weakly modified
by heavy neutrino sector effects.

  In Section~\ref{all} we applied the methods described above
to a specific example.  In this example, the scalar masses have a
common high scale value, up to a hypercharge $D$ term. However,
because of the presence of heavy new physics, this simple
structure does not emerge when the low scale soft scalar masses
are extrapolated up to $M_{GUT}$ using the RG equations of the
MSSM. Based on the low-energy spectrum alone, we were able to
deduce the presence of the hypercharge $D$ term.  The presence of
additional new physics was suggested by the fact that the
splitting between $m_{U_3}^2$ and $m_{U_1}^2$ was considerably
larger than the related splitting between $m_{B_3}^2$ and
$m_{B_1}^2$. By studying the scalar mass combinations of
Eq.~(\ref{hypermasscomb}) and including heavy GUT multiplets and a
right-handed neutrino sector, we were able to reproduce the
low-energy soft spectrum with an underlying mSUGRA model.

  One can also invert this perspective of
overcoming new physics obstacles, and instead view these obstacles
as providing information and opportunities. As the example of
Section~\ref{all} illustrates, analyses of the kind we consider
here probe new physics in indirect ways that can lead to
convincing arguments for its existence or absence.  Such analyses
may be the main way to learn about new physics that cannot be
studied directly.

  If the LHC discovers new physics beyond the standard model,
it will be a challenge to extract the Lagrangian parameters from
the data. It may also be difficult to correctly extrapolate these
parameters to higher scales in order to deduce the underlying
theory that gives rise to the low energy Lagrangian.  In the
present work we have begun to study this second aspect of the
so-called LHC inverse problem, and we have found a few techniques
to address some of the potential obstacles. However, our study is
only a beginning.  We expect that a number of additional
techniques for running up could be discovered with more work.  A
similar set of techniques could also be applied to understanding
the high scale origin of other types of new physics beyond the
standard model.  These techniques deserve further study.

\chapter{Conclusions}\label{conclusions}
The focus of this thesis has been the Deeper Inverse Problem. It
has proposed a pragmatic approach to attacking this important
problem and has studied the various steps which have to be
completed in order to address the deeper inverse problem in a
meaningful way. In this regard, two classes of string/$M$ theory
constructions -- Type IIA intersecting D-brane constructions on
toroidal orientifolds and $M$ theory constructions on singular
$G_2$ holonomy manifolds, were studied with particular emphasis on
their implications for low-energy physics. The effective
four-dimensional theory resulting from both these constructions
fits within the framework of low-energy supersymmetry. Then,
predictions for LHC signatures for many classes of
string-motivated constructions in the literature were studied. The
effective theory of all the studied constructions lies within the
framework of low-energy supersymmetry; this was done because
low-energy supersymmetry seems to provide the most natural and
appealing solution to the hierarchy problem. It was shown that it
is possible to distinguish microscopic constructions on the basis
of their patterns of experimental observables (here LHC
signatures) and that, more importantly, the origin of
distinguishibility can be understood in terms of their underlying
theoretical structure. Finally, a more bottom-up approach of going
to theory from data was studied within the framework of low-energy
supersymmetry and gauge coupling unification and it was shown that
even though many obstacles can make it very difficult to
accurately determine the nature of the high scale theory, it is
still possible to get some insight about it.

The techniques developed and results obtained in this thesis will
be very important in making further progress on the deeper inverse
problem. Moreover, these can be systematically improved to make
the approach more sophisticated. String/$M$ theory constructions
should be made more holistic in the sense that they address all
relevant issues in particle physics and cosmology. Creative
thinking is required to identify more experimental observables
which probe features of the underlying theory. A few such
observables were indeed found in the studies above, many more
should be identified. More techniques for correctly identifying
the high-scale structure of the theory from low scale data should
be developed.

The results obtained in this thesis give us hope and confidence
that string phenomenology is a useful and illuminating exercise.
In the presence of a vast landscape of four dimensional string
vacua and the absence of a dynamical vacuum selection principle,
the most pragmatic approach one can take (and which is proposed by
the thesis) is to compute predictions for low energy experimental
observables for as many classes of realistic string vacua as
possible and try to learn how information from experimental data
may favor some regions of the M-theory amoeba over others. Doing
so will lead to a deeper understanding of string/$M$ theory, and
could be crucial to learning how or if string/$M$ theory can be
related to the real world.

%-------------------------------------------------------------------
%     APPENDICES
%-------------------------------------------------------------------
\startappendices                     % Needed for right format.
\chapter{Some Soft Supersymmetry Breaking Parameters for
Intersecting D-brane Models}\label{idbsoft}

Here, we list the formulas for the soft scalar mass parameters and
the trilinear parameters. It turns out that for the matter content
in Table 2, we have the following independent soft parameters :

\begin{itemize}
\item \underline{Trilinear parameters ($A$)}:
\begin{eqnarray}
A_{Q_LH_uU_R} &=& \frac{-\sqrt{3}m_{3/2}}{4\pi}\,[
2\pi({\Theta}_2+{\Theta}_3)+[({\Theta}_3-{\Theta}_2)
\left( \Psi({\theta}^2_{ab})+\Psi({\theta}^3_{ab})-\Psi({\theta}^2_{ca})-\Psi({\theta}^3_{ca}) \right) \nonumber \\
& & + {\Theta}_1 \, \left(
\Psi({\theta}^2_{ab})-\Psi({\theta}^3_{ab})-\Psi({\theta}^2_{ca})+\Psi({\theta}^3_{ca})
\right)]
\,\sin(2\pi\alpha)\,] \\
A_{Q_LH_dD_R} &=& -\frac{\sqrt{3}m_{3/2}}{4\pi}\,[
2\pi({\Theta}_2+{\Theta}_3)+[({\Theta}_3-{\Theta}_2)
\left( \Psi({\theta}^2_{ab})+\Psi({\theta}^3_{ab})-\Psi({\theta}^2_{c*a})-\Psi({\theta}^3_{c*a}) \right) \nonumber \\
& & + {\Theta}_1 \, \left(
\Psi({\theta}^2_{ab})-\Psi({\theta}^3_{ab})-\Psi({\theta}^2_{c*a})+\Psi({\theta}^3_{c*a})
\right)]
\,\sin(2\pi\alpha)\,] \\
A_{Q_LH_dD_R} &=& A_{LH_dE_R};\;\; A_{Q_LH_uU_R} = A_{LH_uN_R}
\nonumber
\end{eqnarray}
\newpage
\item \underline{Scalar Mass parameters (${\tilde{m}}^2$)}:
\begin{eqnarray}
{\tilde{m}}^2_{Q_L} &=&
\frac{-m_{3/2}^2}{16{\pi}^2}\;[-12\pi\,\{{{\Theta}_1}^2({\Psi}({\theta}^2_{ab})
-{\Psi}({\theta}^3_{ab})) - ({{\Theta}_2}^2-{{\Theta}_3}^2)
({\Psi}({\theta}^2_{ab})+{\Psi}({\theta}^3_{ab}))\}\times
\nonumber\\ & &\sin(2\pi\alpha) + 3\,\{-2 {\Theta}_1
({\Theta}_2-{\Theta}_3)
({\Psi}'({\theta}^2_{ab})-{\Psi}'({\theta}^3_{ab})) +
{{\Theta}_1}^2 ({\Psi}'({\theta}^2_{ab})+{\Psi}'({\theta}^3_{ab})) + \nonumber \\
& & ({\Theta}_2-{\Theta}_3)^2
({\Psi}'({\theta}^2_{ab})+{\Psi}'({\theta}^3_{ab}))\}
\, {\sin}^2(2\pi\alpha) + \pi \,[(-4\pi) + \nonumber \\
& & 3 \,\{{\Theta}_1^2
({\Psi}({\theta}^2_{ab})-{\Psi}({\theta}^3_{ab})) +
({\Theta}_2-{\Theta}_3)^2
({\Psi}({\theta}^2_{ab})-{\Psi}({\theta}^3_{ab}))
- \nonumber \\
& & 2 {\Theta}_1 ({\Theta}_2-{\Theta}_3) ({\Psi}({\theta}^2_{ab})+{\Psi}({\theta}^3_{ab}))\}\,\sin(4\pi\alpha) \,]\;] \\
{\tilde{m}}^2_{U_R} &=&
\frac{-m_{3/2}^2}{16{\pi}^2}\;[-12\pi\,\{{{\Theta}_1}^2({\Psi}({\theta}^2_{ac})
-{\Psi}({\theta}^3_{ac})) - ({{\Theta}_2}^2-{{\Theta}_3}^2)
({\Psi}({\theta}^2_{ac})+{\Psi}({\theta}^3_{ac}))\}\times
\nonumber \\ & &\sin(2\pi\alpha) + 3\,\{-2 {\Theta}_1
({\Theta}_2-{\Theta}_3)
({\Psi}'({\theta}^2_{ac})-{\Psi}'({\theta}^3_{ac})) +
{{\Theta}_1}^2 ({\Psi}'({\theta}^2_{ac})+{\Psi}'({\theta}^3_{ac})) + \nonumber \\
& & ({\Theta}_2-{\Theta}_3)^2
({\Psi}'({\theta}^2_{ac})+{\Psi}'({\theta}^3_{ac}))\}
\, {\sin}^2(2\pi\alpha) + \pi \,[(-4\pi) + \nonumber \\
& & 3 \,\{{\Theta}_1^2
({\Psi}({\theta}^2_{ac})-{\Psi}({\theta}^3_{ac})) +
({\Theta}_2-{\Theta}_3)^2
({\Psi}({\theta}^2_{ac})-{\Psi}({\theta}^3_{ac}))
- \nonumber \\
& & 2 {\Theta}_1 ({\Theta}_2-{\Theta}_3) ({\Psi}({\theta}^2_{ac})+{\Psi}({\theta}^3_{ac}))\}\,\sin(4\pi\alpha) \,]\;] \\
{\tilde{m}}^2_{D_R} &=&
\frac{-m_{3/2}^2}{16{\pi}^2}\;[-12\pi\,\{{{\Theta}_1}^2({\Psi}({\theta}^2_{ac*})
-{\Psi}({\theta}^3_{ac*})) - ({{\Theta}_2}^2-{{\Theta}_3}^2)
({\Psi}({\theta}^2_{ac*})+{\Psi}({\theta}^3_{ac*}))\}\times
\nonumber\\ & & \sin(2\pi\alpha) + 3 \,\{-2 {\Theta}_1
({\Theta}_2-{\Theta}_3)
({\Psi}'({\theta}^2_{ac*})-{\Psi}'({\theta}^3_{ac*})) +
{{\Theta}_1}^2 ({\Psi}'({\theta}^2_{ac*})+{\Psi}'({\theta}^3_{ac*})) + \nonumber \\
& & ({\Theta}_2-{\Theta}_3)^2
({\Psi}'({\theta}^2_{ac*})+{\Psi}'({\theta}^3_{ac*}))\}
\, {\sin}^2(2\pi\alpha) + \pi \,[(-4\pi) + \nonumber \\
& & 3 \,\{{\Theta}_1^2
({\Psi}({\theta}^2_{ac*})-{\Psi}({\theta}^3_{ac*})) +
({\Theta}_2-{\Theta}_3)^2
({\Psi}({\theta}^2_{ac*})-{\Psi}({\theta}^3_{ac*}))
- \nonumber \\
& & 2 {\Theta}_1 ({\Theta}_2-{\Theta}_3) ({\Psi}({\theta}^2_{ac*})+{\Psi}({\theta}^3_{ac*}))\}\,\sin(4\pi\alpha) \,]\;] \\
{\tilde{m}}^2_{Q_L} &=&  {\tilde{m}}^2_{L}; \;\;
{\tilde{m}}^2_{U_R} = {\tilde{m}}^2_{N_R}; \;\;
{\tilde{m}}^2_{D_R} =  {\tilde{m}}^2_{E_R} \nonumber
\end{eqnarray}
\end{itemize}

\noindent The above formulas are subject to the constraint
$|{\Theta}_1|^2+|{\Theta}_2|^2+|{\Theta}_3|^2 = 1$, as can be seen
clearly below equation (\ref{idb:eq:Fu}).

\chapter{K\"{a}hler metric for visible chiral matter in $M$
theory}\label{kahlermetricM}

As stated in section \ref{others}, we will generalize the result
obtained for the K\"{a}hler metric for visible sector chiral
matter fields in toroidal orientifold constructions in IIA
\cite{Bertolini:2005qh} to that in $M$ theory. The result obtained
for the K\"{a}hler metric for the (twisted) chiral matter fields
($\phi_{\alpha}$) in the supergravity limit ($\alpha' \rightarrow
0$) in \cite{Bertolini:2005qh} is:
\begin{eqnarray}\label{metric1}
\tilde{K}^{0}_{\bar{\alpha}\beta} &=& \prod_{i=1}^3
\left(\frac{\Gamma(1-{\theta}^{\alpha}_i)}{\Gamma(\theta^{\alpha}_i)}
\right)^{1/2}
\,{\delta}_{\bar{\alpha}\beta} \nonumber \\
\tan(\pi{\theta}^{\alpha}_i)&=&
\frac{U_2^i}{U_1^i+q^{\alpha}_i/p^{\alpha}_i};\;\; U^i\equiv
U^{i}_1+iU^{i}_2
\end{eqnarray}
where $i=1,2,3$ denote the number of moduli in the specific IIA
example, $\{p,q\}$ are integers and $U^i$ are the complex
structure moduli in type IIA in the geometrical basis. As
mentioned before, we will restrict to factorized rectangular tori
with commuting magnetic fluxes, for which $U_1^i=0$. Then \ba
\tan(\pi\theta^{\alpha}_i)=\frac{p^{\alpha}_i}{q^{\alpha}_i}\,U_2^i\ea

The first step towards the generalization is to identify the $G_2$
moduli in terms of IIA toroidal moduli by imposing consistency
between results of IIA and $M$ theory. The consistency check is
the formula for ${\kappa}^2$ - the physically measured four
dimensional gravitational coupling. We have:
\begin{eqnarray}\label{kappa4}
{\kappa}^2 &=& \frac{{\kappa}^2_{11}}{\mathrm{Vol}(X_7)} ,\;\;\;\;M\;Theory\;.\nonumber\\
{\kappa}^2 &=&
\frac{{\kappa}^2_{10}\,g_s^2}{\mathrm{Vol}(X_6)},\;\;\;\;IIA\;String\;Theory\;.
\end{eqnarray}
where $X_7$ and $X_6$ are the volumes of the internal 7-manifold
and 6-manifold (in IIA) respectively. The $M$ theory gravitational
coupling ${\kappa}^2_{11}$ and the string theory gravitational
coupling ${\kappa}^2_{10}$ can be written in terms of the string
coupling in IIA ($g_s\equiv e^{{\phi}_{10}^A}$) and $\alpha'$:
\ba \label{kappa11}{\kappa}^2_{11}&=&\frac{1}{2}(2\pi)^8g_s^3(\alpha')^{9/2}\,[173] \nonumber \\
{\kappa}^2_{10}&=&\frac{1}{2}(2\pi)^7{\alpha'}^4\; (\mathrm{eqn.\,
13.3.24\,of\, [174]})\ea Also, the volumes of $X_7$ and $X_6$ (for
a IIA toroidal orientifold) can be written as: \ba
\label{volumes}\mathrm{Vol}(X_7) &=& V_X\,l_M^7\;\;\;\;
\mathrm{where}\;\; V_X=\prod^N_{i=1}(s_i)^{a_i};\;\;\;
\frac{l^9_M}{4\pi} = {\kappa}^2_{11}\;\nonumber \\
\mathrm{Vol}(X_6) &=& (2\pi R^{(1)}_1)(2\pi R^{(1)}_2)(2\pi
R^{(2)}_1)(2\pi R^{(2)}_2)(2\pi R^{(3)}_1)(2\pi R^{(3)}_2)\ea
Using (\ref{kappa11}), we get \be \label{Mlength} l_M =
2\pi{\alpha'}^{1/2}g_s^{1/3}\ee which gives rise to the following
expression for $\kappa$ in $M$ theory: \be
\label{kappa4-M}{\kappa}^2 = \frac{\pi{\alpha'}g_s^{2/3}}{V_X}\ee

In IIA String theory, the definition of the IIA moduli $T_i$ and
$U_i$ in terms of the geometry of the torus is given below. We
will stick to the case of a factorized $T^6$, rectangular tori,
and commuting magnetic fluxes (in the type IIB dual) for
simplicity, in which case only the imaginary parts of the moduli
are important:
\ba \label{moduli} Im(T)^i &\equiv& T^{(i)}_2 = \frac{R^{(i)}_2}{R^{(i)}_1} \nonumber \\
Im(U)^i &\equiv& U^{(i)}_2 =
\frac{R^{(i)}_1\,R^{(i)}_2}{{\alpha}'};\;\;i=1,2,3.\ea where
$R^{(i)}_1,R^{(i)}_2$ are the radii of the $i^{th}$ IIA torus
along the $x$ and $y$ axes respectively.

\noindent Now, from (\ref{kappa4}), (\ref{kappa11}),
(\ref{volumes}) and (\ref{moduli})we get: \be \label{kappa4-II}
{\kappa}^2 =
\frac{\pi{\alpha'}g_s^{2}}{U^{(1)}_2U^{(2)}_2U^{(3)}_2} \ee
Combining (\ref{kappa4-M}) and (\ref{kappa4-II}) gives: \be
\label{Vx} \boxed{V_X =
\frac{U^{(1)}_2U^{(2)}_2U^{(3)}_2}{g_s^{4/3}}}\ee

The above formula (eqn.(\ref{Vx})) is quite general and should
always hold\footnote{within the limits of the IIA setup
considered.}, since it has been derived by requiring consistency
between formulas for the physically measured gravitational
coupling constant. To identify the individual $G_2$ moduli
however, is harder and model-dependent. In the next subsection, we
will do the mapping for the case of a simple toroidal $G_2$
orbifold ($T^7/Z_3$) considered in \cite{Lukas:2003dn}, where it
can be shown that eqn (\ref{Vx}) is satisfied.

\subsection{Particular Case}

In \cite{Lukas:2003dn}, the definitions of the $G_2$ moduli (eqn
2.4) and the K\"{a}hler potential (eqn 2.10) are not given in a
dimensionless form. Therefore, we will make them dimensionless as
is done in \cite{Acharya:2005ez}. So, we have: \ba \label{lukasG2}
s^i \equiv \frac{a^i}{l^3_M} = \frac{\int_{C^i}
\tilde{\phi}}{l^3_M};\;\;\; K = -3\ln
\left(\frac{\mathrm{Vol}(X_7)}{l^7_M}\right) + \mathrm{constant}
\ea The volume of the manifold and the moduli in this
compactification are explicitly given as \cite{Lukas:2003dn}: \ba
\label{explicit}
\mathrm{Vol}(X_7) = \prod_{i=1}^{7} R_i;\;a^1 = R_1R_2R_7;\;a^2 = R_1R_3R_6;\;a^3 = R_1R_4R_5;\nonumber\\
a^4 = R_2R_3R_5;\;a^5 = R_2R_4R_6;\;a^6 = R_3R_4R_7;\;a^7 =
R_5R_6R_7. \ea From (\ref{explicit}), we can write
$\mathrm{Vol}(X_7)=(a^1a^2a^3a^4a^5a^6a^7)^{1/3}$, which implies
that $V_X=\prod^7_{i=1}(s_i)^{1/3}$. Therefore, in the notation of
\cite{Acharya:2005ez}, $a_i=1/3;\,i=1,2,...,7$. We will identify
the $M$ theory circle radius as $R_7$, the remaining six radii can
just be identified as the $x$ and $y$ radii of the three tori in
IIA: \ba
R_1=(2\pi)R^{(1)}_1;\;R_2=(2\pi)R^{(1)}_2;\;R_3=(2\pi)R^{(2)}_1;\;
R_4=(2\pi)R^{(2)}_2 \nonumber \\
R_5=(2\pi)R^{(3)}_1;\;R_6=(2\pi)R^{(3)}_2;\;R_7=(2\pi)g_s{\alpha'}^{1/2}.\ea
With these identifications, and using (\ref{Mlength}), we can
write the individual $G_2$ moduli in terms of the IIA moduli: \ba
\label{G2moduli}
s^1&=&U^{(1)}_2;\;\;\;s^6=U^{(2)}_2;\;\;\;s^7=U^{(3)}_2; \nonumber \\
s_2&=&\frac 1
{g_s}\left(\frac{T^{(3)}_2U^{(1)}_2U^{(2)}_2U^{(3)}_2}{T^{(1)}_2T^{(2)}_2}\right)^{1/2};\;\;
s_3=\frac 1 {g_s}\left(\frac{T^{(2)}_2U^{(1)}_2U^{(2)}_2U^{(3)}_2}{T^{(1)}_2T^{(3)}_2}\right)^{1/2};\nonumber\\
s_4&=&\frac 1
{g_s}\left(\frac{T^{(1)}_2U^{(1)}_2U^{(2)}_2U^{(3)}_2}{T^{(2)}_2T^{(3)}_2}\right)^{1/2};\;\;
s_5=\frac 1
{g_s}\left({T^{(1)}_2T^{(2)}_2T^{(3)}_2U^{(1)}_2U^{(2)}_2U^{(3)}_2}\right)^{1/2}\,.\nonumber
\ea \noindent Therefore, we get: \ba V_X =
\prod^7_{i=1}(s^i)^{1/3}=\frac{U^{(1)}_2U^{(2)}_2U^{(3)}_2}{g_s^{4/3}}\ea
which is the same as (\ref{Vx}). So, for the particular case, this
suggests the following generalization: \ba \tilde{K}_{\alpha} &=&
\prod_{i=1,6,7}
\left(\frac{\Gamma(1-\theta^{\alpha}_i)}{\Gamma(\theta^{\alpha}_i)} \right)^{1/2}\nonumber \\
\tan(\pi{\theta}^{\alpha}_i)&=&
\frac{p^{\alpha}_i}{q^{\alpha}_i}\,s_i;\;\;i=1,6,7\ea

\subsection{General Case}
For more general $G_2$ manifolds with many moduli, the precise map
of the individual moduli is not completely clear. However, it
seems plausible that the complex structure moduli appearing in the
K\"{a}hler metric in IIA map to a subset of the $G_2$ moduli in a
similar way, as in the particular case.

Therefore, we use the following expression for the K\"{a}hler
metric for visible chiral matter fields in $M$ theory:\ba
\label{metric2} \tilde{K}_{\alpha} &=& \prod_{i=1}^p
\left(\frac{\Gamma(1-\theta^{\alpha}_i)}{\Gamma(\theta^{\alpha}_i)}
\right)^{1/2}; \;\;i=1,2,..,p \leq N (\equiv N)
\nonumber \\
\tan(\pi{\theta}^{\alpha}_i)&=&
c^{\alpha}_i\,(s_i)^l;\;\;c^{\alpha}_i=
\mathrm{constant};\;l=\mathrm{rational\;number\;of\;{\cal
O}(1)}.\ea

The derivatives of the K\"{a}hler metric w.r.t the moduli are very
important as they appear in the soft scalar masses, the trilinears
as well as the anomaly mediated gaugino masses, as seen from
section \ref{others}. The first derivatives appear in the
trilinears and anomaly mediated gaugino masses while the second
derivatives appear in the scalar masses. For the metric in
(\ref{metric2}), these can be written as follows:
\begin{eqnarray}\label{derivatives} & &\partial_n
\ln(\tilde{K}_{\alpha}) = \psi^{\alpha}_n \left(\frac{\partial
{\theta}^{\alpha}_n}{\partial z^n}\right); \;\;\;\;\;
\psi^{\alpha}_n(\theta^{\alpha}_n)\equiv
\frac{1}{2}\frac{d}{d{\theta}^{\alpha}_n}\,\ln(\frac{\Gamma(1-{\theta^{\alpha}_n})}
{\Gamma(\theta^{\alpha}_n)})\\
& &\partial_{\bar{m}}\partial_n \ln(\tilde{K}_{\alpha}) =
\delta_{\bar{m}n}\,\left[{\psi}^{\alpha}_{\bar{n}n}\left(\frac{\partial
{\theta}^{\alpha}_n}{\partial \bar{z}^n}\right)
\left(\frac{\partial{\theta}^{\alpha}_n}{\partial
z^n}\right)+{\psi}^{\alpha}_n\left(\frac{\partial^2{\theta}^{\alpha}_n}
{\partial\bar{z}^n\partial
z^n}\right)\right];\;\;\;{\psi}^{\alpha}_{\bar{m}n}(\theta^{\alpha}_n)\equiv\frac{d}
{d\theta^{\alpha}_m}{\psi}^{\alpha}_n \nonumber
\end{eqnarray} The functions $\psi^{\alpha}_n$ and
$\psi^{\alpha}_{\bar{n}n}$ depend on the angular variable
$\theta^{\alpha}_n$, which in turn depend on the moduli. The first
and second derivatives of ${\theta}^{\alpha}_n$ with respect to
$z_n$ are given by: \ba \label{partial} \frac{\partial
{\theta}^{\alpha}_n}{\partial z^n} &=& \frac{1}{2i}\frac{\partial
{\theta}^{\alpha}_n}{\partial s^n} =
\frac{l}{4\pi i\,s^n}\,\sin(2\pi{\theta}^{\alpha}_n) \\
\frac{\partial^2{\theta}^{\alpha}_n} {\partial\bar{z}^n\partial
z^n} &=& \frac{1}{4}\frac{\partial^2{\theta}^{\alpha}_n} {\partial
({s}^n)^2} =
\frac{l}{16\pi\,(s^n)^2}\,\left[l\sin(4\pi{\theta}^{\alpha}_n)-2\sin(2\pi{\theta}^{\alpha}_n)\right]
\nonumber \ea

The dependence of the soft parameters in section \ref{others} on
$\theta^{\alpha}_n$ is extremely simple. Instead of reexpressing
the dependence on $\theta^{\alpha}_n$ in terms of the moduli, it
is much more convenient to retain the dependence on
$\theta^{\alpha}_n$, as $\theta^{\alpha}_n\in [0,1)$
\cite{Bertolini:2005qh} and the variation of relevant functions
with $\theta^{\alpha}_n$ in the allowed range can be plotted
easily. In particular, the function $F(\theta^{\alpha}_n)\equiv
\frac{1}{2\pi}\{{\psi}^{\alpha}_n\,\sin(2\pi{\theta}^{\alpha}_n)\}$
appears in the expression for the anomaly mediated gaugino masses
(eqn. (\ref{anomalyds13})) and the trilinears (eqn.
(\ref{trides})), the function $G(\theta^{\alpha}_n) \equiv
\ln\left(\frac{\Gamma(1-{\theta}^{\alpha}_n)}{\Gamma({\theta}^{\alpha}_n)}\right)$
appears in the expression for the trilinears (eqn. (\ref{trides}))
and the function $H(\theta^{\alpha}_n) \equiv
\frac{1}{4\pi}\{l^2\,{\psi}^{\alpha}_{\bar{n}n}\,\sin^2(2\pi{\theta}^{\alpha}_n)+l^2\,{\psi}^{\alpha}_n\,
\sin(4\pi{\theta}^{\alpha}_n)-2l\,{\psi}^{\alpha}_n\,\sin(2\pi{\theta}^{\alpha}_n)\}$
appears in the expression for the scalars (eqn.
(\ref{scalarsds2})). The variation of these functions with
$\theta^{\alpha}_n$ is quite mild as seen from Figure \ref{Am2}:
\begin{figure}[hbtp]\label{Am2}
  \centerline{\hbox{ \hspace{0.0in}
    \epsfxsize=2.0in
    \epsfbox{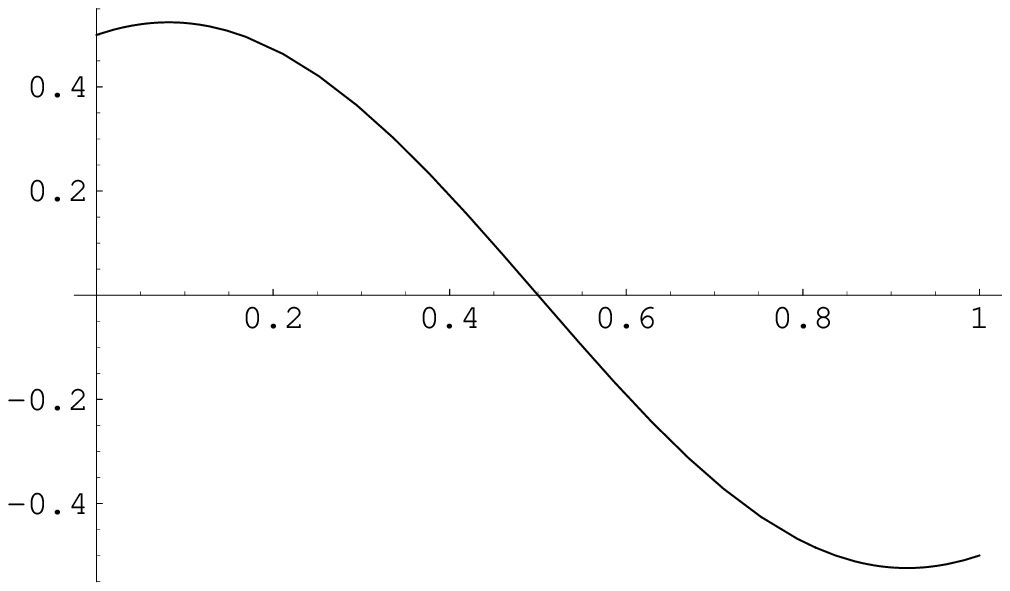}
    \hspace{0.0in}
    \epsfxsize=2.0in
    \epsfbox{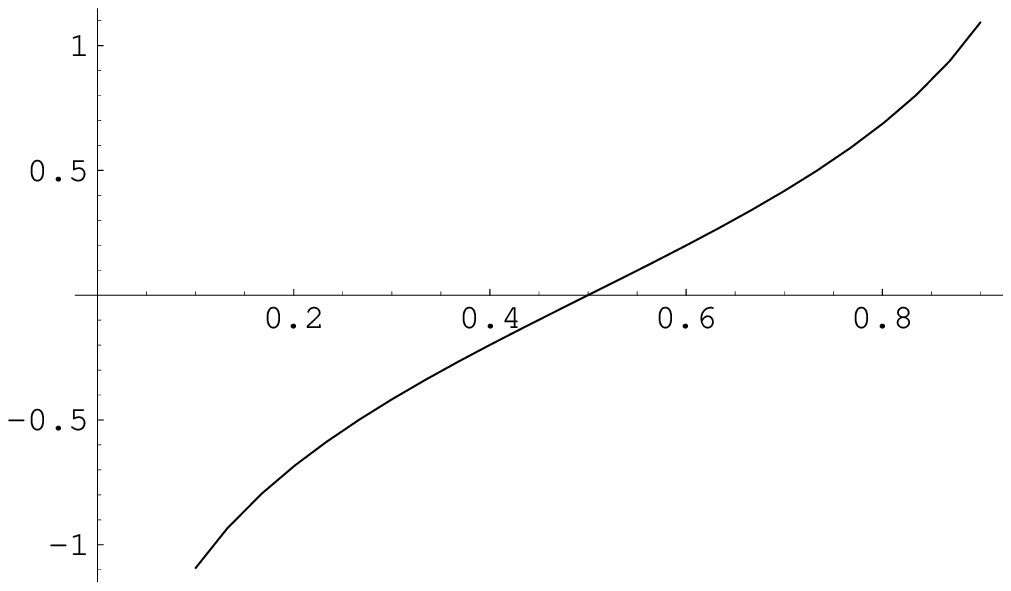}
    \hspace{0.0in}
    \epsfxsize=2.0in
    \epsfbox{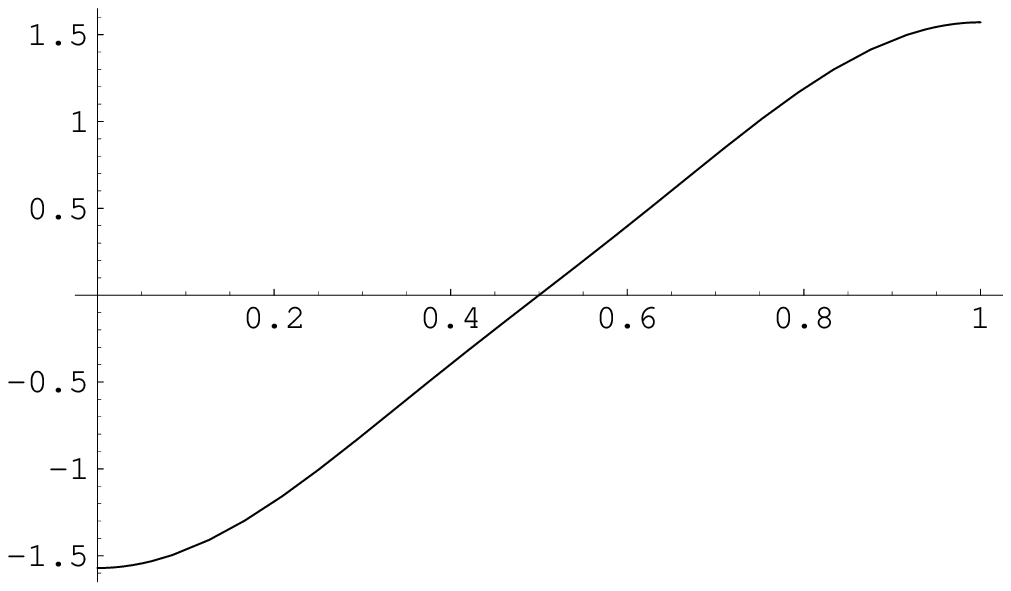}
    }
  }
\caption{Left: $F$ as a function of $\theta^{\alpha}_n$. Middle:
$G$ as a function of $\theta^{\alpha}_n$. Right: $H$ as a function
of $\theta^{\alpha}_n$ (for l=1). From \cite{Bertolini:2005qh},
$\theta_n \in [0,1)$.}
\end{figure}
Since the functions $F$, $G$ and $H$ vary very mildly with
$\theta^{\alpha}_n$ and are ${\cal O}$(1) in the whole range, it
is reasonable to replace the above functions by ${\cal O}(1)$
numbers, as is done in section \ref{others}. This is justified
since we are only interested in estimating the rough overall
scales of various soft parameters.

\chapter{Useful Combinations of Scalar Masses}

Here, we collect some combinations of soft scalar masses that are
particularly useful for running up.

\subsection{$S$ Term Effects}

  The mass differences
\beq \Delta m_{ij}^2 = (Y_j\,m_i^2 - Y_i\,m_j^2)/(Y_j-Y_i)
\label{stermmass} \eeq are useful when there is a non-vanishing
hypercharge $D$ term. A hypercharge $D$ term can shift the low
energy values of the soft scalar masses, and can also modify their
RG running through the $S$ term, as discussed in
Section~\ref{sterm}, and in Refs.~\cite{dterm,dgut,stermrefs}. The
effects of a hypercharge $D$ term cancel out these mass
differences, as well as in the RG equations for them.

  This feature is helpful for running up because the effect
of the $D$ term on the running is determined by the low scale
value of the $S$-term, which depends on all the soft scalar masses
(of hypercharged fields) in the theory. If one of these soft
masses goes unmeasured, there will be a large uncertainty in the
value of the $S$ term, and this in turn will induce a significant
uncertainty in the high scale values of the soft scalar masses
after RG evolution. By focusing on the mass differences of
Eq.~(\ref{stermmass}), these ambiguities cancel each other out.

  On the other hand, if all the MSSM soft scalar masses are measured,
the low scale values of the soft scalar mass combinations \bea
S &=& Tr(Y\,m^2),\\
S_{B-L} &=& Tr[(B\!-\!L)m^2],\nnmb \eea provide useful information
about the high scale theory, and can be used to test possible GUT
embeddings of the MSSM~\cite{dgut}.

\subsection*{Flavor Splitting Effects}

  New physics can obscure the underlying flavor structure of
the soft scalar masses.  Family-universal soft masses derived from
a theory containing new physics can generate a low-energy spectrum
that does not appear to be family-universal after it is evolved
back up to the high scale without including this new physics. This
is true even if the new physics couples in a flavor universal way
to the MSSM.  We presented a particular example of this in
Section~\ref{gutmult}, where the new physics took the form of
complete GUT multiplets having no superpotential couplings with
the MSSM sector.

  There are four pairs of soft mass combinations that
are helpful in this regard~\cite{iblop}. By comparing these pairs
(at any given scale), it is sometimes possible to obtain clues
about the underlying flavor structure of the MSSM soft masses.
These combinations are: \beq
\begin{array}{cclcccl}
m^2_{A_3} &=& 2\,m_{L_3}^2-m_{E_3}^2&\leftrightarrow~~&
m^2_{A_1} &=& 2\,m_{L_1}^2-m_{E_1}^2\\
m^2_{B_3} &=& 2\,m_{Q_3}^2 - m_{U_3}^2 -
m_{D_3}^2&\leftrightarrow~~&
m^2_{B_1} &=& 2\,m_{Q_1}^2 - m_{U_1}^2 - m_{D_1}^2\\
m^2_{X_3} &=& {2}\,m_{H_u}^2 - 3\,m_{U_3}^2&\leftrightarrow~~&
m^2_{X_1} &=& {2}\,m_{L_1}^2 - 3\,m_{U_1}^2\\
m^2_{Y_3} &=& {3}\,m_{D_3}^2 + 2\,m_{L_3}^2 -
2\,m_{H_d}^2&\leftrightarrow~~& m^2_{Y_1} &=& {3}\,m_{D_1}^2
\end{array}
\label{masscombo} \eeq If the high scale soft scalar masses are
family-universal, we expect that each of these pairs, with the
possible exception of the $m_{X_i}^2$, to be roughly equal at the
low scale in the MSSM.  The $m_{X_i}^2$ combinations are expected
to match only if $S=0$ as well.

  To apply the soft mass combinations in Eq.~(\ref{masscombo}),
one should compare them to the splitting between individual soft
masses after running all soft masses up to the high scale within
the MSSM (without new physics).  For instance, an inequality of
the form \beq |m_{B_3}^2-m_{B_1}^2| \ll
max\left\{|m_{Q_3}^2-m_{Q_1}^2|,\,
|m_{U_3}^2-m_{U_1}^2|,\,|m_{D_3}^2-m_{D_1}^2|\right\}, \eeq is
suggestive of high scale family-universality or a particular
relationship between the $Q$, $U$, and $D$ soft masses that has
been obscured by new physics.  This can arise from GUT multiplets
as in Section~\ref{gutmult}, or from a heavy RH neutrino sector as
in Section~\ref{neut}. Note that heavy neutrinos can disrupt the
relationships between the $A$, $X$, and $Y$ pairs.

%-------------------------------------------------------------------
%     BIBLIOGRAPHY
%-------------------------------------------------------------------

\newpage

%-------------------------------------------------------------------
%     ABSTRACT
%-------------------------------------------------------------------
\pagestyle{empty}
\startabstractpage{Connecting String/$M$ Theory
to the Electroweak Scale and to LHC Data}
{Piyush Kumar}{Chairperson: Gordon L. Kane}   % Title in capitals
The Standard Model of particle physics explains (almost) all
observed non-gravitational microscopic phenomena but has many open
theoretical questions. We are on the threshold of unravelling the
mysteries of the Standard Model and discovering its extension.
This could be achieved in the near future with the help of many
experiments in particle physics and cosmology, the LHC in
particular. Assuming that data confirming the existence of new
physics beyond the Standard Model is obtained, one is left with
the very important and challenging task of solving the ``Inverse
Problem", \emph{viz.} ``How can one deduce the nature of the
underlying (perhaps microscopic) theory from data?" This thesis
explores this question in detail, and also proposes an approach to
address the problem in a meaningful way which could prove crucial
to the possible solution to this problem in the future. The
proposed approach has three aspects - a) To systematically study
classes of microscopic (string/$M$ theory) constructions to the
extent that they could be connected to low energy physics
(electroweak scale), b) To find patterns of experimental
observables which are sensitive to the properties of the
underlying theoretical constructions thereby allowing us to
distinguish among different constructions, and c) To try to get
insights about the qualitative features of the theoretical model
from data in a bottom-up approach which complements the top-down
approach and strengthens it as well. This thesis studies all the
above aspects in detail. The methods used and results obtained in
this thesis will hopefully be of great importance in solving the
Inverse Problem.
\end{document}